\documentclass[aps,prl,twocolumn,a4paper,10pt,notitlepage,footinbib,superscriptaddress,longbibliography]{revtex4}
\usepackage[english]{babel}
\usepackage{lmodern}
\usepackage[utf8]{inputenc}
\usepackage{amssymb,amsmath,amsfonts}
\usepackage{textcomp}
\usepackage{braket}
\usepackage[normalem]{ulem}
\usepackage{soul}
\usepackage{graphicx}
\usepackage{dcolumn}
\usepackage{bm}
\usepackage{soul}
\usepackage{comment}
\usepackage{float}
\usepackage{soul}
\usepackage{xcolor}
\usepackage{appendix}
\usepackage{tikz}
\usepackage{enumitem}
\usepackage{longtable}
\usepackage{array}

\usepackage{xcolor}
\definecolor{blueprl}{RGB}{13.0, 18.0, 180.0 }
\usepackage[colorlinks=true, urlcolor=blueprl, allcolors=blueprl]{hyperref}

\newcommand{\comments}[1]{}
\newcommand{\de}{\mathrm{d}}

\begin{document}

\setcounter{secnumdepth}{3}

\title{
Anderson Localization and the Localization Landscape on the Bethe Lattice: \\
Exact Results and Limits of the Landscape Picture
}

\date{\today}

\author{Lorenzo Tonetti}
\affiliation{Sorbonne Universit\'e, Laboratoire de Physique Th\'eorique et Hautes Energies, CNRS-UMR 7589, 4 Place Jussieu, 75252 Paris Cedex 05, France}

\author{Leticia F. Cugliandolo}
\affiliation{Sorbonne Universit\'e, Laboratoire de Physique Th\'eorique et Hautes Energies, CNRS-UMR 7589, 4 Place Jussieu, 75252 Paris Cedex 05, France}

\author{Marco Tarzia}
\affiliation{Sorbonne Universit\'e, Laboratoire de Physique Th\'eorique de la Mati\`ere Condens\'ee, CNRS-UMR 7600, 4 Place Jussieu, 75252 Paris Cedex 05, France}

\begin{abstract}
The Localization Landscape Theory (LLT) offers a classical analogy for understanding Anderson localization through the construction of an effective confining potential, whose percolation threshold has been proposed to mark the mobility edge. While this correspondence demonstrates striking numerical agreement in three dimensions, its theoretical foundations remain an open question. In this work, we extend the analysis of the LLT on the Bethe lattice that we presented in~\cite{Tonetti2026}. This is a setting in which both the Anderson localization transition and the LLT percolation problem admit exact solutions. Our analysis reveals that the two transitions are distinct, exhibiting markedly different critical behaviors. Notably, the percolation transition within the LLT falls into the standard mean-field universality class, which stands in sharp contrast to the unconventional critical behavior characteristic of the Anderson transition on the Bethe lattice.
Nonetheless, the LLT framework succeeds in reproducing a number of exact results, capturing nontrivial features of the very low-disorder regime: it predicts the position of the isolated eigenvalue, the minimal disorder at which  the LLT percolation curve first appears which is compatible with the onset of the mobility edge, and, through the analytic continuation of its cavity equations, the Aizenman--Warzel lower bound for localization. At large connectivity, we show that the mobility edge and the LLT percolation line emerge from the band edge at the same point, $W_{\min}$, and we determine analytically how they separate at stronger disorder. We also study the dependence of the LLT percolation threshold on the energy shift, evaluate the LLT prediction for the Density of States, and derive several results on the statistical properties of the variables controlling the problem. Finally, we develop an extreme-value analysis of the statistics of the cavity Green's functions and of the Localization Landscape variables.
These findings provide an exact analytical benchmark showing that, despite its geometric appeal, the LLT does not reproduce the quantum critical properties of Anderson localization on all geometries, while still offering a powerful tool to understand its very low-disorder regime.
\end{abstract}

\maketitle

\tableofcontents

\vspace{0.25cm}

\section{Introduction}
\label{sec:Introduction}

The phenomenon of Anderson localization -- the complete cessation of wave diffusion in disordered media due to coherent backscattering and destructive quantum interference~\cite{anderson1958absence,lee1985disordered,evers2008anderson,lagendijk2009fifty}--has established itself as a cornerstone of modern condensed matter physics. Originally conceived to explain the absence of electronic conduction in random lattices~\cite{anderson1958absence}, its wave-mechanical nature has enabled experimental realization across remarkably diverse physical platforms. These range from optical configurations~\cite{lagendijk2009fifty, segev2013anderson} and ultracold atomic gases~\cite{aspect2009anderson,roati2008anderson,billy2008direct,kondov2011three,jendrzejewski2012three,semeghini2015measurement} to kicked rotors~\cite{chabe2008experimental} and classical acoustic waves~\cite{hu2008localization}. The spatial dimensionality of the medium plays a decisive role in this physics: while scaling-theory, transfer-matrix, and Green's function methods demonstrate that any infinitesimal disorder unconditionally localizes all eigenstates in one and two dimensions~\cite{mott1961theory,gor1996particle,Abrahams79}, a genuine metal-insulator transition emerges in three or higher dimensions. In these higher-dimensional setups, a critical disorder threshold establishes a mobility edge that cleanly divides the spectrum into localized and extended states~\cite{Kramer93,wegner1979mobility,evers2008anderson}. In recent years, interest in this field has been further re-energized by expanding into the realms of topological phases~\cite{Ludwig_2016,PhysRevLett.81.862,PhysRevLett.81.4704,PhysRevLett.82.4524,PhysRevB.60.4245,PhysRevB.61.10267,PhysRevB.63.235318} and many-body localization in interacting quantum systems~\cite{Basko2006,gornyi2005interacting,Nandkishore2015, abanin-colloquium-2019, alet-many-body-2018, sierant-many-body-2025}.

To investigate these phenomena theoretically, the tight-binding Anderson model (AM) featuring uncorrelated random on-site potentials serves as the standard paradigm~\cite{anderson1958absence}. Despite its apparent simplicity, an exact analytical description of the three-dimensional transition continues to elude analytical approaches. Consequently, our understanding of the corresponding mobility edges and critical exponents relies heavily on sophisticated numerical simulations~\cite{PhysRevLett.105.046403,Kramer93,evers2008anderson,lagendijk2009fifty,tarquini2017critical}. A long-standing conceptual hurdle stems from the fact that localization centers do not simply coincide with local extrema of the random potential. Because the underlying physics is fundamentally rooted in complex, non-local phase interference rather than classical trapping, constructing an intuitively clear or analytically tractable classical limit directly from the bare disorder potential has proven notoriously difficult.

A prominent framework designed to bridge this quantum-classical gap is the Localization Landscape Theory (LLT), put forward by Filoche and Mayboroda~\cite{filoche2012universal}. The core architecture of the LLT relies on an auxiliary scalar field -- the localization landscape -- whose reciprocal is mathematically mapped to an effective confining potential~\cite{filoche2012universal,PhysRevLett.116.056602,arnold2019localization,arnold2019computing,david2021landscape,filoche2024anderson}. In this transformed picture, the spatial trapping of quantum wave functions is predicted to occur within the valleys of this effective potential. This maps the quantum localization-delocalization transition onto a purely geometric classical percolation problem, where the mobility edge is identified as the threshold energy at which classically allowed regions first span the system. While this geometric interpretation yields promising agreement with numerical computations of the mobility edge near the bottom of the spectrum in the three-dimensional AM~\cite{filoche2024anderson}, the broader predictive capability of the LLT as well as its accuracy in reconstructing the electronic Density of States remains a subject of active discussion and debate~\cite{arnold2019computing, david2021landscape, texier2020comment, filoche2020reply}.

The search for a rigorous benchmark of this geometric paradigm motivates its evaluation in a structural environment where both the quantum tight-binding physics and the classical percolation thresholds can be solved analytically. The Bethe lattice -- an infinite regular tree geometry where every node is connected to exactly $K+1$ nearest neighbors~\cite{Bollobas} -- provides an ideal testing ground. Because loops are entirely absent, both the Anderson transition and classical percolation models are exactly tractable on this graph. The Anderson model on the Bethe lattice undergoes a phase transition at a well-defined critical disorder strength~\cite{abou1973selfconsistent}, offering a unique arena characterized by a unconventional mean-field critical behavior~\cite{abou1973selfconsistent,zirnbauer1986localization,verbaarschot1988graded,mirlin1991localization,mirlin1991universality,fyodorov1991localization,fyodorov1992novel,mirlin1994statistical,PhysRevLett.72.526,tikhonov2019statistics,tikhonov2019critical,biroli2010anderson,biroli2022critical,vanoni2024renormalization,sierant2023universality,prado2026anderson}.

Beyond serving as an exact benchmark for comparing quantum localization with the LLT, the Bethe-lattice limit possesses a rich and intricate physics that warrants independent investigation. Far from being a completely settled textbook example, the tree geometry has revealed unexpected subtleties over the past several years, fueling an intense debate that has occasionally challenged the very interpretation of its exact solution~\cite{biroli2012difference,de2014anderson,Kravtsov2018nonergodic,pino2020scaling,bera2018return,de2020subdiffusion}. This ongoing controversy is rooted in two fundamental difficulties. First, the model exhibits extraordinarily severe finite-size corrections that persist deep into the phases, well away from the critical threshold. This was vividly brought to light by numerical studies on Random Regular Graphs (RRGs) -- disordered sparse networks of fixed connectivity $K+1$~\cite{wormald1999models} that structurally converge to the Bethe lattice locally in the thermodynamic limit (see below for more details). Initial numerical findings on finite RRGs appeared strikingly inconsistent with the asymptotic predictions of the analytical solution~\cite{biroli2012difference,de2014anderson}. Second, the critical behavior on the Bethe lattice is highly unconventional, breaking the scaling paradigms established for finite-dimensional spaces. Recent conceptual and quantitative advances~\cite{baroni2024corrections,tikhonov2016anderson,prado2026anderson} have resolved these apparent paradoxes by establishing that the Bethe lattice represents a singular, upper critical dimension ($d_u=\infty$) limit for Anderson localization, validating long-standing historical conjectures~\cite{tarquini2017critical,Castellani1986,PhysRevLett.72.526}. Propelled by these unresolved nuances, research into the disordered Bethe lattice has remained a vibrant and rapidly evolving subfield over the last decade.

\subsection{Main Results}
\label{subsec:main-results}
\begin{figure}[b!]
    \centering
    \includegraphics[width=0.49\textwidth]{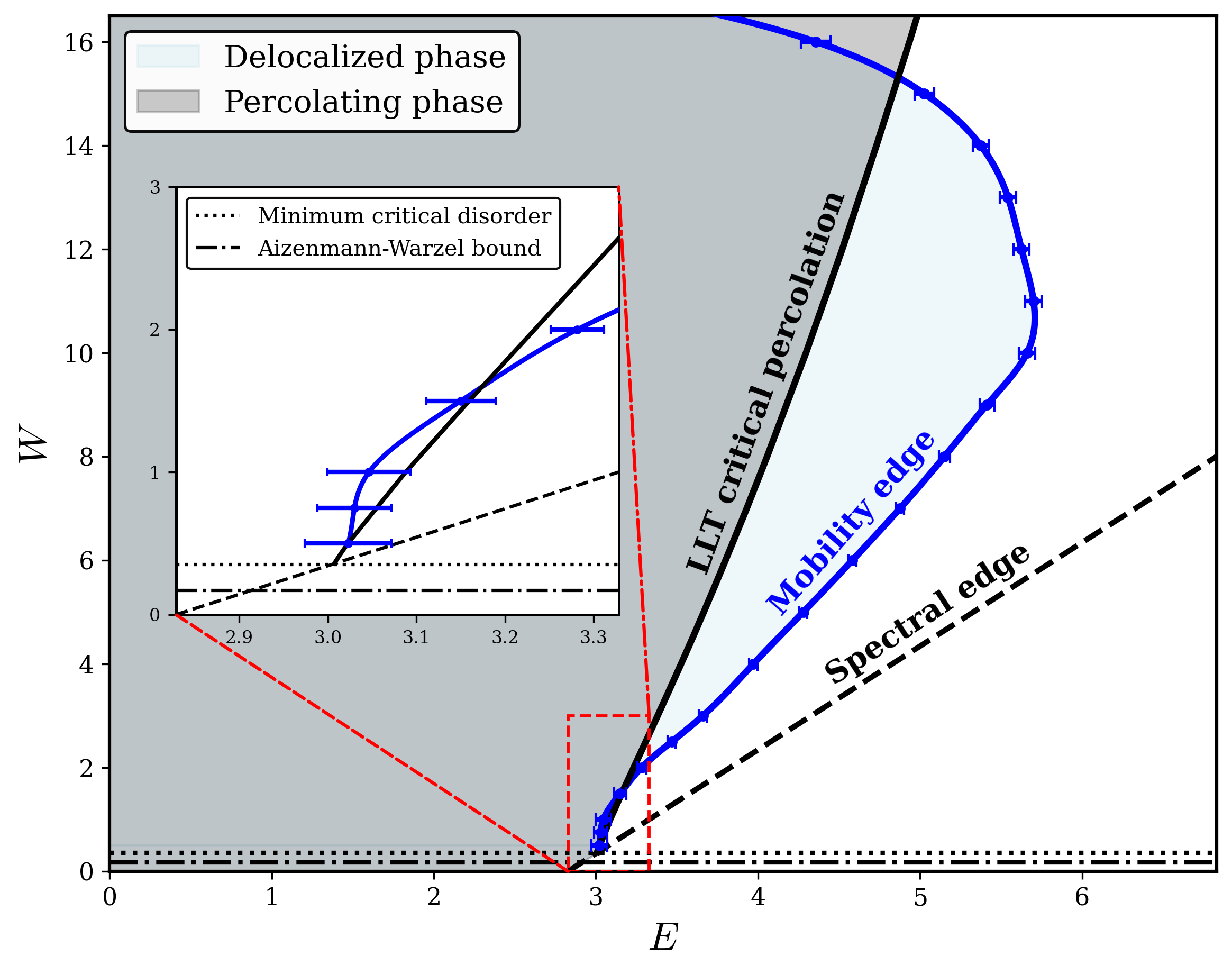}
    \caption{The phase diagram of the Anderson model on the Bethe lattice. 
    The black solid line and the blue data points with errorbars represent the 
    mobility edge found with the LLT ($E_{\rm perc}$) with minimal shift  
    and the actual one ($E_{\rm loc}$), respectively.
    The dotted and dashed-dotted horizontal lines are the minimum critical  
    disorder, $W_{\rm min}$, and the Aizenman-Warzel bound~\cite{warzel2012}, $W_{\rm AW}$, respectively.
 }
    \label{fig:phase-diagram}
    \end{figure}

In Ref.~\cite{Tonetti2026} we investigated the Anderson Model on the Bethe lattice and, using the cavity method~\cite{mezard2001bethe}, we 
studied both the conventional localization and Localization Landscape 
percolation transitions. In so doing, we showed that the LLT percolation threshold does not coincide with the Anderson transition 
on this geometry. In Fig.~\ref{fig:phase-diagram} we show the phase diagram in the energy-disorder strength plane that we 
derived in this reference. While the transition lines are notably close at weak disorder, they separate
significantly at strong disorder. Moreover, differently from the three-dimensional case, the critical percolation curve   
does not provide an upper bound to the Anderson localization's mobility edge, as the two curves cross at a finite value of the disorder.

Furthermore, we analyzed the critical behaviors close to both transitions 
and we found that they are fundamentally different. 
Firstly, we compared the  eigenstates’ inverse
participation rate (IPR) of the Anderson Model, which measures the
inverse volume occupied by an eigenstate, with the inverse
of the average cluster size, $1/S$, in the LLT percolation problem. 
Whereas the IPR jumps from zero to a finite value at the critical energy, 
$E_{\rm loc}$, and then grows as a square root of the distance
from $E_{\rm loc}$~\cite{rizzo2024localized}, $1/S$ grows 
linearly from zero at $E_{\rm perc}$, with the same critical exponent,
$\gamma=1$, as in large-dimensional random percolation.

\begin{figure}[t]
    \centering
    \includegraphics[width=\linewidth]{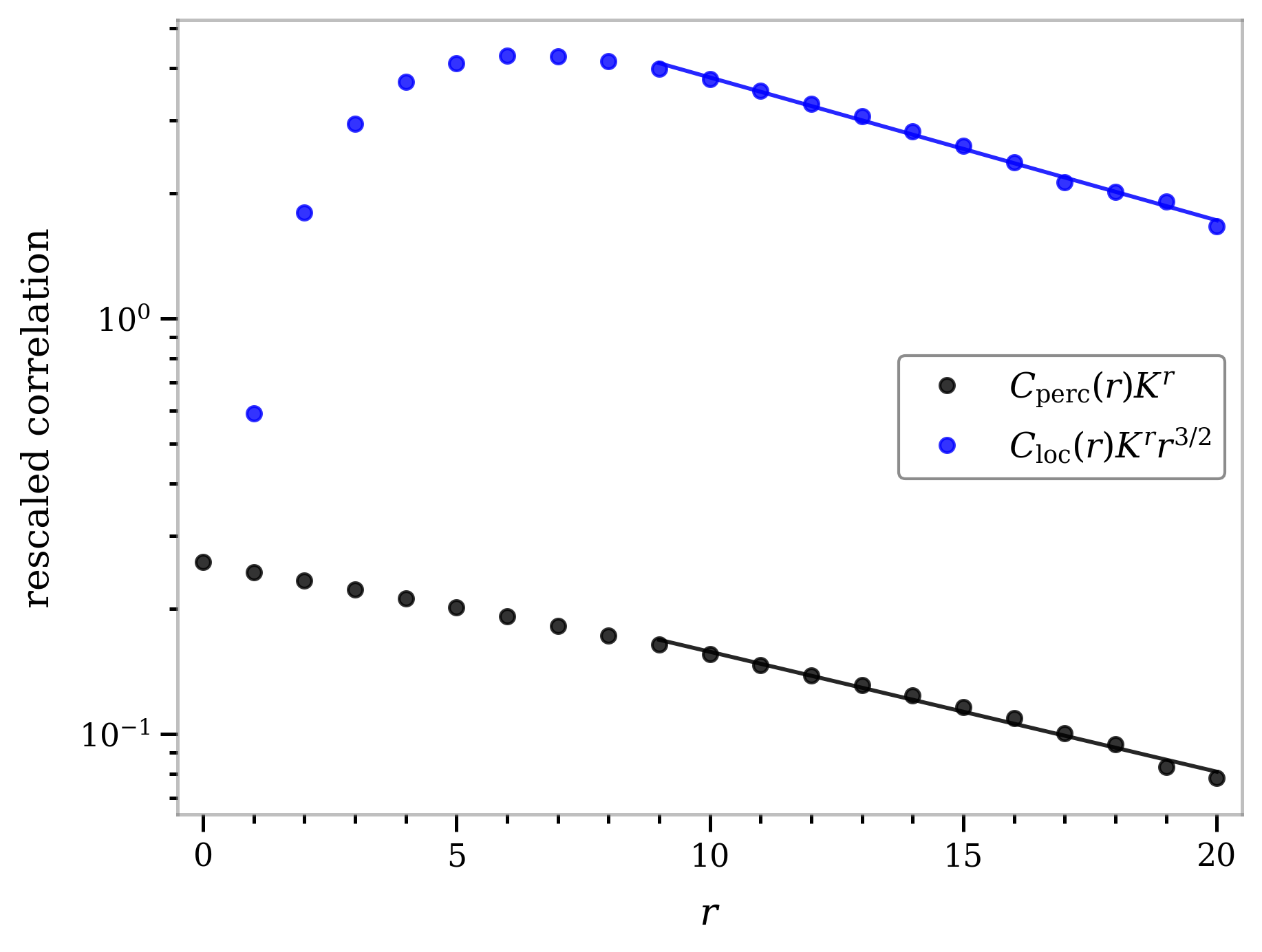}
    \caption{\small{Rescaled correlation functions in the delocalized and non-percolating phases. The rescaling indicated in the 
    key transforms the large-distance decay to the linear form 
    ${\rm const}- r/\xi$. $W=3.5$, while the two energies are chosen separately in order to
    yield correlation lengths which are close in value, $\xi_{\rm loc} \sim 13$ and 
    $\xi_{\rm perc}\sim 15$. 
    Blue dots: Anderson eigenstate correlation
    with a linear tail fit in blue. Black dots: LLT percolation correlation
    with a linear tail fit in black. The two asymptotic behaviors differ by the extra factor $r^{-3/2}$ present in the Anderson case and absent in the percolation one, 
    as explained in Sec.~\ref{subsec:main-results}.}}
    \label{fig:correlations}
\end{figure}

A further insightful comparison can be drawn by examining the spatial dependence of the correlation functions characteristic of each framework. For the quantum problem, the relevant observable is the eigenstate correlation function, defined as $C_{\rm loc}(r)\equiv \mathbb{E}[\vert \psi_n (0) \vert^2 \vert \psi_n (r) \vert^2]$. Within the geometric framework of the LLT, its natural classical counterpart is the percolation connectivity function $C_{\rm perc}(r)$, which measures the probability that two occupied sites separated by a distance $r$ belong to the same cluster. In the large-distance asymptotic limit, these two correlation functions exhibit distinct functional forms due to the different underlying physics. Specifically, the quantum correlation function falls off as $C_{\rm loc}(r) \propto r^{-3/2}K^{-r} e^{-r / \xi_{\rm loc}} $~\cite{zirnbauer1986localization,verbaarschot1988graded,mirlin1991localization,mirlin1991universality,fyodorov1991localization,fyodorov1992novel,mirlin1994statistical,tikhonov2019statistics,prado2026anderson}, carrying a power-law correction that is absent in the purely geometric percolation counterpart, which decays as $C_{\rm perc}(r) \propto K^{-r} e^{-r / \xi_{\rm perc}}$. One example of each is shown in Fig.~\ref{fig:correlations}, where we rescaled the correlations so that their large-distance behavior is ${\rm const} - r/\xi$ in both cases. Careful fits of the tails allowed us to estimate the 
parameter dependencies of the two correlation lengths $\xi_{\rm loc}$ and $\xi_{\rm perc}$
reported in~\cite{Tonetti2026}.

We have directly contrasted the quantum localization length $\xi_{\rm loc}$ approaching the delocalization transition against the corresponding correlation length $\xi_{\rm perc}$ of the critical LLT percolation. As the respective thresholds are approached, both lengths diverge according to a power law, $\xi\simeq (E-E_{\rm c})^{-\nu}$, sharing the same mean-field critical exponent $\nu=1$. Crucially, however, they are characterized by markedly different quantitative prefactors~\cite{Tonetti2026}. Taken together, these results demonstrate that the classically allowed regions satisfying $1/u_i \le E_+= E-E_{\rm sh}$ drastically underestimate the true spatial profile and physical extent of the localized quantum eigenstates at 
energy $E$.

Therefore, we concluded that, 
while the LLT provides a useful and numerically accurate approximation in three dimensions, 
it does not constitute an exact theory of the Anderson transition
in all geometries.

The results recalled so far were presented in Ref.~\cite{Tonetti2026}: the phase diagram of Fig.~\ref{fig:phase-diagram}, the comparison between the IPR and $1/S$, the asymptotic form of the correlation functions, and the comparison between $\xi_{\rm loc}$ and $\xi_{\rm perc}$. Here, we summarize them for completeness, as 
they are used and extended in the rest of the paper. The derivations already given in Ref.~\cite{Tonetti2026} are not repeated. The only additions to this part are the explicit example of the two correlation functions in Fig.~\ref{fig:correlations} and the description of the numerical methods, which was not given in Ref.~\cite{Tonetti2026} (Sec.~\ref{sec:numerics}).

In this paper, we extend the analysis begun in Ref.~\cite{Tonetti2026} and provide a complete account of the definitions, methods, and additional new important results underlying the comparison between Anderson localization and the Localization Landscape Theory on the Bethe lattice.

We first clarify the geometrical setting, by defining the Bethe lattice as the thermodynamic limit of random regular graphs and distinguishing it from the thermodynamic limit of Cayley trees, a point which is often a source of confusion.

We then recall the main features of Anderson localization on the Bethe lattice, emphasizing the Green's function formulation, the cavity equations, and the linear stability criterion that determines the mobility edge.
We next develop the corresponding cavity formulation of the Localization Landscape Theory on the same lattice. In particular, we derive the recursive equations for the Localization Landscape variables and for the associated correlated percolation problem, providing a derivation complementary to the one presented in the End Matter of Ref.~\cite{Tonetti2026}. This allows us to compare, within a common cavity framework, the Anderson localization transition and the LLT percolation transition.

In addition to the results presented in Ref.~\cite{Tonetti2026}, we analyze several aspects of the LLT construction 
that are specific to the Bethe lattice. 
Some of these results were briefly announced, without derivation, in the conclusions of Ref.~\cite{Tonetti2026}: the choice of the spectral shift, the role of the isolated eigenvalue and of $W_{\min}$, the Aizenman--Warzel bound, and the poor LLT estimate of the Density of States. Here we derive and discuss them in detail.

First, we study the dependence of the LLT percolation line on the spectral shift used to make the Hamiltonian positive definite. This is a relevant degree of freedom of the theory, which is not fixed by any first-principle criterion, and we show that changing this shift does not improve the agreement between the LLT percolation threshold and the Anderson mobility edge  in the regime $W\lesssim15$, i.e., before the two curves cross at strong disorder. We also propose a selection criterion based on the LLT bound on the eigenstate amplitudes, which is meaningful in the band tails and selects the bulk-edge shift in the weak-disorder regime, where the LLT percolation line and the mobility edge are close.

Second, we discuss the role of the isolated eigenvalue 
of the single-particle tight-binding Hamiltonian
which, at weak disorder,  lies below the continuous spectrum~\cite{biroli2010anderson}. 
On the Bethe lattice, one must therefore distinguish between the true minimal 
eigenvalue and the lower edge of the continuous spectrum. As the disorder strength $W$ increases, the isolated eigenvalue 
approaches the continuous spectrum and merges with it at a finite disorder, which we denote $W_{\rm min}$. We show that within the LLT, this merging 
controls the onset of the percolation transition: the LLT percolation curve appears exactly at $W_{\rm min}$, which explains why 
it disappears below a finite disorder scale. This suggests that the same scale controls the onset of Anderson localization.
For $K=2$, our numerical data for the mobility edge are compatible with this conjecture, but can be reliably extrapolated only for $W\ge0.5$,
and cannot resolve the onset directly. We also derive an exact lower bound for $W_{\min}$, which lies only a few percent below its actual value.

At large connectivity, the conjecture can be settled analytically. The reason is that a single condition controls both phenomena. 
Indeed, the stability condition of the 
cavity equations which determines the mobility edge and, below the spectrum, the position of the isolated eigenvalue, are given by the same equation.
Following Abou-Chacra, Anderson, and Thouless~\cite{abou1973selfconsistent}, we evaluate this condition 
for $K\gg1$ close to the band edge, and find that the mobility edge is simply the continuation, above $W_{\rm min}$, of the curve of the 
isolated eigenvalue. The mobility edge therefore emerges from the edge of the spectrum exactly at $W_{\min}$, 
where the LLT percolation line also starts. We also obtain the large-$K$ LLT percolation line in closed form.
Beyond this common starting point,  the lines separate: the mobility edge first enters the band faster, by a factor of order $\ln K$, and at stronger disorder
the two lines cross twice, as for $K=2$, with the LLT line missing the re-entrance of the mobility edge.
Third, we show that the analytic continuation of the LLT cavity equations below $W_{\min}$ reproduces the Aizenman--Warzel bound $W_{\rm AW}=t(\sqrt K-1)^2$~\cite{aizenman2011absence,warzel2012}, below which the spectrum is purely absolutely continuous close to the band edges. This result follows from the exact edges of the support of the distributions of the cavity variables, without approximations, but in a regime where the shifted Hamiltonian is not positive definite. It is therefore a property of the analytically continued cavity equations, rather than a prediction of the standard LLT. We also explain the origin of this agreement: both the LLT result and the proof of Aizenman and Warzel bound are controlled by the extreme configuration in which all the on-site energies take their maximal value. That the LLT cavity equations encode such a rigorous result is surprising, and provides a positive counterpoint to the more tempered finding that the LLT percolation transition fails to replicate the Anderson critical behavior.

Fourth, we investigate the  predictions of the LLT for the Density of States and its integrated (or cumulative) form 
on the Bethe lattice, and we compare them with the exact Anderson results. 
We show that the Weyl-type estimate can be derived from an exact decomposition of the shifted Hamiltonian into an effective kinetic operator and the effective potential $1/u_i$, which makes explicit the approximations involved. For no finite choice of the shift does the resulting estimate reproduce the symmetry and the shape of the Anderson Density of States, and at intermediate and strong disorder it is not more accurate than the naive Weyl estimate built with the bare disorder. These discrepancies are failures of the Weyl-type construction, and not necessarily of the LLT itself.

Finally, We describe the numerical methods used throughout the paper, including population dynamics and the numerical determination of the relevant stability criteria. In addition, We present a detailed analysis of the cavity equations in the high-connectivity limit (for which we derive a fully analytical solution), and we provide numerical evidence supporting the assumptions used in this limit.

For the reader's convenience, Table~\ref{tab:new-results} summarizes, for each result, whether it was presented in Ref.~\cite{Tonetti2026}, only announced there, or is new, together with the Section in which it is discussed.

\begin{table*}[t]

\caption{Relation between the present paper and Ref.~\cite{Tonetti2026}. ``Presented'': result shown in Ref.~\cite{Tonetti2026} and recalled here. ``Announced'': result stated in the conclusions of Ref.~\cite{Tonetti2026} without derivation. ``---'': result not contained in Ref.~\cite{Tonetti2026}.
}
\label{tab:new-results}
\footnotesize
\newcommand{\tcA}[1]{\parbox[t]{0.36\textwidth}{\raggedright #1}}
\newcommand{\tcB}[1]{\parbox[t]{0.16\textwidth}{\raggedright #1}}
\newcommand{\tcC}[1]{\parbox[t]{0.40\textwidth}{\raggedright #1}}
\renewcommand{\arraystretch}{1.25}
\begin{ruledtabular}
\begin{tabular}{lll}
\tcA{\textbf{Result}} & \tcB{\textbf{Ref.~\cite{Tonetti2026}}} & \tcC{\textbf{This paper}} \\
\hline
\tcA{Cavity equations for the Localization Landscape (Gaussian representation)} & \tcB{Presented (End Matter)} & \tcC{Recalled for a generic shift; new alternative derivation based on the Green's function chain rule, which does not require positive definiteness (Sec.~\ref{subsec:cavity-derivation})} \\
\tcA{Phase diagram for $K=2$: LLT percolation line vs.\ mobility edge} & \tcB{Presented} & \tcC{Recalled (Fig.~\ref{fig:phase-diagram}); extrapolation of the mobility edge described in Sec.~\ref{sec:mobility}} \\
\tcA{Critical behavior of the IPR vs.\ inverse cluster size $1/S$} & \tcB{Presented} & \tcC{Recalled (Sec.~\ref{subsec:main-results}); numerical method for $E\neq0$ in Sec.~\ref{sec:numerics}} \\
\tcA{Correlation functions; $\xi_{\rm loc}$ vs.\ $\xi_{\rm perc}$} & \tcB{Presented (asymptotic forms and correlation lengths)} & \tcC{Recalled; explicit example (Fig.~\ref{fig:correlations}); numerical method in Sec.~\ref{app:numeval}} \\
\tcA{Isolated eigenvalue, $W_{\min}$, and the onset of localization} & \tcB{Announced} & \tcC{Derived and discussed (Secs.~\ref{subsec:EisoWmin} and \ref{subsubsec:lowW})} \\
\tcA{Aizenman--Warzel bound from the LLT cavity equations} & \tcB{Announced} & \tcC{Derived and discussed (Secs.~\ref{subsec:EisoWmin} and \ref{subsec:ExVal})} \\
\tcA{Dependence of the LLT percolation line on the spectral shift; $X\to\infty$ asymptotes} & \tcB{Choice of the shift announced} & \tcC{New (Sec.~\ref{subsec:dependence-shift}, Fig.~\ref{fig:PDshift})} \\
\tcA{LLT (Weyl-type) estimate of the Density of States and of its integrated form, and its tails} & \tcB{Poor agreement announced} & \tcC{New (Secs.~\ref{subsec:density-of-states} and \ref{subsec:ExVal})} \\
\tcA{High-connectivity solution: marginal distributions, factorization approximation, linear stability, phase diagram for $K=5$ and comparison with the cubic lattice} & \tcB{Comparison with the cubic lattice announced} & \tcC{New (Sec.~\ref{subsec:hicon})} \\
\tcA{{Analytic large-$K$ mobility edge close to the band edge; onset of localization at $W_{\min}$; comparison with the large-$K$ LLT percolation line}} & \tcB{{---}} & \tcC{{New (Sec.~\ref{subsubsec:largeK-loc}, Fig.~\ref{fig:largeK})}} \\
\tcA{Extreme-value statistics of the cavity Green's functions, rescaled fields, and Localization Landscape variables} & \tcB{---} & \tcC{New (Sec.~\ref{subsec:ExVal})} \\
\tcA{Numerical methods: population dynamics, mobility edge at $E\neq0$, IPR in the localized phase, LLT percolation observables} & \tcB{---} & \tcC{New (Sec.~\ref{sec:numerics})} \\
\end{tabular}
\end{ruledtabular}
\end{table*}

\subsection{Layout of the paper}

The paper is organized as follows. In Sec.~\ref{sec:RRG}, we establish the underlying geometric framework of our study. We provide formal definitions for both random regular graphs and the infinite Bethe lattice, placing particular emphasis on a crucial structural distinction: namely, how the Bethe lattice emerges naturally as the local thermodynamic limit of disordered sparse networks, contrasting sharply with the global thermodynamic limit of a boundary-dominated Cayley tree.

In Sec.~\ref{sec:Anderson} we introduce the Anderson model on the Bethe lattice and summarize the Green's function tools used to characterize the localization transition, including the self-consistent distributional equation and its linear stability analysis.

Section~\ref{sec:EQGU} is devoted to the Localization Landscape Theory on the same geometry. We introduce the shifted positive Hamiltonian, the Localization Landscape, and the associated effective potential. We then derive the cavity equations for the Localization Landscape variables, the recursive equations for the LLT percolation probability, and the expressions for the percolation correlation function and the average cluster size.

The main results of this paper
are presented in Sec.~\ref{sec:results}. In Sec.\ref{subsec:EisoWmin}, we discuss the role of the isolated eigenvalue and the resulting lower bounds for the critical disorder. This part clarifies the disorder scale at which the LLT percolation threshold disappears and its relation to the onset of Anderson localization. In Sec.~\ref{subsec:dependence-shift}, we study the phase diagram and analyze how the LLT percolation line depends on the spectral shift used to define the shifted Hamiltonian. In Sec.~\ref{subsec:density-of-states} we analyze the LLT prediction for the Density of States and the Integrated Density of States, comparing it with the corresponding Anderson results.

Section~\ref{sec:numerics} describes the numerical methods used in the paper and in Ref.~\cite{Tonetti2026}. We explain the population dynamics algorithm, the procedure used to determine the Anderson mobility edge, the evaluation of observables in the localized phase, and the computation of the LLT percolation observables, including occupation probabilities, correlation functions, and average cluster sizes.

In Sec.~\ref{sec:analytic} we discuss several analytical limits and distributional properties of the cavity equations. We first introduce the independent-site approximation in Sec.~\ref{subsec:indsites}. We then analyze the high-connectivity limit in Sec.~\ref{subsec:hicon}, including the marginal distributions, the quality of the factorization approximation, the lower bound for the critical disorder, the isolated eigenvalue, the linear stability analysis, and the high-connectivity phase diagram, as well as the analytic large-connectivity mobility edge and its comparison with the LLT percolation line. Finally, in Sec.~\ref{subsec:ExVal} we perform an extreme-value analysis of the relevant fluctuating quantities, including the cavity Green's functions, the rescaled auxiliary fields, the Localization Landscape variables, and the effective potentials.

We conclude in Sec.~\ref{sec:Conclusions}. A notation table collecting the main symbols used throughout the paper is provided in App.~\ref{sec:notation-table}.

\section{Random Regular Graphs}
\label{sec:RRG}

Random regular graphs are an important class of sparse random graphs in which each node has exactly the same degree $K+1$, ensuring uniform connectivity across the network. They are random in the sense that a random regular graph with $N$ vertices is a graph drawn uniformly from the set of all graphs with $N$ nodes and fixed degree~\cite{Bollobas,wormald1999models}. These graphs serve as useful models of space in statistical physics since, in the thermodynamic limit, they represent the infinite dimensional limit for Euclidean lattices (since the number of nodes at a fixed distance from a given node grows exponentially with the distance), while preserving a finite local connectivity. In contrast, standard mean-field approximations achieved by taking the fully connected limit yield an infinite connectivity in the thermodynamic limit. This geometry entirely destroys any notion of spatial distance, as all pairs of sites are trivially separated by a path of length $1$. Furthermore, for several problems -- including Anderson localization~\cite{EversMirlin} -- the phase transition is either completely absent in the fully connected limit or exhibits a fundamentally different nature from the physical transition. Consequently, the random-regular graph provides the only true mean-field framework that preserves a well-defined local tree structure, making it the indispensable starting point.

\begin{figure}[h!]
\hspace{-5cm} (a)
 \\
    \includegraphics[width=0.49\textwidth]{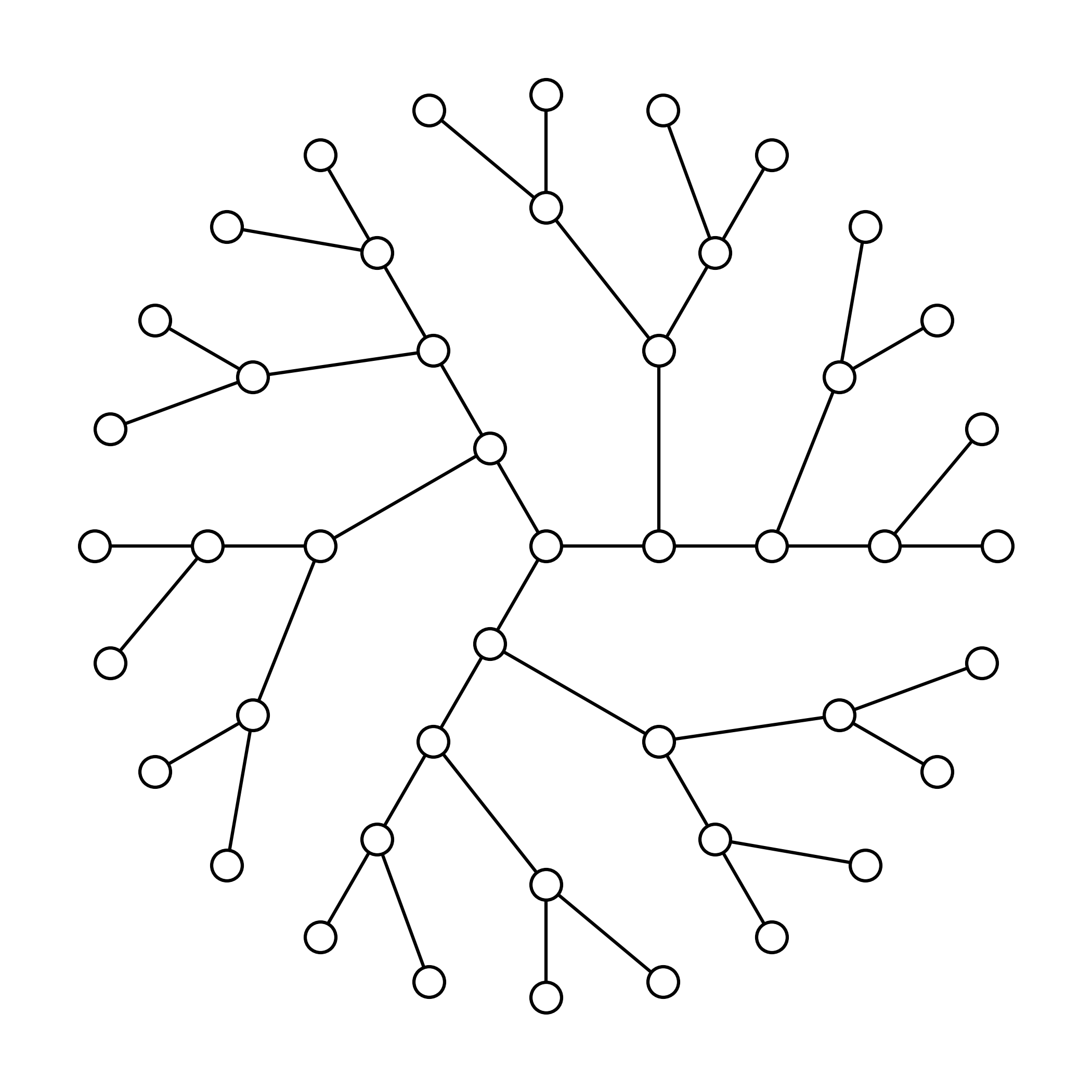}
    \\
 \hspace{-5cm}    (b)
    \\
    \includegraphics[width=0.49\textwidth]{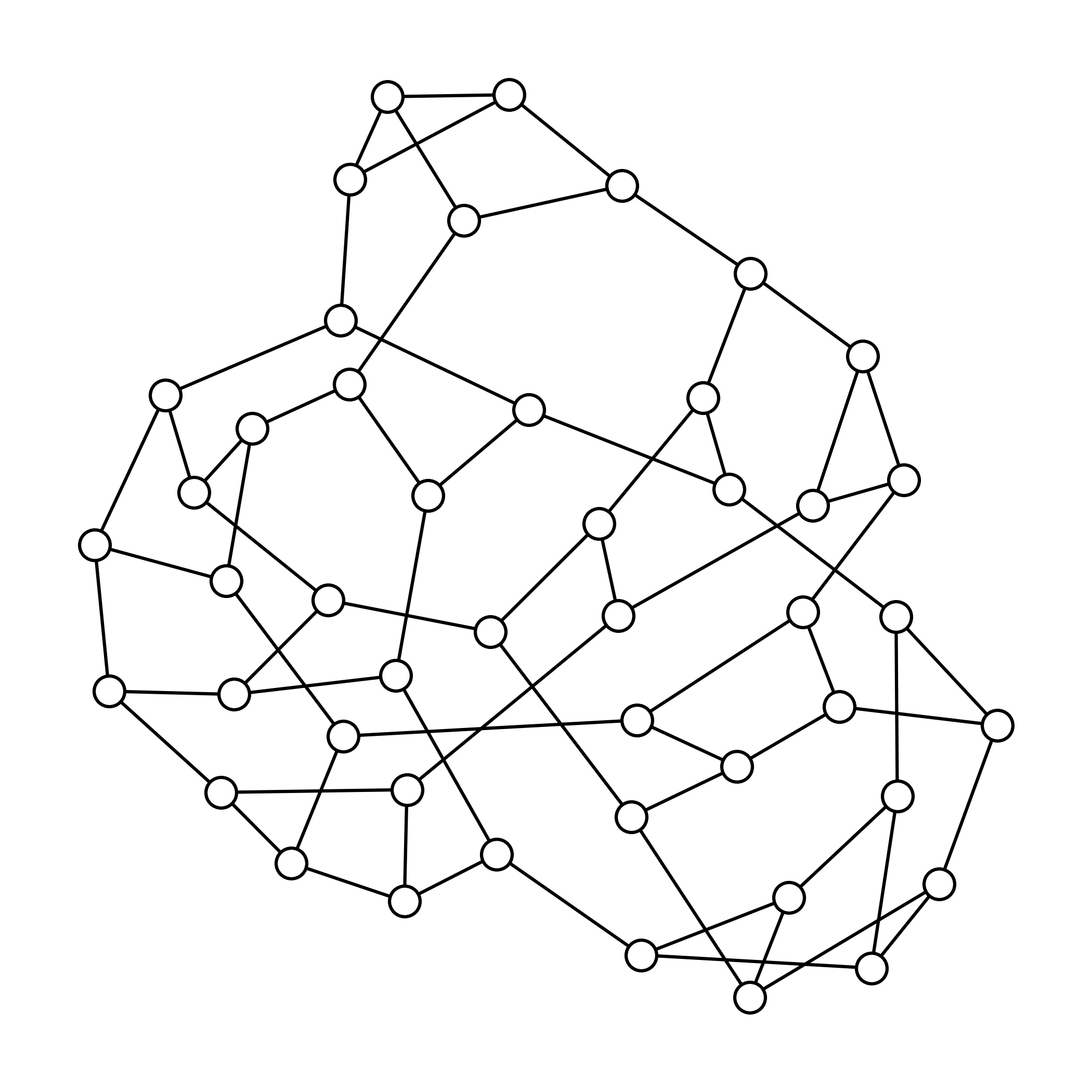}
    \vspace{-0.5cm}
    \caption{\small{Schematic representation of a Cayley tree and a random regular graph. (a) Cayley 
    tree with $4$ generations and coordination number $K+1 = 3$ (thus having $46$ nodes). (b) A random regular graph also 
    with $K+1=3$ and the same number of nodes.}}
    \label{fig:RRGCT}
\end{figure}

An important feature of random regular graphs is that they are locally tree-like. Indeed, it can be shown that the typical loop length scales as 
$\ln N/\ln K$~\cite{wormald1999models}.
The infinite graph associated with the thermodynamic limit of a random regular graph is the so-called Bethe lattice. Since the loop length diverges as 
$N\to\infty$, from the perspective of a generic reference site the lattice appears as an infinite tree with fixed connectivity.
Here, its definition as the thermodynamic limit of a random regular graph is essential. Because random regular graphs exhibit no intrinsic hierarchy among nodes, taking the limit $N\to\infty$ implies that each site sees the remainder of the lattice as an infinite tree and the system becomes translationally invariant.

A frequent source of confusion lies in conflating the infinite Bethe lattice with the thermodynamic limit of a finite Cayley tree. A Cayley tree of $n$ generations is a loopless hierarchical structure whose root node is connected to $K+1$ offspring in the first generation, while all subsequent nodes up to generation $n-1$ branch into $K$ offspring in the next generation. The total number of nodes $N$ within an $n$-generation Cayley tree is given by
\begin{equation}
N = 1 + \frac{K+1}{K-1}(K^{n}-1).
\end{equation}
Crucially, the Cayley tree lacks translational invariance and is inherently bipartite. Furthermore, it possesses a pathological boundary that contains a macroscopic fraction of the total number of sites, even as $n \to \infty$. Recent studies have demonstrated that this boundary dominance causes Anderson localization on Cayley trees to exhibit profound qualitative and quantitative differences compared to random regular graphs -- particularly within the delocalized metallic phase -- discrepancies that persist robustly in the thermodynamic limit~\cite{biroli2020anomalous,sonner2017multifractality,tikhonov2016fractality}. To visually contrast these geometric paradigms, Fig.~\ref{fig:RRGCT} illustrates both a Cayley tree and a random regular graph for a small system size of $N=46$.

The most important properties of the Bethe lattice, which add to the ones of random regular graphs, can be summarized as:
\begin{itemize}[noitemsep]
    \item It is statistically translational invariant in the thermodynamic limit.
    \item Conditioning on one of the vertices, the $K+1$ branches of the remaining graph become statistically independent in the large-$N$ limit~\cite{mezard2009information}.
    \item There is only one simple path of finite length between two vertices.
\end{itemize}
These properties simplify the analysis of the statistical properties of models defined on Bethe lattices.

\section{Anderson Localization on the Bethe Lattice}
\label{sec:Anderson}

In this Section we summarize the analysis of the  Anderson model on the Bethe lattice and we 
present some results relevant to the comparison to predictions of the Localization Landscape Theory. A pedagogical introduction to Anderson localization on the Bethe lattice, which covers in detail the results summarized below, can be found in the recent review~\cite{prado2026anderson}.
In Sec.~\ref{sec:Anderson-def} we recall the definition of the Anderson model and the local Density of States (LDoS), 
in Sec.~\ref{sec:Anderson-Bethe}  we present the Green's function formalism used to analyze the LDoS and phase transition,
in Sec.~\ref{sec:SCanderson} we describe the distribution analysis of the Green's function,  
and in Sec.~\ref{sec:ALeigenval} a linear analysis of the  distribution equation which allows one to 
locate the phase transition.

\subsection{Model and main observables}
\label{sec:Anderson-def}

The tight-binding Anderson model~\cite{anderson1958absence} reads
\begin{equation}
\label{eq:ham}
    \hat{\mathcal{H}} = \sum_{i} \epsilon_i \hat{c}_i^\dagger \hat{c}_i - t \sum_{\langle i,j \rangle} \left( \hat{c}_i^\dagger \hat{c}_j + \hat{c}_j^\dagger \hat{c}_i \right)
\end{equation}
where $\hat{c}_i^\dagger$ and $\hat{c}_i$ are 
creation and annihilation (spinless) fermion operators at site $i$ in the second-quantization formalism. The parameters 
$\epsilon_i$ are the on-site disorder potentials, and they are independent and identically distributed random
variables drawn from the uniform distribution
\begin{equation}
\label{eq:gammas}
    \gamma(\epsilon)=\begin{cases}
        1/W &\text{if} \,\,\epsilon\in [-W/2,W/2]
        \;,
        \\
        0 &\text{else} \; .
    \end{cases}
\end{equation}
 The hopping amplitude between neighboring sites is denoted $t$, and $\sum_{\langle i,j \rangle}$ runs over nearest-neighbor pairs.
 
 The mobility edge is the disorder strength dependent critical energy, $E_{\rm loc} (W)$,  which 
 separates localized from extended states. Below the mobility edge, the wave functions are spatially 
 localized due to interference from disorder, inhibiting transport; above it, states remain delocalized 
 and can carry current. As disorder increases, the mobility edge shifts until all states become localized, 
 marking the Anderson transition to an insulating phase. 

A key mathematical tool for analyzing the localization properties of this (and similar) models 
is the resolvent or Green's function, defined as
\begin{equation}
\label{eq:defG}
    \hat{\mathcal{G}}(z) \equiv (\hat{\mathcal{H}}-z\hat{\mathcal{I}} )^{-1},
\end{equation}
where $\hat{\mathcal{H}}$ is the Hamiltonian of the system, and $z$ is a complex number whose imaginary part is useful for regularization reasons.
The imaginary part of the diagonal elements of $\hat{\mathcal{G}}(z)$, 
\begin{equation}
\label{eq:diagGF}
    \mathcal{G}_{ii}(z)=\langle i | \hat{\mathcal{G}}(z)|i\rangle
    \, ,
\end{equation}
evaluated in the orthonormal basis of states $|i\rangle$ localized at site $i$, is directly linked to the local density of states to be defined below. Its decay properties characterize the spatial extent of the wavefunctions.

The Local Density of States (LDoS) is a position-resolved version of the Density of States, and 
describes how quantum states are distributed in both energy $E$ and space, parametrized by the site index $i$, 
\begin{equation}
\label{eq:rhoi-def}
    \rho_i(E) =\sum_n \; |\psi_i^{(n)}|^2 \; \delta(E-E_n) 
    \; , 
\end{equation}
where $E_n$ are the eigenvalues and $\psi^{(n)}$ the 
eigenstates of the Hamiltonian $\hat{\mathcal H}$.
 In terms of the imaginary part of the Green's function, the LDoS is given by
\begin{eqnarray}
    \rho_i(E) 
    &=&
    \frac{1}{\pi} \; {\rm Im} \, \mathcal{G}_{ii}(E+ {\rm i} 0^+) 
    \label{eq:rhoi}
    \; ,
\end{eqnarray}
where $\mathcal{G}_{ii}(E+{\rm i}\alpha)$ is the diagonal element of the Green's function of Eq.~(\ref{eq:defG}) evaluated at $z=E+{\rm i}\alpha$, with $\alpha\to0^+$. 
The expression of $\rho_i$ in terms of $\mathcal{G}_{ii}$ is a well-known result, and 
the derivation can be found, for example,  in Ref.~\cite{economou1972existence}.

The LDoS serves as the order parameter for the delocalization-localization transition. Physically, it represents the inverse lifetime of a particle of energy $E$ created on 
site $i$. In the localized phase, $\rho_i(E)$ is exponentially small on $O(N)$ sites and of order unity on only $O(1)$ sites. In contrast, in the delocalized phase, $\rho_i(E)$ remains nonzero on $O(N)$ sites, enabling transport across the system.

Another order parameter for the delocalization-localization transition is the Inverse Participation Ratio or IPR, which is defined as
\begin{eqnarray}
\label{eq:IPR}
    I_2(E)
    &\equiv &
    \mathbb{E}\Bigg[\frac{\sum_n \sum_i \big|\psi^{(n)}_i\big|^4 \delta(E_n-E)}{\sum_n\delta(E_n-E)}\Bigg] 
    \nonumber\\
    &=&
    \mathbb{E}\Bigg[\lim_{\alpha \to 0^+} \frac{\alpha \sum_i |\mathcal{G}_{ii}(E+{\rm i}\alpha)|^2}{ \sum_i {\rm Im} \, \mathcal{G}_{ii}(E+{\rm i}\alpha)  }\Bigg]\,, 
\end{eqnarray}
and is essentially a measure of the average inverse volume occupied by an eigenstate. 
Henceforth, the expectation value ${\mathbb E}[ \dots]$ is taken with respect to the disorder distribution. 
In the delocalized phase $I_2(E) \sim O(1/N)$, while in the localized one $I_2(E) \sim O(1)$.

Finally, the last quantity on which we are interested in is the eigenstate correlation function, which is defined as 
\begin{eqnarray}
\label{eq:corrfuncAL}
    C_{\rm loc}(|i-j|\,;E)
    &\equiv &
    \mathbb{E}\Bigg[\frac{\sum_n  \big|\psi^{(n)}_i\big|^2 \big| \psi^{(n)}_j\big|^2 \delta(E-E_n)}{\sum_n\delta(E-E_n)}\Bigg] 
    \nonumber\\
    &=&
    \mathbb{E} \Bigg[ \lim_{\alpha \to 0^+} \frac{\alpha |\mathcal{G}_{ij}(E+{\rm i}\alpha)|^2}{\sum_i {\rm Im} \, \mathcal{G}_{ii}(E+{\rm i}\alpha)} \Bigg]
    \; .
\end{eqnarray}
It represents the average correlation between the amplitude of the eigenstates on two sites at distance $|i-j|\delta$, with $\delta$ the 
lattice spacing. 
The algebraic steps leading to the final expression in Eq.~(\ref{eq:IPR}) are detailed in Ref.~\cite{rizzo2024localized}
 (see also Ref.~\cite{mirlin1994statistical}), and the last term in Eq.~(\ref{eq:corrfuncAL}) is derived in an analogous way.
For completeness, we sketch the derivation here. Using the spectral decomposition of the Green's function at $z=E+{\rm i}\alpha$,
\begin{equation}
\label{eq:Gspectral}
\mathcal G_{ij}(E+{\rm i}\alpha)=\sum_n\frac{\psi^{(n)}_i\,\psi^{(n)*}_j}{E_n-E-{\rm i}\alpha}\,,
\end{equation}
one has
\begin{eqnarray}
 {\rm Im}\,\mathcal G_{ii}(E+{\rm i}\alpha)
=
\sum_n\big|\psi^{(n)}_i\big|^2\,\frac{\alpha}{(E_n-E)^2+\alpha^2}\,,
\end{eqnarray}
which reduces to $\pi\rho_i(E)$ for $\alpha\to0^+$ [Eq.~(\ref{eq:rhoi})], and
\begin{eqnarray}
&& 
\alpha\,|\mathcal G_{ij}(E+{\rm i}\alpha)|^2=
\nonumber\\
&& \qquad\qquad
\sum_{n,m}\frac{\alpha\,\psi^{(n)}_i\psi^{(n)*}_j\,\psi^{(m)*}_i\psi^{(m)}_j}{(E_n-E-{\rm i}\alpha)(E_m-E+{\rm i}\alpha)}\,.
\end{eqnarray}
If $\alpha$ is much smaller than the spacing between the eigenvalues, the terms with $n\neq m$ vanish for $\alpha\to0^+$. Each term with $n=m$ contains the Lorentzian $\alpha/[(E_n-E)^2+\alpha^2]\to\pi\delta(E-E_n)$, so that
\begin{eqnarray}
&& \lim_{\alpha\to0^+}\alpha\,|\mathcal G_{ij}(E+{\rm i}\alpha)|^2
=
\nonumber\\
&& 
\qquad\qquad
\pi\sum_n\big|\psi^{(n)}_i\big|^2\big|\psi^{(n)}_j\big|^2\,\delta(E-E_n)
\;.
\label{eq:Gspectral-limit}
\end{eqnarray}
Taking $i=j$ and summing over $i$, the ratio with $\sum_i{\rm Im}\,\mathcal G_{ii}$ gives the last expression in Eq.~(\ref{eq:IPR}). Taking $i\neq j$ gives the last expression in Eq.~(\ref{eq:corrfuncAL}). The order of limits is essential: $\alpha\to0^+$ must be taken before $N\to\infty$, i.e., with $\alpha$ smaller than the level spacing. Otherwise, in the delocalized phase, each Lorentzian overlaps many levels, and the expression in Eq.~(\ref{eq:IPR}) is of order $\alpha$ instead of giving $I_2(E)=O(1/N)$. In the localized phase on the infinite lattice the spectrum is pure point, and the same expressions can be used directly in the thermodynamic limit. This is the property exploited by the numerical method of Sec.~\ref{sec:numerics}.
From Eqs.~(\ref{eq:rhoi})-(\ref{eq:corrfuncAL}),  it is clear  that the critical 
properties of Anderson localization on a generic lattice are fully determined by $\hat{\mathcal{G}}(z)$.

\subsection{The Green's Function on the Bethe Lattice}
\label{sec:Anderson-Bethe}

The Anderson model can be solved exactly on the Bethe lattice thanks its hierarchical structure, which allows for an exact recursive  computation of the components 
$\mathcal{G}_{ij}(z)$. With methods explained in~\cite{abou1973selfconsistent,biroli2010anderson,prado2026anderson}, one obtains that  the diagonal elements  are given by
\begin{equation}
\label{eq:gAL}
    \mathcal{G}_{ii}(z)=\frac{1}{\epsilon_i-z-t^2\sum\limits_{k\in \partial i}\mathcal{G}_{k\rightarrow i}(z)}
    \; ,
\end{equation}
where the symbol $\partial i$ represents the set of nearest neighbors of site $i$, and the terms $\mathcal{G}_{k\rightarrow i}(z)$ are the cavity Green's functions on site $k$ in the absence of site $i$. The latter satisfy an analogous recursion relation, 
\begin{equation}
\label{eq:gcavAL}
    \mathcal{G}_{k\rightarrow i}(z)=\frac{1}{\epsilon_k-z-t^2\sum\limits_{l\in {\partial k \setminus i}} \mathcal{G}_{l\rightarrow k}(z)}
    \; ,
\end{equation}
where $\partial  {k \! \setminus \! l}$ is a shorthand notation for $\partial {k \! \setminus \! \{l\}}$ and  ``$\setminus$" is the symbol of difference between sets, and hence denotes the set of all neighbors of $k$ except $l$.

The imaginary part of the Green's function constitutes the central quantity of interest for the localization transition since, according to Eq.~\eqref{eq:rhoi}, its probability distribution governs the behavior of the LDoS. Crucially, the mobility edge defines the boundary in the $(E,W)$-plane that separates a localized phase, where the typical value of the imaginary part vanishes, from a delocalized phase, where it remains finite. While the physical implications of the imaginary part are well established, a deep interpretation of the real part of the Green's function has remained comparatively elusive.  Localization Landscape Theory offers a powerful framework to bridge this conceptual gap. Indeed, as we demonstrate in Sec.~\ref{sec:EQGU}, the localization landscape $\mathbf{u}$ is directly related to the resolvent operator $\hat{\mathcal{G}}(z)$ evaluated entirely below the lower spectral edge, $z \in \mathbb{R}$ with $z \le E_{\rm min}$. In this specific energy regime, the cavity recursive equations~\eqref{eq:gAL}--\eqref{eq:gcavAL} yield strictly real, non-negative values, providing a clear geometric meaning to the real component of the Green's function.

\subsection{Self-consistent distributional equation}
\label{sec:SCanderson}

The derivation that we present in this Section is a summary of the one in Ref.~\cite{rizzo2024localized}, with the only difference that we consider here the case with finite energy $E$.
Let us focus on the cavity Green's functions $\mathcal{G}_{k\to i}(z)$, with $z=E+{\rm i} \alpha$ where $\alpha \ll 1$ serves as a regularization parameter. From now on, and to lighten the notation, we only 
write $\mathcal{G}_{k\to i}$ without the explicit $z$-dependence. 

We decompose the cavity Green's function in the localized phase as
\begin{equation}
    \mathcal{G}_{k\to i} = \mathcal{G}^R_{k\to i} + {\rm i} \mathcal{G}^I_{k\to i}
    \; .
\end{equation}

The cavity equation (\ref{eq:gcavAL}) can be separated into real and imaginary parts. By expanding for small $\alpha$ and $\mathcal{G}^I_{k\to i}$, and retaining the leading order, we obtain the two cavity equations governing the critical properties:
\begin{align}
\label{eq:ReGcavcrit}
    (\mathcal{G}_{k\to i}^R)^{-1} &= \epsilon_k- E - t^2 \sum_{l \in \partial k \setminus i} \mathcal{G}_{l\to k}^R \; ,  \\
    \label{eq:ImGcavcrit}
    \mathcal{G}_{k\to i}^I &= t^2 (\mathcal{G}_{k\to i}^R)^2 \sum_{l \in \partial k \setminus i} \mathcal{G}_{l\to k}^I \; .
\end{align}
We can now exploit the statistical independence of the cavity variables by noting that, after averaging over disorder, the Hamiltonian becomes translationally invariant. Consequently, the joint probability distribution of the on-site variables is identical on every site, and the same applies to the cavity variables. This allows us to write a recursive equation that relates the joint probability distribution of their real and imaginary parts [Eqs.~(\ref{eq:ReGcavcrit}) and (\ref{eq:ImGcavcrit})] to $K$ copies of itself evaluated on the nearest neighboring cavity sites~\cite{rizzo2024localized,abou1973selfconsistent,tikhonov2019critical,prado2026anderson}:
\begin{eqnarray}
\label{eq:SCanderson}
    P(g, \hat g) &=& 
    \int d\epsilon \, \gamma(\epsilon) \int \prod_{l=1}^K \left[dg_l d\hat g_l \, 
    P( g_l, \hat g_l )\right]  \; 
    \nonumber\\
    &&
   \quad 
   \times \;
   \delta \bigg( g - \frac{1}{\epsilon-E - t^2 \sum_l g_l} \bigg) 
   \nonumber\\
   && 
   \quad 
   \times \;
   \delta \bigg( \hat g - t^2 g^2 \sum_l \hat g_l \bigg)
   \; ,
\end{eqnarray}
where the disorder distribution $\gamma$ is the one in Eq.~(\ref{eq:gammas}) and the functions $P$ in the left-hand-side and 
within the integrals in the right-hand-side are the same. This follows from the fact that the cavity variables are statistically independent, thus the joint probability distribution of $K$ cavity variables on the right-hand-side have been decomposed as
\begin{eqnarray}
    P(\{ g_l, \hat g_l \}_{l=1,\dots,K})=\prod_{l=1}^K P( g_l, \hat g_l )
    \; .
\end{eqnarray}
Eq.~\eqref{eq:SCanderson} should be compared with Eq.~(\ref{eq:stochSC}) below, which represents its counterpart in the Localization Landscape percolation problem. This comparison will let us highlight the difference between the Anderson localization and the Localization Landscape percolation transitions on the Bethe lattice. 

\subsection{Linear stability analysis}
\label{sec:ALeigenval}

Equation (\ref{eq:SCanderson}) can be rewritten as an integral eigenvalue equation using the following procedure.  
Starting from Eq.~(\ref{eq:SCanderson}), we use the integral representation of the delta function for the variable $\hat g$, 
\begin{equation}
    \delta\bigg( \hat g - t^2 g^2\sum_l \hat g_l \bigg)
    = \int_{-\infty}^{\infty} 
    \frac{d\lambda'}{2\pi} \,e^{ - {\rm i} \lambda' \left(\hat g - t^2 g^2\sum_l \hat g_l \right)} \, ,
\end{equation}
to  integrate explicitly over all the $\hat g_l$'s in the right-hand-side of Eq.~(\ref{eq:SCanderson}). The result is
\begin{eqnarray}
    P(g,\hat g) \! & \! = \!\!  & \!  \int \frac{d\lambda'}{2\pi}  \, e^{ -{\rm i} \lambda' \hat g} 
    \! \int d\epsilon \, \gamma(\epsilon) 
\int \prod_{l=1}^K \left[dg_l  \, \hat P( g_l , \lambda' t^2g^2)\right] \; 
\nonumber\\
&&
\quad 
\times \;
    \delta \bigg( g - \frac{1}{\epsilon-E - t^2 \sum_l g_l} \bigg)
    \label{eq:Pgghat1}
\end{eqnarray}
with 
\begin{equation}
 \hat P( g_l, \lambda' t^2g^2) \equiv 
 \int  d\hat g_l \; P( g_l, \hat g_l) \; e^{ {\rm i} \lambda' t^2 g^2 \hat g_l} 
 \; .
\end{equation}
Then, Fourier transforming Eq.~(\ref{eq:Pgghat1}) 
over $\hat g$ on both sides, and recognizing the integral representation of $\delta(\lambda-\lambda')$, 
we obtain 
\begin{eqnarray}  
    \hat P(g,\lambda)
    &=&
    \int d\epsilon \, \gamma(\epsilon) \int \prod_{l=1}^K \left[dg_l \; \hat P( g_l ,\lambda t^2g^2)\right]
    \; \nonumber\\
    &&
    \quad
    \times \;
    \delta \bigg( g - \frac{1}{\epsilon -E - t^2 \sum_l g_l} \bigg)\,.  
\end{eqnarray}  
Next, we expand this characteristic function around the solution with zero imaginary part to lowest order in $\lambda$, assuming an algebraic form for the distribution~\cite{rizzo2024localized,abou1973selfconsistent,tikhonov2019critical,prado2026anderson},  
\begin{equation}  \label{eq:ansatz}
    \hat P(g,\lambda) \approx P(g)+f(g)|\lambda|^\beta\,,  
\end{equation}  
which gives  
\begin{equation} 
\label{eq:fAL}
    f(g)= \int d g'\,\mathcal{K}_{\rm loc}^\beta(g,g')f(g') 
\end{equation}  
with  the kernel 
\begin{eqnarray}
\label{eq:kernelAL}
    && \mathcal{K}_{\rm loc}^\beta(g,g')
    =
    K|tg|^{2\beta} \int d\epsilon \, \gamma(\epsilon) \; \int d \tilde g \;  R_{\rm loc}(\tilde g) 
    \qquad\quad
    \; \nonumber\\
    && 
    \qquad\qquad
    \times \;
    \delta \bigg (g- \frac{1}{\epsilon-E-t^2 (\tilde g+g')}\bigg)\,,  
\end{eqnarray}  
where  
\begin{equation}  
    R_{\rm loc}(\tilde{g})=\int \prod_{l=1}^{K-1}\left[dg_l\,P(g_l )\right] \; \delta\bigg(\sum_{l=1}^{K-1}g_l-\tilde{g}\bigg)
    \; .  
\end{equation}  
The function $R_{\rm loc}$ represents the distribution of the sum of the $K-1$ real parts of the cavity Green functions. 
It can be evaluated numerically by computing the marginal distribution of the $\mathcal{G}_{k\to i}^R$'s through the population dynamics algorithm
explained in Sec.~\ref{subsec:population} and sampling from it sums of $K-1$ variables.

Given this equation, the critical line can be identified as the curve in the 
$(E,W)$ plane where the largest eigenvalue of the kernel equals one. In the localized phase, the only stable solution of the self-consistent distributional equation~\eqref{eq:SCanderson} is the one with a vanishing imaginary part of the Green's function; therefore, any perturbation $f(g)$ of this solution must decay under iteration of the integral operator. 
Conversely, in the delocalized phase, such perturbations grow under iteration. If the leading eigenvalue of 
${\mathcal K}_{\rm loc}$ is smaller than one, all components of $f$, when decomposed on the eigenbasis of 
${\mathcal K}_{\rm loc}$, decay upon iteration. If it exceeds one, any component not orthogonal to the corresponding eigenfunctions diverges. Hence, the boundary separating these two regimes is precisely the curve where the largest eigenvalue of the kernel equals one.

This integral operator can be diagonalized numerically with high precision for all $E$, as it has been done in Refs.~\cite{parisi2019anderson} and~\cite{tikhonov2019critical} for $E=0$. Because of the high computational cost of this procedure, in order to obtain the full curve (sacrificing accuracy) we used the population dynamic approach. 
This method will be explained in detail in Secs.~\ref{sec:numerics}, and the resulting phase diagram is discussed in the 
companion Letter~\cite{Tonetti2026}.

The integral eigenvalue equation \eqref{eq:fAL} is fundamentally different from its counterpart in the Localization Landscape percolation problem. In Sec. \ref{subsubsec:linstab}, we will show this explicitly in the high-connectivity limit. This difference provides an additional motivation for the mismatch between the critical properties of AL and the ones of LLT percolation proved in Ref.~\cite{Tonetti2026}. 

\section{The Localization Landscape Theory on the Bethe Lattice}
\label{sec:EQGU}

In this Section we derive the relevant equations to study the critical behavior of the Localization Landscape percolation problem.
We begin by recalling the definition of the Localization Landscape in Sec.~\ref{subsec:localization-landscape}. 
In Sec.~\ref{subsec:cavity-derivation}, 
we derive the set of coupled equations that let us compute the Localization Landscape on a generic site $i$ of the Bethe lattice as a function of independent quantities defined on the cavity lattices rooted at the nearest neighbors of $i$.
Here, we present an alternative derivation to that in the End Matter of the companion Letter~\cite{Tonetti2026}, 
employing known results for the Green's function of Anderson localization on the Bethe lattice~\cite{biroli2010anderson,rizzo2024localized}.
Next, in Sec. \ref{sec:EQP} we obtain the equation governing the probability that a generic site belongs to an infinite cluster. Then, in Sec.~\ref{sec:SCP} we show how this equation and the cavity equations of Sec.~\ref{subsec:cavity-derivation} can be recast into a self-consistent distributional equation.
Finally, in Sec.~\ref{sec:CFP} we derive the expressions for the percolation correlation function and average cluster size.

\subsection{The Localization Landscape}
\label{subsec:localization-landscape}

The Localization Landscape Theory (LLT)~\cite{filoche2012universal}, here treated in its discrete formulation as in Ref.~\cite{filoche2024anderson}, introduces an $N$-dimensional positive vector ${\mathbf u}$ as the solution of the equation
\begin{equation}
\hat {\mathcal H}_+ {\mathbf u} = {\mathbf 1}
\end{equation}
with ${\mathbf 1}$ the vector with identical components equal to one, and defines a real-space effective potential, $1/u_i$.
Here, $\hat{\mathcal H}_+$ denotes a shifted version of the original Anderson Hamiltonian, chosen so that the LLT construction is well defined. In its usual formulation LLT requires $\hat{\mathcal{H}}_+$ to be positive definite, however, as we discuss later, the question is slightly more complex for the Bethe lattice. Given an energy $E_+$ associated with $\hat{\mathcal H}_+$, we define the set of nearest-neighboring sites on the lattice with lower effective potential than the particle energy as
\begin{equation}
    \Omega_{E_+}
    =
    \left\{
    i \; \Bigg| \; \frac{1}{u_i}\leq E_+
    \right\}.
\end{equation}
According to the LLT, these are the spatial regions where a particle is classically confined.
Within this framework, when a macroscopic (``giant'') cluster of this kind exists, that is, when the set $\Omega_{E_+}$ forms a connected path spanning the system, the quantum particle delocalizes.

In principle, the spectral shift of $\hat{\cal H}$ to ensure that the shifted Hamiltonian is positive definite is arbitrary. We write the shifted Hamiltonian as
\begin{equation}
\label{eq:Hplus}
\hat{\cal H}_+
=
\hat{\cal H}
-
E_{\rm sh}\hat{\cal I}
\; ,
\end{equation}
where $E_{\rm sh}$ is the spectral reference subtracted from $\hat{\cal H}$. The energy $E_+$ in the shifted Hamiltonian is therefore related to the energy $E$ in the original Anderson Hamiltonian by
\begin{equation}
\label{eq:EplusE}
    E_+
    =
    E-E_{\rm sh}
    \; ,
    \qquad
    E
    =
    E_{\rm sh}+E_+ 
    \; .
\end{equation}
As discussed in Ref.~\cite{filoche2017localization}, the minimal shift, equal to minus the minimal eigenvalue of $\hat{\cal H}$, is often the optimal choice.
Therefore, in standard applications of the LLT one chooses the spectral reference $E_{\rm sh}$ to coincide with the ground state energy.

~We stress that the Localization Landscape is not covariant under a change of the shift: the solution of $(\hat{\mathcal H}-E_{\rm sh}\hat{\mathcal I})\mathbf u=\mathbf 1$ is not simply translated when $E_{\rm sh}$ is changed. As a consequence, the LLT percolation threshold depends on $E_{\rm sh}$ even after the energies are translated back to those of $\hat{\mathcal H}$ [Eq.~\eqref{eq:Epercpositive} below], and no first-principle criterion uniquely selects the shift beyond the requirement that $\hat{\mathcal H}_+$ be positive definite. For any admissible shift, $E_{\rm sh}\le E_{\min}$, $\hat{\mathcal H}_+$ is a non-singular M-matrix, so that $\hat{\mathcal G}(E_{\rm sh})=\hat{\mathcal H}_+^{-1}$ has non-negative entries (see Sec.~\ref{subsec:cavity-derivation}). The original LLT bound~\cite{filoche2012universal} then holds. If $\hat{\mathcal H}\psi=E\psi$, then $\psi=(E-E_{\rm sh})\hat{\mathcal G}(E_{\rm sh})\psi$ and, therefore,
\begin{equation}
\label{eq:LLTbound}
\frac{|\psi_i|}{\|\psi\|_\infty}\le(E-E_{\rm sh})\,u_i(E_{\rm sh})\,,\quad\|\psi\|_\infty=\max_j|\psi_j|\,.
\end{equation}
The amplitude of an eigenstate of energy $E$ is thus constrained on the sites where $(E-E_{\rm sh})u_i<1$, i.e., where $1/u_i>E_+$: these are the classically forbidden regions of the LLT. In Sec.~\ref{subsec:dependence-shift} we use this bound to discuss the choice of the shift.

\begin{figure*}[t]
\centering
\includegraphics[width=0.85\textwidth]{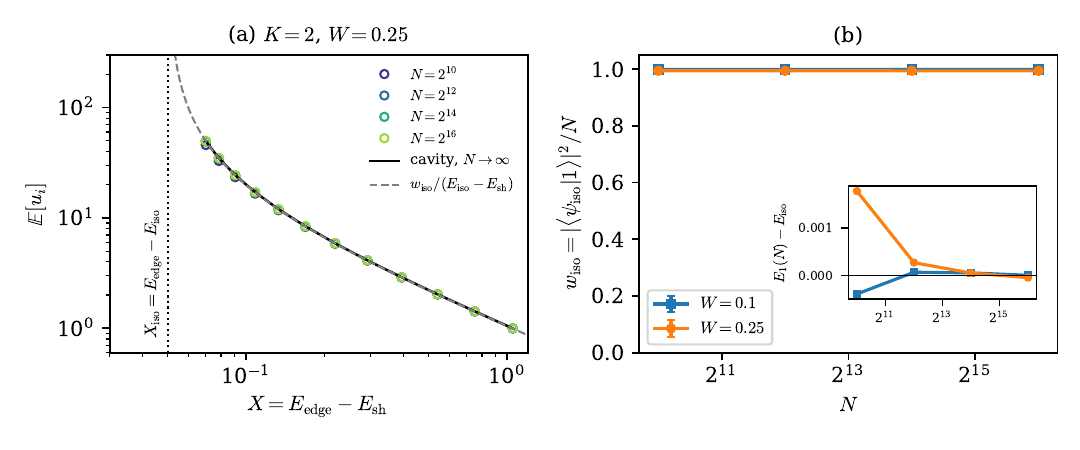}
\caption{Isolated eigenvalue and mean Localization Landscape on finite RRGs with $K=2$ and $t=1$. (a)~$\mathbb E[u_i]$ as a function of the shift $X=E_{\rm edge}-E_{\rm sh}$ for $W=0.25$, and RRGs with $N=2^{10},\dots,2^{16}$ sites (symbols; averages over $16$, $16$, $8$, and $2$ samples, respectively). These are compared with the cavity result in the thermodynamic limit (solid line, population dynamics with $10^6$ elements) and with the contribution $w_{\rm iso}/(E_{\rm iso}-E_{\rm sh})$ of the isolated eigenvalue alone [dashed line, see Eq.~\eqref{eq:Eu-spectral}]. $\mathbb E[u_i]$ diverges at $X_{\rm iso}=E_{\rm edge}-E_{\rm iso}>0$ (dotted vertical line, cavity value), i.e.\ at $E_{\rm sh}=E_{\rm iso}$, and not at the edge $E_{\rm edge}$ of the continuous spectrum. (b)~Weight $w_{\rm iso}$ of the isolated eigenvector in the spectral measure of the uniform vector, Eq.~\eqref{eq:Eu-spectral}, for $W=0.1$ and $W=0.25$. It is independent of $N$ and of order one. Inset: difference between $E_1(N)$, the lowest eigenvalue of a RRG with $N$ sites (averaged over samples), and the cavity value of $E_{\rm iso}$.}
\label{fig:finiteN-u}
\end{figure*}

A further subtlety arises when applying the LLT on the Bethe lattice. Unlike Euclidean lattices, the Hamiltonian of the Anderson model on the Bethe lattice has an isolated eigenvalue which lies below the lower edge of the continuous spectrum for disorder strengths $W\leq W_{\min}$. It is therefore useful to distinguish explicitly between the lower edge of the continuous spectrum,
\begin{equation}
\label{eq:Eedge}
    E_{\rm edge}(W)
    =
    -2t\sqrt{K}-\frac{W}{2},
\end{equation}
and the true minimal eigenvalue. In absence of disorder the isolated eigenvalue is located at $E_{\rm iso}(W=0) = - t(K+1)$, corresponding to the uniform eigenvector $\psi_{\rm iso} = 1/\sqrt{N} ( 1, \ldots, 1)$. When disorder is turned on the spectrum is deformed. The bulk of the spectrum broadens and the energy of the isolated eigenvalue is modified. The distance between $E_{\rm iso}(W)$ and the lower edge of the continuous part of the spectrum is progressively reduced upon increasing $W$. The disorder $W_{\min}$ is defined by
\begin{equation}
    E_{\rm iso}(W_{\min})
    =
    E_{\rm edge}(W_{\min}) .
\end{equation}
Thus, the minimal eigenvalue is
\begin{equation}
\label{eq:Emin}
E_{\rm min}(W) =
    \begin{cases}
        E_{\rm edge}(W) &\text{if} \quad W>W_{\rm min}\; ,\\
        E_{\rm iso}(W) &\text{if} \quad W\leq W_{\rm min}\; .
    \end{cases}
\end{equation}
For $W>W_{\min}$, the true bottom of the spectrum and the lower edge of the continuous spectrum coincide. For $W\leq W_{\min}$, instead, they are distinct.
On a finite RRG with $N$ sites, the weight of the isolated eigenvector in the local spectral measure of Eq.~\eqref{eq:rhoi-def} is $|\psi_{\rm iso}(i)|^2=O(1/N)$. The isolated eigenvalue therefore disappears from the local (and average) Density of States in the thermodynamic limit, and it does not belong to the spectrum of $\hat{\mathcal H}$ on the infinite tree. Nonetheless, it plays a central role in the LLT. The reason is that the Localization Landscape is a non-local quantity: as shown in Sec.~\ref{subsec:cavity-derivation} [Eq.~\eqref{eq:ui}], $u_i=\sum_j\mathcal G_{ij}(E_{\rm sh})$ is a row sum of the resolvent. Its average over the sites of a finite graph reads
\begin{eqnarray}
\label{eq:Eu-spectral}
\mathbb E[u_i]&=& \frac1N\sum_i u_i=\frac1N\langle\mathbf 1|\hat{\mathcal G}(E_{\rm sh})|\mathbf 1\rangle
\nonumber\\
&=& 
\sum_n\frac{w_n}{E_n-E_{\rm sh}}\,,
\end{eqnarray}
with $w_n=|\langle\psi_n|\mathbf 1\rangle|^2/N$ and $\sum_n w_n=1$.
Hence $\mathbb E[u_i]$ is the Stieltjes transform of the spectral measure associated with the normalized uniform vector $|\mathbf 1\rangle/\sqrt N$, and not with a single site. Writing $\psi_{\rm iso}(i)=a_i/\sqrt N$, with $a_i=O(1)$ and $\sum_ia_i^2=N$, the weight of the isolated eigenvalue in this measure is $w_{\rm iso}=\bar a^2$, with $\bar a=N^{-1}\sum_ia_i=O(1)$. At $W=0$, $\psi_{\rm iso}$ is the uniform vector, $w_{\rm iso}=1$, and $u_i=1/(E_{\rm iso}-E_{\rm sh})$ exactly. For $W>0$, $w_{\rm iso}$ remains of order one as $N\to\infty$, as we have verified on finite RRGs [Fig.~\ref{fig:finiteN-u}(b)].

Throughout this paper, the thermodynamic limit is taken first, at fixed $E_{\rm sh}$, by means of the cavity method, and the dependence on $E_{\rm sh}$ is studied afterwards. In this limit, the isolated eigenvalue is recovered as follows. For $E_{\rm sh}<E_{\rm edge}$, $\hat{\mathcal G}(E_{\rm sh})$ is a bounded operator on square-summable vectors, and the cavity recursions for the Green's functions converge. However, the uniform source $\mathbf 1$ is not square summable. The row sum $u_i$ can thus diverge even though $\mathcal G_{ij}$ decays with the distance $r$ between $i$ and $j$, because the number of sites at distance $r$ grows as $(K+1)K^{r-1}$. This is transparent at $W=0$. By the chain rule of Eq.~\eqref{eq:chain-rule}, $\mathcal G_{ij}=\mathcal G_{ii}(tg)^r$, where $g$ is the cavity Green's function, solution of $g=1/(-E_{\rm sh}-Kt^2g)$. Therefore
\begin{equation}
u_i=\mathcal G_{ii}\Big[1+(K+1)\sum_{r\ge1}K^{r-1}(tg)^r\Big]\,,
\end{equation}
which is finite if and only if $Ktg<1$, i.e., $E_{\rm sh}<-t(K+1)=E_{\rm iso}(0)$, whereas the spectrum on square-summable vectors begins at $-2t\sqrt K$. In other words, $E_{\rm edge}$ is the bottom of the spectrum probed by local observables, while $E_{\rm iso}$ is the threshold below which the response to a uniform source, which is bounded but not square summable, is finite. The same mechanism holds for $W>0$. We define $E_{\rm iso}(W)$ in the thermodynamic limit as the largest value of $E_{\rm sh}$ for which $\mathbb E[u_i]$ is finite. It coincides with the $N\to\infty$ limit of the lowest eigenvalue of finite RRGs~\cite{biroli2010anderson} [see Fig.~\ref{fig:finiteN-u} and Sec.~\ref{subsubsec:lowW}].

A simple exact bound on $E_{\rm iso}(W)$ follows from averaging $\hat{\mathcal H}_+\mathbf u=\mathbf 1$ over the sites, which gives $\mathbb E[u_i(\epsilon_i-E_{\rm sh}-t(K+1))]=1$. Since $\partial u_i/\partial\epsilon_j=-\mathcal G_{ij}u_j\le0$, $u_i$ is a non-increasing function of all the (independent) on-site energies, and the Harris inequality gives $\mathbb E[u_i\epsilon_i]\le\mathbb E[u_i]\,\mathbb E[\epsilon_i]=0$. Hence
\begin{equation}
\label{eq:Eu-lowerbound}
\mathbb E[u_i]\ge\frac{1}{-t(K+1)-E_{\rm sh}}\,,
\end{equation}
which implies $E_{\rm iso}(W)\le-t(K+1)$ for every $W$, with equality at $W=0$.

The strictly minimal LLT prescription corresponds to choosing
\begin{equation}
    E_{\rm sh}(W)=E_{\rm min}(W) 
    \; .
\end{equation}
However, in order to follow continuously the bulk-edge prescription below $W_{\min}$, and to study the dependence of the LLT predictions on the choice of the shift, we introduce the one-parameter family
\begin{equation}
\label{eq:Eshift}
    E_{\rm sh}(W,X)
    =
    E_{\rm edge}(W)-X
    \; ,
    \qquad X\geq 0 
    \; .
\end{equation}
The case $X=0$ corresponds to the bulk-edge shift. For $W\le W_{\min}$, the true isolated-eigenvalue shift is obtained by choosing
\begin{equation}
\label{eq:Xiso}
    X=X_{\rm iso}(W)
    \equiv
    E_{\rm edge}(W)-E_{\rm iso}(W)
    \; ,
\end{equation}
since then $E_{\rm sh}(W,X_{\rm iso})=E_{\rm iso}(W)$. Values $0\leq X<X_{\rm iso}(W)$ correspond to an under-shifted continuation, for which the isolated eigenvalue lies below the reference energy and $\hat{\cal H}_+(X)$ is not positive definite. In the under-shifted regime, the LLT recursion defining the probability distribution of the Localization Landscape variables no longer converges. Nevertheless, the recursive equations for the edges of its support remain well defined, and their behavior allows the LLT framework to recover known low-disorder results for Anderson localization on the Bethe lattice, including the onset scale of localization and the Aizenman--Warzel lower bound~\cite{warzel2012}~(Sec.~\ref{subsec:EisoWmin}). We stress that these are properties of the analytically continued cavity equations, and not predictions of the standard, positive-definite LLT.  Values $X>X_{\rm iso}(W)$ correspond to an over-shifted, strictly positive definite Hamiltonian. The value $X=X_{\rm iso}(W)$ is the boundary value associated with the true minimal spectral reference.

In the following, we use Eq.~\eqref{eq:Eshift} as a convenient way to define both the standard bulk-edge continuation and the shifted LLT family. More precisely, for each value of $X$ we define the landscape field through
\begin{equation}
\label{eq:uXdef}
    \hat{\cal H}_+(X)\mathbf u
    =
    \mathbf 1 
    \; ,
    \qquad
    \hat{\cal H}_+(X)
    =
    \hat{\cal H}
    -
    E_{\rm sh}(W,X)\hat{\cal I}
    \; .
\end{equation}
When $\hat{\cal H}_+(X)$ is positive definite, this is the usual LLT definition. When $W\le W_{\min}$ and $0\leq X<X_{\rm iso}(W)$, the same equations define an analytical continuation of the bulk-edge LLT branch. This analytical continuation is not the standard positive-definite LLT prescription, but it is useful because it preserves the structure of the cavity equations obtained above $W_{\min}$ and allows us to test how much information about the onset of localization is already encoded in the bulk-edge prescription.

From Eqs.~\eqref{eq:g}-\eqref{eq:caveta} and~\eqref{eq:pcavlin}-\eqref{eq:uk} below, we determine the LLT critical percolation parameter $E_{+,\rm perc}(W,X)$ for the shifted Hamiltonian $\hat{\cal H}_+(X)$, as well as the exact position of the mobility edge $E_{\rm loc}(W)$ for the statistically symmetric Anderson Hamiltonian $\hat{\cal H}$~\cite{abou1973selfconsistent,biroli2010anderson,prado2026anderson}.
The LLT threshold $E_{+,\rm perc}$ is measured in the shifted spectrum. Therefore, since the Anderson spectrum is statistically symmetric around zero, we can compare the LLT prediction with the positive-energy mobility edge using the reflected value
\begin{equation}
\label{eq:Epercpositive}
    E_{\rm perc}(W,X)
    =
    -\left[E_{+,\rm perc}(W,X)+E_{\rm sh}(W,X)\right].
\end{equation}
In Sec.~\ref{subsec:dependence-shift} we present the phase diagram and analyze the effect of choosing different shifts, namely different values of $X$ in Eq.~\eqref{eq:Eshift}.

\subsection{Recursive derivation of the cavity equations}
\label{subsec:cavity-derivation}

We now derive the cavity equations for the Localization Landscape. Since the role of the spectral shift is only to translate the diagonal part of the Hamiltonian, it is useful to keep the reference energy $E_{\rm sh}$ arbitrary throughout the derivation. In this way, the same equations apply both to the standard LLT prescription, where $E_{\rm sh}=E_{\rm min}$, and to the shifted family introduced in Sec.~\ref{subsec:localization-landscape}, where $E_{\rm sh}=E_{\rm edge}-X$. In the following, we omit the explicit dependence of $E_{\rm sh}$ on $W$ and $X$.

The shifted Hamiltonian
\begin{equation}
    \hat{\mathcal H}_+
    =
    \hat{\mathcal H}
    -
    E_{\rm sh}\hat{\mathcal I}
\end{equation}
has the same functional form as the Hamiltonian in Eq.~\eqref{eq:ham}, except that the on-site energies are redefined as
\begin{equation}
\label{eq:eps-shifted}
    \varepsilon_i
    =
    \epsilon_i-E_{\rm sh}.
\end{equation}
Therefore, if the original on-site energies are drawn from the distribution $\gamma$ in Eq.~\eqref{eq:gammas}, the shifted variables are distributed according to
\begin{eqnarray}
\label{eq:gamma}
    \gamma_+(\varepsilon)
    &=&
    \gamma(\varepsilon+E_{\rm sh})
    \nonumber\\
    &=&
    \begin{cases}
        1/W \,\,\qquad &\text{if }\,\, \varepsilon \in[\varepsilon_{\rm min},\varepsilon_{\rm max}]
        \; ,\\
        0 \,\, \qquad &\text{else}
        \; ,
    \end{cases}
\end{eqnarray}
with
\begin{eqnarray}
[\varepsilon_{\rm min},\varepsilon_{\rm max}]
\equiv
\left[
-\frac{W}{2}-E_{\rm sh},
\frac{W}{2}-E_{\rm sh}
\right]
\label{eq:epsminmax}
\; .
\end{eqnarray}
Thus, the different choices of the LLT shift enter the cavity equations only through the shifted disorder distribution $\gamma_+$.

From the inversion of the Green's function definition,
\begin{equation}
    (\hat{\mathcal H}-z\hat{\mathcal I})\hat{\mathcal G}(z)
    =
    \hat{\mathcal I}
    \; ,
\end{equation}
one obtains
\begin{equation}
    \hat{\mathcal H}_+
    =
    \hat{\mathcal H}-E_{\rm sh}\hat{\mathcal I}
    =
    \hat{\mathcal G}^{-1}(E_{\rm sh}) 
    \; .
\end{equation}
Therefore, the Localization Landscape associated with the shifted Hamiltonian is
\begin{eqnarray}
\label{eq:ui}
\hat{\mathcal H}_+{\mathbf u}
&=&
{\mathbf 1} \quad  \Longleftrightarrow \quad
{\mathbf u}
=
\hat{\mathcal G}(E_{\rm sh}){\mathbf 1}
\end{eqnarray}
and, in components,  
\begin{eqnarray}
u_i
&=&
\sum_{j=1}^N
\mathcal G_{ij}(E_{\rm sh}) 
\; .
\end{eqnarray}
When $E_{\rm sh}$ lies at the lower edge of the continuous spectrum, or below it, the Density of States vanishes at this energy and the Green's functions are real. Equation~\eqref{eq:ui} can then be written in the form adopted in Ref.~\cite{Tonetti2026},
\begin{equation}
\label{eq:uiReG}
    u_i
    =
    \sum_{j=1}^N
    \mathrm{Re}\,
    \mathcal G_{ij}(E_{\rm sh}) 
    \; .
\end{equation}
Henceforth, in the derivation of the cavity equations, we omit the explicit $E_{\rm sh}$ dependence and the $\mathrm{Re}$ symbol.

Equation~\eqref{eq:uiReG} also provides a compact definition of  the analytical continuation discussed in Sec.~\ref{subsec:localization-landscape}. Namely, for the shifted family
\begin{equation}
    E_{\rm sh}=E_{\rm edge}-X
    \; ,
\end{equation}
we write

\begin{equation}
\label{eq:uacont}
      u_i
      =
      \sum_{j=1}^N
      \mathcal G_{ij}(E_{\rm edge}-X)
      \; ,
      \qquad X\geq 0 
      \; .
\end{equation}
For $W>W_{\min}$ and $X=0$, this coincides with the standard bulk-edge, and hence minimal, LLT prescription. For $W\le W_{\min}$, instead, Eq.~\eqref{eq:uacont} allows one to follow the bulk-edge branch below the point where the isolated eigenvalue separates from the continuous spectrum. In this regime, the usual positive-definite LLT prescription is recovered only when the reference energy is chosen at, or below, the isolated eigenvalue, while $0\leq X<X_{\rm iso}$ corresponds to the under-shifted analytical continuation.

This distinction will be important in Sec.~\ref{subsec:EisoWmin}. For the derivation below, however, no further modification is needed: the cavity equations have the same form for all choices of $E_{\rm sh}$, and the shift only fixes the support of the distribution $\gamma_+$.

In the End Matter of Ref.~\cite{Tonetti2026} we showed that
\begin{equation}
    \label{eq:etau}
    u_i = \mathcal{G}_{ii}\eta_i 
\end{equation}
where
\begin{eqnarray}
    \label{eq:g}
    \mathcal{G}_{ii}^{-1} & =  & \varepsilon_i - t^2 \sum_{k \in \partial i} \mathcal{G}_{k \to i} 
    \; , 
    \\
    \label{eq:eta}
    \eta_{i} &= &1 + t \sum_{k \in \partial i} \mathcal{G}_{k \to i} \eta_{k \to i}
    \; , \\
    \label{eq:cavg}
    \mathcal{G}_{k \to i}^{-1} & = & \varepsilon_k - t^2 \sum_{l \in \partial k \setminus i} \mathcal{G}_{l \to k}
     \;  , \\
    \label{eq:caveta}
    \eta_{k \to i}  &= & 1 + t \sum_{l \in \partial k \setminus i}
    \mathcal{G}_{l \to k} \eta_{l \to k}   \; .
\end{eqnarray}
We obtained this result by representing the Green's functions of the system as
\begin{equation}
    \label{eq:gGaussian}
    \mathcal{G}_{ij} =  \frac{1}{Z_0}\int {\cal D}{\mathbf x} \;  x_i x_j \; e^{-S_0[{\mathbf x}]}
    \equiv \langle x_i x_j\rangle_{0}
    \; ,
\end{equation}
where
\begin{eqnarray}
    S_J[{\mathbf x}]  &=& \frac{1}{2}  \sum_i \varepsilon_i x_i^2 - t \sum_{\langle i,j \rangle} x_i x_j -J\sum_j x_j 
    \nonumber\\
    &=&
    \frac{1}{2}\mathbf{x^t}{\mathcal H}_+\mathbf{x}-J\sum_j x_j
    \; ,
\\
    Z_J  &=& \int {\cal D} x \, e^{-S_J[{\mathbf x}]}
    \; ,
\end{eqnarray}
which is justified in the standard and over-shifted case by the fact that, $\mathcal{H}_+$ is a positive definite matrix, thus $e^{-S_0[{\mathbf x}]}/Z_0$ is a properly normalized Gaussian measure. More precisely, as already stated in Ref.~\cite{Tonetti2026}, $\mathcal{H}_+$ is a non-singular M-matrix, so its inverse $\mathcal G$ has non-negative components. In Eq.~\eqref{eq:gGaussian} the symbol $\langle \,\cdot \, \rangle_J$ indicates the expectation value defined by the Gaussian measure with action $S_J[\mathbf{x}]$.
The component $u_i$ is given by the Gaussian expectation
\begin{equation}
    u_i   = \sum_j \mathcal{G}_{ij} = \langle x_i \sum_j x_j \rangle_0 = \frac{\partial \langle x_i \rangle_J}{\partial J} \Bigg\vert_{J=0},
\end{equation}
parametrizing the on-site marginal distributions on the normal and cavity lattice as
\begin{eqnarray}
    \mu_i(x_i) & \propto & e^{-\frac{x_i^2}{2 \, \mathcal{G}_{ii}} + J\eta_i x_i} \,, 
    \\
    \mu_{k \to i}(x_k) &\propto & e^{-\frac{x_k^2}{2 \, \mathcal{G}_{k \to i}} + J\eta_{k \to i} x_k} \,,
\end{eqnarray}
and finding the self-consistent recursive equations~\eqref{eq:g}-\eqref{eq:caveta}.

To validate this result, and extend it to the under-shifted case, we compute $u_i$ recursively by 
using the standard cavity method on a tree, see Ref.~\cite{mezard2009information}. This derivation relies only on the properties of the Green's function, and it does not require $\hat{\mathcal{H}}_+$  to be positive definite. 

For $\alpha>0$, we define
\begin{equation}
\label{eq:complex-action}
    S^{(\alpha)}[\mathbf{x};z]
    =
    \frac{1}{2}
    \mathbf{x}^{t}
    \left[
    \alpha \hat{\mathcal I}
    -
    {\rm i}
    \left(
    \hat{\mathcal H}
    -
    z\hat{\mathcal I}
    \right)
    \right]
    \mathbf{x}
    \; .
\end{equation}
The real part of this quadratic form is positive for every $\alpha>0$, so the Gaussian integral is convergent. We denote the corresponding covariance by
\begin{equation}
\label{eq:Calpha}
    \mathcal C_{ij}^{(\alpha)}(z)
    =
    \frac{1}{Z^{(\alpha)}}
    \int \mathcal D\mathbf{x}
    \;
    x_i x_j
    \, 
    e^{-S^{(\alpha)}[\mathbf{x};z]} 
    \; .
\end{equation}
Equivalently,
\begin{equation}
\label{eq:Cmatrix}
    \hat{\mathcal C}^{(\alpha)}(z)
    =
    \left[
    \alpha \hat{\mathcal I}
    -
    {\rm i}
    \left(
    \hat{\mathcal H}
    -
    z\hat{\mathcal I}
    \right)
    \right]^{-1}
    \; .
\end{equation}
Hence, whenever $z$ is outside the spectrum,
\begin{equation}
\label{eq:CtoG}
    \lim_{\alpha\to0^+}
    \hat{\mathcal C}^{(\alpha)}(z)
    =
    {\rm i}\hat{\mathcal G}(z)
    \; ,
    \qquad
    \hat{\mathcal G}(z)
    =
    \left(
    \hat{\mathcal H}
    -
    z\hat{\mathcal I}
    \right)^{-1}.
\end{equation}
Thus, 
\begin{equation}
    \mathcal G_{ij}(z)
    =
    -{\rm i}
    \lim_{\alpha\to0^+}
    \mathcal C_{ij}^{(\alpha)}(z)
    \; .
\end{equation}

The $i$-th component of the Localization Landscape can be expressed as
\begin{equation}
\label{eq:ubari}
    u_i = \sum_j \mathcal{G}_{ij} = \mathcal{G}_{ii}+\sum_{k\in \partial i}\sum_{j\in B_{k\rightarrow i}}\mathcal{G}_{ij},
\end{equation}
where $B_{k \rightarrow i} \equiv \left\{ j \,\middle|\, j \right.$ 
belongs to the cavity lattice rooted at  $k\in \partial i$ after the removal of site  $\left. i \right\}$.
The second sum in the second term can be rewritten as
\begin{equation}
\label{eq:sumBranch}
    \sum_{j\in B_{k\rightarrow i}}\mathcal{G}_{ij} = \mathcal{G}_{ik}+\sum_{j\in B_{k\rightarrow i}\setminus k}\mathcal{G}_{ij}.
\end{equation}

Take two sites on the Bethe lattice, say $0$ and $r$, connected by a non-intersecting path of length $r$.
After integrating out all variables outside this path, the remaining one-site cavity marginals have the Gaussian form
\begin{eqnarray}
\label{eq:complex-marginal-normal}
    \mu_i^{(\alpha)}(x_i)
    &\propto&
    e^{
    -\frac{x_i^2}{2\mathcal C_{ii}^{(\alpha)}}},
    \\
\label{eq:complex-marginal-cavity}
    \mu_{k\to i}^{(\alpha)}(x_k)
    &\propto&
    e^{
    -\frac{x_k^2}{2{\mathcal C}_{k\to i}^{(\alpha)}}
   },
\end{eqnarray}
where $\mathcal C_{k\to i}^{(\alpha)}$ is the cavity covariance on the branch rooted at $k$ after removing site $i$. 

Using the action~\eqref{eq:complex-action}, the coupling between two neighboring sites on the path contributes the factor
\begin{equation}
    \exp\left[-{\rm i} t x_s x_{s+1}\right].
\end{equation}
Therefore, the regularized covariance between the endpoints can be written as
\begin{eqnarray}
\mathcal C_{0r}^{(\alpha)}
&=&
\int
\prod_{s=0}^{r-1}
\left[
dx_s\,\mu_{s\to s+1}^{(\alpha)}(x_s)
\right]
dx_r\,\mu_r^{(\alpha)}(x_r)
\;
x_0x_r
\nonumber\\
&&
\qquad\qquad\times
\exp\left[
- {\rm i} t
\sum_{s=0}^{r-1}
x_sx_{s+1}
\right].
\label{eq:Cpath-integral}
\end{eqnarray}
The first integration gives
\begin{eqnarray}
&&
\int dx_0\,
\mu_{0\to 1}^{(\alpha)}(x_0)\,
x_0\,
e^{- {\rm i} tx_0x_1}
\nonumber\\
&&
\qquad
=
- {\rm i} t\,\mathcal C_{0\to 1}^{(\alpha)}\,x_1\,
\exp\left[
-\frac{t^2}{2}
\mathcal C_{0\to 1}^{(\alpha)}
x_1^2
\right].
\label{eq:first-complex-integration}
\end{eqnarray}
The exponential factor produced by this integration is precisely the Gaussian factor that renormalizes the marginal of the next site. Repeating the same step along the path, one obtains
\begin{equation}
\label{eq:C-chain-rule}
    \mathcal C_{0r}^{(\alpha)}
    =
    \mathcal C_{rr}^{(\alpha)}
    \prod_{s=0}^{r-1}
    \left(
    - {\rm i} t\,\mathcal C_{s\to s+1}^{(\alpha)}
    \right).
\end{equation}
Using Eq.~\eqref{eq:CtoG} and taking $\alpha\to0^+$, this becomes
\begin{equation}
\label{eq:chain-rule}
    \mathcal G_{0r}
    =
    \mathcal G_{rr}
    \prod_{s=0}^{r-1}
    \left(
    t\,\mathcal G_{s\to s+1}
    \right).
\end{equation}
Analogously, starting the integration from the endpoint $r$ gives the equivalent form
\begin{equation}
\label{eq:chain1}
    \mathcal G_{0r}
    =
    \mathcal G_{00}
    \prod_{s=1}^{r}
    \left(
    t\,\mathcal G_{s\to s-1}
    \right)
    =
    \mathcal G_{rr}
    \prod_{s=0}^{r-1}
    \left(
    t\,\mathcal G_{s\to s+1}
    \right).
\end{equation}

Here, the cavity Green's functions in the products are the ones connecting the sites along the path going from site $r$ to site $0$ (in the first equality) or vice versa (in the second). Note that the recursive equations defining these normal and cavity Green's functions are the same as those for the real part of the normal and cavity Green's functions in the Anderson localization context, after substituting $E=E_{\rm sh}$ [Eqs.~\eqref{eq:gAL} and \eqref{eq:gcavAL}].

Inserting Eq.~(\ref{eq:chain1}) in Eq.~(\ref{eq:sumBranch}) we obtain
\begin{eqnarray}
    \mathcal{G}_{ij} &= & \mathcal{G}_{jj}\prod_{(s,s')\in \rho_{ij}}t \, \mathcal{G}_{s \rightarrow s'}
    \nonumber\\
    &=&
    t \, \mathcal{G}_{i \rightarrow k}\underbrace{\mathcal{G}_{jj}
    \prod_{(s,s')\in \rho_{kj}}t \, \mathcal{G}_{s \rightarrow s'}}_{\mathcal{G}_{kj}}=t \, \mathcal{G}_{i \rightarrow k}\mathcal{G}_{kj}
    \; ,
\end{eqnarray}
where $\rho_{ij}$ represents the directed path connecting site $i$ and site $j$.
Plugging this expression and $\mathcal{G}_{ik} = t \, \mathcal{G}_{i \rightarrow k}\mathcal{G}_{kk}$ in Eq.~(\ref{eq:ubari}) we immediately obtain
\begin{eqnarray}
\label{eq:u}
u_i &=& \mathcal{G}_{ii} + t \sum_{k \in \partial i} \mathcal{G}_{i \rightarrow k} u_{k \rightarrow i} \; ,
\\
\label{eq:uktoi}
u_{k \rightarrow i} &=& \mathcal{G}_{kk} + t \sum_{l \in \partial k \setminus i} \mathcal{G}_{k \rightarrow l} u_{l \rightarrow k} \;  ,
\end{eqnarray}
where we have defined
\begin{equation}
    u_{k \rightarrow i}\equiv \sum_{j\in B_{k\rightarrow i}}\mathcal{G}_{kj}
    \; .
\end{equation}

These equations are equivalent to Eqs.~(\ref{eq:eta}) and (\ref{eq:caveta}). In fact, if we replace
\begin{equation}
    u_{k\to i}=\mathcal{G}_{kk}\eta_{k\to i}
\end{equation}
in Eqs.~(\ref{eq:eta}) and  (\ref{eq:caveta}), we recover the recursive Eqs.~(\ref{eq:u}) and (\ref{eq:uktoi}):
\begin{eqnarray}
    u_i &= & \mathcal{G}_{ii}\eta_{i}= \mathcal{G}_{ii}+t\sum_{k \in \partial i}\mathcal{G}_{ii} \eta_{k \to i}  \mathcal{G}_{k \to i}
    \nonumber\\
    &=& 
    \mathcal{G}_{ii}+t\sum_{k\in \partial i}\mathcal{G}_{i \rightarrow k}u_{k \rightarrow i}
    \,,\\
    u_{k \to i} &=& \mathcal{G}_{kk}\eta_{k \to i}= \mathcal{G}_{kk}+t\sum_{l \in \partial k \setminus i}\mathcal{G}_{kk} \eta_{l \to k}  \mathcal{G}_{l \to k}
    \nonumber\\
    &=& 
    \mathcal{G}_{kk}+t\sum_{l\in \partial k \setminus i}\mathcal{G}_{k \rightarrow l}u_{l \rightarrow k}\,.
\end{eqnarray}
The second equalities in the two equations follow from Eqs.~(\ref{eq:chain1}). Indeed,
\begin{eqnarray}
    \mathcal{G}_{kk} \, t \, \mathcal{G}_{l \to k} \, \eta_{l \to k} &=& \mathcal{G}_{kl} \,  \eta_{l \to k}=
    \mathcal{G}_{ll} \, t \, \mathcal{G}_{k \to l} \, \eta_{l \to k}
    \nonumber\\
    &=& t \, \mathcal{G}_{k \to l} \, u_{l \to k}\,.
\end{eqnarray}
Conversely, it is easy to check that by defining $\eta_i =u_i/\mathcal{G}_{ii}$ and $\eta_{k\to i} =u_{k\to i}/\mathcal{G}_{kk}$, Eqs.~(\ref{eq:u}) and (\ref{eq:uktoi}) 
boil down to Eqs.~(\ref{eq:eta}) and (\ref{eq:caveta}).

As we can see, Eqs.~(\ref{eq:u}) and (\ref{eq:uktoi}) depend both on the normal and the cavity Green's functions. Therefore, Eqs.~(\ref{eq:g}), (\ref{eq:cavg}), (\ref{eq:u}) and (\ref{eq:uktoi}) form a set of coupled recursive equations that can be solved self-consistently. However, for computing the percolation critical curve it is more convenient to work with the set of equations (\ref{eq:g})-(\ref{eq:caveta}).

\subsection{The percolation probability}
\label{sec:EQP}

In order to study the percolation problem for the Hamiltonian $\hat{\cal{H}}_+$ at energy $E_+$, one has to consider the standard cavity equation for the random site percolation on the Bethe lattice \cite{stauffer2018introduction}, where the probability of a node being occupied is replaced by the condition $u_i \geq 1/E_+$. 
In this setting, we define
\begin{equation}
    p_i\equiv \Pr\{\text{site $i$ belong to the giant cluster}\}.
\end{equation}

Since the value of the Localization Landscape $u_i$ at site $i$ is correlated to the ones of its nearest neighboring sites, 
$u_{k \in \partial i}$,
the probability of a site being occupied is site dependent. 

The recursive equations read
\begin{align}
\label{eq:pi}
    p_{i} &= \theta(u_i-1/E_+) \Big[ 1 - \prod_{k \in \partial i } (1 -  p_{k \to i}) \Big] , 
   \\
    \label{eq:cavpi}    
    p_{k\to i} &= \theta(u_k-1/E_+) \Big[ 1 - \prod_{l \in \partial k \setminus i } (1 -  p_{l \to k}) \Big] .
\end{align}
From now on we will refer to the $p_i$'s as the percolation probabilities.
An important observation is that $\{p_{k \to i}\}_{k\in \partial i}$ in Eqs.~(\ref{eq:pi})-(\ref{eq:cavpi}) are not statistically independent, 
because for each of them we need to compute the $\theta(u_k-1/E_+)$'s, which also depend 
on quantities evaluated at site $i$. For this reason, it is useful to switch to a different set of cavity variables that we define as
\begin{eqnarray}
    \bar p_{k \to i} \! & \! \equiv \! & \! \Pr\{\text{cavity site $k$ (in absence of $i$) belong}
   \nonumber  \\
     && \;\; \quad {\text{  to the giant cluster if occupied}\} } \,,
\end{eqnarray}
thus:
\begin{equation}
   p_{k \to i} =\theta(u_k-1/E_+)\,\bar p_{k \to i}\,.
\end{equation}
In terms of these variables, Eqs.~(\ref{eq:pi})-(\ref{eq:cavpi}) become 
\begin{align}
    \label{eq:pi2}
    p_{i} &= \theta(u_i-1/E_+) 
    \nonumber\\
    & \;\;\quad  \times \, \Big[ 1 - \prod_{k \in \partial i } \Big(1 - \theta(u_k-1/E_+)\bar p_{k \to i} \Big ) \Big]
    \; , \\
    \label{eq:cavpbari}
    \bar p_{k \to i} &=1 - \prod_{l \in \partial k \setminus i} \Big(1 - \theta (u_l -1/E_+)\bar p_{l \to k} \Big)
    \; .
\end{align}
Since in the non-percolating phase all these probabilities must be exactly zero, and the critical percolation
transition is expected to be continuous,  close to the percolation critical curve we can 
linearize the second equation above. Under this assumption it becomes
\begin{eqnarray}
\label{eq:pcavlin}
    \bar p_{k \to i} = \sum_{l \in \partial k \setminus i} \theta (u_l -1/E_+)\bar p_{l \to k} \, .
\end{eqnarray}
Here $u_l=\mathcal{G}_{ll}\eta_l$, see Eq.~(\ref{eq:etau}), and it can be computed in terms of independent cavity variables once we realize that
\begin{eqnarray}
    \mathcal{G}_{ll}^{-1} &=& \mathcal{G}_{l \to k}^{-1} - t^2\mathcal{G}_{k \to l} 
    \; , 
   \\
    \eta_l &=& \eta_{l \to k} + t\mathcal{G}_{k \to l} \eta_{k \to l} 
    \; ,
\end{eqnarray}
and 
\begin{eqnarray}
    \mathcal{G}_{k \to l}^{-1} &=& \varepsilon_k - t^2\sum_{m \in \partial k \setminus l} \mathcal{G}_{m \to k} \; , 
    \\
    \eta_{k \to l}  &=& 1 +t \sum_{m \in \partial k \setminus l} \mathcal{G}_{m \to k} \eta_{m \to k} \; , 
\end{eqnarray}
which follow from Eqs.~(\ref{eq:g})-(\ref{eq:caveta}).
The final expression for the Localization Landscape for a site $l\in \partial k\setminus i$ is 
\begin{align}
        \label{eq:uk}
        u_l &= \left( \mathcal{G}_{l \to k}^{-1}-\frac{t^2}{\varepsilon_k-t^2\sum\limits_{m \in \partial_k \setminus l} \mathcal{G}_{m \to k}} \right)^{-1} \nonumber\\
        &\qquad  \times \left(\eta_{l \to k}+t \; \frac{1+t\sum\limits_{m\in \partial_k \setminus l}\mathcal{G}_{m \to k}\eta_{m\to k}}{\varepsilon_k-t^2\sum\limits_{m \in \partial_k \setminus l} \mathcal{G}_{m \to k}}\right) 
        \nonumber\\
        & \equiv U_l(\{\mathcal{G}_{l\to k}, \eta_{l\to k}\}_{l \in \partial k}).
\end{align}

\subsection{Self-consistent distributional equation}
\label{sec:SCP}

We can now exploit the statistical independence of the cavity variables by observing that, after averaging over disorder, the Hamiltonian becomes translational invariant. Therefore, the joint probability distribution of on-site variables must be the same on every site (and the same must hold for the cavity variables). This means that we can write a recursive equation that relates the joint probability distribution of $\mathcal{G}_{k\to i}, \eta_{k\to i},\bar p_{k \to i} $ on a cavity site, to independent copies of itself. In other words, $\mathcal{G}_{k\to i}, \eta_{k\to i},\bar p_{k \to i} $ are written in terms of independent sets of cavity variables $\{\mathcal{G}_{l\to k}, \eta_{l\to k},\bar p_{l \to k}\}_{l \in \partial k}$ drawn from the same probability distribution $P(g,\eta,\bar p)$.
Because of translational invariance $P(g,\eta,\bar p)$ must also be the probability distribution of $\mathcal{G}_{k\to i}, \eta_{k\to i},\bar p_{k \to i} $ themselves. Therefore, using all the cavity equations which define the cavity variables, i.e. Eqs.~(\ref{eq:eta}), (\ref{eq:cavg}),  (\ref{eq:pcavlin}) and (\ref{eq:uk}), we can write the self-consistent distributional equation 
\begin{align}
\label{eq:stochSC}
    & P(g,\eta,\bar p) = \int d\varepsilon\, \gamma_+ (\varepsilon)
    \prod_{l=1}^{K+1} \left[dg_ld\eta_l d\bar p_l  \,P( g_l,\eta_l,\bar p_l ) \right] 
      \nonumber\\
    & 
  \qquad \times  \delta \bigg(\bar p-\sum\limits_{l=1}^K\theta\Big(U_l\big(\{g_l,\eta_l\}_{l\leq K+1}\big)-1/E_+\Big)\bar p_l \bigg)
    \nonumber\\
    & \qquad \times 
    \delta \bigg ( g-\frac{1}{\varepsilon-t^2\sum\limits_{l=1}^K g_l}\bigg ) 
    \nonumber\\
    &\qquad \times \delta \bigg (  \eta- 1-t\sum\limits_{l=1}^K g_l \eta_l \bigg ) 
  .
\end{align}
This equation holds close to criticality.  If one wants to obtain the distribution for a generic pair $(E_+,W)$, the equation must be solved exchanging the delta function enforcing Eq.~(\ref{eq:pcavlin}) with a delta enforcing the full relation in 
Eq.~(\ref{eq:cavpbari}). 

The critical curve of this percolation problem is given by the points in the $(E_+,W)$-plane where $\bar p_{k \to i}$ passes from having zero to non-zero expectation value.

Equation~(\ref{eq:stochSC}) can be solved by using population dynamics, which is  a standard numerical technique used to solve self-consistent distributional equations, 
and is explained in Sec~\ref{subsec:population}.

\subsection{Correlation function and average cluster size}
\label{sec:CFP}

The correlation function $C_{\rm perc}(r)$ for a percolation problem  is defined 
as the probability that two sites at distance $r$ belong to the same cluster. 
Denoting as $0,\dots,r$ the sites on the lattice along a non-intersecting path of length $r$,  we have
\begin{equation}
\label{eq:corrfuncnobond}
     C_{\rm perc}(r) = \Pr \{O_0=1, \dots, O_{r}=1\}=\mathbb{E}\bigg[\prod_{i=0}^r O_i\bigg]
     \; , 
\end{equation}
with the occupation variables $O_i$ defined as
\begin{equation}
    \label{eq:Oiperc}
    O_{i} = \begin{cases}
        1 & \text{if } u_{i} \geq 1/E_+ \; ,  \\
        0 & \text{otherwise} \; 
    \end{cases} =\theta(u_i -1/E_+)
    \; . 
\end{equation}

Since the Bethe lattice is statistically translational invariant, 
the occupation probability is site-independent. Accordingly, we can define
\begin{equation}
\label{eq:q}
    q \equiv \Pr\{O_i=1\} = \Pr \{u_i\geq 1/E_+\}
    \; . 
\end{equation}
Since the $O_i$ depend only on the set of independent cavity quantities $\big\{(\mathcal{G}_{k\to i},\eta_{k \to i})\,|\,i\in\{0,\dots, r\}\,,k\in \partial i \cap\partial \{0,\dots, r\}\}$, given the joint probability distribution $P(g,\eta,\bar p)$ we can compute $C_{\rm perc}(r)$. The expected scaling of the correlation function for large distances is
\begin{equation} \label{eq:Cperc}
    C_{\rm perc}(r)\sim \frac{e^{-r/\xi_{\rm perc}}}{K^r}\,.
\end{equation}
The factor $K^{-r}$ is standard on tree-like geometries and compensates for the exponential growth, $\sim K^r$, of the number of sites located at distance $r$ from a given node. Indeed, the volume integral of $C_{\rm perc}(r)$ -- obtained by summing $C_{\rm perc}(r)$ over all sites at distance $r$ -- is directly related to the mean cluster size $S$ (see Eq.~\eqref{eq:avgclsize} below), which remains finite in the non-percolating phase and diverges at the critical point. Since the number of sites at distance $r$ grows as $K^r$, convergence of this sum requires the correlation function to decay at least as fast as $K^{-r}$. This behavior is precisely what emerges from the solution of the recursive equations for the percolation problem. To define a meaningful correlation length, it is therefore convenient to absorb this purely geometric factor into the definition of $C_{\rm perc}(r)$, as done in Eq.~\eqref{eq:Cperc}. With this convention, the characteristic length scale $\xi_{\rm perc}$ becomes critical at the transition and diverges with the standard mean-field exponent $\nu=1$. An analogous definition is commonly adopted in the context of Anderson localization as well (see below). 
The procedure for the numerical computation of $C_{\rm perc}(r)$ and $\xi_{\rm perc}$ is explained in Sec.~\ref{app:numeval}. 

The average cluster size $S$ represents the average size of the connected component to which a generic occupied site belongs. This can be computed directly from the correlation function, by observing that the average number of sites at distance $r$ belonging to the same cluster as site $i$ is $(K+1)K^{r-1}C_{\rm perc}(r)/q$ (the factor $1/q$ is needed to condition the probability on the occupation of the reference site). Therefore, the average number of sites belonging to the same cluster of site $i$, regardless of their distance, is 
just
\begin{equation}
\label{eq:avgclsize}
    S=1+\sum_{r=1}^\infty (K+1)K^{r-1}\frac{C_{\rm perc}(r)}{q}
    \; . 
\end{equation}
$q$ can be computed numerically from $P(g,\eta)$ as explained in Sec.~\ref{subsubsec:qqbar}.

At criticality, the correlation length $\xi_{\rm perc}$ diverges, together with the average cluster size $S$. 
Using Eqs.~\eqref{eq:Cperc} and~\eqref{eq:avgclsize}, one finds asymptotically, to leading order close to the percolation threshold,
\begin{equation}
S \propto \left(1 - e^{-1/\xi_{\rm perc}}\right)^{-1}
\simeq \xi_{\rm perc}\, .
\end{equation}
Therefore, in order to determine the position of critical curve for the Localization Landscape percolation problem [black line in Fig.\ref{fig:phase-diagram}], we evaluate numerically the pairs $(E_+,W)$ at which $\xi_{\rm perc}\to +\infty$.

\section{Results}
\label{sec:results}

This section is organized as follows. In Sec.~\ref{subsec:EisoWmin}, we discuss the properties of the Localization Landscape variables below $W_{\min}$, namely the LLT prediction for the position of the isolated eigenvalue and its relation to the onset of localization. Moreover, we show how the Aizenman-Warzel disorder bound~\cite{warzel2012} can be derived 
from the analytic continuation of the LLT cavity equations.
 In Sec.~\ref{subsec:dependence-shift}, we present the phase diagram obtained from the generic-connectivity equations for $K=2$ and analyze how the percolation critical energy depends on the choice of the shift. Finally, in Sec.~\ref{subsec:density-of-states}, we compute the Density of States predicted by the LLT-improved Weyl approximation~\cite{arnold2019computing,david2021landscape,Fefferman1983theUP} and compare it with the known Density of States of the Anderson model on this geometry.

\subsection{Isolated eigenvalue, Localization Landscape and lower bounds for critical disorder}
\label{subsec:EisoWmin}

An important result of the LLT percolation framework is that it lets us understand two features of the phase diagram of the AM on the Bethe lattice that have remained somewhat mysterious so far. First, it  suggests that the localized phase of the Anderson model should start to exist when the isolated eigenvalue $E_{\rm iso}(W)$ enters the bulk of the spectrum. In fact, as we show in Sec.~\ref{subsubsec:lowW}, decreasing the disorder strength, the expected value of the Localization Landscape variables $u_i$ diverges exactly at the value $W_{\min}$ where
\begin{equation}
\label{eq:EisoWmin}
    E_{\rm iso}(W_{\min})=-2t \sqrt K -\frac{W_{\rm min}}{2}\,,
\end{equation}
indicating that a typical realization of $u_i$ at $W=W_{\min}$ corresponds to a vanishing effective potential. Therefore, the system is in the percolating phase for any value of the energy. 

The high-connectivity solution (see Sec.~\ref{subsubsec:lowW}) provides estimates of $E_{\rm iso}$ and $W_{\rm min}$ that 

are very accurate also for $K=2$ [Eqs.~\eqref{eq:Eiso} and \eqref{eq:bound}]:
\begin{eqnarray}
     E_{\rm iso}(W)&=& -t-\frac{W}{2}\coth\left(\frac{W}{2Kt}\right)\,,\\
     W_{\rm\min}&=&Kt \ln\left| 1+\frac{W_{\rm min}}{t(2\sqrt K-1)} \right|\,.\label{eq:Wmin-HC}
\end{eqnarray}
For $K=2$ and $t=1$, the second equation gives $W_{\min}\simeq0.354$. The exact cavity solution, obtained by locating the disorder at which the average of the $y$-variables of Sec.~\ref{subsubsec:lowW} stops growing for $E_{\rm sh}=E_{\rm edge}$, gives $W_{\min}\simeq0.357$. Similarly, for $W=0.1$ and $W=0.25$ the high-connectivity and exact values of $E_{\rm iso}$ differ by less than $10^{-3}$ (see also Fig.~\ref{fig:finiteN-u}).

The value of $W_{\min}$ is also constrained by an exact bound. Since $E_{\rm iso}(W)\le-t(K+1)$ [Eq.~\eqref{eq:Eu-lowerbound}], the condition $E_{\rm iso}(W_{\min})=E_{\rm edge}(W_{\min})$ implies $2t\sqrt K+W_{\min}/2\ge t(K+1)$, i.e.,
\begin{equation}
\label{eq:Wmin-bound}
W_{\min}\ge2t(\sqrt K-1)^2\,,
\end{equation}
which gives $W_{\min}\ge0.343$ for $K=2$ and $t=1$, only about $4\%$ below the actual value. This is twice the Aizenman--Warzel threshold $W_{\rm AW}$ discussed below. The bound has a simple meaning: $2t(\sqrt K-1)^2$ is the disorder at which the edge of the continuous spectrum, $2t\sqrt K+W/2$, reaches the energy $t(K+1)$, i.e., the modulus of the isolated eigenvalue of the clean lattice. This energy also plays a special role for Anderson localization at weak disorder. Aizenman and Warzel showed that extended states appear well beyond the clean band, in the Lifshitz-tail regime, and that, as $W\to0$, localization is only proven at energies that approach $|E|=t(K+1)$~\cite{aizenman2011extended}. This suggests that localized states at weak disorder appear only once the band edge exceeds $t(K+1)$, i.e., for $W>2t(\sqrt K-1)^2$, close to $W_{\min}$.

We cannot, however, determine the onset of Anderson localization directly. As discussed in Sec.~\ref{sec:mobility}, reliable extrapolations of the mobility edge are only possible for $W\ge0.5$. Our data are compatible with an onset close to $W_{\min}$: at $W=0.5$, the mobility edge is already very close to $t(K+1)$ [see Fig.~\ref{fig:PDshift}(b)]. They do not, however, establish that the two scales coincide, which therefore remains a conjecture supported by the LLT.

At large connectivity, however, the question can be settled analytically. In Sec.~\ref{subsubsec:largeK-loc} we show that the mobility edge obtained from the stability criterion of Sec.~\ref{sec:ALeigenval} is given, close to the band edge, by the same equation as the isolated eigenvalue, Eq.~\eqref{eq:Eiso}, continued to $W>W_{\min}$. Within this approximation, the mobility edge therefore emerges from the edge of the spectrum exactly at $W_{\min}$, at the same point as the LLT percolation line.

For $W\le W_{\rm min}$ it is useful to study the statistical properties of the analytical continuation of the Localization Landscape, obtained by generalizing its definition as follows.
Using the notation introduced in Sec.~\ref{subsec:localization-landscape}, we consider the family of reference energies
\begin{equation}
    E_{\rm sh}
    =
    E_{\rm edge}-X = -2t\sqrt K -W/2 -X
    \; ,
    \quad X\geq 0
    \;,
\end{equation}
and define
\begin{equation}
    u_i = \sum_{j=1}^N \mathcal{G}_{ij}(E_{\rm sh})
    \;.
\end{equation}
The case $E_{\rm sh}=E_{\rm iso}$ corresponds to the standard LLT prescription, while the cases $E_{\rm sh}<E_{\rm iso}$ and $E_{\rm iso}< E_{\rm sh}<E_{\rm edge}$ correspond to the over-shifted and under-shifted regimes respectively.

With this definition, a trivial extension of the high-connectivity solution of Sec.~\ref{subsubsec:lowW}, predicts that the expected value of $u_i$ diverges at $E_{\rm sh}=E_{\rm iso}$. Therefore,
\begin{equation}
    \mu_u\equiv \mathbb{E}[u_i]=+\infty \iff E_{\rm sh}\ge E_{\rm iso}\,,
\end{equation}
so the under-shifted regime has $\mu_u=+\infty$.

Aizenman and Warzel proved that, for bounded disorder, the spectrum of the Anderson model on the Bethe lattice is purely absolutely continuous in a neighborhood of the band edges if the disorder is weak enough~\cite{aizenman2011absence,warzel2012}. In their notation, the Hamiltonian is $T+\lambda V$, with $T$ the adjacency matrix and $V_i$ uniformly distributed in $[-1,1]$, i.e., $\lambda=W/(2t)$ in our units, and the condition reads $\lambda<(\sqrt K-1)^2/2$. In our convention, this gives
\begin{equation}
\label{eq:WAW}
    W<W_{\rm AW}\equiv t(\sqrt K-1)^2\,.
\end{equation}
Hence, for $W<W_{\rm AW}$, the states at the spectral edges are delocalized, and localized states cannot appear there. Aizenman and Warzel also conjectured that, at weak enough disorder, localization does not occur anywhere in the spectrum~\cite{aizenman2011absence,warzel2012}. The same threshold is obtained from the analytic continuation of the LLT cavity equations, as we now show.
In Sec.~\ref{subsubsec:utails}, we show that the effective potential 
variables $ v_i\equiv 1/ u_i$ concentrate on the deterministic value
\begin{eqnarray}
    v_i=\mathbb{E}[v_i]=0\;,\forall i
    &\iff&
    W+X\le X_c \equiv t(\sqrt K -1)^2 \nonumber\\ &\iff& E_{\rm sh}\ge  \frac{W}{2}-t(K+1)\,.
\end{eqnarray}
This means that, for
\begin{equation}
    W\le W_{\rm AW}(X) \equiv t(\sqrt{K}-1)^2-X\,,
\end{equation}
 the effective potentials vanish 
identically, so that $E_+ > v_i$ holds with probability one and the system is 
necessarily in the percolating (delocalized) phase. 
For $X=0$, this threshold coincides with the Aizenman--Warzel bound of Eq.~\eqref{eq:WAW}, while for $X>0$ it is smaller.

This result requires careful consideration.
For $W\le W_{\rm AW}<W_{\min}$ and $X=0$, the reference energy lies above the isolated eigenvalue, i.e., we are in the under-shifted regime. There, $\hat{\mathcal H}_+$ is not positive definite, the distribution of the Localization Landscape variables does not converge to a stationary solution, and $1/u_i$ does not have its usual interpretation as an effective potential. The result is therefore a property of the analytic continuation of the LLT cavity equations. More precisely, it is a property of the deterministic recursions that fix the edges of the support of the distributions of the cavity variables, which remain well defined in this regime. It is not a prediction of the standard, positive-definite LLT: for $E_{\rm sh}\le E_{\rm iso}$, the bound $E_{\rm iso}\le-t(K+1)$ of Eq.~\eqref{eq:Eu-lowerbound} implies $X\ge X_c-W/2$, hence $W+X>X_c$ for any $W>0$, and the effective potentials never vanish identically.

The reason why the analytically continued equations reproduce the Aizenman--Warzel bound can be understood by comparing the two derivations. The lower edge of the support of $P_u$ is determined by the extreme configuration in which all the on-site energies take their maximal value, $\varepsilon_i=\varepsilon_{\max}=2t\sqrt K+W+X$ (Sec.~\ref{subsubsec:gtails}). In this configuration all the cavity Green's functions are equal to $m_g$, the cavity Green's function of the clean tree at energy $-\varepsilon_{\max}$, and the Green's function decays along a path of length $r$ as $(tm_g)^r$. Since the number of paths of length $r$ grows as $K^r$, $u_i$ diverges if and only if $Ktm_g\ge1$, i.e., if the decay rate of the Green's function, $-\ln(tm_g)$, does not exceed $\ln K$. The proof of Aizenman and Warzel relies on the same two ingredients~\cite{aizenman2011absence}. First, the spectrum is absolutely continuous at the energies where the Lyapunov exponent $L(E)$, which measures the decay rate of the Green's function along a path, is smaller than $\ln K$. Second, at the band edge $E_\lambda$, the Lyapunov exponent of the disordered tree is bounded by that of the clean tree at an energy shifted by the full width of the disorder distribution, $L_\lambda(E_\lambda)\le L_0(E_0-2\lambda)$, with $E_0=-2\sqrt K$. In our units, this is the clean tree at energy $-2t\sqrt K-W$, i.e., the extreme configuration above for $X=0$. In both cases, the condition reads $2t\sqrt K+W<t(K+1)$, i.e., $W<X_c=W_{\rm AW}$. The LLT cavity equations therefore recover $W_{\rm AW}$ as a worst-case bound controlled by the extreme realizations of the disorder, rather than through the typical properties of the landscape. We also note that this result follows from the exact edges of the supports, and does not rely on the high-connectivity approximation.

The difference between the threshold $W_{\rm AW}$ and the minimal disorder 
$W_{\rm min}$  can be understood by 
considering what happens as $W$ is decreased from above $W_{\rm min}$. Two distinct 
transitions occur.
At $W = W_{\rm min}$, the expectation value of the cavity Green's functions reaches 
$1/(Kt)$, causing $\mu_u \to +\infty$. A typical $u_i$ diverges, 
driving the system into the percolating phase, even though locally 
finite values of $u_i$ remain possible with nonzero probability.
At $W = W_{\rm AW}$, the lower edge of the support of $P_{u}$ 
diverges as well. Below this threshold, no finite $u_i$ is realizable, 
not even with exponentially small probability, and the percolating phase becomes the only possibility.
Thus, $W_{\rm min}$ and $W_{\rm AW}$ mark two successive offsets of percolation, distinguished by whether the divergence of $ u_i$ is 
typical or global.

\begin{figure}
    \centering
    \includegraphics[width=0.50\textwidth]{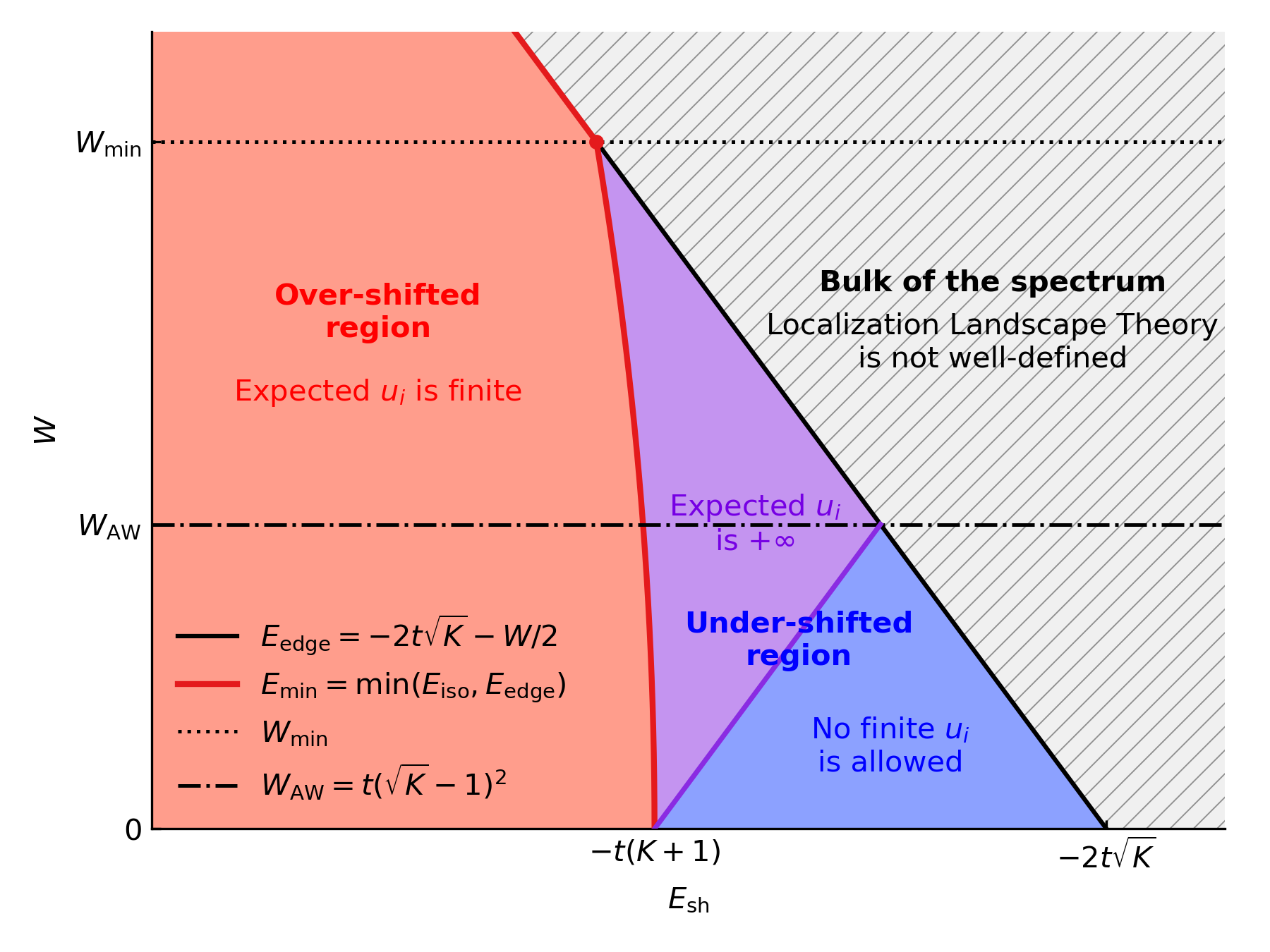}
    \caption{\small{Shifting regimes. The statistical properties of the Localization Landscape depend uniquely on the value of the reference energy $E_{\rm sh}$ and the disorder strength $W$. Red region: over-shifted regime, the expectation value of the Localization Landscape components $u_i$ is always finite. Purple region: under-shifted regime with finite lower bound. Here the system is always in the percolating phase, but atypical finite components $u_i$ can exist. Blue region: under-shifted regime with diverging lower bound. No finite $u_i$ solution is possible; therefore, the effective potential is identically equal to zero on the whole lattice.}}
    \label{fig:shift-regions}
\end{figure}

The different phases and bounds for the disorder presented in this Section are sketched in Fig.~\ref{fig:shift-regions}. 

\subsection{Phase diagram and optimal shift choice}
\label{subsec:dependence-shift}

In this Section we present the LLT percolation critical curves obtained for different choices of the spectral shift. 
With the same convention for $E_{\rm sh}$ used in the previous Section, the LLT percolation curve for $X=0$ starts at $W=W_{\min}$, where the isolated eigenvalue merges with the lower edge of the continuous spectrum, see Sec.~\ref{subsec:EisoWmin}.
For each value of $X$ we solved the cavity equations derived in Sec.~\ref{subsec:cavity-derivation}
and determined the corresponding percolation threshold $E_{\rm perc}(W,X)$. 

We compared the bulk-edge family with shifts defined relative to the isolated eigenvalue, namely with reference energies of the form $E_{\rm sh}=E_{\rm iso}-X$ for $W\le W_{\min}$. The resulting critical curves are only weakly affected by this choice. The reason is that, close to $W_{\min}$, the isolated eigenvalue is still very close to the lower edge of the continuous spectrum,
    $E_{\rm iso}(W)\simeq E_{\rm edge}(W).$
Therefore, the two prescriptions differ only by a small offset along the short portion of the curve extending below $W_{\min}$.

However, since the 
Anderson mobility edge can be reliably determined only for $W\ge0.5$, and it is expected to disappear close to $W_{\min}$, 
this comparison (not shown) does not provide a useful criterion for selecting the optimal shift.
A more informative indication comes instead from the weak-disorder analysis discussed in Sec.~\ref{subsec:EisoWmin}, where we show that the bulk-edge prescription with $X=0$, analytically continued below $W_{\min}$, allows the LLT framework to recover the Aizenman--Warzel bound $W_{\rm AW}$~\cite{warzel2012}. This suggests that the bulk-edge prescription, rather than the isolated-eigenvalue prescription, is the natural choice for extracting the low-disorder information encoded in the LLT.

In both the general case and the high-connectivity case (discussed in detail in Sec.\ref{subsec:hicon}), the introduction of a finite shift $X>0$ drives the critical curves towards different asymptotic lines, which can be computed analytically. The physical effect of the shift is to reduce the correlations on the lattice. 
Indeed, as $X\to+\infty$, both the cavity Green's functions and the cavity rescaled fields become deterministic, and the transition is described by an effective independent-site percolation problem.

The large-$X$ limit can be derived as follows. In the presence of an extra shift $X>0$, the support of the cavity Green's function marginal $P_g$ is $[m_g,M_g]$. As shown in Sec.~\ref{subsubsec:gtails}, 
in the limit $X\to\infty$ at fixed $K$ and $W$, defining $\mu_g$ as the expected value of cavity Green's functions, one has
\begin{eqnarray}
\mu_g\sim M_g\sim m_g
&\sim&
\frac{1}{X}
+O\left(\frac{1}{X^2}\right),
\\
M_g-m_g
&=&
\frac{W}{X^2}
+O\left(\frac{1}{X^3}\right)\,.
\end{eqnarray}
Thus, the cavity Green's functions concentrate around $1/X$ as $X\to+\infty$.
\begin{figure*}[t]
\centering
\includegraphics[width=0.45\textwidth,trim=0.6cm 0cm 0.8cm 0cm,clip]{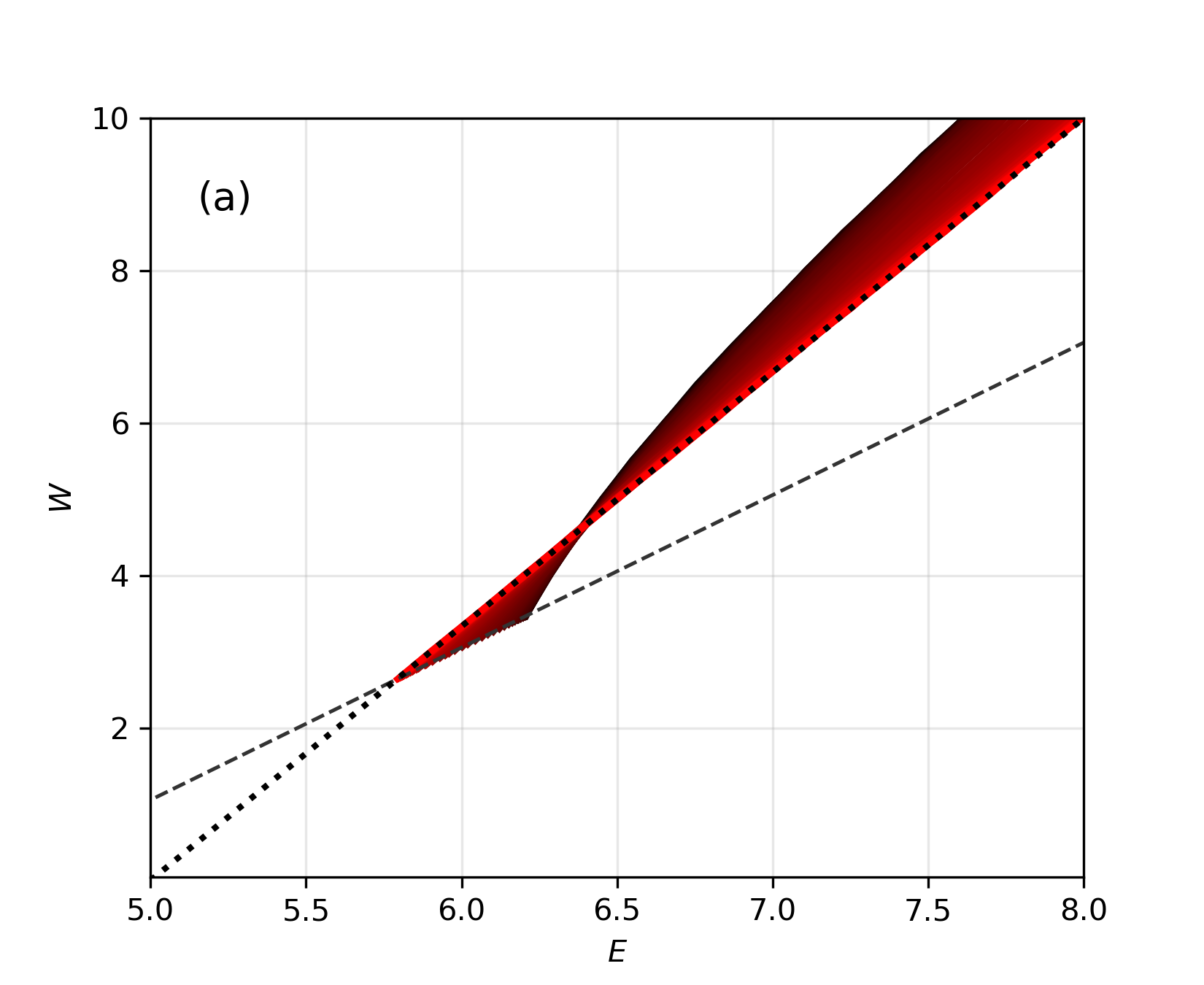}
\includegraphics[width=0.49\textwidth,trim=0.8cm 0cm 0.8cm 0cm,clip]{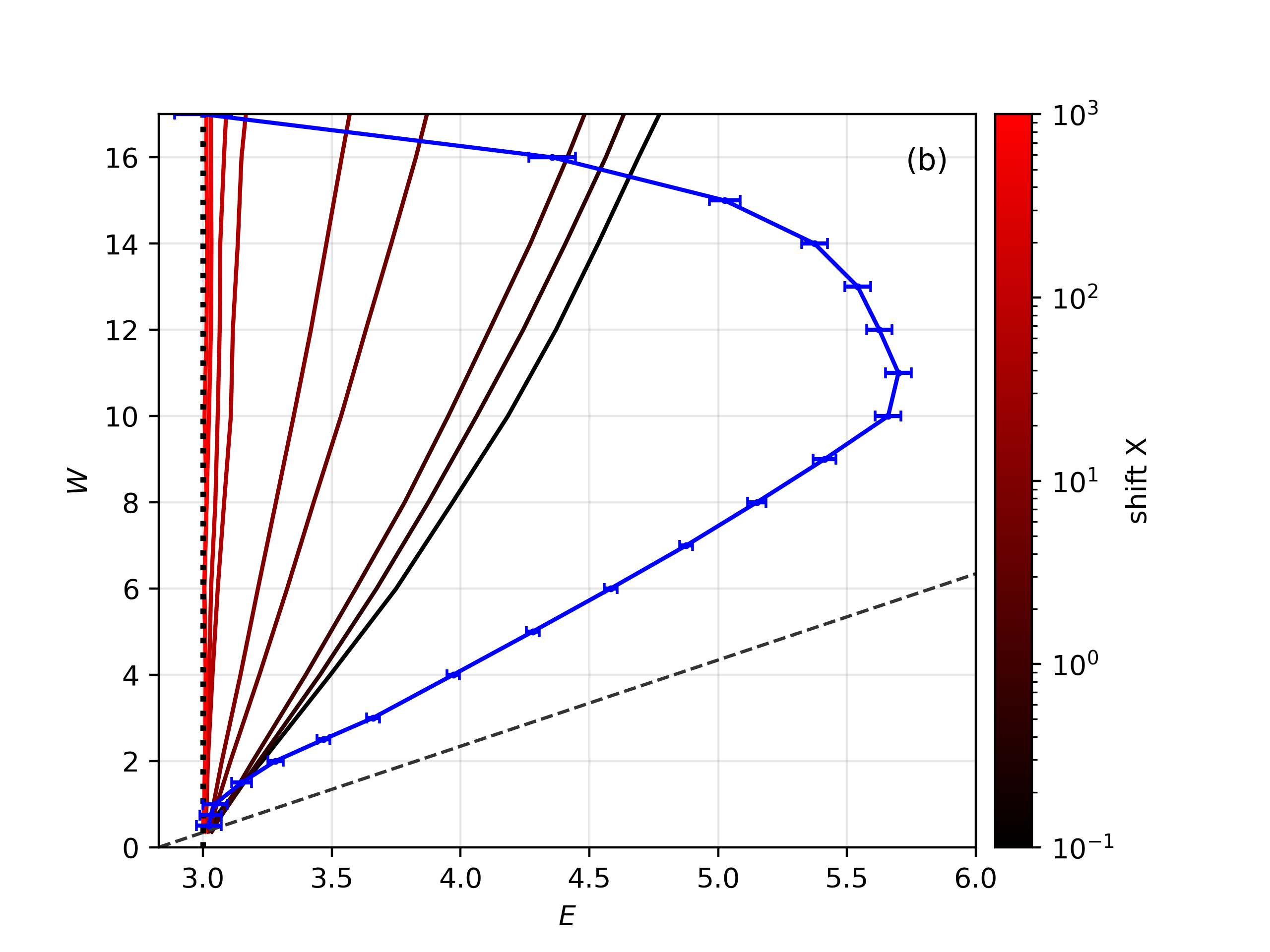}
    \caption{\small {Dependence of the percolation critical curve on the shift $-E_{\rm edge}(W)+X$. (a) Critical curves for $K=5$ and $t=1$, obtained from the high-connectivity equations~\eqref{eq:gsys},  
    \eqref{eq:etasys} and \eqref{eq:psys} (using the method described in Sec.~\ref{subsubsec:qqbar} to evaluate the curve $\bar q=1/K$) for several values of $X$, reported in the legend of panel (b). As $X\to\infty$, the curves approach the asymptotic line given by Eq.~\eqref{eq:EcXinftyHC}} (black dotted line). (b) Critical curves for $K=2$ and $t=1$, obtained from the full equations~\eqref{eq:g}, \eqref{eq:eta}, \eqref{eq:pi}, and \eqref{eq:uk} (computed using the criterion of divergence of the correlation length as described in Sec.~\ref{subsubsec:CcorrS}) for several values of $X$, reported in the legend. The black dotted line denotes the asymptotic percolation threshold given by Eq.~\eqref{eq:EcXinftyEXACT}. 
    In both panels, the black dashed line denote the spectral boundaries and the blue curve in (b) is 
    the mobility edge.}
    \label{fig:PDshift}
\end{figure*}
By inserting $\mathcal{G}_{l \to k}\sim 1/X$ into Eq.~\eqref{eq:caveta}, one obtains
\begin{equation}
    \eta_{k \to i}
    =
    1+\frac{t}{X}
    \sum_{l \in \partial k \setminus i} \eta_{l \to k}
    +O\left(\frac{1}{X^2}\right),
\end{equation}
which becomes a deterministic recursion in the large-$X$ limit. Its translationally invariant solution is
\begin{equation}
\mu_\eta
=
1+\frac{Kt}{X}
+O\left(\frac{1}{X^2}\right).
\end{equation}
As a consequence, the normal and cavity Localization Landscape variables become asymptotically independent random variables of the form
\begin{eqnarray}
u_i^{-1}
&=&
\varepsilon_i-t(K+1)
+O\left(\frac{1}{X}\right)
\, ,
\\
u_{i\to j}^{-1}
&=&
\varepsilon_i-Kt
+O\left(\frac{1}{X}\right)\; ,
\end{eqnarray}
where $\varepsilon_i=\epsilon_i-E_{\rm sh}$ is the shifted on-site energy of Eq.~\eqref{eq:eps-shifted}, uniformly distributed in $[2t\sqrt K+X,\,2t\sqrt K+W+X]$.

The limiting percolation critical curve is then obtained, in the general case, by imposing
\begin{equation}
\label{eq:EcXinfty}
\Pr\left\{
u_i^{-1}<-(E_{\rm perc}(W,X)+E_{\rm sh}(W,X))
\right\}
=
\frac{1}{K}
\; .
\end{equation}
This gives
\begin{equation}
E_{\rm perc}(W,X\to \infty)
=
W\left(\frac{1}{2}-\frac{1}{K}\right)
+t(K+1)
\; .
\label{eq:EcXinftyEXACT}
\end{equation}

In the high-connectivity approximation, the only difference is that condition~\eqref{eq:EcXinfty} is imposed on the cavity Localization Landscape variables (see Sec.~\ref{subsec:hicon}).
Therefore,
\begin{equation}
E_{\rm perc}(W,X\to \infty)
=
W\left(\frac{1}{2}-\frac{1}{K}\right)
+Kt
\; .
\label{eq:EcXinftyHC}
\end{equation}
We stress that Eqs.~\eqref{eq:EcXinftyEXACT} and~\eqref{eq:EcXinftyHC} are not two successive orders of a $1/K$ expansion. Both are exact $X\to\infty$ limits, at fixed $K$ and $W$, of two different sets of equations. The first follows from the full cavity equations. The second follows from the high-connectivity equations of Sec.~\ref{subsec:hicon}, in which the normal variables are replaced by the cavity ones. For this reason, all the terms are kept in both expressions, including $W/K$. The difference between the two, $t$, is of order $K^0$, i.e., of relative order $1/K$ with respect to the leading terms, and it measures the error of the high-connectivity approximation in this limit. Equation~\eqref{eq:EcXinftyHC} is the relevant asymptote for Fig.~\ref{fig:PDshift}(a), where the critical curves are computed from the high-connectivity equations.
For the case shown in Fig.~\ref{fig:PDshift}(b), namely $K=2$, Eq.~\eqref{eq:EcXinftyEXACT} reduces to
\begin{equation}
E_{\rm perc}(W,{X\to\infty})
=
t(K+1)
=
3t
\; .\label{eq:EcXinftyK2}
\end{equation}

In order to evaluate the effect of a more generic extra shift on the phase diagram, we computed the LLT critical curves for several values 
of $X$, using the high-connectivity criticality condition [Sec.~\ref{subsec:hicon}, Eq.~\eqref{eq:hiconcrit}] for $K=5$ [Fig.~\ref{fig:PDshift}(a)] and in the general case as the value of the energy $E_+$ at which the correlation length $\xi_{\rm perc}$ diverges for the first time [see Sec.~\ref{sec:CFP}], for $K=2$ 
[Fig.~\ref{fig:PDshift}(b)]. In panel (b), we also show the exact mobility edge for comparison, plotted as a blue line.
We omit the mobility edge for $K=5$ in panel (a) as achieving the necessary precision for the Anderson transition  
is computationally extremely heavy (see Sec.~\ref{sec:mobility} for details); consequently, these calculations were restricted to the $K=2$ case.

The analytical results described above clarify the behavior observed numerically. 
The displacement of the critical line as $X$ increases is controlled by the fact that the two approaches converge towards different asymptotic limiting curves, whose position depends explicitly on $K$. In particular, the high-connectivity and exact equations do not converge to the same large-$X$ percolation line. This explains why the corresponding critical curves drift in different directions when the shift is varied. For $K=2$, increasing the shift clearly increases the distance between the percolation critical curve and the Anderson mobility edge for $W\lesssim15$, i.e., before the two curves cross. Beyond the crossing, the curves with $X>0$ are closer to the mobility edge, although none of them reproduces its re-entrant shape. Therefore, for the purpose of estimating the mobility edge as the LLT percolation critical curve, the most favorable extra shift choice 
among those considered, on the basis of the observed agreement, is $X=0$, as discussed in literature~\cite{filoche2017localization}, and as suggested by the results of Sec.~\ref{subsec:EisoWmin}.

~The LLT bound of Eq.~\eqref{eq:LLTbound} suggests a criterion to select the shift. To make the bound as tight as possible on average, one can choose the shift that minimizes the average of its right-hand side,
\begin{equation}
\label{eq:fshift}
f(E_{\rm sh})\equiv(E-E_{\rm sh})\,\mathbb E[u_i(E_{\rm sh})]\,,
\end{equation}
among the admissible shifts $E_{\rm sh}\le E_{\min}$. Since $\partial u_i/\partial E_{\rm sh}=\sum_j\mathcal G_{ij}u_j$, one has $\partial\,\mathbb E[u_i]/\partial E_{\rm sh}=\mathbb E[u_i^2]$, and an interior minimum satisfies $\mathbb E[u_i]/\mathbb E[u_i^2]=E-E_{\rm sh}$.

This criterion, however, is meaningful only in the tails of the band. Indeed, Eq.~\eqref{eq:Eu-lowerbound} implies $f>1$ for every admissible shift if $E>-t(K+1)$. Moreover, the Jensen inequality $\mathbb E[u_i^2]\ge\mathbb E[u_i]^2$ gives $\partial f/\partial E_{\rm sh}\ge\mathbb E[u_i]\,(f-1)$, so that, for $E>-t(K+1)$, $f$ is an increasing function of $E_{\rm sh}$. Its minimum is then reached for $X\to\infty$, where $f\to1$ and the bound becomes trivial. In this limit, the condition $(E-E_{\rm sh})u_i<1$ reduces to $\epsilon_i-t(K+1)>E$, i.e., to the independent-site percolation problem discussed above. A nontrivial minimum can therefore exist only for $|E|>t(K+1)$. This is precisely the region where the LLT percolation line lies for $W>W_{\min}$ (Fig.~\ref{fig:phase-diagram}), and, at weak disorder, the region where localized states can appear~\cite{aizenman2011extended}.

We have evaluated $f$ along the $X=0$ critical line of Fig.~\ref{fig:phase-diagram}, i.e., for $E=-E_{\rm perc}(W,X=0)$, using population dynamics for $K=2$ and $t=1$. For $W\lesssim4$, $f$ is an increasing function of $X$: the unconstrained minimizer lies above $E_{\min}$, and the minimum over the admissible shifts is reached at $X=0$. This includes the whole weak-disorder region in which the LLT percolation line and the mobility edge are close. For larger disorder, the minimizer moves to $X^*>0$, and it grows with $W$: $X^*\approx0.5$, $1$, $3$, $4$, $6$, and $10$ for $W=5$, $6$, $8$, $10$, $12$, and $14$, respectively. In this regime, however, a larger shift moves the LLT percolation line further away from the mobility edge [Fig.~\ref{fig:PDshift}(b)], so the criterion is not a useful guide there. We conclude that the criterion confirms the bulk-edge prescription $X=0$ precisely in the weak-disorder regime, where the LLT percolation line provides a good approximation of the mobility edge, but it should not be regarded as a general principle.

Another interesting remark is that, in the limit $X\to+\infty$, and in terms of the energy $E$ of the statistically symmetric Hamiltonian $\hat{\mathcal{H}}$, the effective potential variables $v_i\equiv 1/u_i$ become
\begin{equation}
    v_i = \varepsilon_i-t(K+1)\,.
\end{equation}
Thus, in this limit, the picture proposed by the Localization Landscape Theory reduces to the independent-site percolation in an effective potential energy landscape obtained by correcting the Anderson on-site disorders $\epsilon_i$ by a constant term $-t(K+1)$. For $K=2$, the corresponding asymptote, $E_{\rm perc}$ [Eq.~\eqref{eq:EcXinftyK2}], coincides with the energy $t(K+1)$ that controls the onset of localization at weak disorder (Sec.~\ref{subsec:EisoWmin}), at which the mobility edge starts in Fig.~\ref{fig:PDshift}(b).

\subsection{Integrated Density of States}
\label{subsec:density-of-states}

The integrated Density of States (IDoS) is defined as,
\begin{equation}
   \mathcal N(E)
   =
   \int_{-\infty}^{E} dE'\,\rho(E')
   \, ,
\end{equation}
where
\[
\rho(E)=\frac{1}{N}\sum_i \rho_i(E),
\]
and the \(\rho_i\)'s are the local densities of states defined in Eq.~\eqref{eq:rhoi-def}.

Obtaining \(\mathcal N(E)\) is in general a difficult task, and for this reason several approximate methods have been developed for its evaluation. One of the earliest and most important ones is the approach introduced by Hermann Weyl in 1911, which is known to provide an accurate description of the high energy asymptotics of the IDoS in finite dimensional systems for certain classes of potentials~\cite{Ivrii_2016}.

In finite dimension, the Weyl approximation relies on a simple idea: if the potential varies slowly in space, then the local number of quantum states below energy \(E\) can be approximated by the number of free states with kinetic energy smaller than \(E-V(\mathbf x)\), and this local contribution is then averaged over the whole system. This leads to the estimate
\begin{eqnarray}
   &&  \!\!\!\!\!\!\!\!\!\!
   \mathcal N^{(d)}(E)\approx 
    \nonumber\\
   && \!\!\! \frac{1}{|\Omega|}\int_{\Omega} d^dx \, \int_{\mathbb R^d} \frac{d^d\xi}{(2\pi)^d} \; \theta \left( E-V({\mathbf x})-|\bm\xi|^2\right)\,.
\end{eqnarray}
The right-hand-side can also be interpreted as the spatial average of the integrated Density of States of a free system evaluated at the locally shifted energy \(E'=E-V(\mathbf x)\), namely
\begin{equation}
\mathcal N^{(d)}(E)\approx \frac{1}{|\Omega|}\int_{\Omega} d^dx \; \mathcal N^{(d)}_0\!\left(E-V(\mathbf x)\right),
\end{equation}
with
\begin{equation}
\mathcal N_{0}^{(d)}(E')
=
\int_{\mathbb R^d} \frac{d^d\xi}{(2\pi)^d} \; \theta \left( E'-|\bm \xi|^2\right).
\end{equation}
Written in this way, the physical meaning of the Weyl approximation is very transparent: at each point \(\mathbf x\), one counts the free states available at the effective local energy \(E-V(\mathbf x)\), and then averages this quantity over space.

The same reasoning can be extended naturally to the Bethe lattice. In this case, however, there is no momentum variable. 
The roles of plane waves and their associated wave-vectors are assumed by the eigenvectors and eigenvalues of the adjacency matrix for the  lattice in absence of disorder. Consequently, it is natural to replace the free phase-space counting with the integrated Density of States of the adjacency matrix of the clean graph. Since the Bethe lattice has no semiclassical momentum variable and no slowly varying limit, this replacement is heuristic. Below we show how it can be derived from an exact rewriting of the shifted Hamiltonian in terms of the Localization Landscape, which also makes explicit the approximations involved.

Whenever the potential fluctuates on spatial scales smaller than \(O(E^{-1/2})\), the Weyl approximation becomes progressively less accurate. For this reason, it is not well suited to describe the IDoS of systems with strongly disordered potentials, such as the Anderson model.

Several refinements of the Weyl approximation in finite dimension have been proposed.
An important example is the method developed by Fefferman and Phong~\cite{Fefferman1983theUP}, which yields upper and lower bounds on \(\mathcal N^{(d)}(E)\), again with very good performance in the high energy regime. More recently, the Localization Landscape framework was used to further improve these estimates. References~\cite{arnold2019computing,david2021landscape} provided 
numerical evidence for such an improvement, together with a rigorous derivation 
of upper and lower bounds on \(\mathcal N^{(d)}(E)\). In this Section, we compute the IDoS predicted by the Localization Landscape extension of Weyl's approximation
adapted to be applied to the Bethe lattice. The key idea of Refs.~\cite{arnold2019computing,david2021landscape} 
that we will use here is to replace the original 
potential \(V(\mathbf x)\) by the effective potential \(1/u(\mathbf x)\).

If the Localization Landscape is defined by
\begin{equation}
\hat{\mathcal H}_+\mathbf{u}
=
(\hat{\mathcal H} - E_{\rm sh}\hat{\mathcal I}) \mathbf{u}
=
\mathbf{1}
\; ,
\end{equation}
the associated effective potential is \(1/u_i\). 

The first point to note is that, for $W\le W_{\rm min}$, the result depends on the choice of the shift prescription. For
$E_{\rm iso}\leq E_{\rm sh}<W/2-t(K+1)$, the typical values of the Localization Landscape variables diverge. Therefore, in order to obtain the LLT estimate of the IDoS, one should in principle study the probability of rare finite values of the Localization Landscape variables. However, up to very small corrections, the LLT prediction reduces to the IDoS associated with the hopping term of the Hamiltonian, which we denote by $\mathcal N_0^{(\infty)}$. In the regime
$W/2-t(K+1)\leq E_{\rm sh}\leq E_{\rm edge}$, instead, no finite value of $u_i$ is possible, and the LLT prediction reduces exactly to $\mathcal N_0^{(\infty)}$. Finally, in the over-shifted regime, we observe numerically that the LLT prediction for the IDoS below $W_{\min}$ becomes increasingly different from the exact one, with the agreement worsening as the shift is moved deeper into the over-shifted region.
At finite disorder below $W_{\rm min}$, the LLT Weyl approximation is almost unable to distinguish the disordered case from the ordered one, since the on-site effective potentials vanish either typically or identically, depending on the shift regime. More generally, any approximation method that uses only the effective potential $1/u_i$ to estimate the spectral properties of the Anderson Hamiltonian below $W_{\rm min}$ will have very limited sensitivity to the presence of disorder on the lattice.

Using the definition of the Localization Landscape, $\hat{\mathcal H}_+\mathbf u=\mathbf 1$, i.e., $\varepsilon_iu_i-t\sum_{j\in\partial i}u_j=1$, the shifted on-site energies can be written exactly as $\varepsilon_i=1/u_i+t\sum_{j\in\partial i}u_j/u_i$. Hence
\begin{equation}
\label{eq:Hdecomposition}
\hat{\mathcal H}_+=\hat{\mathcal K}_{\rm eff}+\hat{\mathcal V}_{\rm eff}\,,
\end{equation}
with
\begin{eqnarray}
\label{eq:Keff}
\hat{\mathcal K}_{\rm eff} \! & \! = \! & \! t\sum_i\sum_{j\in\partial i}\frac{u_j}{u_i}\,\hat c_i^\dagger\hat c_i-t\sum_{\langle i,j\rangle}\left(\hat c_i^\dagger\hat c_j+\hat c_j^\dagger\hat c_i\right), \;\; \\
\label{eq:Veff}
\hat{\mathcal V}_{\rm eff} \! & \! = \! & \! \sum_i\frac{1}{u_i}\,\hat c_i^\dagger\hat c_i\,.
\end{eqnarray}
This is the lattice analogue of the exact transformation of the continuum LLT~\cite{PhysRevLett.116.056602}, in which the Hamiltonian is rewritten as a weighted kinetic operator, $-u^{-2}\nabla\cdot(u^2\nabla)$, plus the effective potential $1/u$. Indeed, $\hat{\mathcal K}_{\rm eff}\mathbf u=0$, and $\hat{\mathcal K}_{\rm eff}=\hat U^{-1}\hat L_u\hat U^{-1}$, where $\hat U={\rm diag}(u_i)$ and $\hat L_u$ is the graph Laplacian with positive weights $tu_iu_j$ on the links $\langle i,j\rangle$. Hence $\hat{\mathcal K}_{\rm eff}$ is positive semi-definite, with $\mathbf u$ as zero mode, and $\hat{\mathcal V}_{\rm eff}$ plays the role of a potential.

The Weyl construction requires two approximations. First, the landscape is assumed to vary slowly in space, $u_j/u_i\simeq1$, so that the kinetic operator is replaced by $t$ times the graph Laplacian of the clean lattice,
\begin{eqnarray}
\label{eq:Ksv}
\hat{\mathcal K}_{\rm eff} &\simeq& \hat{\mathcal K}_0
\\
\hat{\mathcal K}_0 &\equiv& t(K+1)\sum_i\hat c_i^\dagger\hat c_i-t\sum_{\langle i,j\rangle}\left(\hat c_i^\dagger\hat c_j+\hat c_j^\dagger\hat c_i\right).
\qquad
\end{eqnarray}
On the infinite tree, the spectrum of $\hat{\mathcal K}_0$ is $[t(\sqrt K-1)^2,\,t(\sqrt K+1)^2]=[X_c,\,X_c+4t\sqrt K]$. Its zero mode, the uniform vector, is not normalizable there; on finite RRGs it corresponds to the isolated eigenvalue. Second, the number of states below $E_+$ is estimated by counting the eigenstates of $\hat{\mathcal K}_0$ below the local energy $E_+-1/u_i$, using only the one-site distribution $P_u(u)$ and neglecting the spatial correlations of the landscape. Neither approximation is justified a priori on the Bethe lattice, where $u_j/u_i$ fluctuates by an amount of order one. A failure of the resulting estimate should therefore be regarded as a limitation of this Weyl-type construction, and not necessarily of the LLT itself.

Restricting to the case $W> W_{\rm min}$, the Weyl approximation for the IDoS is
\begin{equation}
\label{eq:IDoSWgeWmin}
\mathcal N_{\mathrm{LLT,+}}^{(\infty)}(E_+)= \int du \, P_u(u)\,
\mathcal N_{0,+}^{\infty}\!\left(E_+-\frac{1}{u}\right),
\end{equation}
where \(P_u(u)\) denotes the distribution of the Localization  Landscape variable at a typical site, and
\(\mathcal N_{0,+}^{(\infty)}\) is the integrated Density of States of the operator $\hat{\mathcal K}_0$ of Eq.~\eqref{eq:Ksv}.

\(\mathcal N_{0,+}^{(\infty)}\) is directly related to the IDoS of the adjacency matrix (multiplied by $t$) of an RRG in the thermodynamic limit $\mathcal{N}_0^{(\infty)}$, as $\mathcal N_{0,+}^{(\infty)}(E_+)=\mathcal{N}_0^{(\infty)}(E_+-t(K+1))$, where
\begin{equation}
\mathcal N_0^{(\infty)}(E)
=
\int_{-\infty}^{E} dE'\,\rho_0^{(\infty)}(E')
\end{equation}
and
\begin{widetext}
\begin{equation}
\label{eq:DoSRRG}
\rho_0^{(\infty)}(E)
=
\frac{K+1}{2\pi}\,
\frac{\sqrt{4Kt^2-E^2}}{(K+1)^2 t^2-E^2}\,
\mathbf 1_{\{|E|\le 2t\sqrt K\}} \, .
\end{equation}
Performing the integral explicitly, one obtains
\begin{equation}
\mathcal N_0^{(\infty)}(E)=
\begin{cases}
0 \qquad & 
\quad 
\text{if}\quad E<-2t\sqrt K\,,\\[1em]
\displaystyle
\frac{1}{2}
+\frac{K+1}{2\pi}\arcsin\!\left(\frac{E}{2t\sqrt K}\right)
-\frac{K-1}{2\pi}
\arctan\!\left(
\frac{K-1}{K+1}\,
\frac{E}{\sqrt{4Kt^2-E^2}}
\right)
\qquad
&
\quad
\text{if}\quad |E|\le 2t\sqrt K\,,\\[1.2em]
1 \qquad &
\quad
\text{if}\quad E>2t\sqrt K\,.
\end{cases}
\end{equation}

Finally, rewriting as usual for $W>W_{\min}$
\begin{equation}
    E = E_++E_{\rm sh}=E_+-2t\sqrt K-W/2-X
\end{equation}
 in Eq.~\eqref{eq:IDoSWgeWmin}, and writing $\mathcal N_{+,0}^\infty$ in terms of $\mathcal N_0^{(\infty)}$, one has
\begin{equation}
\label{eq:IDOSLLT}
\begin{cases}
\mathcal N^{(\infty)}_{\mathrm{LLT}}(E)
    \approx\mathcal N_0^{(\infty)}(E)\quad&
    \quad
    \text{if}\quad W\le W_{\rm min}
    \; , \\[1em]
     \mathcal N^{(\infty)}_{\mathrm{LLT}}(E) = \displaystyle\int du \, P_u(u)\,
    \mathcal N_{0}^{(\infty)}\!\left(E+\frac W2+X-X_c-\frac{1}{u}\right) \quad&
    \quad
    \text{if}\quad W>W_{\rm min}
    \; .
\end{cases}
\end{equation}
In conclusion, the finite dimensional formula and its Bethe lattice analogue have exactly the same structure: in both cases, one evaluates the free integrated Density of States at an energy shifted by the effective potential, and then averages the result, over space in finite dimension, and over the distribution of \(u\) in the Bethe lattice case.

From this derivation it follows that the Localization Landscape prediction of the DoS is
\begin{equation}
\label{eq:DOSLLT}
\begin{cases}
    \rho^{(\infty)}_{\mathrm{LLT}}(E)\approx
    \rho_0^{(\infty)}(E)\quad&
    \quad
    \text{if}\quad W\le W_{\rm min}
    \; , \\[1em]
     \rho^{(\infty)}_{\mathrm{LLT}}(E) = \displaystyle\int du \, P_u(u)\,
    \rho_{0}^{(\infty)}\!\left(E+\frac W2+X-X_c-\frac{1}{u}\right)\quad&
    \quad
    \text{if}\quad W>W_{\rm min} 
    \; .
\end{cases}
\end{equation}
Note that, in the top expressions of  Eqs.~\eqref{eq:IDOSLLT},\eqref{eq:DOSLLT} we do not consider the dependence on the shift, as numerical simulations show that the best LLT-Weyl estimates are effectively indistinguishable from the ones of the clean lattice for $W<W_{\rm min}$.
In the first version of this work, the kinetic operator was taken as $\hat{\mathcal K}_+=(2t\sqrt K+X)\sum_i\hat c_i^\dagger\hat c_i-t\sum_{\langle i,j\rangle}(\hat c_i^\dagger\hat c_j+\hat c_j^\dagger\hat c_i)$. Its diagonal constant was chosen so that its spectrum starts at $X$, but it was not derived. This leads to $\rho_0^{(\infty)}(E+W/2-1/u)$ in Eq.~\eqref{eq:DOSLLT}. The two constructions coincide for $X=X_c$. Moreover, the exact relation $1/u_i=\varepsilon_i-t\sum_{j\in\partial i}u_j/u_i$ implies
\begin{equation}
\frac{1}{u_i(X)}-\frac{1}{u_i(0)}=X-t\sum_{j\in\partial i}\left[\frac{u_j(X)}{u_i(X)}-\frac{u_j(0)}{u_i(0)}\right],
\end{equation}
where $u_i(X)$ denotes the landscape for the shift $E_{\rm sh}=E_{\rm edge}-X$. Since the landscape is not covariant under a change of the shift, the original estimate at $X=X_c$ and the revised one at $X=0$ are not identical. They differ only through the dependence of the ratios $u_j/u_i$ on the shift, which is weak. Numerically, for $K=2$ and $W=1$, $4$, and $8$, the quantiles of the two effective potentials differ by a few hundredths of $t$, and the integrated absolute difference between the two DoS estimates is about $1\%$. The two estimates therefore practically coincide in the bulk of the spectrum. They differ near the lower edge: since the smallest value of $1/u_i$ is zero both for $X=0$ and for $X=X_c$, the revised estimate at $X=0$ has a gap $X_c$ above $E_{\min}$, while the original one at $X=X_c$ extends down to $E_{\min}$.

\begin{figure}
    \centering \hspace{-1.1cm}
    \includegraphics[width=0.39\linewidth]{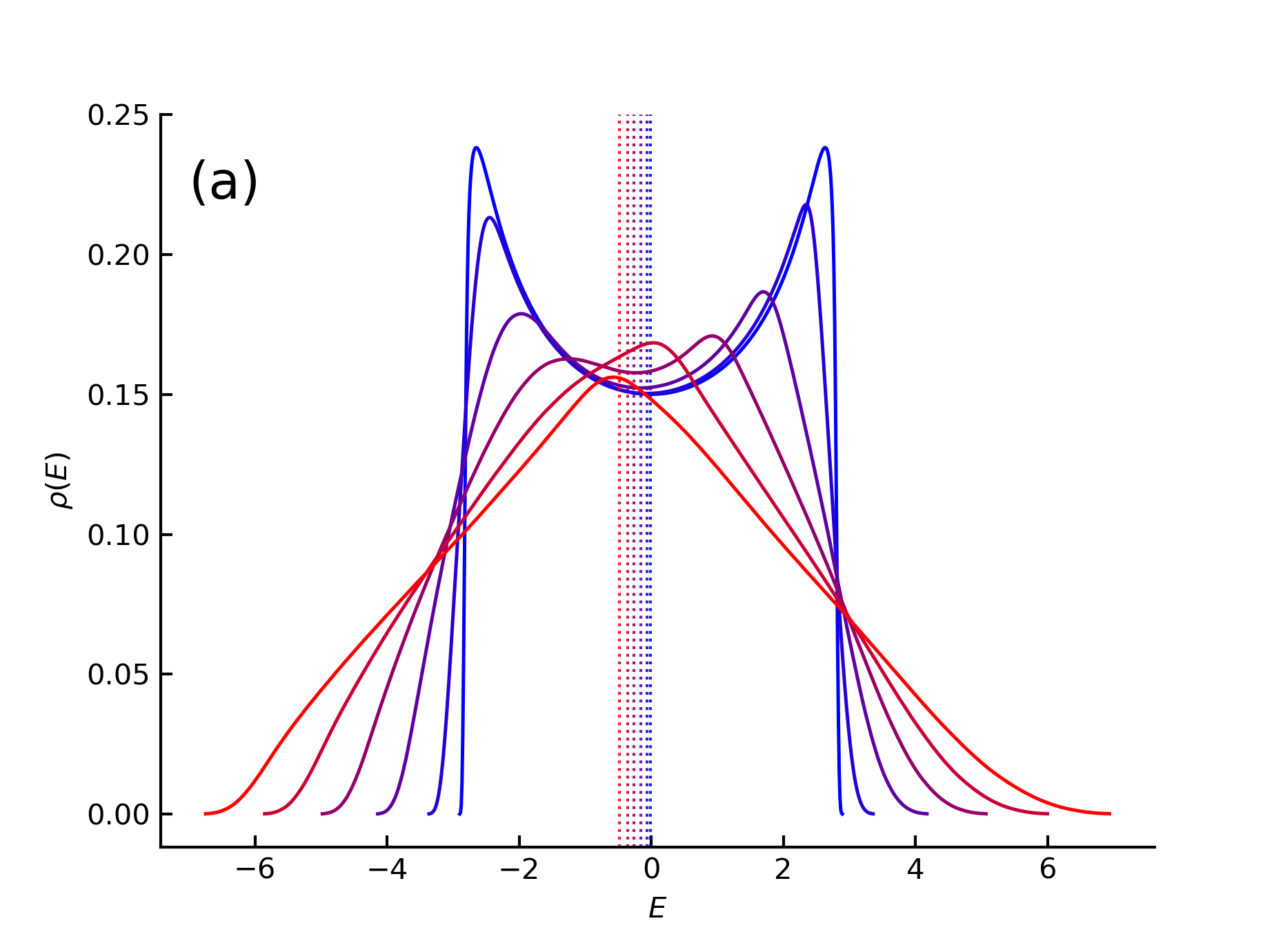}\hspace{-1.3cm}
    \includegraphics[width=0.39\linewidth]{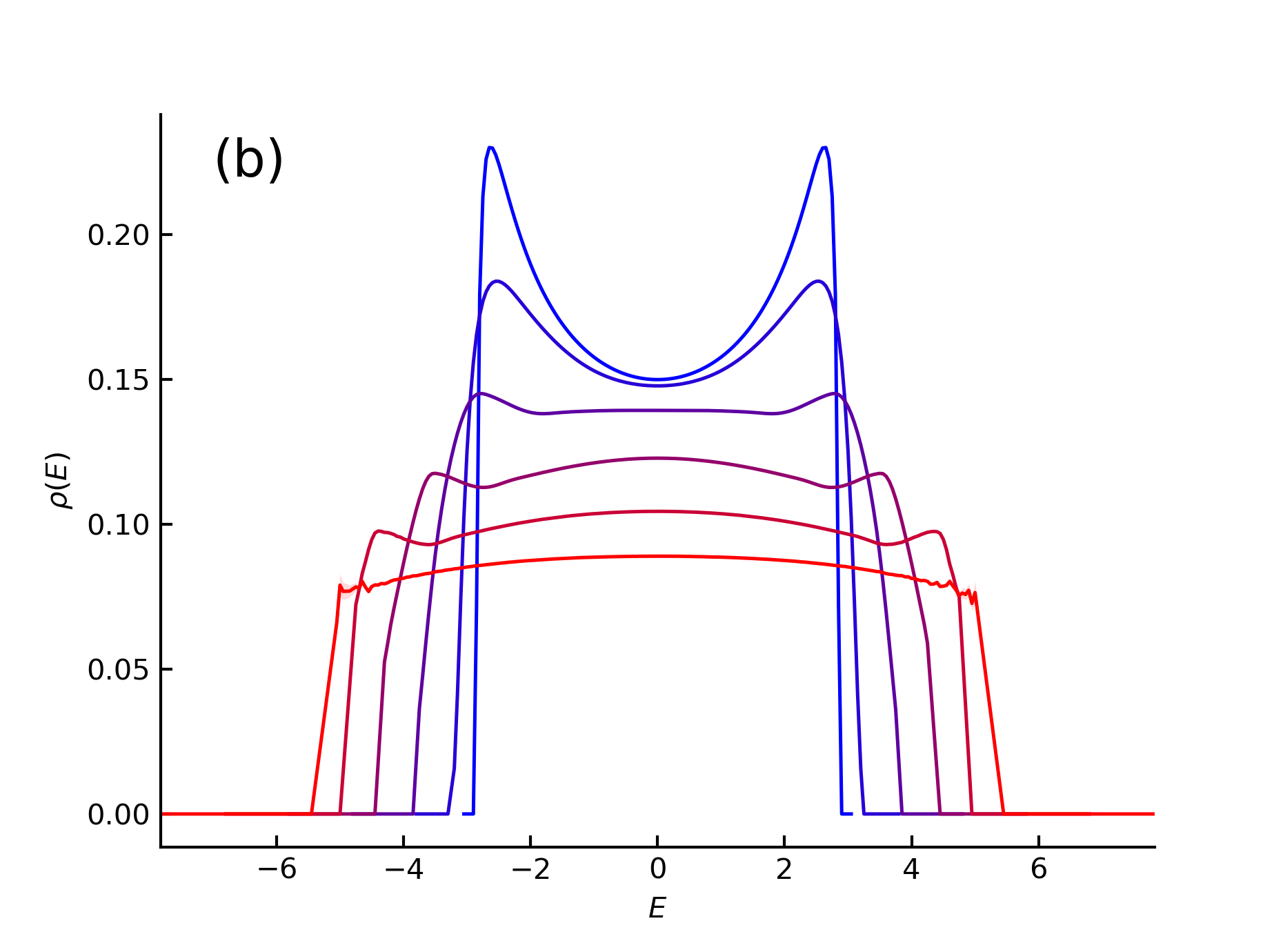}\hspace{-1.3cm}
    \includegraphics[width=0.39\linewidth]{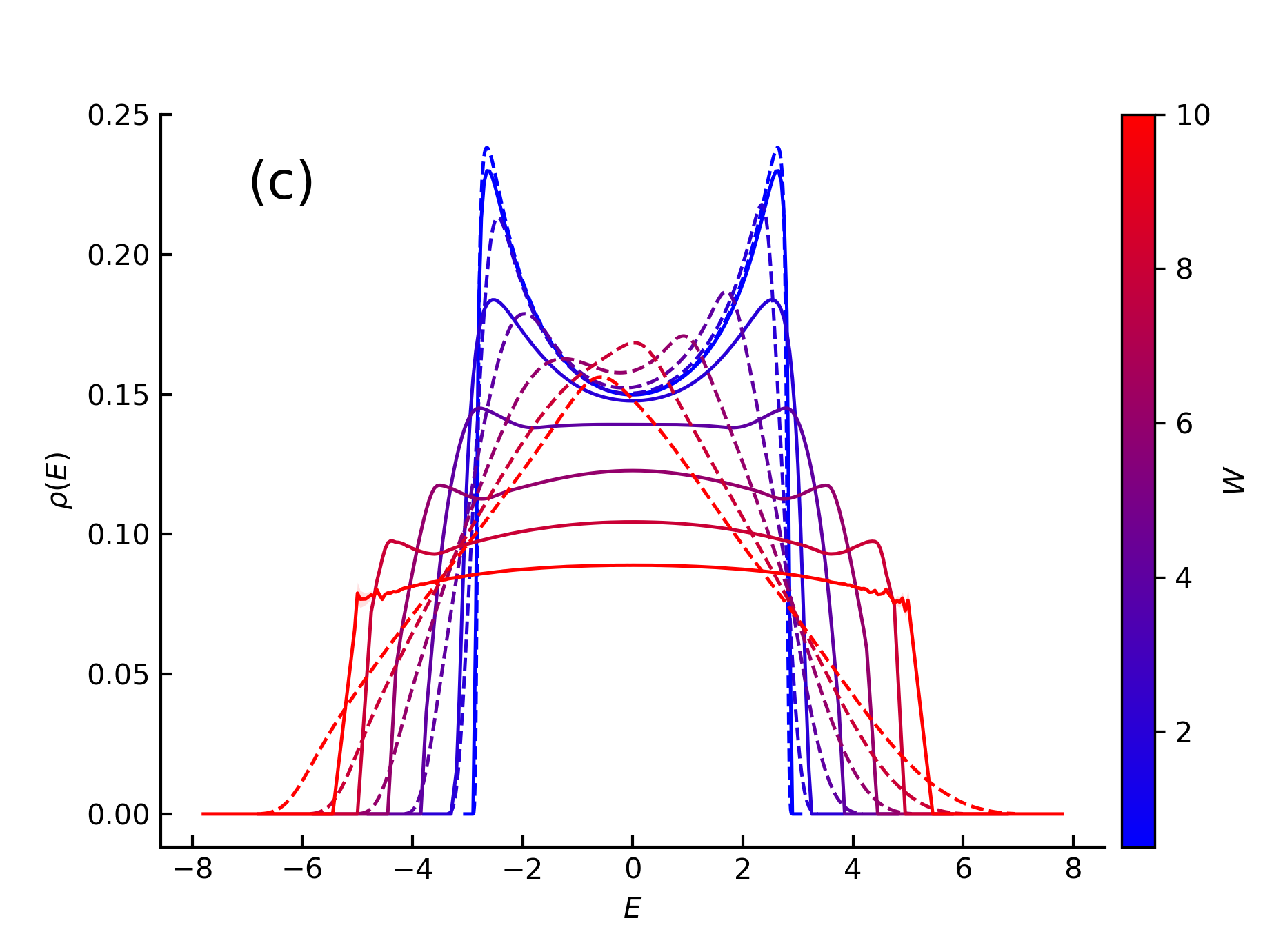}\hspace{-1.0cm}
    \caption{\small{Density of States (DoS) of the Anderson model on the Bethe lattice with  $K=2$ and $t=1$. (a) LLT prediction for the DoS at $X=X_c$ [Eq.~\eqref{eq:DOSLLT}], the smallest shift for which its support coincides with the exact one; in the bulk it practically coincides with the estimate at $X=0$}. The vertical dotted lines represent the median of the distribution. (b) Exact DoS [Eq.~\eqref{eq:DOSAL}]. 
    (c) Comparison between the LLT prediction with dashed lines and the exact result with solid lines. The discrepancy becomes more pronounced for increasing values of $W$.
    }
    \label{fig:DoS}
\end{figure}

The LLT estimate of the DoS can be compared with the exact result for the Anderson model on the Bethe lattice, namely
\begin{equation}
\label{eq:DOSAL}
\rho^{(\infty)}_{\rm AM}(E)
\equiv
\mathbb{E}[\rho_i(E)]
=
\frac{1}{\pi}
\mathbb{E}\!\left[
{\rm Im}\,\mathcal{G}_{ii}(E+{\rm i} 0^+)
\right]
\; ,
\end{equation}
which we evaluate numerically  
using the population dynamics algorithm to determine the distributions of the Anderson Green's functions. Further details about the Anderson model on the Bethe lattice and the numerical methods are given in Secs.~\ref{sec:Anderson} and \ref{sec:numerics}.

The first important feature of Eq.~\eqref{eq:DOSLLT} is that its support 
matches that of the exact DoS only for $X\ge X_c$. In fact for $W>W_{\min}$ (and using that $W_{\min}>X_c$), $\rho^{(\infty)}_{\mathrm{LLT}}$ is supported on
\begin{eqnarray}
    [E_{\rm min}^{\rm LLT}\,,\, E_{\rm max}^{\rm LLT}]
    &=&
    [E_{\rm min}+X_c-X +\tilde m_v\,,\,E_{\rm max} -W-X+X_c + \tilde M_{v}]\,\nonumber\\
    &=& \begin{cases}
    [E_{\rm min}+X_c-X\,,\, E_{\rm max}] \quad &\text{if} \quad X< X_c\,,\\
    [E_{\rm min}\,,\, E_{\rm max}] \quad &\text{if} \quad X\ge X_c\,,
    \end{cases}\label{eq:LLTedges}
\end{eqnarray}
where $\tilde M_{v}$ and $\tilde m_{v}$ are respectively the maximum and minimum possible values for the effective potentials $v_i=1/u_i$ [Eqs.~\eqref{eq:tildeMv} and \eqref{eq:tildemv}], namely
\begin{equation}
\widetilde M_v
=
\begin{cases}
0
\; ,
\quad &\text{if} \quad W+X\le X_c 
\; ,\\[6pt]
W+X-X_c
\; ,
\quad &\text{if} \quad W+X>X_c
\; ,
\end{cases}
\qquad\qquad
\widetilde m_v
=
\begin{cases}
0
\; ,
\quad &\text{if} \quad X\le X_c 
\; ,\\[6pt]
X-X_c
\; ,
\quad &\text{if} \quad X>X_c
\; .
\end{cases}
\label{eq:vmaxmin}
\end{equation}

 Even for $X\ge X_c$, where the supports of the two DoS match, $\rho_{\rm LLT}^{(\infty)}$ does not reproduce the statistical symmetry of the exact DoS in Eq.~\eqref{eq:DOSAL}. Indeed, since the distribution of $1/u$ is not symmetric around $W/2$, the usual energy transformation $E = E_++E_{sh}$ does not restore the symmetry of the DoS. As shown in Fig.~\ref{fig:DoS}, the LLT prediction and the exact DoS do not even agree qualitatively, and their discrepancy increases for stronger disorder.

In Fig.~\ref{fig:DoS} we use $X=X_c$, the smallest shift for which the support of the LLT estimate coincides with the exact one. For this value, the revised estimate coincides exactly with the one obtained with the original construction discussed above. In the bulk of the spectrum, it practically coincides with the estimate at $X=0$, the shift used for the mobility edge: the two prescriptions differ appreciably only in the interval $[E_{\min},E_{\min}+X_c]$.

The qualitative discrepancy with the exact DoS is robust with respect to the shift. We have computed the LLT estimate for $0\le X\le4$ and $W=1$, $4$, $8$, and $12$ ($K=2$, $t=1$). For every finite $X$, it is asymmetric and does not reproduce the flat plateau of the exact DoS at intermediate and strong disorder. Its integrated absolute deviation from the exact DoS remains large: between $0.21$ and $0.26$ at $W=4$, between $0.30$ and $0.40$ at $W=8$, and between $0.32$ and $0.43$ at $W=12$, decreasing slowly as $X$ increases. This dependence has a simple interpretation. As $X\to\infty$, the effective potential reduces to $\varepsilon_i-t(K+1)$ (Sec.~\ref{subsec:dependence-shift}), and Eq.~\eqref{eq:DOSLLT} reduces to the naive Weyl estimate built with the bare disorder, $\int d\epsilon\,\gamma(\epsilon)\,\rho_0^{(\infty)}(E-\epsilon)$, which is symmetric and contains no information on the landscape. Increasing $X$ therefore interpolates between the LLT--Weyl estimate and the naive Weyl estimate. The latter deviates from the exact DoS by $0.10$, $0.25$, $0.24$, and $0.20$ for $W=1$, $4$, $8$, and $12$, respectively. Compared with the naive Weyl estimate, the LLT improves the estimate of the DoS at weak disorder ($0.055$ at $W=1$, $X=0$), but not at intermediate and strong disorder.

\end{widetext}

One may also ask whether the LLT estimate reproduces the Lifshitz tails of the Anderson DoS close to the edge of the spectrum. For the Anderson model on the Bethe lattice these tails are double exponential, $\exp[-e^{c\,\delta^{-1/2}}]$ with $\delta=E-E_{\min}$, as shown by the rigorous bounds of Refs.~\cite{bapst2011lifshitz,hoecker2014} (see also Ref.~\cite{biroli2010anderson}). Close to $E_{\min}$, Eq.~\eqref{eq:DOSLLT} only involves the rare large values of the Localization Landscape variables. Its behavior, however, depends on the shift. For $X<X_c$ the estimate has a gap of width $X_c-X$ above $E_{\min}$, and therefore no tail at all at the edge of the spectrum. For $X\ge X_c$ it extends down to $E_{\min}$, and its asymptotic form is controlled by the extreme-value statistics of the landscape variables discussed in Sec.~\ref{subsec:ExVal}. That analysis provides a stretched-exponential lower bound for the large-$u$ tail of $P_u$ for $X<X_c$. This bound degenerates at $X=X_c$, and for $X>X_c$ the landscape variables are bounded. We have therefore not determined the asymptotic form of $\rho^{(\infty)}_{\rm LLT}$ close to $E_{\min}$ for $X\ge X_c$, and we do not draw quantitative conclusions on the comparison between the tails of the LLT estimate and the Lifshitz tails of the Anderson model. We only note that, for $X<X_c$, the LLT--Weyl estimate misses the Lifshitz tail entirely.

\section{Numerical Methods}
\label{sec:numerics}

In this Section we explain the numerical methods that we used to solve the critical percolation properties of the 
Localization Landscape. The numerical procedures for computing the  critical properties of the Anderson model on the Bethe lattice are described in detail in Refs.~\cite{rizzo2024localized,tikhonov2019critical}. Therefore, in this Section we only explain how they generalize for $E\ne 0$, and we devise the extrapolation method used to evaluate the critical energies close to the spectral boundary.

\subsection{Population dynamics}
\label{subsec:population}

The population dynamics algorithm is a method used to solve self-consistent distributional integral equations. It is accurately described in the context of message-passing algorithms on random graphical models in Ref.~\cite{mezard2009information}, and in the context of Anderson localization in Ref.~\cite{prado2026anderson}.
Consider the Bethe lattice with connectivity $K+1$ and the cavity equation that relates a cavity variable $X$ to $K$ independent copies of itself on the nearest neighboring sites $\{X_l\}_{l=1,\dots,K}$, 
along with a random variable $Y$ drawn from a distribution $\gamma_+$:
\begin{equation}  
\label{eq:cavgeneric}  
    X = \Psi\left(\{X_l\}_{l=1,\dots,K}\,; Y\right).  
\end{equation}  
It is important to note that the variables in this equation are not necessarily scalars. For example, in the set of cavity equations given by Eqs.~(\ref{eq:cavg})-\eqref{eq:caveta}, we 
have $X = (\mathcal{G}_{k \to i}, \eta_{k \to i})$, $Y=\varepsilon_k$,
and $\{X_l\}_{l=1,\dots,K} = \{\mathcal{G}_{l \to k}, \eta_{l \to k}\}_{l\in \partial k\setminus i}$, 
and the function $\Psi$ relating these. For the case of the full set of equations~(\ref{eq:cavg}),~\eqref{eq:caveta}, and~\eqref{eq:cavpbari} the number of neighbors to be drawn is $K+1$, but apart from that, the algorithm remains unchanged.

We aim to find the probability distribution $P$ such that, when $\{X_l\}_{l=1,\dots,K}$ are independent random variables drawn from $P$, the equality in Eq.~(\ref{eq:cavgeneric}) holds in distribution. This means that $P$ is the solution to the equation  
\begin{eqnarray}  
\label{eq:popdyneq}  
    P(x) &=& \int dy \, P_y(y) \int \prod_{l=1}^{K} \left[dx_l \, P( x_l )\right]  
    \nonumber\\
    & &
 \quad \times \, \delta \bigg( x - \Psi \left( \{x_l\}_{l=1,\dots,K} \,; y \right) \bigg)
  \; ,  
\end{eqnarray}  
where the statistical independence of the $X_l$ variables has let us factorize the joint probability distribution within the integral of 
the right-hand-side,
\begin{equation}
    P(\{x_l\}_{l=1,\dots,K}) = \prod_{l=1}^{K} P(x_l)
    \; .
\end{equation}
The algorithm is based on defining a ``pool'' (or ``population'') of $N$ variables $\{X_i\}_{i=1,\dots,N}$, which approximately represents the probability distribution:
\begin{equation}
    P(x) \simeq \frac{1}{N} \sum_{i=1}^N \delta(x - X_i) \, .
\end{equation}
Each element of the pool is initialized independently at random. 
We select an integer $T$ as the number of iterations. The empirical distribution of $\{X_i^{(\tau)}\}_{i=1,\dots,N}$ at iteration $\tau$ is denoted $\tilde P^{(\tau)}$. At each iteration step $\tau$, we sample a variable $y^{(\tau)}$ from $P_y$ and randomly select $K$ indices $\{i_1, \dots, i_{K}\}$ from $\{1, \dots, N\}$. 
 We then replace $X_{i_1}^{(\tau-1)}$ in the population with  
\begin{equation}  
    X_{i_1}^{(\tau)} = \Psi \left( \{X_{i_l}^{(\tau-1)}\}_{l=1,\dots,K} \,; y^{(\tau)} \right),  
\end{equation}  
while keeping all other variables unchanged.

Assuming Eq.~(\ref{eq:popdyneq}) has a solution, it can be argued that for sufficiently large $T$ and $N$, the empirical distribution $\tilde P^{(\tau)}$ will be a good approximation of $P$.

\subsection{Numerical determination of the mobility edge} \label{sec:mobility}

Here, we describe the numerical procedure used to determine the localization transition. The starting point is the set of recursion equations for the cavity Green’s functions, Eq.~\eqref{eq:gcavAL}. As discussed in Sec.~\ref{sec:SCanderson}, in the localized phase the imaginary part of the Green’s function vanishes proportionally to the regulator~$\alpha$. Anderson localization can therefore be investigated through the linear stability of Eqs.~\eqref{eq:gcavAL} with respect to a small imaginary component, which leads to the simplified recursion equations~\eqref{eq:ReGcavcrit}-\eqref{eq:ImGcavcrit}. Equation~\eqref{eq:ReGcavcrit} determines the distribution of the real parts alone, independently of the (small) imaginary components, while Eq.~\eqref{eq:ImGcavcrit} is linear in the imaginary parts. As a consequence, the typical value of the imaginary parts either grows exponentially (in the delocalized phase) or decreases exponentially (in the localized phase) under iteration. The corresponding growth (or decay) rate is given by the largest eigenvalue~$\lambda$ of the linear integral operator~\eqref{eq:fAL},
\begin{equation}
\mathcal{G}^I_{\rm typ} \propto \lambda^{T} \, ,
\end{equation}
and can be computed numerically using the population dynamics algorithm.

The numerical method proceeds as follows. We initialize a pool of $N$ pairs of cavity Green’s functions ${ \mathcal{G}_i^R, \mathcal{G}_i^I }$. We first iterate Eq.~\eqref{eq:ReGcavcrit} using population dynamics for the real parts only, until their distribution converges to a stationary form. We then start updating the entire pool, real and imaginary parts together, according to Eqs.~\eqref{eq:ReGcavcrit}–\eqref{eq:ImGcavcrit}. The only modification compared to the standard algorithm is the following: since our goal is to monitor the exponential growth or decay of the typical imaginary part, at each iteration step $\tau+1$ we generate an entirely new pool of $N$ elements from the pool at “time’’~$\tau$, replace the old pool, and iterate again. After a transient of about $200$ iterations, the typical value of $\mathcal{G}_i^I$ begins to grow or decrease exponentially, depending on whether the parameters $(E,W)$ lie in the delocalized or localized phase. We extract the growth exponent through a fit of the resulting data, and repeat the whole procedure several times to improve the statistical accuracy.

The main difficulty of this approach stems from the fact that the probability distributions of $(\mathcal{G}_i^R, \mathcal{G}_i^I)$ develop power-law tails at large arguments, as extensively discussed in the literature~\cite{abou1973selfconsistent,tikhonov2019critical,rizzo2024localized,prado2026anderson}. Consequently, finite-size effects associated with the finite pool size $N$ used to represent these distributions are very strong. These finite-size effects have been investigated in detail and with high precision in Ref.~\cite{tikhonov2019critical} for $K=2$ and $E=0$, where the localization transition is driven by increasing $W$. In that specific case, the operator~\eqref{eq:fAL} has been diagonalized numerically with very high accuracy in Refs.~\cite{parisi2019anderson,tikhonov2019critical}, allowing for a careful characterization and control of the finite-$N$ corrections, since the $N \to \infty$ asymptotic result is known exactly. We take advantage of this analysis to carry out an accurate determination of the mobility edge at $E>0$.

\begin{figure}[h!]
    \begin{center}
         \includegraphics[width=0.5\textwidth]{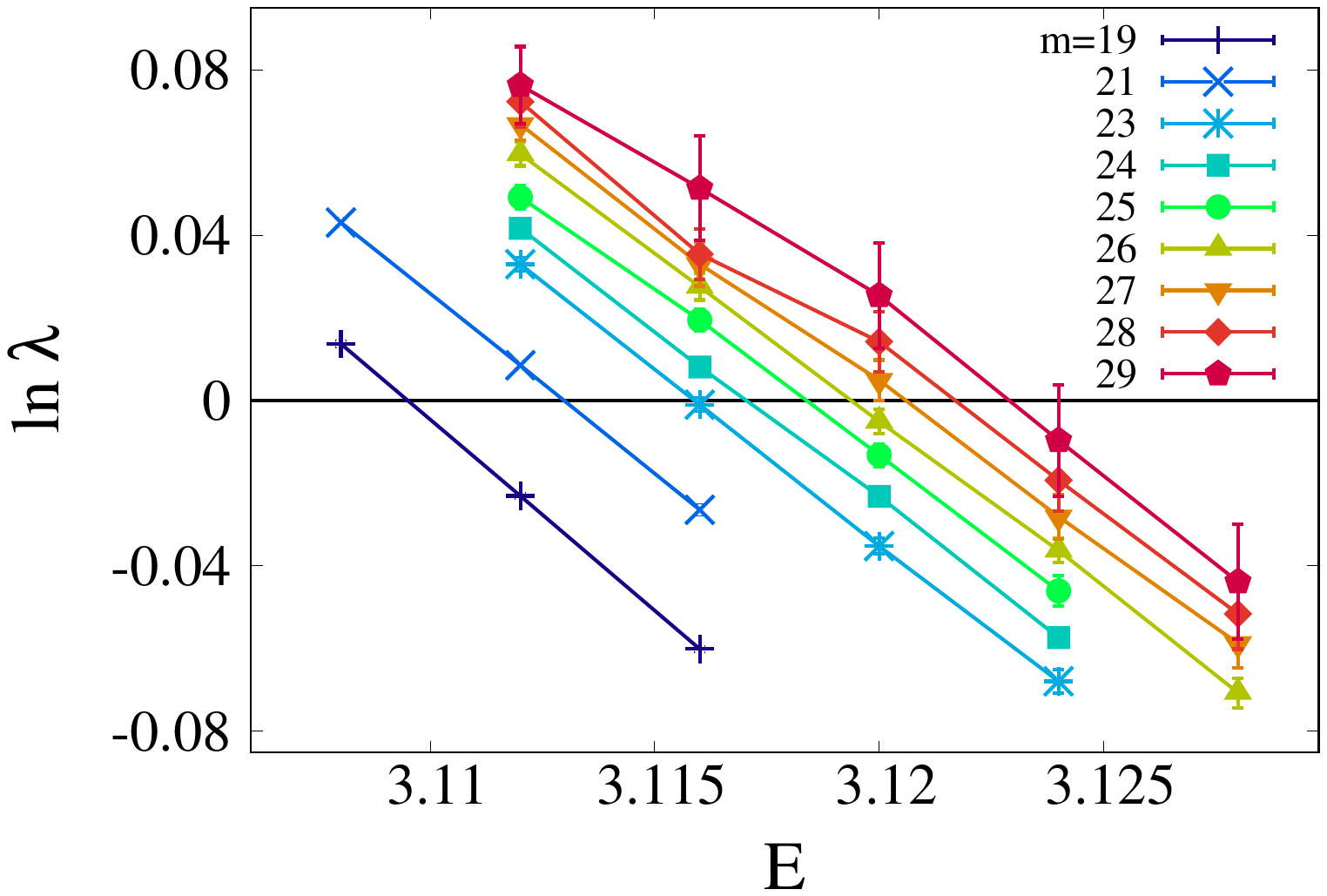}
    \end{center}
       \vspace{-0.5cm}
    \caption{\small{Exponent governing the exponential growth or decay of the typical imaginary part of the Green’s function in the linearized regime, $\ln \lambda$, as a function of the energy $E$ for $W = 1.5$ and $K = 2$, shown for several population sizes $N = 2^m$ with $m$ ranging from $19$ to $29$. The finite-$N$ estimate of the mobility edge, $E_{\rm loc}(N)$, is obtained from a linear fit of the data in the region where $\ln \lambda$ is small.}}
    \label{fig:critical-W1}
\end{figure}

\begin{figure*}[t] 
    \centering
    \includegraphics[width=0.33\textwidth]{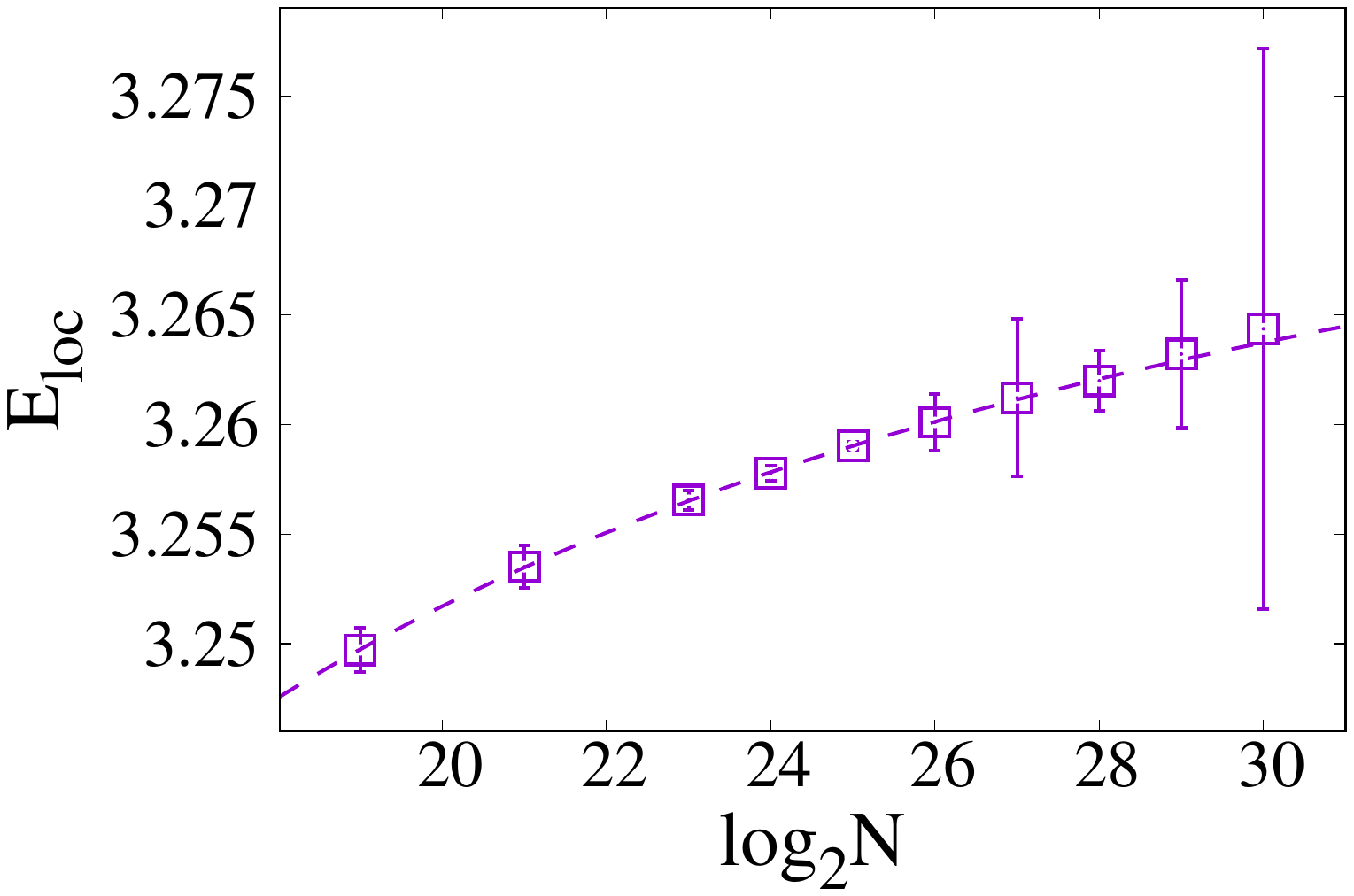} \put(-127,91){\small (a)}
    \includegraphics[width=0.33\textwidth]{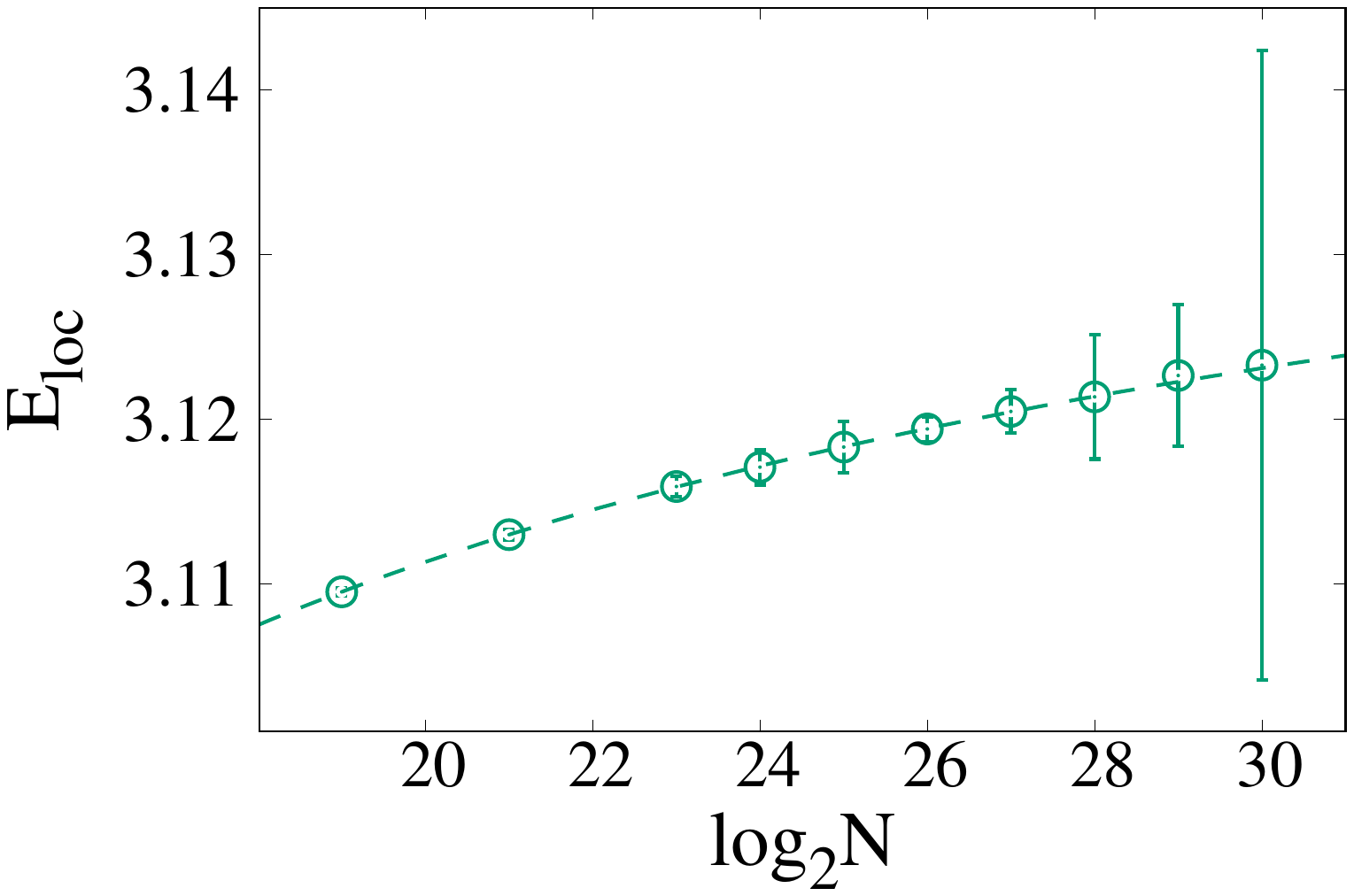} \put(-127,91){\small (b)}
    \includegraphics[width=0.33\textwidth]{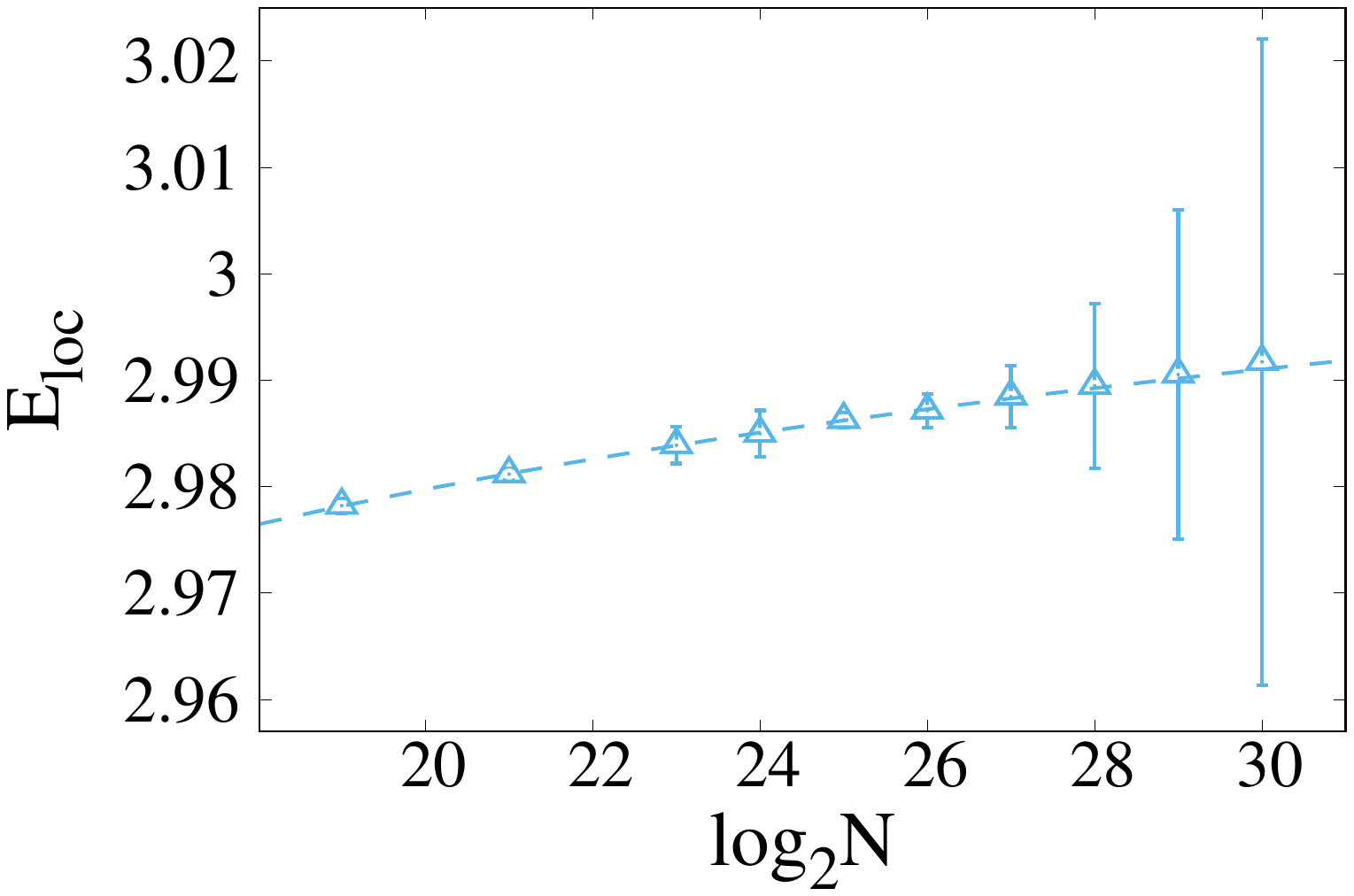} \put(-127,91){\small (c)}
    
    \vspace{0.2cm} 
    
    \includegraphics[width=0.33\textwidth]{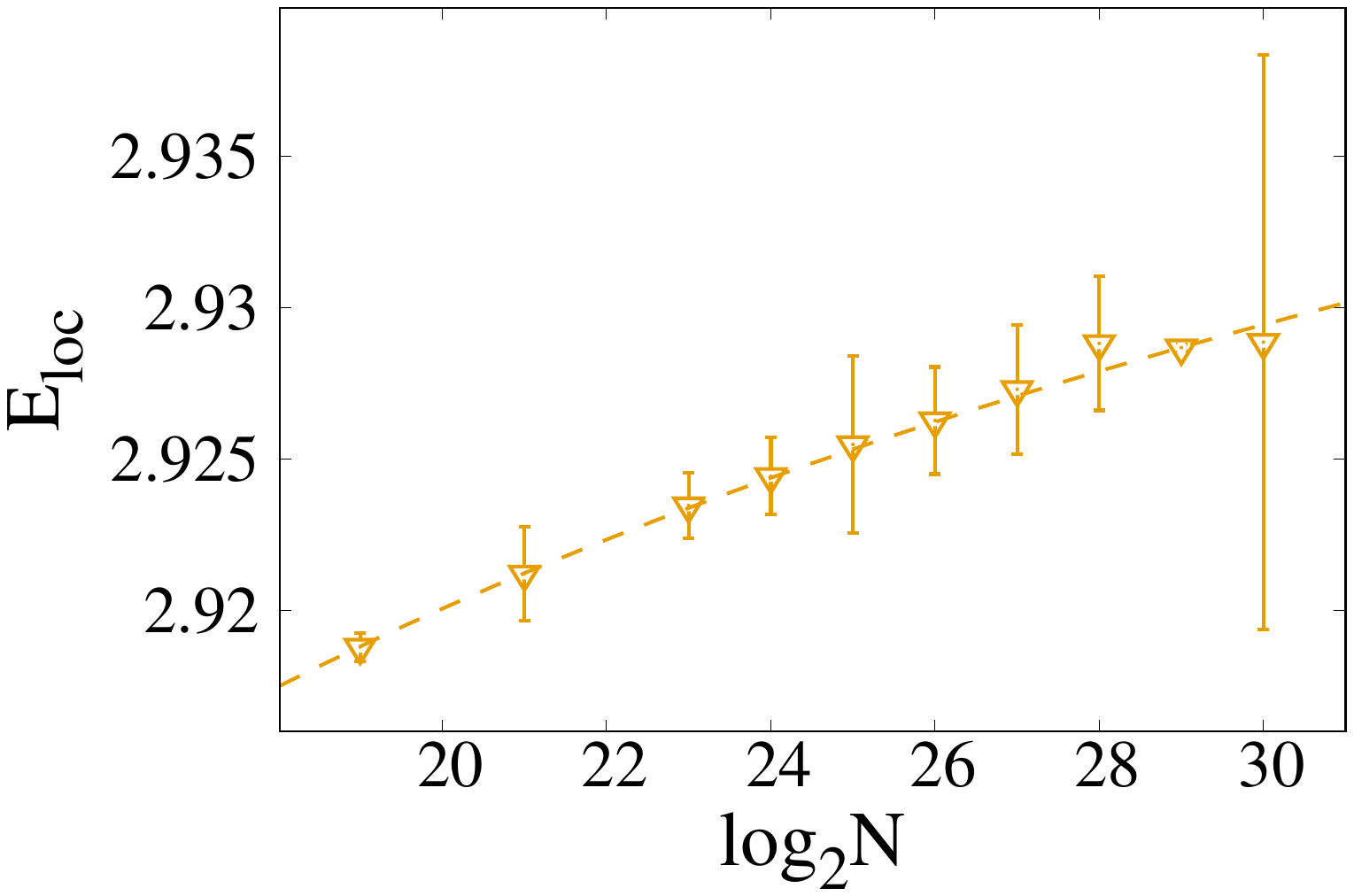} \put(-127,91){\small (d)}
    \includegraphics[width=0.33\textwidth]{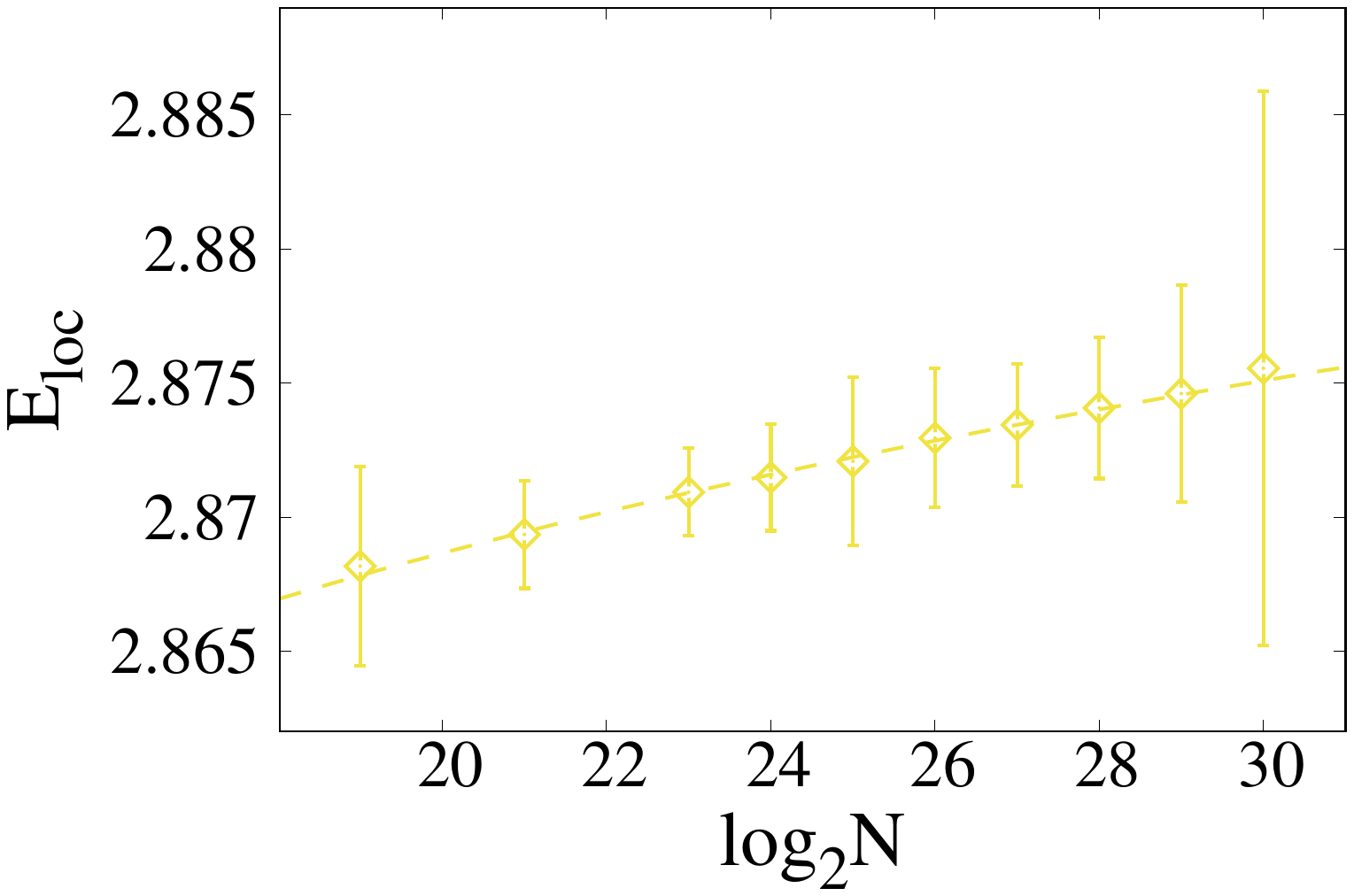} \put(-127,91){\small (e)}
    \includegraphics[width=0.32\textwidth]{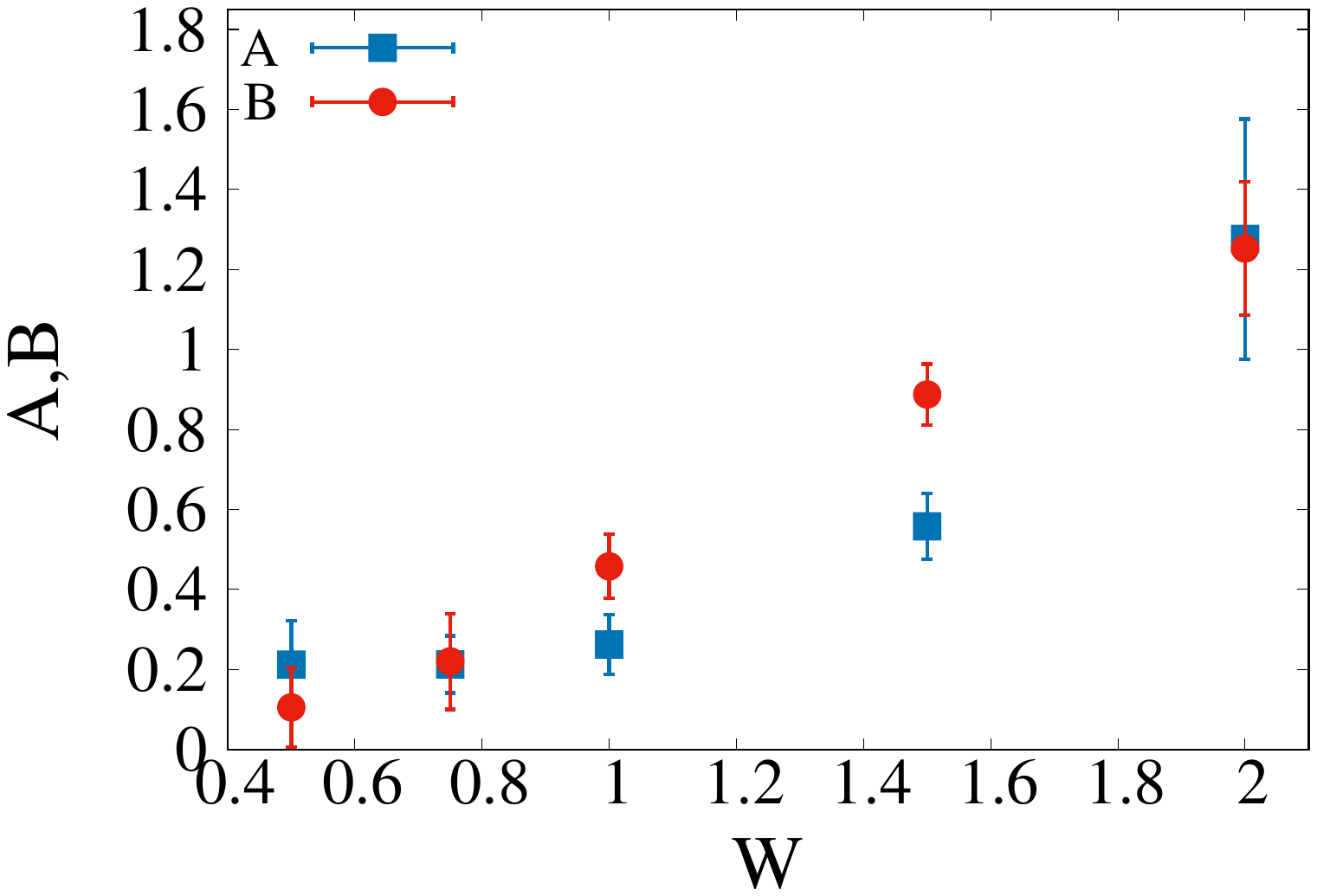} \put(-78,88){\small (f)}
    
    \vspace{-0.2cm}
    \caption{\small{  Fitting procedure used to determine the asymptotic critical energy $E_{\rm loc}$ according to Eq.~\eqref{eq:fit_Ec} for various disorder strengths: (a) $W=2$, (b) $W=1.5$, (c) $W=1$, (d) $W=0.75$, and (e) $W=0.5$. Panel (f) shows the $W$-dependence of the fitting parameters $A$ and $B$.}}
    \label{fig:fits_Ec}
\end{figure*}

In Fig.~\ref{fig:critical-W1} we show $\ln \lambda$ as a function of $E$ for $W = 1.5$ (with $K = 2$) for several pool sizes $N = 2^m$, with $m$ ranging from $19$ to $29$. The figure clearly demonstrates that the energy at which $\ln \lambda$ crosses zero, $E_{\rm loc}(N)$, obtained from a linear fit of the data and corresponding to the estimated position of the mobility edge, drifts slowly but systematically to larger energies as the pool size increases. At $E=0$, where the exact $N \to \infty$ asymptotic value of the localization transition is known with high precision, the analysis of Ref.~\cite{tikhonov2019critical} shows that convergence to the thermodynamic limit is logarithmically slow in $N$. Motivated by this, we extrapolate the asymptotic mobility edge assuming the same type of finite-size corrections, and fit $E_{\rm loc}(N)$ with
\begin{equation} \label{eq:fit_Ec}
E_{\rm loc}(N) = E_{\rm loc} - \frac{A}{(\ln N)^B} \, ,
\end{equation}
where $A$ and $B$ are disorder-dependent fitting parameters of order $1$.
In Figs.~\ref{fig:fits_Ec}(a)–(e) we show the results of these fits for several disorder strengths. As $W$ decreases, the concavity of $E_{\rm loc}(N)$ becomes flatter, corresponding to a decrease of the exponent $B$. This indicates that finite-size effects become increasingly severe at smaller disorder. Combined with the limited accessible range of $\ln N$ -- due to increasing computational cost -- and with the increasing statistical uncertainty for the largest $N$, this results in rather large error bars in the extrapolated asymptotic value of the mobility edge, especially for small $W$. Consequently, we are unable to obtain reliable extrapolations for $W < 0.5$.

A further important point is that, because the accessible range of $\ln N$ is limited and the error bars for the largest $N$ are substantial, a relatively broad range of values of the fitting parameters $E_{\rm loc}$, $A$, and $B$ provides fits of comparable quality for each disorder strength. Within this range, we select the fitting parameters such that their dependence on $W$ is smooth (for instance, by imposing that $E_{\rm loc}$, $A$, and $B$ are decreasing functions of $W$), as illustrated in Fig.~\ref{fig:fits_Ec}(f).

The result of this procedure finally yields the estimation of the asymptotic position of the mobility edge which is plotted in the phase diagram of Ref.~\cite{Tonetti2026}.

\subsection{Evaluation of relevant observables in the localized phase}
In this Section, we show how the linearized cavity recursion equations~\eqref{eq:ReGcavcrit}-\eqref{eq:ImGcavcrit} can be efficiently used to compute relevant observables in the localized phase. For clarity, we illustrate the method using the IPR. The extension to the computation of the two-point correlation function is straightforward and will not be discussed in detail. The approach is a direct generalization of the algorithm introduced in Ref.~\cite{rizzo2024localized} to the case $E \neq 0$.

The IPR is related to the second moment of $|\mathcal{G}_{ii}|$, which is broadly distributed according to Eq.~\eqref{eq:ansatz}. The power-law tails of its probability distribution would lead, in principle, to divergent expressions for $\mathbb E\big[ |\mathcal{G}_{ii}|^2 \big]$. In practice, this divergence is avoided because the power-law behavior is cut off at large values of the imaginary part around $\alpha^{-1}$, which marks the limit of validity of the linearized equations. Nevertheless, computing $\mathbb E\big[ |\mathcal{G}_{ii}|^2 \big]$ requires accessing the region of large imaginary parts of the Green's function, rather than the region of typical finite values. At first sight, this seems to suggest that the linearized equations might not be useful.

Fortunately, this is not the case. To see this, we define:
\begin{equation} \label{eq:mii1}
{\cal M}_{ii} \equiv \frac{1}{{\cal G}_{ii}} = \epsilon_{i} - E - {\rm i} \alpha - \!\! \sum_{m \in \partial i} {\cal G}_{m \to i} = m_{ii} - {\rm i} \alpha \, \hat{m}_{ii} \, , 
\end{equation}
from which  one immediately obtains that 
\begin{equation}
\mathbb E\big[ |\mathcal{G}_{ii}|^2 \big]  = \int  \de m \, \de \hat{m} \; Q(m,\hat{m}) \frac{1}{m^2+\hat{m}^2 \alpha^2}   \,  \, ,
\end{equation}
where $Q(m,\hat{m})$ is the joint pdf of $m$ and $\hat m$.
Similarly, using the fact that ${\cal G}_{ii}^I = \alpha \hat{m}_{ii}/(m_{ii}^2 + \alpha^2 \hat{m}_{ii}^2)$, $\mathbb E \big [ {\rm Im} {\cal G}_{ii} \big ]$ is expressed as
\begin{equation}
\mathbb E \big [ {\rm Im} {\cal G}_{ii} \big ] = \int  \de m\, \de \hat{m} \; 
Q(m,\hat{m}) \frac{\alpha {\hat{m}}}{m^2+\hat{m}^2 \alpha^2}   \,  \, ,
\end{equation}
Given that $\hat{m}$ is strictly positive we can make the change of variables $m = \alpha \hat{m} x$, which  leads to
\begin{eqnarray} 
\mathbb E \big [ |{\cal G}_{ii}|^2 \big ] &=& \int \de x \, \de \hat{m} \;  Q(\alpha \hat{m} x  ,\hat{m}) 
\frac{(\alpha \, \hat{m})^{-1}}{1 + x^2}   \,  \, , \label{eq:gii} \\
\label{eq:img} \mathbb E \big [ {\rm Im} {\cal G}_{ii} \big ] &=&  \int  \de x \, \de \hat{m} \; Q(\alpha \hat{m} x  ,\hat{m}) \frac{1}{1 + x^2}   
 \, .
\end{eqnarray}
In the $\alpha \to 0^+$ limit we can approximate $Q(\alpha \hat{m}  x,\hat{m}) \approx Q(0,\hat{m})$ and perform the integration over $x$ explicitly. Plugging Eqs.~\eqref{eq:gii} and \eqref{eq:img} into Eq.~\eqref{eq:IPR}, one finally obtains~\cite{mirlin1994statistical,rizzo2024localized,prado2026anderson}:
\begin{equation} \label{eq:IpQ}
    I_2 = 
    \frac{ \int  \de \hat{m} \; Q(0  ,\hat{m}) \, \hat{m}^{-1} }
    {\int  \de \hat{m} \; Q(0 ,\hat{m}) } \; .
\end{equation}
To sum up, although the moments of the local Green's functions are controlled by the fact that  $|{\cal G}_{ii}|$ is $O(1/\alpha)$ with probability $O(\alpha)$, they can be computed in terms of the {\it typical} values of ${\cal M}_{ii}$, whose real part is typically $O(1)$ and whose imaginary part is typically $O(\alpha)$. The fact that in the localized phase one can use the linearized equations to compute the relevant observables, such as the IPR, facilitates the adoption of highly efficient computational methods that strongly reduce 
the effect of the finite size of the population.

We thus introduce a modification of the population dynamics algorithm [see Ref.~\cite{rizzo2024localized} for more details] which allows us to perform the extrapolation of $Q(m,\hat{m})$ to $m=0$ very efficiently, thereby allowing one to evaluate the numerator and the denominator of Eq.~\eqref{eq:IpQ} with arbitrary accuracy.  In fact, from Eq.~\eqref{eq:mii1} we have that $m_{ii} = \epsilon_{i} - E - \sum_{m \in \partial i} {\cal G}_{m \to i}^R$. Hence, the probability that $m_{ii}=0$ is equal to the probability that $\epsilon_i = E + \sum_{m \in \partial i} g_{m \to i}$. This occurs with probability density $1/W$  if $|E + \sum_{m \in \partial i} g_{m \to i}|<W/2$, and with zero probability otherwise. Based on this observation, we thus proceed in the following way:
For a given pair of values $(E,W)$ in the localized phase, we implement the standard population dynamics algorithm described in Sec.~\ref{subsec:population} and obtain the stationary probability distribution of the real and imaginary parts of the cavity Green's functions $P(g,\hat g)$ in the linearized regime [see Eq.~\eqref{eq:SCanderson}], corresponding to the solution to Eqs.~\eqref{eq:ReGcavcrit}–\eqref{eq:ImGcavcrit}. We extract $K+1$ elements from the pool and compute $m$ and $\hat{m}$ from Eq.~\eqref{eq:mii1}. We define $S = E + \sum_{i=1}^{K+1} {\cal G}^R_{i}$. If (and only if) $|S|<W/2$ we add $\hat{m}^{-1}/W$ to the numerator and $1/W$ to the denominator of $I_2$. We repeat this process several times and divide the numerator and the denominator by the total number of attempts. We renew the elements of the pool of the cavity Green's function by performing a few steps of the standard population dynamics algorithm and repeat the whole process until the desired accuracy on $I_2$ is reached. 
The algorithm described here can be straightforwardly extended to the computation of generic two-point correlation functions.

\subsection{Evaluation of Localization Landscape percolation observables}
\label{app:numeval}

We now summarize the method that we used to compute numerically 
many of the quantities related to the Localization Landscape percolation. 
The following procedures have in common that they involve the sampling of pairs of 
cavity Green's functions and cavity rescaled fields from the $P(g,\eta)$ that solves
\begin{eqnarray}
    P(g,\eta) &=& 
    \int d\varepsilon\, \gamma_+(\varepsilon) \int \prod_{k=1}^K\left[dg_k d\eta_k \,P( g_k,\eta_k ) \right]\;
     \nonumber\\
    && \qquad
    \times \;  \delta\bigg(g-\frac{1}{\varepsilon-t^2\sum_k g_k}\bigg)
    \nonumber\\
    && \qquad \times \; \delta \bigg(\eta -1-t\sum_kg_k\eta_k\bigg)\,.
\end{eqnarray}
 $P(g,\eta)$ can be  computed from this equation via population dynamics, see Sec.~\ref{subsec:population}. 

 As argued in Sec.~\ref{subsubsec:joint}, if the connectivity of the lattice if high enough, with a 
 negligible error we can substitute $P(g,\eta)$ 
with the product $P_g(g)P_\eta(\eta)$ where the marginal distributions $P_g(g)$ and $P_\eta(\eta)$ are the ones derived in Sec.~\ref{subsubsec:marginals}.  This avoids repeated population dynamics simulations when varying $W$. Numerically, we found that the Pearson correlation coefficient $r_{g\eta}$ and the mutual information $I(g,\eta)$ are negligible for the parameter ranges that we used, confirming that the factorization approximation proposed in Eq.~\eqref{eq:MF} introduces negligible errors in high enough connectivity.

 \subsubsection{The occupation probabilities}
\label{subsubsec:qqbar}

The occupation probability $q$ for the Localization Landscape percolation problem [see Eqs.~\eqref{eq:Oiperc} and \eqref{eq:q}], 
can be written as
\begin{equation}
    q \equiv \Pr\{O_i=1\}= \mathbb{E}[O_i]
\end{equation}
and it can be computed explicitly through Monte Carlo sampling:
\begin{equation}
    q = \frac{1}{M}\sum_{m=1}^M O_i^{(m)}\,,
\end{equation}
where for each sample $m$ we draw $\{\mathcal{G}_{k\to i}^{(m)}, \eta_{k \to i}^{(m)}\}_{l \in \partial k}$ from $P(g,\eta)$ to compute the occupation variable $O_{i}^{(m)}$. Alternatively, in the high-connectivity limit, we have an analytical expression of 
the marginal distribution of $u_{i}$ [see Eq.~\eqref{eq:P_u}] and we can just calculate $q= \int du \;P_u(u) \, \theta(u-1/E_+)$.

The conditional probability $\bar{q}$ is defined in Eq.~\eqref{eq:qbar}. It reads
\begin{equation}
  \bar{q} \equiv \Pr \{ O_{k\to i} =1 \mid O_{i\to j}=1 \} 
  \; , 
\end{equation}
$\forall i, \; \forall j \in \partial i, \; \forall k \in \partial i \setminus j $, 
with the cavity occupation variables defined in Eq.~\eqref{eq:cavOi}, 
and governs the high-connectivity transition. It can computed numerically by rewriting it as
\begin{equation}
\label{eq:ProbOsol}
    \bar{q} = \frac{\Pr \{ O_{k \to i} =1, O_{i \to j}=1 \}}{\Pr \{ O_{i \to j} =1 \}} =  \frac{\mathbb{E}[O_{k\to i}O_{i\to j}]}{\mathbb{E}[O_{i\to j}]}\,.
\end{equation}
Again, the expectation values of this expression can be evaluated using Monte Carlo sampling, i.e.
\begin{eqnarray}
    \bar{q} &=& \frac{\sum_{m=1}^M O_{i \to j}^{(m)} O_{k \to i}^{(m)}}{\sum_{m=1}^M O_{i \to j}^{(m)}} 
    \nonumber\\
    & \overset{d}{=}& \frac{1}{K} \frac{\sum_{m=1}^M O_{i \to j}^{(m)} \sum_{k \in \partial i \setminus j} O_{k \to i}^{(m)}}{\sum_{m=1}^M O_{i \to j}^{(m)}}\,,
    \label{eq:qbarmonte}
\end{eqnarray}
where for each of the $M$ samples, $\{\mathcal{G}_{k\to i}^{(m)}, \eta_{k \to i}^{(m)}\}_{k \in \partial i \setminus j}$ are drawn from $P(g,\eta)$ to compute the occupation variables $O_{i \to j}^{(m)}$ and $\{O_{k \to i}^{(m)}\}_{k \in \partial i}$. The last equality in Eq.~(\ref{eq:qbarmonte}) follows from the statistical equivalence of the variables $\{O_{k \to i}\}_{k \in \partial i}$.

\subsubsection{The percolation correlation function and the average cluster size}
\label{subsubsec:CcorrS}
From its definition in Eq.~(\ref{eq:corrfuncnobond}), we compute the correlation function numerically by approximating the expectation value with the average over $M$ samples as
\begin{equation}
    C_{\rm perc}(r)=\frac{1}{M}\sum_{m=1}^M\prod_{s=0}^r O_s^{(m)}\,,
\end{equation}
where the $O_s$ are the ones defined in Eq.~(\ref{eq:Oiperc}).  
Each $u_s$ is computed in terms of the on-site Green's function  and the  rescaled field  as $u_s=\mathcal{G}_{ss}\eta_s$, where $\mathcal{G}_{ss}$ and $\eta_s$ are computed as functions of their cavity counterparts on the nearest-neighboring sites of the path between site $0$ and site $r$ (i.e. $\big\{(\mathcal{G}_{k\to i},\eta_{k \to i})\,|\,i\in\{0,\dots, r\}\,,k\in \partial i \cap\partial \{0,\dots, r\}\}$)  according to Eqs.~(\ref{eq:g})-(\ref{eq:caveta}). 
Thus, the $O_s^{(m)}$ of a given realization are all strongly correlated to each other and, in order to compute 
their product $\prod_{s=0}^r O_s^{(m)}$, we need to draw {\it all} $K(r+1)+2$ independent identically distributed pairs of 
cavity Green's functions and cavity rescaled fields from $P(g,\eta)$ incoming to all the nodes along the path.

The correlation length of $C_{\rm perc}(r)$ is evaluated by fitting the functional form $e^{-r/\xi_{\rm perc}}/K^r$ to the curve $C_{\rm perc}(r)$ in the non-percolating phase, for $r\gg 1$.

For the computation of $S$ we just use the definition in Eq.~(\ref{eq:avgclsize}). Therefore, we need to compute the correlation function $C_{\rm perc}(r)$ up to a cutoff distance $r_{max}$ and perform explicitly the sum in Eq.~(\ref{eq:avgclsize}). 

\section{Distributions and interesting limits}
\label{sec:analytic}

In this section, we provide the technical details for the derivation of several analytical results, some of which were already used in the preceding sections. In Sec.~\ref{subsec:indsites} we present the equation determining the parameter dependence of the percolation critical curve in the independent-site approximation. Next, in Sec.~\ref{subsec:hicon} we define the high-connectivity limit, we write the equations governing the system in this regime, 
we derive many exact and approximate results which provide very good approximations for finite and not too small $K$, and we obtain the exact solution in the high connectivity limit for the critical properties of the Localization Landscape percolation problem.  
In Sec.~\ref{subsec:ExVal} we study the exact edges of the supports of the marginal distributions of the Green's functions, the rescaled fields and the Localization Landscape variables, we summarize the approximate results for their bulks, and we perform an asymptotic analysis to determine their tail behaviors.

\subsection{The independent-site approximation}
\label{subsec:indsites}

The uncorrelated site percolation model on a lattice assumes that each site is independently occupied with probability $q$ (and not occupied with probability $1-q$). On the Bethe lattice with connectivity $K+1$ the critical value of the occupation probability $q_c$ above which the system is in the percolating phase is $1/K$ \cite{stauffer2018introduction}. The independent-site approximation for the Localization Landscape percolation problem consists of treating the components of the Localization Landscape $u_i$ on different sites as statistically independent. Under this assumption, the problem maps directly onto an uncorrelated percolation problem where the probability $q$ that a generic site $i$ is occupied, is given by the probability of $u_i$ being greater or equal than $1/E_+$ [see Eq.~\eqref{eq:q}]. Thus, the equation determining the critical curve $E^{\rm perc}_c(W)$ in the $(E,W)$-plane in the independent-site approximation reads 
\begin{equation} 
\label{eq:indsitescrit}
    \Pr\{u_i\geq1/E^{\rm perc}_{+,c}(W)\}\equiv q_c=1/K.
\end{equation}

The probability $q$ can be computed numerically as explained in Sec.~\ref{subsubsec:qqbar}.
The critical curve in the high-connectivity limit has been plotted in Fig.~\ref{fig:critcurve1} with a solid red line. The independent-site approximation 
curve deviates significantly from the exact one in the high-connectivity limit, indicating that correlations are not negligible 
while determining the critical behavior of the system. 
The independent-site critical curve matches with the exact one in a unique point, corresponding to $W=W_{\rm min}$, and for which the expected value of the Localization Landscape variables diverge (see Sec.~\ref{subsubsec:PD}).

\subsection{The high-connectivity limit}
\label{subsec:hicon}

 In the high-connectivity limit  $K \gg1$, many of the  quantities describing the transition 
 can be obtained either analytically, under a few approximations, or with little numerical effort. 
 Moreover, many of these results remain accurate even for low values of $K$, 
 making the high-connectivity limit a very convenient 
 regime to study the main features of 
 the transition. 
 
Here, we derive the equations that describe the critical curve in the high-connectivity limit in two ways. 
The first one is based on defining an integral eigenvalue equation analogous to the one that we derived 
in Sec.~\ref{sec:ALeigenval} for Anderson localization. This derivation lets us 
understand that the equations describing the percolation transition are different from the ones describing the localization transition. 
The second method is more straightforward, and it is based on enforcing that the average cluster size diverges at the transition.

The main idea behind the high-connectivity limit is that normal and cavity variables can be considered to be 
statistically equivalent, i.e.
\begin{align}
    \mathcal{G}_{ii} &\overset{d}{=} \mathcal{G}_{i\to j} \,,
      \\
       \eta_i & \overset{d}{=} \eta_{i\to j} \; ,
         \\
    u_i & \overset{d}{=} u_{i\to j}\overset{d}{=}\mathcal{G}_{i\to j}\eta_{i\to j} \,,
       \\
    p_i &  \overset{d}{=} p_{i\to j} \; , 
\end{align}
$\forall i \; {\rm and} \; \forall j \in \partial i$,  for $K\gg 1$, 
where the symbol $\overset{d}{=}$ represents equality in distribution.
 This is because recursive equations for normal and cavity variables differ by just one of the $O(K)$ terms inside sums of the type $\sum_{k\in \partial i}$ or $\sum_{k\in \partial i \setminus j}$, and according to large deviation theory, under the hypothesis that the 
 cavity variables are random variables with finite mean and variance, we can safely neglect one of the terms in the sums. 

 The sets of cavity variables $\{\mathcal{G}_{k\to i}, \eta_{k\to i},p_{k\to i}\}_{k \in \partial i \setminus j}$ are not independent in the general case, because for each of the $p_{k \to i}$ one has to compute $u_k$, and all the $u_k$'s are statistically dependent, since they depend on quantities evaluated at their common neighbor $i$. 
 Now, since $u_k\overset{d}{=}\mathcal{G}_{k \to i}\eta_{k \to i}$, the cavity percolation probabilities become independent cavity variables. For this reason, the three cavity equations that determine the transition in the high-connectivity limit are
 \begin{align}
     \label{eq:gsys}
    \mathcal{G}_{i\to j}^{-1} & =  \varepsilon_i - t^2 \sum_{k \in \partial i \setminus j} \mathcal{G}_{k \to i} \, ,\\
    \label{eq:etasys}
    \eta_{i \to j} & = 1 + t \sum_{k \in \partial i \setminus j} \mathcal{G}_{k \to i } \eta_{k \to i}   \, ,\\
    \label{eq:psys}
    p_{i \to j} &= \theta(\mathcal{G}_{i\to j}\eta_{i\to j} -1/E_+)\sum_{k \in \partial i \setminus j} p_{k \to i} \, ,
 \end{align}
 where the last one has been obtained by expanding Eq.~\eqref{eq:cavpi} for small cavity percolation probabilities close to the critical curve. The self-consistent distributional equation that we have to solve is much simpler compared to Eq.~(\ref{eq:stochSC}), and reads
 \begin{align}
    P(g,\eta,p)&=\int d\varepsilon \,\gamma_+(\varepsilon)
    \int \prod_{k=1}^K \left[dg_kd\eta_kdp_k \,P( g_k,\eta_k,p_k ) \right]
    \; 
    \nonumber\\
    &
    \qquad \times
    \delta \bigg ( g-\frac{1}{\varepsilon-t^2\sum_kg_k}\bigg ) 
    \nonumber\\ 
    & \qquad \times \delta \bigg (  \eta- 1-t\sum_k g_k \eta_k \bigg ) 
    \nonumber\\
    &
    \qquad \times
    \delta \bigg (  p -\theta(g\eta-1/E_+)\sum_k p_k \bigg )
    \; . 
\end{align}
 
 In Sec.~\ref{subsubsec:marginals} we show that the marginal distributions of $\mathcal{G}_{i \to j}$, $\eta_{i \to j}$ and $u_{i\to j}$ admit solutions with finite mean and variance, 
 and that these solutions are very close to the ones obtained with population dynamics simulations also for $K \sim O(1)$. 
 In Sec.~\ref{subsubsec:joint} we show numerically
  that the joint distribution $P(g,\eta)$ can effectively be approximated with the product of their marginals. Section~\ref{subsubsec:lowW} determines the position of the isolated eigenvalue as a function of disorder and the minimal disorder needed to have a percolation transition.
In Sec.~\ref{subsubsec:linstab} we perform a linear stability analysis and in Sec.~\ref{subsubsec:divergence-cluster-size}
we study the  average cluster size to locate the critical percolation curve.

\subsubsection{Marginal distributions}
\label{subsubsec:marginals}

 In order to obtain a self-consistent solution for the marginal distribution of the cavity Green's function $P_g(g)= \int d\eta dp \, P(g,\eta,p)$, we observe that Eq.~(\ref{eq:gsys}) 
 is closed, therefore it can be solved independently from the other cavity equations.
If $K$ is large enough, assuming that $P_g$ has finite mean and variance,  we can approximate the sum over the 
nearest neighbors of site $i$ in absence of $j$ as  
\begin{equation}
\label{eq:CLTG}
\sum_{k\in \partial i \setminus j} \mathcal{G}_{k\to i} \approx K \mu_g +\sqrt{K\sigma_g^2}Y
\; ,
\end{equation}
where
\begin{equation}
    Y  \sim \mathcal{N}(0,1) \; , \quad
    \mu_g=\mathbb{E}[\mathcal{G}_{k\to i}] \; , \quad   \sigma_g^2 = \mathbb{V}[\mathcal{G}_{k\to i}] \; ,
\end{equation}
and $\mathbb{V}[ \dots ]$ is the variance of the random variable. 
This follows directly from the central limit theorem. Now, the roughest approximation that we can make is to neglect Gaussian fluctuations.
In this limit the probability distribution of the cavity Green's functions is simply obtained by changing variables in the probability distribution of $\varepsilon_i$.
Thus, the probability distribution of the cavity Green's functions becomes
\begin{equation}
\label{eq:roughappr}
    Q_g(g;\mu_g) = 
    \begin{cases}
        \displaystyle \frac{1}{W g^2} & \qquad \text{if } g \in D_g
        \; , \\
        0 & \qquad \text{else}
        \; ,  
    \end{cases}
\end{equation}
with 
\begin{displaymath}
D_g \equiv\left[\frac{1}{W/2-E_{\rm sh}-K t^2 \mu_g },\; \frac{1}{ -W/2-E_{\rm sh}-K t^2 \mu_g}\right] 
\end{displaymath}
From this expression, $\mu_g$ and $\sigma^2_g$ are easily obtained self-consistently by solving
\begin{equation}
\label{eq:gavgapprox}
\mu_g = \frac{1}{W} \ln\left| 1-\frac{W}{W/2+E_{\rm sh} +K t^2 \mu_g} \right|
\; ,
\end{equation}
and computing
\begin{equation}
\label{eq:gvarapprox}
    \sigma^2_g = \frac{1}{(K t^2 \mu_g+E_{\rm sh})^2-W^2/4}-\mu_g^2
    \; .
\end{equation}
More generally, the $n$-th moment $\mu_{g,r}$ of the cavity Green's function can be expressed in a nicer form as
\begin{equation}
    \mu_{g,n}=\mu_g^nV_n\left(\frac{W\mu_g}{2}\right)\,,
\end{equation}
where we have defined
\begin{equation}
V_n(x)=
\frac{\sinh\!\big((n-1)x\big)}{(n-1)x}
\left(\frac{\sinh x}{x}\right)^{n-1},
\qquad n> 1\,.
\end{equation}
In terms of $V_2$ the variance reads
\begin{equation}
    \sigma_g^2 = \mu_g^2\,\left[V_2\left(\frac{W\mu_g}{2}\right)-1\right]\,.
\end{equation}
Another useful quantity for characterizing the physics of the system is the logarithmic mean, which is given by
\begin{equation}
\mathbb E\left[\ln\mathcal{G}_{i \to j}\right] = \ln \mu_g - T\!\left(\frac{W\mu_g}{2}\right),
\end{equation}
with
\begin{equation}
T(x)=x\,\coth x-1-\ln\!\left(\frac{\sinh x}{x}\right)\,,
\end{equation}
from which one can define compute the typical value of a cavity Green's function as
\begin{equation}
    \mu_g^{\rm typ}\equiv e^{\mathbb E [\ln\mathcal{G}_{r \to j}]}\,.
\end{equation}

From population dynamics simulations we have observed that these approximations of mean and typical value are very accurate already for $K=2$ for any value of $W$, while the estimate of the variance becomes accurate at low $W$ only for slightly higher $K$, namely $K>3$. 

A more precise expression for the full probability distribution of the cavity Green's functions can be found keeping the Gaussian fluctuations. Starting from
\begin{equation}
    \mathcal{G}_{i\to j}^{-1} \approx \varepsilon-t^2\Big( K \mu_g +\sqrt{K\sigma_g^2}Y\Big)
    \; ,
\end{equation}
we obtain
\begin{eqnarray}
\label{eq:Pgacc}
    &&
    P_g(g;\mu_g,\sigma_g^2) = 
    \nonumber\\
    &&
    \frac{1}{W g^2} \int_{-W/2-E_{\rm sh}}^{W/2-E_{\rm sh}} \!\!  \!\! d\varepsilon \, \mathcal{N} \Big( \varepsilon \,; Kt^2 \mu_g +1/g \,,\, Kt^4\sigma_g^2\Big) 
    \, ,
    \qquad\;\;\;
\end{eqnarray}
where  mean and variance have to be obtained self-consistently from 
\begin{eqnarray}
\label{eq:SCgind1}
\mu_g &=& \int dg\,P_g(g\,;\mu_g,\sigma_g^2)g\,,
\\
\label{eq:SCgind2}
\sigma^2_g &=& \int dg\,P_g(g\,;\mu_g,\sigma_g^2)(g-\mu_g)^2 \,.
\end{eqnarray}

We compute now the marginal distribution of the cavity rescaled fields by employing a similar approximation. We assume that
\begin{eqnarray}
    && \eta_{i\to j} \approx 
    \nonumber
    \\
    && \ \  1+t\Big(  K\mathbb{E}[\mathcal{G}_{k \to i }\eta_{k \rightarrow i}] + \sqrt{K\mathbb{V}[\mathcal{G}_{k \to i }\eta_{k \rightarrow i}]}Y\Big)\,,  \qquad
    \label{eq:approxsumeta}
 \end{eqnarray}   
  with 
  \begin{equation}
   Y  \sim \mathcal{N}(y\,;0,1)
    \; . 
\end{equation}

We will argue later that this Gaussian approximation does not hold, even at high connectivity, when the disorder is too strong. This is because the distribution of the cavity Green’s function becomes highly asymmetric in the strong-disorder regime. Consequently, this asymmetry propagates to the distribution of the cavity rescaled fields, which thereby loses its Gaussian form. Nevertheless, as we will see, it remains sufficiently accurate to provide an excellent estimate of the marginal distribution of the cavity Localization Landscape variables, even at low connectivity.

The variables $\eta_{k \to i}$ and $\mathcal{G}_{k \to i }$ are not independent. Therefore, computing the expectation value and the variance of their product in Eq.~(\ref{eq:approxsumeta}) is not trivial. However, for high-connectivity, we can argue that they are weakly correlated. 
Therefore, we will  consider them as independent, i.e.
\begin{equation}
\label{eq:MF}
    P(g,\eta) = \int dp \; P(g,\eta,p)\approx P_g(g)P_{\eta}(\eta) \; .
\end{equation}

Under this approximation,  the mean and variance in Eq.~(\ref{eq:approxsumeta}) now read
\begin{eqnarray}
\label{eq:meanvaretag}
    \mathbb{E}[\mathcal{G}_{k \to i }\eta_{k \rightarrow i}] &=& \mu_g\mu_{\eta}\,, 
    \\
\label{eq:meanvaretag2}
    \mathbb{V}[\mathcal{G}_{k \to i }\eta_{k \rightarrow i}] &=& \sigma_g^2\sigma_{\eta}^2+\sigma_g^2\mu_{\eta}^2+\sigma_{\eta}^2\mu_g^2
    \; , 
\end{eqnarray}
where
\begin{equation}
    \mu_\eta=\mathbb{E}[\eta_{k\to i}]\,, \qquad  \sigma_{\eta}^2 = \mathbb{V}[\eta_{k\to i}]\,.
\end{equation}
This implies that the probability distribution of the cavity rescaled fields is
\begin{equation}
\label{eq:Peta}
    P_{\eta}(\eta;\mu_{\eta},\sigma_{\eta}^2) = \mathcal{N}(\eta \,;\mu_{\eta},\sigma_{\eta}^2) \,, 
\end{equation}
with
\begin{eqnarray}
\label{eq:etastats}
\mu_{\eta} &=& \frac{1}{1-K t \mu_g }\,, 
\\
\label{eq:etastats2}
 \sigma_{\eta}^2 &=& \frac{Kt^2\sigma^2_g}{[1 - Kt^2 (\sigma^2_g+\mu_g^2)](1-Kt\mu_g)^2}\,.   
\end{eqnarray}
These expressions have been obtained by enforcing that both sides of Eq.~(\ref{eq:approxsumeta}) have the same mean and variance.

The factorization of $P(g, \eta)$ in Eq.~(\ref{eq:MF})  also simplifies the calculation of the marginal distribution of the cavity
 Localization Landscape variables, 
which in distribution are given by 
 $u_{i \to j} \overset{d}{=} \mathcal{G}_{i\to j} \eta_{i \to j}$. The result is
\begin{align}
\label{eq:P_u}
    P_u(u) &= \int dg \,P(g\, ,\eta =u/g )\left | \frac{d(u/g)}{du}\right| 
    \nonumber\\
    & = \int \frac{dg}{|g|} \, P_g(g)P_{\eta} (u/g)
    \nonumber\\ 
    & = \int dg \, P_g(g) \, \mathcal{N}(u \, ; g\mu_{\eta} \, ,g^2\sigma^2_{\eta})
    \; .
\end{align}

\begin{figure}[!ht]
    \begin{center}
         \includegraphics[width=0.48\textwidth]{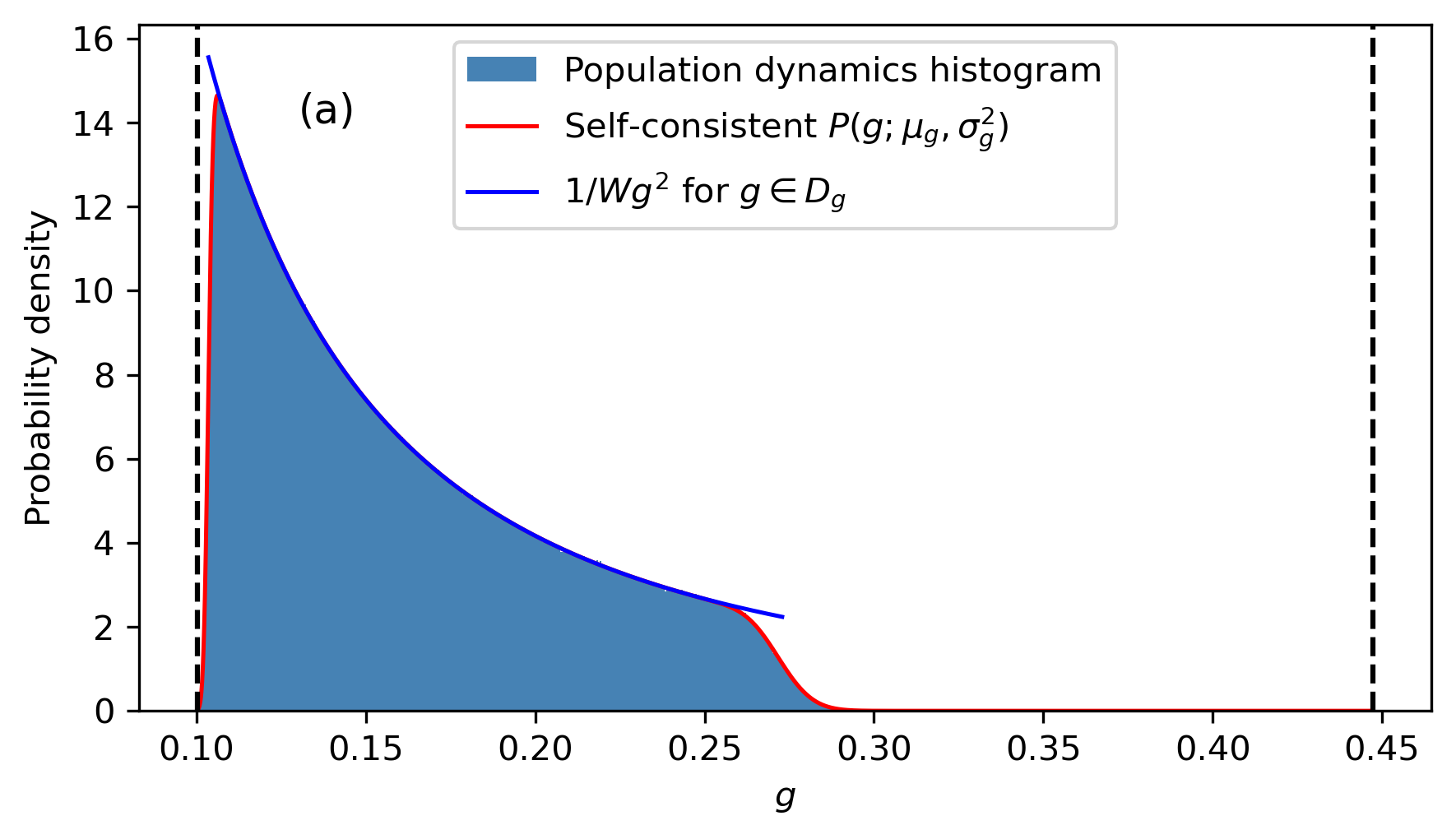}
        \includegraphics[width=0.48\textwidth]{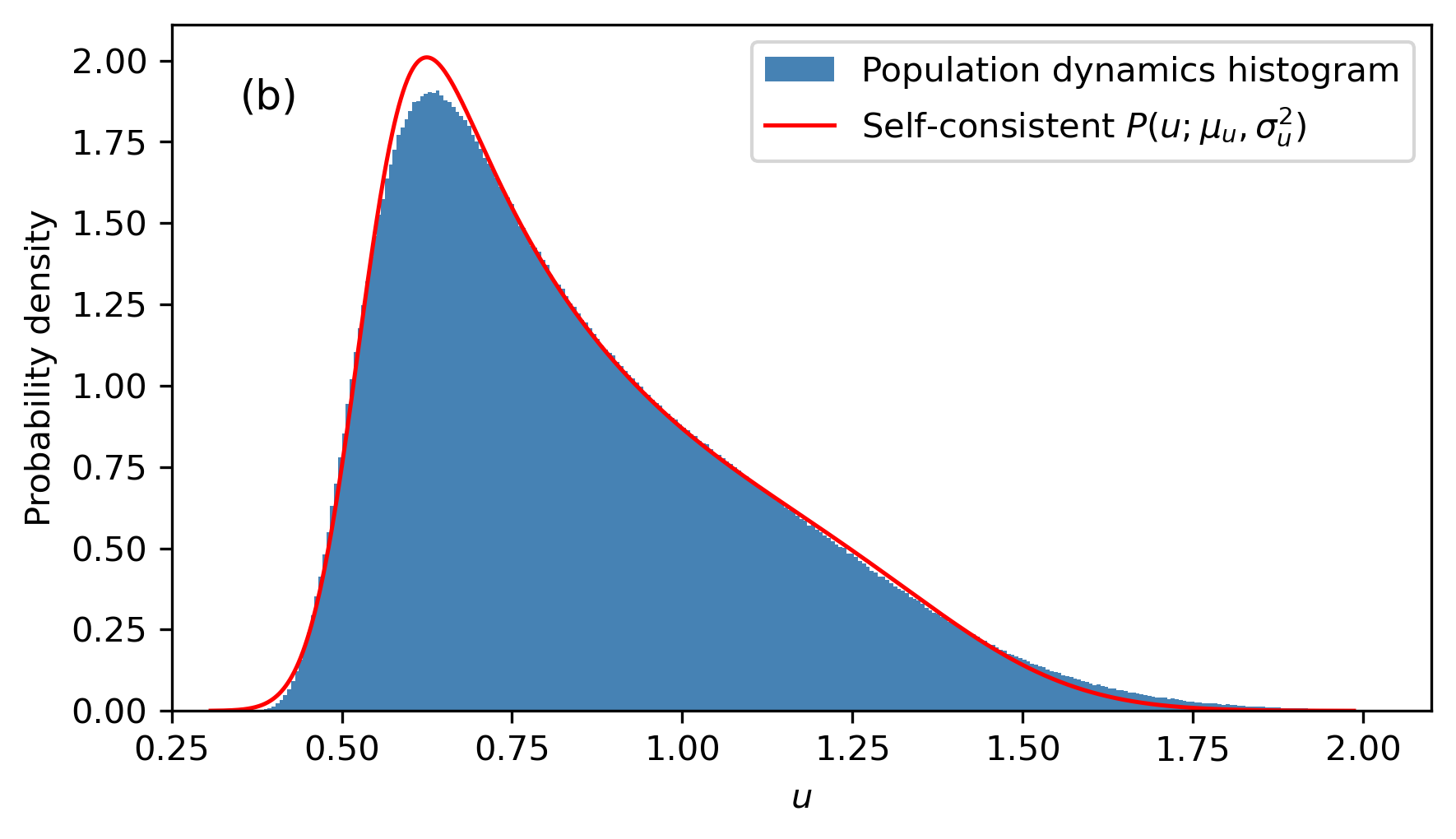}
    \end{center}
       \vspace{-0.5cm}
    \caption{\small{Marginal distributions of cavity Green's functions $\mathcal{G}_{i \to j}$ (a) and cavity Localization Landscape variables $u_{i\to j}$ (b) for  $K=5$, $t=1$, and $W=6$. 
      Light blue histograms: empirical distributions obtained with population dynamics with $N=10^7$. 
    (a) Red curve: Probability distribution in Eq.~(\ref{eq:Pgacc}) with parameters $\mu_g$ and $\sigma_g^2$ computed solving by iteration the self-consistent 
    Eqs.~(\ref{eq:SCgind1}) and (\ref{eq:SCgind2}). Blue curve: rougher approximation for the probability distribution given in 
    Eq.~(\ref{eq:roughappr}) with  $\mu_g$ computed self-consistently by iteration of Eq.~(\ref{eq:gavgapprox}). Vertical dashed lines: exact upper and lower bounds of the support of the distribution valid at any connectivity, 
    Eqs.~\eqref{eq:Mg} and \eqref{eq:mg}.
     (b) Red curve: probability distribution $P_u$ in Eq.~(\ref{eq:P_u}), with parameters $\mu_u$ and $\sigma^2_u$ obtained self-consistently from Eqs.~(\ref{eq:ustats}).}
     }
    \label{fig:PgPu}
\end{figure}

We can observe that $u_{i\to j}$ is the quantity with mean and variance 
 in Eqs.~(\ref{eq:meanvaretag}) and (\ref{eq:meanvaretag2}). Thus, using the results in Eqs.~(\ref{eq:etastats}) and \eqref{eq:etastats2} we have
\begin{eqnarray}
    \label{eq:ustats}
    \mu_u \! &\! = \! & \!  
    \mu_{\eta}\mu_g = \frac{\mu_g}{1-Kt\mu_g}\,,
    \\
    \sigma^2_u \! &\! = \! & \!  
    \frac{\sigma^2_{\eta}}{Kt^2}=\frac{\sigma^2_g}{[1 - Kt^2 (\sigma^2_g+\mu_g^2)](1-Kt\mu_g)^2}\,.
\end{eqnarray}

In Fig.~\ref{fig:PgPu} we plotted the distribution of the cavity Green's functions  $\mathcal{G}_{i \to j}$ (a) 
and the distribution of $u_{i \to j}=\mathcal{G}_{i\to j}\eta_{i \to j}$ (b) from population dynamics simulations
in a model with $K=5$, $t=1$, and $W=6$. 
We compare the histograms (light blue) to the large $K$ approximations (red curves) and found that the agreement is very good.

\subsubsection{Quality of the factorization approximation}
\label{subsubsec:joint}

The joint probability distribution of the cavity Green's function and the cavity auxiliary fields $P(g,\eta)$ is the one satisfying the self-consistent distributional equation
\begin{eqnarray}
\label{eq:PgetaSC}
     P(g,\eta) &=&
     \int d\varepsilon\, \gamma_+(\varepsilon) \int \prod_{k=1}^K\left[dg_k d\eta_k \; P( g_k,\eta_k )\right]
     \nonumber\\
     && 
     \times \; \delta\bigg(g-\frac{1}{\varepsilon-t^2\sum_k g_k}\bigg)
      \nonumber\\
     && 
     \times \;
     \delta \bigg(\eta -1-t\sum_kg_k\eta_k\bigg)\,.
\end{eqnarray}
As we have anticipated in the previous Section,  $P(g,\eta)$ is well approximated by the product of the two marginals $P_g(g)$ and $P_\eta(\eta)$ as in Eq.~(\ref{eq:MF}), and checked numerically. The quality of the approximation can be further quantified by computing the mutual information.

The mutual information of two continuous random variables $X,Y$ distributed with $P$ is defined as
\begin{equation}
    I(X,Y) \equiv \int dxdy\; P(x,y) \, \ln_2 \frac{P(x,y)}{P(x)P(y)}\,,
\end{equation}
and it represents the information theoretical measure of the dependence of two random variables. More precisely, it is the information about the variable $X$ that one obtains after measuring $Y$, quantified in terms of Shannon's entropy.
The mutual information of two random variables is always greater or equal than zero. $I(X,Y)=0$ corresponds to the case in which $X$ and $Y$ are statistically independent.

\begin{figure}[!ht]
\vspace{0.5cm}
    \begin{center}
    \includegraphics[width=\linewidth]{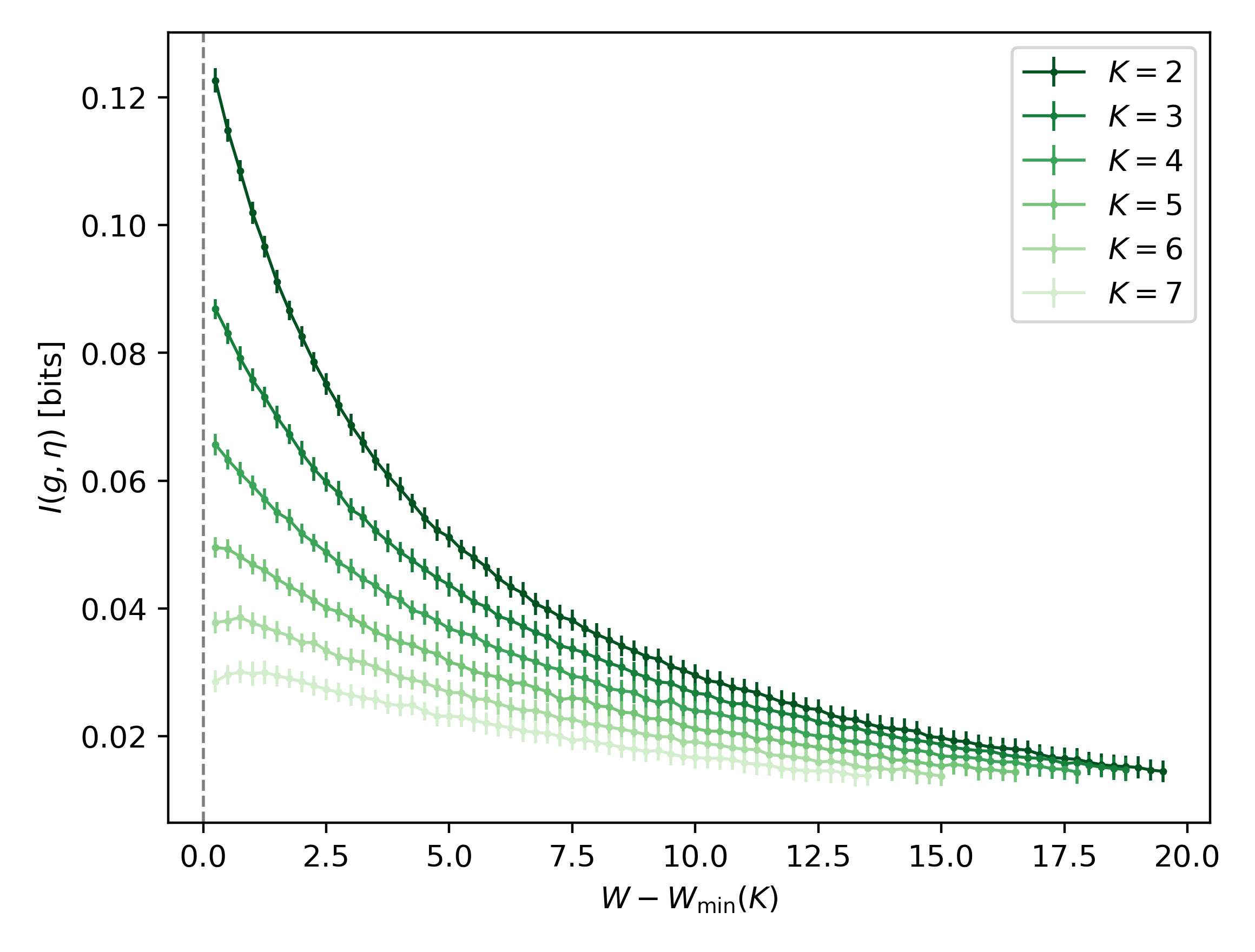}
    \end{center}
    \vspace{-0.5cm}
    \caption{\small{Mutual information of cavity Green's functions and cavity auxiliary fields. The mutual information, computed numerically following the procedure devised in Ref.~\cite{kraskov2004estimating} from the population dynamics distribution with a population of $N=10^6$ cavity variable pairs, is plotted for many values of the disorder strength above the lower critical disorder $W_{\rm min}$ (see Sec.~\ref{subsubsec:lowW}) for multiple values of $K$ and $t=1$. As it can be seen from the plot, the mutual information decreases monotonically with $W$, remaining always below $6\cdot 10^{-2}\, \mbox{\rm bits}$ for $K\ge5$. This indicates that the factorization approximation of Eq.~\eqref{eq:MF} represents the exact joint distribution with good accuracy.}}
    \label{fig:MI}   
\end{figure}

The estimation of the mutual information can be done using population dynamics. 
Assuming that the population of $N$ pairs of cavity Green's functions and cavity rescaled fields represents a set of $N$ samples drawn from the exact distribution, we can compute $I(\mathcal{G}_{i\to j},\eta_{i\to j})$ empirically through the procedure devised in Ref.~\cite{kraskov2004estimating}.
We computed the mutual information for $K=2,\dots,7$, and  many values of $W$ in the range that will be used later to compute the critical curve. The results are plotted in Fig.~\ref{fig:MI}. As we can see from the plot, the mutual information decreases with $K$ and $W$, meaning that for high enough $K$ the factorization approximation is justified. 

The mutual information analysis indicates that the quantities $\mu_\eta$ and $\sigma^2_\eta$ can be computed self-consistently with minimal error through Eqs.~\eqref{eq:meanvaretag}, and that $\mu_u$ and $\sigma^2_u$ can then be obtained from Eqs.~\eqref{eq:ustats}.

It should be noted that the mutual information result does not guarantee that the analytical expression of the cavity rescaled field distribution in Eq.~\eqref{eq:Peta} -- derived under the assumption of independence between $g$ and $\eta$ -- accurately reproduces the true marginal. This expression relies on the additional approximation introduced in Eq.~\eqref{eq:approxsumeta}.
However, the self-consistent distribution of Eq.~\eqref{eq:P_u}, which incorporates the Gaussian approximation of Eq.~\eqref{eq:Peta}, closely matches the marginal distribution of the cavity Localization Landscape variables obtained from population dynamics.

As discussed in Sec.~\ref{subsubsec:divergence-cluster-size}, the percolation transition at high connectivity is governed by the statistics of the cavity Localization Landscape variables. Since the self-consistent prediction of Eq.~\eqref{eq:P_u} accurately reproduces the numerical results, we computed the high-connectivity critical curve by sampling from the factorized distribution $P_g(g)P_\eta(\eta)$, determined self-consistently from Eqs.~\eqref{eq:Pgacc} and \eqref{eq:Peta}. The resulting critical energies (solid black line in Fig.~\ref{fig:critcurve1}) show excellent agreement with the exact ones obtained from sampling the full joint distribution (black dots in Fig.~\ref{fig:critcurve1}).

\begin{figure*}[t]
    \centering
    \includegraphics[width=0.95\textwidth, height=0.4\textheight, keepaspectratio]{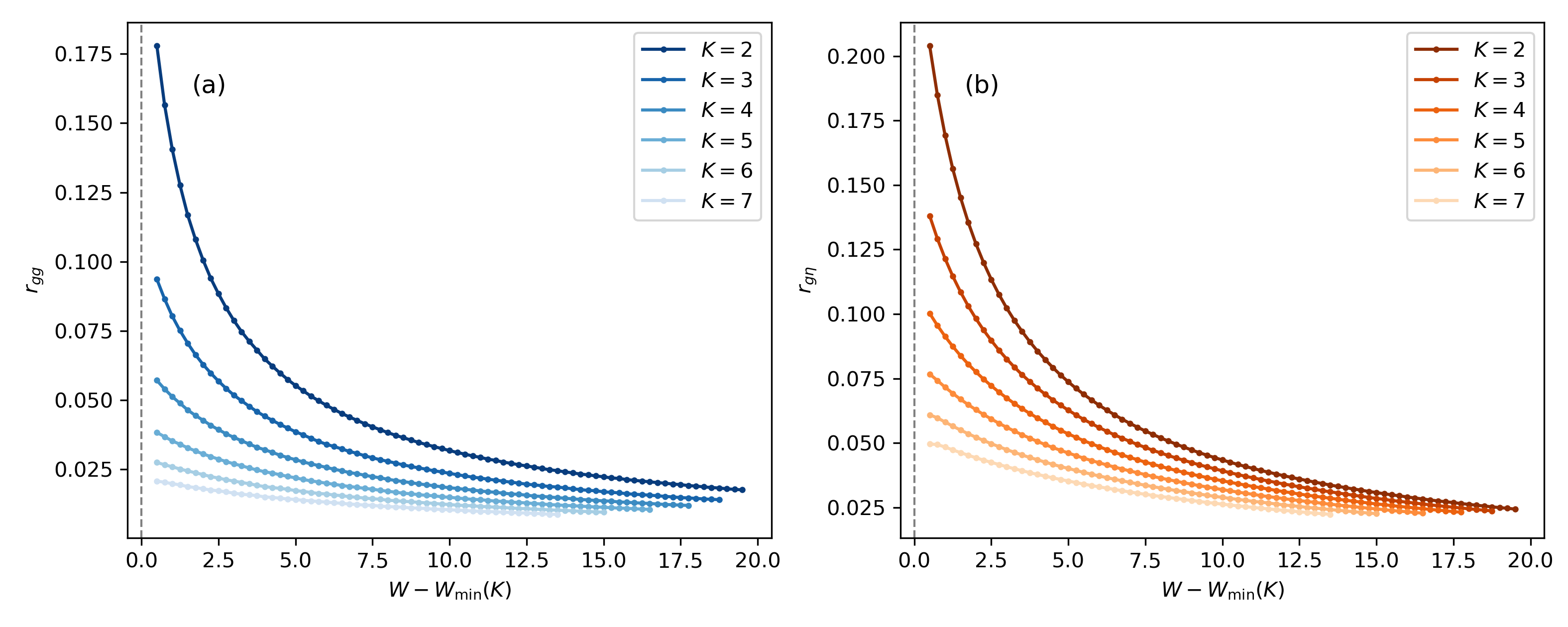}
    \caption{\small{Correlations of cavity variables. (a) Correlation between $\mathcal{G}_{i \to j}$ and one of its neighbors $\mathcal{G}_{k\to i}$, where $k \in \partial i$ [see Eq.~\eqref{eq:corrgg}]. (b) Correlation between $\mathcal{G}_{i \to j}$ and $\eta_{i \to j}$ [see Eq.~\eqref{eq:corrgeta}].}}
    \label{fig:corrgeta}
    
    \vspace{0.5cm} 

    \includegraphics[width=0.95\textwidth, height=0.4\textheight, keepaspectratio]{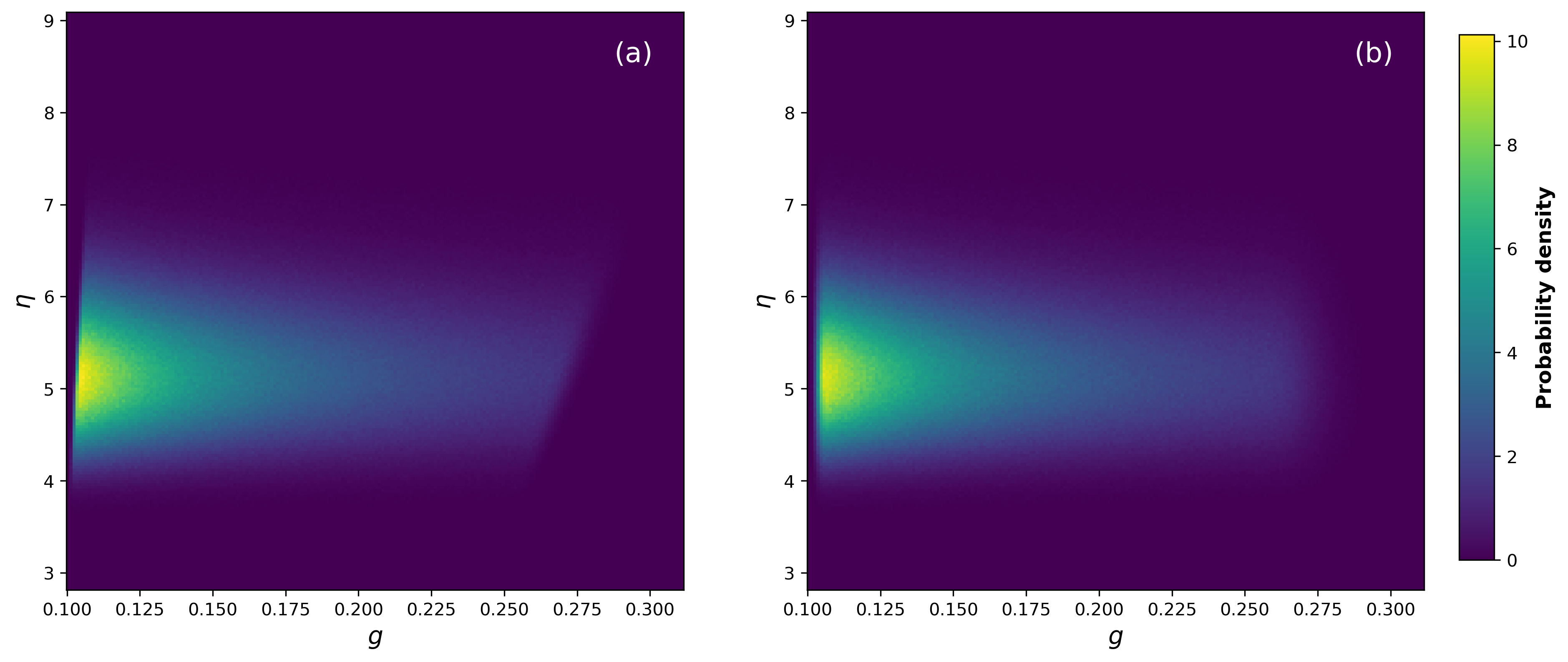}
    \caption{\small{Plots of $P(g,\eta)$ and its mean-field approximation for $K=5$, $t=1$, and $W=6$. (a) Histogram of the joint probability distribution obtained through population dynamics (see Sec.~\ref{subsec:population}) with $N=10^7$. (b) The mean-field approximation obtained by shuffling the cavity Green's functions of the pairs $(g,\eta)$ from panel (a).}}
    \label{fig:Pgetas}
\end{figure*}

A further confirmation of the validity of the factorization approximation is given by the numerical evaluation of the correlation between cavity Green's functions and cavity rescaled fields, i.e.
\begin{equation}
\label{eq:corrgeta}
    r_{g\eta}=\frac{\mathbb E[\mathcal{G}_{i\to j}\eta_{i \to j}]-\mu_g\mu_\eta}{\sigma_g\sigma_\eta}\,,
\end{equation}
computed numerically through Monte Carlo sampling of the joint distribution obtained from population dynamics [see plot (b) in Fig.~\ref{fig:corrgeta}].
Since the difference between $\mathbb E [g\eta]$ and $\mu_g\mu_\eta$ gets smaller as the connectivity increases, then for the purpose of computing expectation values of cavity variables the factorization approximation is justified [see Eqs.~\eqref{eq:meanvaretag} and \eqref{eq:ustats}].

In Fig.~\ref{fig:Pgetas} we represented the colormap plots of the histograms of the joint probability distribution $P(g,\eta)$, and its factorized approximation $P_g(g)P_\eta(\eta)$. The first one has been computed numerically through population dynamics, while the second one has been obtained by shuffling the $g$-variables between the pairs $(g,\eta)$ in the latter distribution, erasing the dependencies of the two variables. As we can see from the plots the two distributions are qualitatively similar. The main difference between the two is that the tails of the distributions have a slightly different shape in the $(g,\eta)$ plane. This is due to the fact that the values that $\eta_{i \to j}$ can take once $\mathcal{G}_{i \to j }$ is given are restricted by the coupled equations \eqref{eq:gsys} and \eqref{eq:etasys}. As revealed from the mutual information measurements, this feature does not result in significant effects on the global dependency of the two variables. 

In Section~\ref{subsubsec:lowW}, we will rely on the assumption that the 
cavity variables which depend on each other along a chain are weakly correlated, in particular, for the cavity Green's functions in Eq.~\eqref{eq:FactorAverage1}. In order to verify this assumption numerically, it is sufficient to show that the correlation between $\mathcal{G}_{i \to j}$ and one of its neighbors $\mathcal G_{k \to i}$, i.e.
\begin{equation}
\label{eq:corrgg}
    r_{gg}= \frac{\mathbb E[\mathcal{G}_{i\to j}\mathcal{G}_{k \to i}]-\mu_g^2}{\sigma_g^2}\,,
\end{equation}
is small enough, as this implies that longer distance correlations are surely negligible. For this reason, we computed $r_{gg}$ numerically for $K=1,\dots,7$ by Monte Carlo sampling from the joint distribution obtained with population dynamics. We observe that, as expected, as $K$ and $W$ increase the correlations decreases. Already for $K=5$ they turn out to be negligible for any value of $W$, see Fig.~\ref{fig:corrgeta}(a).

\subsubsection{Lower bound for the critical disorder and isolated eigenvalue}
\label{subsubsec:lowW}

It is important to note that, for physical consistency, the expectation value of the cavity rescaled fields must be non-negative, since the opposite would imply negative Localization Landscape variables, which is impossible, as explained in the End Matter of Ref.~\cite{Tonetti2026}. 
Therefore, Eqs.~(\ref{eq:etastats}) are valid only in the regime where
\begin{equation}
\label{eq:condboundW}
    0\leq \mu_\eta<+\infty \iff \mu_g <  \frac{1}{Kt}
    \; .
\end{equation}
A more formal derivation of this condition will be given below.

In the high-connectivity limit, Eq.~(\ref{eq:roughappr}) indicates that $\mu_g$ is a decreasing function of the disorder strength for $W>W_{\rm min}$. For $W\le W_{\rm min}$, however, the value of $\mu_g$ depends explicitly on the reference energy $E_{\rm sh}(W)$. For $W<W_{\rm min}$, the standard definition of LLT requires the reference energy to be equal to the isolated eigenvalue $E_{\rm iso}(W)$. Its value was computed numerically as a function of $W$ in Ref.~\cite{biroli2010anderson} by exploiting the fact that the Gaussian integral
\begin{equation}
    \label{eq:gaussy}
    \langle x_i\rangle = \frac{1}{Z_0}\int {\cal D}{\mathbf x} \;  x_i \; e^{-S_0[{\mathbf x}]}\,,  
\end{equation}
where
\begin{equation}
    S_0[\mathbf{x}]=\frac{1}{2}\mathbf{x^t}\left(\hat{\mathcal{H}}-E_{\rm sh}\hat{\mathcal{I}}\right)\mathbf{x}\,,
\end{equation}
and $Z_0$ is a normalization constant, must diverge when $E_{\rm sh}>E_{\rm iso}$. By introducing a source term $Jx_i$ in the action, one can follow a procedure analogous to that discussed in the End Matter of~\cite{Tonetti2026} to compute this expectation value. Parametrizing a generic cavity marginal distribution as
\begin{equation}
    \mu_{i\to j}(x_i) \propto e^{-\frac{x_i^2}{2 \, \mathcal{G}_{i \to j}} + y_{i \to j} x_i} \, ,
\end{equation}
the cavity equations to be enforced are:
\begin{eqnarray}
    \mathcal{G}_{i\to j}^{-1} &=& \varepsilon_i - t^2 \sum_{k \in \partial i \setminus j} \mathcal{G}_{k \to i}    
    \; , 
    \nonumber\\
    y_{i \to j}  &=&  t \sum_{k \in \partial i \setminus j} 
    \mathcal{G}_{k \to i} y_{k \to i}   \, . 
    \label{eq:ycav}
\end{eqnarray}
Here, the $E_{\rm sh}$-dependence is contained as usual in the random variables $\varepsilon_i$, which are uniformly distributed in the interval $[-W/2-E_{\rm sh}, W/2-E_{\rm sh}]$ (we denote their distribution by $\gamma_+$).
The recursion for the variables $y_{i \to j}$ has only two fixed points: $0$ and $\infty$. The fixed point at $0$ corresponds to the case in which the Gaussian integral in Eq.~\eqref{eq:gaussy} converges. By identifying the value of $E_{\rm sh}$ at which this recursion first diverges, one can determine $E_{\rm iso}$ (see also Ref.~\cite{biroli2010anderson} for an interpretation of this results in terms of Bose-Einstein condensation in the ground-state energy).

In terms of distributions, we are interested in understanding when the solution of the integral equation
\begin{eqnarray}
\label{eq:Pgy}
     P(g,y) &=&
     \int d\varepsilon\, \gamma_+(\varepsilon) \int \prod_{k=1}^K\left[dg_k dy_k \; P( g_k,y_k )\right]
     \nonumber\\
     && 
     \qquad \times \; \delta\bigg(g-\frac{1}{\varepsilon-t^2\sum_k g_k}\bigg)
      \nonumber\\
     && 
     \qquad \times \;
     \delta \bigg(y -t\sum_kg_ky_k\bigg)\,,
\end{eqnarray}
transitions from having zero to non-zero expectation value of $y$.
Determining the value of $E_{\rm sh}$ at which this happens is rather simple once a few assumptions are made, assumptions that can be verified numerically (see Sec.~\ref{subsubsec:joint}). 

We present here an argument based on studying directly the iteration of the cavity equation~\eqref{eq:ycav}. For a given cavity root site $i$, in the absence of site $j$, let $\mathcal{P}(r)$ denote the set of all directed paths of length $r$ starting from $i$ and going away from the root. A path $\rho \in \mathcal{P}(r)$ is written as
\[
\rho = (\rho_0,\rho_1,\dots,\rho_r),
\]
with $\rho_0=i$. We then define the truncated cavity field
\begin{equation}
\label{eq:yrdef}
y^{(r)}_{i \to j}
=
t^r \sum_{\rho \in \mathcal{P}(r)}
\prod_{s=1}^{r}
\mathcal{G}_{\rho_s \to \rho_{s-1}} \, .
\end{equation}
By construction, the full cavity variable is obtained in the limit
\begin{equation}
\label{eq:yfullfromyr}
y_{i \to j}
=
\lim_{r \to \infty} y^{(r)}_{i \to j} \, .
\end{equation}
It is useful to rewrite Eq.~\eqref{eq:yrdef} in exponential form. For each path $\rho \in \mathcal{P}(r)$ we define
\begin{equation}
\label{eq:Srdef}
S_r(\rho)
=
\sum_{s=1}^{r} \ln  \mathcal{G}_{\rho_s \to \rho_{s-1}} \, .
\end{equation}
Then, the truncated field can be written as
\begin{equation}
\label{eq:yrpartition}
y^{(r)}_{i \to j}
= t^r
\sum_{\rho \in \mathcal{P}(r)}
e^{ S_r(\rho)}
\, .
\end{equation}
This representation shows that $y^{(r)}_{i \to j}$ can be naturally interpreted as a partition sum over all paths of length $r$ on the rooted tree.
The expectation value of $y_{i \to j}^{(r)}$ can be computed as
\begin{equation}
\label{eq:muyr}
    \mu_y^{(r)}=(Kt)^r\mathbb{E}\left[e^{S_r(\rho)}\right]\,,
\end{equation}
since all paths on the tree are statistically equivalent.
Then, by H\"older's inequality and Jensen's inequality, one obtains respectively an upper and a lower bound for $\mathbb{E}\left[e^{S_r(\rho)}\right]$:
\begin{eqnarray}
\label{eq:eSbounds}
    \mu_{g,r}\ge\mathbb{E}\left[e^{S_r(\rho)}\right]\ge e^{ \mathbb{E}\left[S_r(\rho)\right]}=(\mu_g^{\rm typ})^r
\end{eqnarray}
where $\mu_g^{\rm typ}\equiv e^{\mathbb E[\ln \mathcal {G}_{i \to j}]}$ and $\mu_{g,r}$ is the $r$-th moment of the cavity Green's functions. In particular, the second inequality allows us to obtain an exact upper bound for the value of $E_{\rm iso}$. Indeed, from Eqs.~\eqref{eq:muyr},\eqref{eq:eSbounds} it follows that
\begin{eqnarray}
    \mu_y \equiv \lim _{r\to \infty} \mu_y^{(r)}=
        \infty \quad&\text{if}\quad Kt\mu_g^{\rm typ}>1 \,,
\end{eqnarray}
therefore $E_{\rm iso}(W)$ must be smaller than or equal to the value for which $\mu_g^{\rm typ}=1/(Kt)$ (recall that $E_{\rm iso}$ is the largest value of $E_{\rm sh}$ for which the average Localization Landscape is finite, see Sec.~\ref{subsec:localization-landscape}).
 In practice, we can obtain an estimate of $E_{\rm iso}$ using the properties of the high-connectivity limit. In Sec.~\ref{subsubsec:joint}, we showed that correlations between nearest-neighbor cavity Green's functions along a path are negligible for sufficiently large $K$ and for all disorder values (see Fig.~\ref{fig:corrgeta}), which means that
\begin{eqnarray}
    \mathbb{E}\left[e^{S_r(\rho)}\right]
    &=&
    \mathbb{E}\left[\prod_{s=1}^{r} \mathcal{G}_{\rho_s \to \rho_{s-1}}\right]
    \nonumber\\
    &=&
    \prod_{s=1}^{r} \mathbb{E}\left[\mathcal{G}_{\rho_s \to \rho_{s-1}}\right]=\mu_g^r\,,
    \label{eq:FactorAverage1}
\end{eqnarray}
therefore
\begin{equation}
\label{eq:itery}
    \mu_y = \begin{cases}
        0 \quad &\text{if} \quad \mu_g< 1/(Kt)\,,
        \\
        \infty \quad &\text{if} \quad \mu_g \geq 1/(Kt) \,.
    \end{cases}
\end{equation}
Finally, $E_{\rm iso}$ must be the value for which the fixed point of the recursion for the $y$-variables switches from $0$ to $+\infty$, namely the value for which the expectation value of the cavity Green's function for $W\le W_{\rm min}$ is equal to $1/(Kt)$.  
We can now use the high-connectivity results of Sec.~\ref{subsubsec:marginals} to obtain the value of $E_{\rm iso}$ from the roughest approximation for $\mu_g$ in Eq.~\eqref{eq:roughappr}.
We have
\begin{equation}
    \frac{1}{Kt}=\frac{1}{W}\ln \left(1-\frac{W}{W/2+E_{\rm iso}(W)+t}\right)\,,
\end{equation}
thus
\begin{equation}
\label{eq:Eiso}
    E_{\rm iso}(W)= -t-\frac{W}{2}\coth\left(\frac{W}{2Kt}\right)\,.
\end{equation}
It is important to note that this equation correctly reproduces the zero-disorder value of the isolated eigenvalue, namely $E_{\rm iso}(0)=-t(K+1)$. Figure~\ref{fig:critcurve1} shows the resulting prediction for $-E_{\rm iso}(W)$ in the case $K=5, \,t=1$.

Another important consequence appears when we return to the Localization
Landscape percolation equations for the translated Hamiltonian introduced in
Sec.~\ref{subsec:cavity-derivation}. For $W\leq W_{\rm min}$, the cavity
Green's function reaches the critical value
\begin{equation}
    \mu_g(W)=\frac{1}{Kt}.
\end{equation}
At this point, the self-consistent equation for the average rescaled field no
longer admits a finite positive expectation value.
This also implies
\begin{equation}
\label{eq:ubelowWmin}
    \mu_u(W)\equiv \mathbb E[u_i]=+\infty
    \qquad
    \text{for}
    \qquad
    W\leq W_{\rm min}.
\end{equation}

In the high-connectivity limit, at low-disorder, the typical and average values of cavity variables are very close, so the divergence of the rescaled fields signals that the typical scale of the Localization Landscape becomes arbitrarily large. Consequently, the corresponding effective
potentials
\begin{equation}
    v_i\equiv \frac{1}{u_i}
\end{equation}
has vanishing typical scale. Therefore, for any fixed positive energy $E_+>0$, the inequality $E_+>v_i$ is satisfied on a macroscopic number of sites. Therefore, in this regime, the LLT percolation problem is necessarily in the percolating phase.

Consequently, according to the LLT, the percolation transition ceases to exist precisely at $W=W_{\rm min}$. This means that the minimal critical disorder for the Localization Landscape percolation transition, below which the system is always in the percolating phase, coincides with the disorder strength at which the isolated eigenvalue enters the bulk of the spectrum. The only shifting prescription for which the non-percolating phase can extend below $W_{\min}$ is the over-shifted regime, namely $E_{\rm sh}<E_{\rm iso}$, as in this case the expected value of the Localization Landscape variables never diverge.

A further confirmation of the equivalence between $W_{\rm min}$ and the lower critical disorder for percolation is obtained by observing that, according to the cavity equations~\eqref{eq:gsys} and \eqref{eq:etasys}, a cavity rescaled field is given exactly by the series
\begin{eqnarray}
\label{eq:etatree}
    \eta_{i\to j} &=& \sum_{r=0}^\infty y_{i\to j}^{(r)}\,, \quad y_{i \to j}^{(0)}=1\,.
\end{eqnarray}
After taking the expectation value, we have
\begin{eqnarray}
    && \mu_\eta \ge \sum_{r=0}^{\infty} (Kt\mu_g^{\rm typ})^r
    \nonumber\\
    && \;\;\;\;\; = 
    \begin{cases}
        1/(1-Kt\mu_g^{\rm typ}) \quad &\text{if} \quad \mu_g^{\rm typ}< 1/(Kt)\,,
        \\
        \infty \quad &\text{if} \quad \mu_g^{\rm typ} \geq 1/(Kt) \,.
    \end{cases}
    \qquad
\end{eqnarray}
and in the high-connectivity limit
\begin{eqnarray}
    \mu_\eta =
    \begin{cases}
        1/(1-Kt\mu_g) \quad &\text{if} \quad \mu_g< 1/(Kt)\,,
        \\
        \infty \quad &\text{if} \quad \mu_g \geq 1/(Kt) \,.
    \end{cases}
\end{eqnarray}

The minimal critical disorder $W_{\rm min}$ can be computed from Eq.~\eqref{eq:gavgapprox} as
\begin{equation}
\label{eq:bound}
    W_{\rm\min}=Kt \ln\left| 1+\frac{W_{\rm min}}{t(2\sqrt K-1)} \right|\,,
\end{equation}
which can be solved self-consistently by iteration.
Remarkably, the prediction for $W_{\rm min}$ in Eq.~\eqref{eq:bound} for $K=5,t=1$ matches perfectly the disorder at which the high-connectivity LLT percolation transition ceases to exist. 
For $K=2$ and $t=1$, Eq.~\eqref{eq:bound} gives $W_{\min}\simeq0.354$, very close to the exact cavity value $W_{\min}\simeq0.357$ and to the disorder at which the LLT percolation transition ceases to exist. It is also compatible with our data for the mobility edge, which however cannot be reliably extrapolated below $W=0.5$ (Sec.~\ref{sec:mobility}). Note also the exact bound $W_{\min}\ge2t(\sqrt K-1)^2$ of Eq.~\eqref{eq:Wmin-bound}. This suggests that the onset of the AL transition is closely linked to the position of the isolated eigenvalue, although this connection deserves further investigation. This connection is made explicit at large connectivity in Sec.~\ref{subsubsec:largeK-loc}: the growth rate of $\mu_y^{(r)}$ in Eq.~\eqref{eq:muyr} is the largest eigenvalue of the Anderson kernel~\eqref{eq:kernelAL} at $\beta=1/2$, evaluated below the spectrum, and the same eigenvalue determines the mobility edge inside the spectrum.

\subsubsection{Linear stability analysis}
\label{subsubsec:linstab}

Following the same idea of \cite{abou1973selfconsistent}, and already used in the analysis of the Anderson problem in Sec.~\ref{sec:ALeigenval}, 
we start from the self-consistent distributional equation (\ref{eq:stochSC}). Using the integral representation of the Dirac delta
\begin{widetext}
\begin{equation}
    \delta\bigg( p - \theta(g \eta - 1/E_+) \sum_k p_k \bigg)
    = \int_{-\infty}^{\infty} 
    \frac{d\lambda'}{2\pi} \,e^{-{\rm i} \lambda' \left( p - \theta(g\eta - 1/E_+) \sum_k p_k \right)} \, ,
\end{equation}
we can rewrite
\begin{align}
    P(g,\eta,p)&=\int \frac{d\lambda'}{2\pi} \, e^{i\lambda' p}\int d\varepsilon \, \gamma_+(\varepsilon)
    \int \prod_{k=1}^K \left[dg_kd\eta_kdp_k \,\hat{P}\big ( g_k,\eta_k ,\theta(g\eta-1/E_+)\lambda'\big )\right]
    \nonumber\\ 
    & \quad\qquad\qquad \times\delta \bigg ( g-\frac{1}{\varepsilon-t^2\sum_k g_k}\bigg ) 
     \delta \bigg (  \eta- 1-t\sum_k g_k \eta_k \bigg ).
\end{align}
Then, taking the Fourier transform of the latter with respect to $p$ and recognizing the integral representation of $\delta(\lambda-\lambda')$, we derive
\begin{align}
    \hat{P}(g,\eta,\lambda)&=\int d\varepsilon \,\gamma_+(\varepsilon) 
    \int \prod_{k=1}^K \left[dg_kd\eta_k \, \hat{P}\big( g_k,\eta_k ,\; \theta(g\eta-1/E_+)\lambda\big)  \right]
 \delta \bigg ( g-\frac{1}{\varepsilon-t^2\sum_k g_k}\bigg )\delta \bigg (  \eta- 1-t\sum_k g_k \eta_k \bigg ).
\end{align}
The theta function in the argument of the characteristic function of the distribution distinguishes between two 
cases. For $g \eta>1/E_+$ we have to solve the integral equation
\begin{equation}
    \label{eq:chfunc}
    \hat{P}(g,\eta,\lambda)=\int d\varepsilon \,\gamma_+(\varepsilon)\int \prod_{k=1}^K \left[dg_kd\eta_k \, \hat{P}\big(g_k,\eta_k,\lambda\big) \right] \;
    \delta \bigg ( g-\frac{1}{\varepsilon-t^2\sum_k g_k}\bigg )\delta \bigg (  \eta- 1-t\sum_k g_k \eta_k \bigg ),
\end{equation}
while for $g\eta<1/E_+$ we simply have
\begin{equation}
    \hat{P}\big( g_k,\eta_k,\theta(g\eta-1/E_+)\lambda)=
    \hat{P}( g_k,\eta_k,0)=P( g_k,\eta_k ),
\end{equation}
thus
\begin{equation}
    P(g,\eta,p)=P(g,\eta)\delta(p),
\end{equation} 

In order to find an equation for the critical curve, we are interested in finding a solution for the probability distribution that admits a finite expectation value of the cavity percolation probability. Since its distribution necessarily has finite mean and variance (because $p \in [0,1]$), we can assume that at criticality the characteristic function for $g\eta>1/E_+$ can be expanded in powers of $\lambda$, and that the term of the smallest order is exactly of order one:
\begin{equation}
    \hat{P}(g,\eta,\lambda)=P(g,\eta)+{\rm i} \lambda f(g,\eta) +O(\lambda^2).
\end{equation}
Substituting in the integral Eq.~(\ref{eq:chfunc}), and discarding terms of order higher than one we obtain a self-consistent equation for $f(g,\eta)$ that reads
\begin{align}
    f(g,\eta)&=K\int d\varepsilon\, \gamma_+(\varepsilon) \,dg'd\eta' \,f(g',\eta') 
    \int \prod_{k=1}^{K-1}\left[dg_kd\eta_k \, P( g_k,\eta_k )\right] 
    \nonumber \\ &
   \qquad\qquad\quad  \times\delta \bigg ( g-\frac{1}{\varepsilon-t^2 \big(g'+\sum_{k=1}^{K-1} g_k\big)}\bigg )
    \delta \bigg (  \eta- 1-t \Big (g'\eta'+ \sum_{k=1}^{K-1} g_k \eta_k \Big ) \bigg )\,,
\end{align}
and can be recast as
\begin{equation}
\label{eq:eigintperc}
    f(g,\eta)=\int dg'd\eta'\,\mathcal{K}_{\rm perc}(g,\eta,g',\eta')f(g',\eta'),
\end{equation}
with
\begin{equation}
    \mathcal{K}_{\rm perc}(g,\eta;g',\eta')=\int d\varepsilon d\tilde{g}d\tilde{u} \,\gamma_+(\varepsilon)R_{\rm perc}(\tilde{g},\tilde{u})\delta \bigg ( g-\frac{1}{\varepsilon-t^2 (g'+\tilde{g})}\bigg )\delta \bigg (  \eta- 1-t \Big (g'\eta'+ \tilde{u} \Big ) \bigg )\,, 
\end{equation}
where we have introduced the joint probability distribution of the sums $\sum_{l=1}^{K-1}g_l$ and $\sum_{l=1}^{K-1}g_l\eta_l$:
\begin{equation}
    R_{\rm perc}(\tilde{g},\tilde{u})
    \equiv
    \int \prod_{k=1}^{K-1}\left[dg_kd\eta_k \,P( g_k,\eta_k )\right]
    \;  \delta\bigg(\sum_{k=1}^{K-1}g_k-\tilde{g}\bigg)\delta\bigg(\sum_{k=1}^{K-1}g_k\eta_k-\tilde{u}\bigg).
\end{equation}

Performing the integrals over the tilde variables, the kernel simplifies to 
\begin{equation}
\label{eq:kernel_perc}
    \mathcal{K}_{\rm perc}(g, \eta;\, g', \eta') =
    \frac{K}{g^2\, t^3} \int d\varepsilon \, \gamma_+(\varepsilon) \;
    R_{\rm perc}\left(-g' - \frac{1 - \varepsilon g}{g t^2},\;
    -g'\eta'-\frac{1-\eta}{t}\right).
\end{equation}
\end{widetext}
Now, in the non-percolating phase, the solution with zero percolation probability, i.e. $P(g,\eta,p)=P(g,\eta)\delta(p)$,  is the stable one. This means that any infinitesimal perturbation of its characteristic function, under iteration of the recursive equation defining $\hat{P}(g,\eta,\lambda)$ will vanish. Instead, in the percolating phase, we expect that the first-order perturbative term will increase under iteration. Accordingly, the critical curve will be 
identified as the curve in the space of parameters where the first order correction remains stable under iteration, i.e. where the kernel $\mathcal{K}_{\rm perc}$ has top eigenvalue equal to $1$ (following exactly the same argument at the end of Sec. \ref{sec:ALeigenval}). 

The kernels of Eqs.~\eqref{eq:kernelAL} and \eqref{eq:kernel_perc} are significantly different, and this explains quantitatively the difference between the two critical behaviors. 

The integral operator can be computed numerically, using population dynamics to evaluate the distribution $R_{\rm perc}(\tilde{g}, \tilde{u})$, and it can be diagonalized numerically to obtain the critical curve with very high precision. 
However, this procedure is highly computationally expensive.
Therefore, in order to compute the critical curve in the high-connectivity limit we used a completely different technique that we present in the next Section, where we also present an argument to show that the linear stability analysis produces the same result.

\subsubsection{Divergence of the average cluster size}
\label{subsubsec:divergence-cluster-size}

An alternative method to determine the critical curve involves deriving an expression for the average cluster size $S$ and identifying the values $(E, W)$ for which $S$ diverges.

In the general case, a cluster consists of a connected component of lattice sites $i$ where $u_i \geq 1/E_+$. The statistical dependence between $u_i$ and $u_k$ (for $k \in \partial i$) is complex, but in the high-connectivity limit the problem simplifies significantly. Here, we can treat $u_k \overset{d}{=} u_{k \to i} \overset{d}{=} \mathcal{G}_{k \to i} \eta_{k\to i}$, effectively breaking the mutual dependence between sites $i$ and $k$: $u_i$ depends on all its neighbors, while $u_{k\to i}$ depends on all its neighbors except $i$.

In this high-connectivity limit, we can replace the occupation variable $O_i$ from Eq.~\eqref{eq:Oiperc} with
\begin{equation}
    \label{eq:cavOi}
    O_{i\to j} = \begin{cases}
        1 & \text{if } u_{i \to j} \geq 1/E_+\, \\
        0 & \text{otherwise}
    \end{cases}\,.
\end{equation}
By substituting in the definition of Eq.~(\ref{eq:corrfuncnobond}) the set of occupation variables $\{O_1,\dots,O_{r}\}$ with $\{O_{1 \to 0},\dots,O_{r \to r-1}\}$, we can factorize Eq.~\eqref{eq:corrfuncnobond} as
\begin{align}
    C_{\rm perc}(r) 
    & = 
    \Pr \{O_0=1, O_{1 \to 0}=1, \dots, O_{r \to r-1}=1\} 
    \nonumber \\
           & = \left[ \prod_{s=2}^r \Pr \{O_{s\to s-1} = 1 
           \mid O_{s-1 \to s-2}=1 \} \right] 
           \nonumber\\
           & \  \times
           \Pr \{O_{1\to 0}=1 \mid O_0=1\} \Pr \{O_0=1
           \} 
           \,.
           \label{eq:corrfunchicon}
\end{align}

Due to translational invariance, the conditional probability that a site is occupied given that one of its nearest neighbors is occupied is independent of the specific pair of sites considered. Thus, we define
\begin{equation}
\label{eq:qbar}
    \bar{q} \equiv \Pr \{ O_{k\to i} =1 \mid O_{i\to j}=1 \}
    \end{equation}
   $ \forall i, \forall j \in \partial i,\forall k \in \partial i \setminus j $.
Similarly, the last factor in Eq.~\eqref{eq:corrfunchicon} is given by the unconditional occupation probability $q$ [see Eq.~\eqref{eq:q}].
Using these definitions, $C_{\rm perc}(r)$ takes the form
\begin{equation}
    C_{\rm perc}(r) = q \bar{q}^r = \frac{q}{K^r} e^{-r/\xi_{\rm perc}},
\end{equation}
where the correlation length is given by
\begin{equation}
    \xi_{\rm perc} = - \frac{1}{\ln K\bar q}.
\end{equation}

As in the independent-site percolation problem~\cite{stauffer2018introduction}, the correlation function decays exponentially, so the average cluster size, defined in Eq.~\eqref{eq:avgclsize}, simplifies to
\begin{equation}
\label{eq:S}
    S =  \frac{1+\bar q}{1 - K \bar{q}}.
\end{equation}
We immediately see that $S$ diverges when
\begin{equation}
\label{eq:hiconcrit}
    \bar{q} = \bar{q}_c = 1/K.
\end{equation}
This condition is analogous to that of the independent percolation problem, except that the occupation probability is now conditioned on the occupation of a neighboring site.
By numerically evaluating the pairs $(E, W)$ that satisfy $\bar q = 1/K$, we obtain the critical curve. 
The numerical procedure to compute $\bar q$ is explained in detail in Sec.~\ref{app:numeval}. 
The solution of the integral eigenvalue equation~\eqref{eq:eigintperc} amounts to determining the critical energy at which the recursion for $p_{i\to j}$, defined in its linearized form in Eq.~\eqref{eq:psys}, diverges. In the following we show, using an argument analogous to that of Sec.~\ref{subsubsec:lowW}, to show that this divergence occurs when $\bar q =1/K$.

Let $\mathcal{P}(r)$ denote the set of all directed paths of length $r$ starting from $i$ and going away from the root. A path $\rho\in\mathcal{P}(r)$ is written as
\[
\rho=(\rho_0,\rho_1,\dots,\rho_r)\,,
\]
with $\rho_0=i$. As in Sec.~\ref{subsubsec:lowW}, we introduce a truncated version of the cavity variable by iterating the linearized cavity equation up to depth $r$, namely
\begin{equation}
\label{eq:prdef}
    p^{(r)}_{i\to j}
    =
    O_{i\to j}
    \sum_{\rho\in\mathcal{P}(r)}
    \prod_{s=1}^{r-1}
    O_{\rho_s\to\rho_{s-1}}\,,
\end{equation}

By construction, the full cavity variable is obtained in the limit
\begin{equation}
    p_{i\to j}=\lim_{r\to\infty}p^{(r)}_{i\to j}\,.
\end{equation}

The expectation value of $p^{(r)}_{i\to j}$ is then
\begin{equation}
\label{eq:mupr}
    \mu_p^{(r)}
    =
    K^r \,
    \mathbb{E}\!\left[
    O_{i\to j}
    \prod_{s=1}^{r-1}
    O_{\rho_s\to\rho_{s-1}}
    \right]\,,
\end{equation}
since all paths on the tree are statistically equivalent. At high connectivity, the same property used to factorize the correlation function in Eq.~\eqref{eq:corrfunchicon} implies that the statistical dependence of the occupation variables propagates only from the leaves to the root. Therefore, along a path $\rho$ one has
\begin{widetext}
\begin{eqnarray}
     \Pr \{O_{i\to j}, O_{\rho_{1}\to\rho_{0}}, \dots, O_{\rho_{r-1}\to\rho_{r-2}}\} = \Pr \{O_{i\to j}\mid O_{\rho_{1}\to\rho_{0}}\}
         \prod_{s=2}^{r-1}
         \Pr \{O_{\rho_{s-1}\to\rho_{s-2}} \mid O_{\rho_{s}\to\rho_{s-1}}\}
         \Pr \{O_{\rho_{r-1}\to\rho_{r-2}}\}\,.
           \nonumber
\end{eqnarray}
\end{widetext}

Using the definitions~\eqref{eq:q} and~\eqref{eq:qbar}, this gives
\begin{equation}
    \mathbb{E}\!\left[
    O_{i\to j}
    \prod_{s=1}^{r-1}
    O_{\rho_s\to\rho_{s-1}}
    \right]
    =
    q\,\bar q^{\,r-1}\,,
\end{equation}
and therefore
\begin{equation}
\label{eq:muprfinal}
    \mu_p^{(r)}
    =
    K^r q\,\bar q^{\,r-1}
    =
    (Kq)\,(K\bar q)^{r-1}\,.
\end{equation}

Finally, taking the limit $r\to\infty$, one obtains
\begin{equation}
     \mu_p \equiv \lim_{r\to \infty} \mu_p^{(r)} = \begin{cases} 
        0 \quad&\text{if}\quad \bar q<1/K\,,\\
        +\infty \quad&\text{if}\quad \bar q\ge 1/K\,,
    \end{cases}
\end{equation}
which shows that the integral recursion of Eq.~\eqref{eq:eigintperc} diverges when $\bar q = 1/K$, thus marking the transition to the percolating phase.
 
\subsubsection{The high-connectivity phase diagram}
\label{subsubsec:PD}

\begin{figure}[!h]
\vspace{0.3cm}
        \centering
        \includegraphics[width=0.45\textwidth]{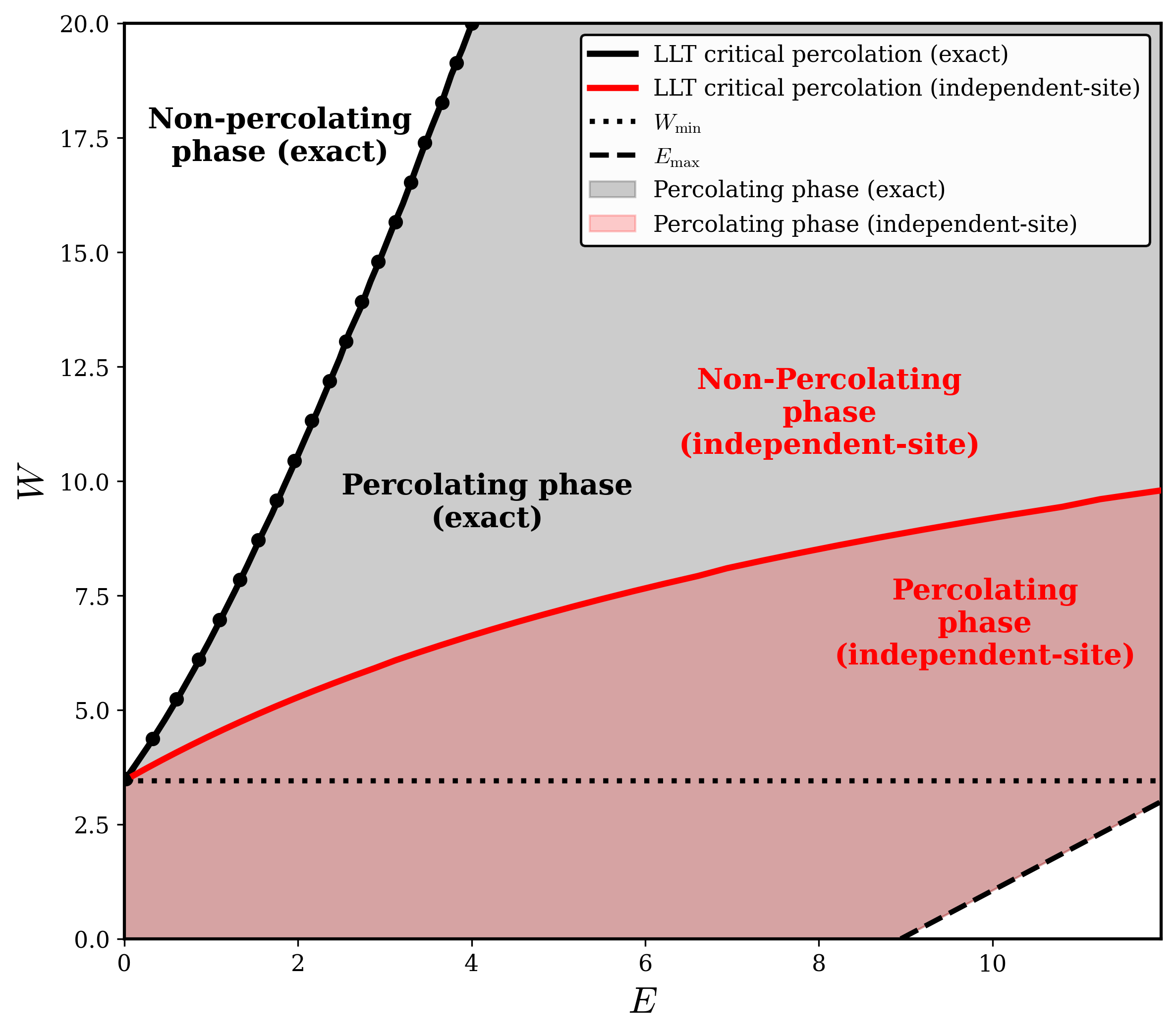}
    \caption{Percolation phase diagrams  for $K=5$ and $t=1$. Positive definite Hamiltonian of Eq.~\eqref{eq:Hplus}.  Red curve: critical curve separating percolating and non-percolating phases for high-connectivity in the independent-site approximation [Eq.~(\ref{eq:indsitescrit})].
    Black line: exact critical curve in the high-connectivity limit [Eq.~(\ref{eq:hiconcrit})]. Dashed line: upper boundary of the bulk of spectrum (see Sec.~\ref{subsec:cavity-derivation}). Dotted line: analytically predicted lower critical disorder [Eq.~(\ref{eq:bound})].
    }
    \label{fig:critcurve1}
\end{figure}

Figure~\ref{fig:critcurve1} presents the phase diagram obtained in the high-connectivity limit. 
It displays both the curve derived from the independent-site approximation condition [Eq.~\eqref{eq:indsitescrit}] for the edge-shift prescription ($E_{\rm sh}(W)=E_{\rm edge}(W)$) and the one obtained using the exact criterion [Eq.~\eqref{eq:hiconcrit}].

\begin{figure}[!h]
\vspace{0.3cm}
    \centering     
    \includegraphics[width=0.45\textwidth]{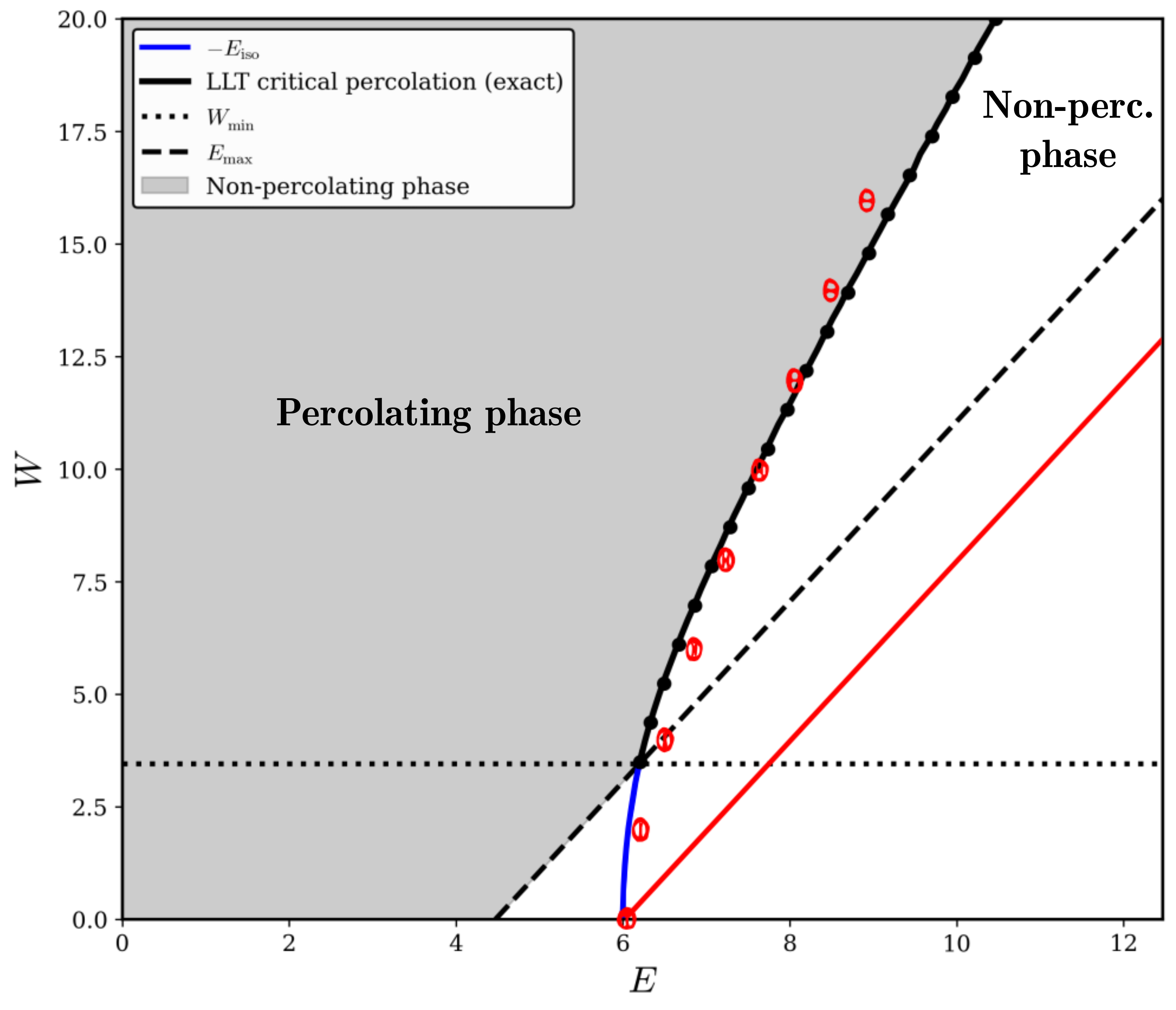}
    \caption{\small{Percolation phase diagram for $K=5$ and $t=1$. Hamiltonian with the statistically symmetric spectrum of Eq.~\eqref{eq:ham}. All the curves  have been obtained from the ones in Fig.~\ref{fig:critcurve1}
    by means of a translation and inversion of sign as explained in Sec.~\ref{subsubsec:PD}. The independent-site critical curve has not been plotted as it loses its physical interpretation in the symmetric case. Red dots: Localization Landscape percolation critical energies for the cubic lattice from \cite{filoche2024anderson}. Blue line: opposite of the isolated eigenvalue $-E_{\rm iso}$ for $W\leq W_{\rm min}$ from Eq.~\eqref{eq:Eiso}. }}
    \label{fig:critcurve2}
\end{figure}

The phase diagram in Fig.~\ref{fig:critcurve1} is plotted in the $(E_+,W)$-plane for the positive definite Hamiltonian $\hat{\cal{H}}_+$ (see Sec.~\ref{subsec:cavity-derivation}). Figure~\ref{fig:critcurve2} shows the positive-energy side of the phase diagram associated with the original Anderson Hamiltonian of Eq.~\eqref{eq:ham}. This was obtained from the plot 
in Fig.~\ref{fig:critcurve1}
by translating the energies by $E_{\rm sh}(W)$ to restore the statistically symmetric spectrum, and then switching $E \mapsto -E$ to display only the positive-energy region. Due to this transformation, the percolating and non-percolating phases are exchanged in the right image.

The exact curve for a given parameter pair $(K,t)$ requires sampling from the joint distribution $P(g,\eta)$, which properly accounts for correlations between cavity Green's functions and rescaled fields (see Sec.~\ref{sec:numerics} for numerical details). However, as anticipated from the results of Sec.~\ref{subsubsec:joint}, the phase diagram computed using the factorized approximation of Eq.~(\ref{eq:MF}) agrees excellently with that obtained from the joint distribution via population dynamics. In Fig.~\ref{fig:critcurve1} and Fig.~\ref{fig:critcurve2}, the solid black line corresponds to the curve derived from the self-consistent marginal distributions of Sec.~\ref{subsubsec:marginals}, while the black dots represent critical energies computed by sampling the joint distribution from population dynamics.

A key observation is that the independent-site curve deviates significantly from the exact one in 
Fig.~\ref{fig:critcurve1}, indicating that correlations between nearest-neighbor sites are essential to describe the physics of the system. Neglecting these correlations leads to incorrect results. This discrepancy is so pronounced that the transformed independent-site curve becomes physically meaningless in the symmetric case, and thus it is not plotted in the right panel.
The only case for which the independent-site critical energy corresponds to the exact one is when $W=W_{\rm min}$, i.e. when all the Localization Landscape variables diverge, becoming effectively independent.
The solid blue curve in Fig.~\ref{fig:critcurve2} shows the prediction of $-E_{\rm iso}(W)$ for $W \leq W_{\rm min}$ derived in Sec.~\ref{subsubsec:lowW}. As expected, it intersects the spectral boundary $-E_{\rm edge}(W) = 2t\sqrt{K} + W/2$ at the point where the percolation transition ceases to exist. The disorder value at which this occurs is indicated by a horizontal dotted line.
Although the high-connectivity equations are formally valid for $K \gg 1$, many probability distributions from Sec.~\ref{subsubsec:marginals} are accurate even at $K=5$ (e.g., the marginals of the cavity Localization Landscape variables and cavity Green's functions, as shown in Fig.~\ref{fig:PgPu}). Therefore, we plot both curves for $K = 5$, arguing that the exact high-connectivity condition captures many important features of the real curve at this connectivity, despite neglecting the influence of one neighbor out of six. Moreover, the solution on the Bethe lattice with $K = 5$ represents the Bethe approximation for the cubic lattice, allowing for direct comparison with the numerical results from Ref.~\cite{filoche2024anderson} (plotted as red circles in ~\ref{fig:critcurve2}). The Bethe and cubic lattice transitions have a qualitative agreement. The main difference is that on the Bethe lattice, the spectral boundary prevents the curve from reaching zero disorder, as discussed in Sec.~\ref{subsubsec:lowW}.

\subsubsection{Large-connectivity mobility edge and comparison with the LLT percolation}
\label{subsubsec:largeK-loc}

In this Section we compute the Anderson mobility edge for $K\gg1$, following the approach of Abou-Chacra, Anderson, and Thouless~\cite{abou1973selfconsistent}, and we compare it with the high-connectivity LLT percolation line of Sec.~\ref{subsubsec:PD}. We focus on the region close to the edge of the spectrum, where both lines originate.

\paragraph{Stability criterion and isolated eigenvalue.}
The mobility edge is the line on which the largest eigenvalue $\Lambda_\beta(E,W)$ of the kernel~\eqref{eq:kernelAL} equals one at the symmetric point $\beta=1/2$~\cite{abou1973selfconsistent,tikhonov2019critical}. For $\beta=1/2$ the kernel contains the factor $Kt|g|$, and $\Lambda_{1/2}$ is the growth rate of $K^r\,\mathbb E\big[\prod_{s=1}^{r}t\,|\mathcal G_{\rho_s\to\rho_{s-1}}|\big]$ along a path of length $r$. Below the spectrum all the Green's functions are positive, and this is exactly the growth rate of $\mu_y^{(r)}$ in Eq.~\eqref{eq:muyr}. The isolated eigenvalue is therefore fixed by the same condition, $\Lambda_{1/2}(E_{\rm iso},W)=1$, evaluated below the spectrum. In other words, $-E_{\rm iso}(W)$ and $E_{\rm loc}(W)$ are the parts of the same level line $\Lambda_{1/2}=1$ that lie outside and inside the spectrum, respectively. Whether the two parts join at the band edge depends on the continuity of $\Lambda_{1/2}$ across it, which is explicit in the large-$K$ evaluation below.

\paragraph{Large-$K$ evaluation.}
We consider the lower edge of the spectrum, $E<0$; the upper edge follows by symmetry. For $K\gg1$ two simplifications hold, as in the analysis of Ref.~\cite{abou1973selfconsistent} at $E=0$. First, the real part of the self-energy, $t^2\sum_{l\in\partial k\setminus i}\mathcal G_{l\to k}$, is a sum of $K$ terms of order $1/W$; it concentrates on its average $\Sigma=Kt^2\mu_g$, with fluctuations that vanish for $K\to\infty$. Second, the cavity Green's function $g'$ of the next site along the path enters the denominator only through the single term $t^2g'$, which is small compared to the denominator itself. The kernel then becomes independent of $g'$, i.e., of rank one, and its eigenvalue reduces to
\begin{eqnarray}
\label{eq:Lambda-largeK}
\Lambda_{1/2} &\simeq& Kt\,\mathbb E|g|\,,
\\
g &=& \frac{1}{\epsilon+x}\,,
\\
x &\equiv& -E-\Sigma\,,
\end{eqnarray}
with $\epsilon$ uniformly distributed in $[-W/2,W/2]$.

\paragraph{Edge branch.}
For $x>W/2$ the denominators $\epsilon+x$ are positive for every $\epsilon$. Resonances, i.e., vanishing denominators, are possible only through large fluctuations of the self-energy, whose probability vanishes for $K\to\infty$ (this is the Lifshitz-tail region of the spectrum). Hence $g>0$, $\Lambda_{1/2}=Kt\mu_g$, and
\begin{equation}
\label{eq:Lambda-edge}
\Lambda_{1/2}=\frac{Kt}{W}\ln\frac{x+W/2}{x-W/2}\,.
\end{equation}
On the critical line, $\Lambda_{1/2}=Kt\mu_g=1$ implies $\Sigma=t$, and $x-W/2=W/(e^{W/Kt}-1)$, i.e.,
\begin{equation}
\label{eq:Eloc-edge}
E_{\rm loc}(W)=t+\frac W2\coth\left(\frac{W}{2Kt}\right)\,.
\end{equation}
This is exactly $-E_{\rm iso}(W)$ of Eq.~\eqref{eq:Eiso}, now for $W>W_{\min}$. For $W<W_{\min}$ the solution lies outside the spectrum and gives the isolated eigenvalue; for $W>W_{\min}$ it lies inside the spectrum and gives the mobility edge. It crosses the band edge, $E_{\max}=2t\sqrt K+W/2$, when $x-W/2=2t\sqrt K-t$, i.e., exactly at the value of $W_{\min}$ given by Eq.~\eqref{eq:bound}. Hence, at large connectivity, the mobility edge emerges from the edge of the spectrum exactly at $W_{\min}$, i.e., at the point where the LLT percolation line starts. Contrary to the case $E=0$, no resonance cutoff enters here: $g$ is bounded, the logarithm in Eq.~\eqref{eq:Lambda-edge} is free of ambiguities, and the result is controlled up to corrections that vanish for $K\to\infty$. It holds as long as $x-W/2\gg t$, i.e., $W/(e^{W/Kt}-1)\gg t$. Even for $K=2$, Eq.~\eqref{eq:Eloc-edge} gives $E_{\rm loc}\simeq3.010$, $3.023$, and $3.042$ for $W=0.5$, $0.75$, and $1$, in agreement, within the error bars, with the numerical mobility edge of Fig.~\ref{fig:phase-diagram}.

Close to the onset, setting $W=W_{\min}+a$ and $E_{\rm loc}=E_{\max}-b$, Eq.~\eqref{eq:Eloc-edge} gives
\begin{equation}
\label{eq:b-of-a}
b(a)=\big(2t\sqrt K-t\big)-\frac{W_{\min}+a}{e^{(W_{\min}+a)/Kt}-1}\,.
\end{equation}
For $a=O(t)$, $b\simeq(2/\sqrt K)\,(1-Kt/W_{\min})\,a$, which vanishes for $K\to\infty$, since $W_{\min}\simeq(Kt/2)\ln K$. The nontrivial scaling regime is $a=Kt\,\alpha$ and $b=2t\sqrt K\,\beta$, with $\beta(\alpha)\to1-e^{-\alpha}$ for $K\to\infty$ (up to corrections of order $1/\ln K$): the mobility edge detaches from the band edge over a disorder range of order $Kt$, and moves inwards by an amount of order $2t\sqrt K$, i.e., the width of the clean band.

\paragraph{Bulk branch.}
For $W/2-|x|\gg t$ the denominators vanish at $\epsilon=-x$, and the logarithmic divergence of $\mathbb E|g|$ is cut off by resonances. As in Ref.~\cite{abou1973selfconsistent}, this is handled by the symmetry $\Lambda_\beta=\Lambda_{1-\beta}$. For $\beta<1/2$ the regular part of the kernel gives $\Lambda_\beta\simeq(Kt^{2\beta}/W)\,[(W/2-x)^{1-2\beta}+(W/2+x)^{1-2\beta}]/(1-2\beta)$. Symmetrizing and taking $\beta\to1/2$ gives
\begin{eqnarray}
\label{eq:Lambda-bulk}
\Lambda_{1/2}
&\simeq&
\frac{2Kt}{W}\ln\frac{W^2/4-x^2}{t^2}\,,
\\
\Sigma &=& \frac{Kt^2}{W}\ln\frac{W/2+x}{W/2-x}\,.
\end{eqnarray}
At $E=0$, this reduces to the classical result $W_c=4Kt\ln(W_c/2t)$~\cite{abou1973selfconsistent}, which gives $W_c\simeq17.2$ for $K=2$, to be compared with the exact value $W_c\simeq18.17$~\cite{tikhonov2019critical,parisi2019anderson}. The bulk branch of the mobility edge is $E_{\rm loc}=x+\Sigma$, with $x=[W^2/4-t^2e^{W/2Kt}]^{1/2}$. It is re-entrant and vanishes at $W_c$. It holds for $W/2-x\gg t$, i.e., $W\gg2Kt\ln(W/t)$. Contrary to the edge branch, Eq.~\eqref{eq:Lambda-bulk} is only valid to leading logarithmic order, since the cutoff of order $t$ inside the logarithm is not fixed by this argument.

\begin{figure*}[t]
\centering
\includegraphics[width=0.85\textwidth]{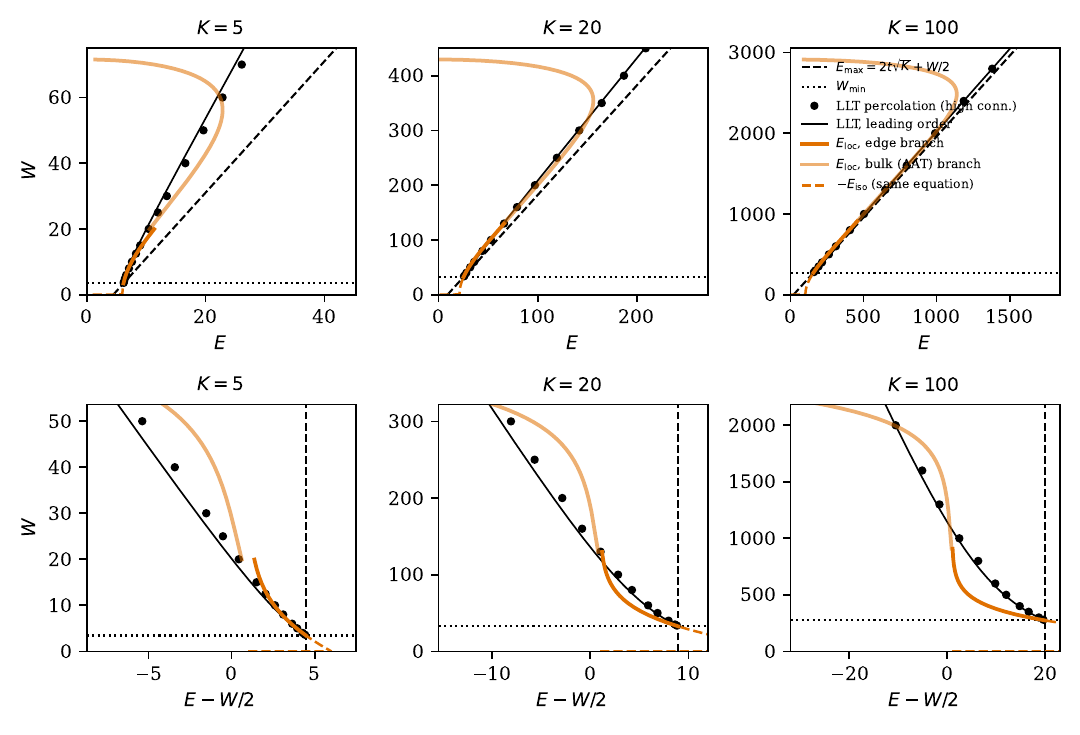}
\caption{Large-connectivity phase diagram for $K=5$, $20$, and $100$ ($t=1$), positive-energy side. Top row: $W$ vs.\ $E$. Bottom row: zoom on the weak-disorder region, $W$ vs.\ $E-W/2$. Black dots: LLT percolation line from the full high-connectivity criterion $\bar q=1/K$ [Eq.~\eqref{eq:hiconcrit}, cavity population dynamics with $X=0$]. Thin black line: its leading-order large-$K$ expression, Eq.~\eqref{eq:Eperc-largeK}. Dark orange line: edge branch of the mobility edge, Eq.~\eqref{eq:Eloc-edge}. Light orange line: bulk (Abou-Chacra--Anderson--Thouless) branch, Eq.~\eqref{eq:Lambda-bulk}. The two branches are joined where their distances from $W/2$ coincide; in this crossover region the leading-order expressions are accurate only up to $O(t)$. Orange dashed line: $-E_{\rm iso}(W)$ for $W<W_{\min}$, Eq.~\eqref{eq:Eiso}, i.e., the same equation as the edge branch. Dashed black line: band edge $E_{\max}=2t\sqrt K+W/2$. Dotted line: $W_{\min}$, Eq.~\eqref{eq:bound}.}
\label{fig:largeK}
\end{figure*}

In the intermediate region, $Kt\ln(W/t)\lesssim W\lesssim2Kt\ln(W/t)$, the mobility edge stays within a distance of order $t$ from the edge of the single-site band, $|E_{\rm loc}|-W/2=O(t)$, and the leading-order expressions above determine it only up to $O(t)$. In Fig.~\ref{fig:largeK} we join the two branches at the point where their distances from $W/2$ coincide.

\paragraph{LLT percolation line at large $K$.}
The same simplifications apply to the LLT cavity equations with $X=0$, i.e., $E_{\rm sh}=E_{\rm edge}$. Let $\kappa\equiv Kt\mu_g$ at $E_{\rm sh}=E_{\rm edge}$, be the solution of $\kappa=(Kt/W)\ln[(W+2t\sqrt K-t\kappa)/(2t\sqrt K-t\kappa)]$. The rescaled fields concentrate on $\mu_\eta=1/(1-\kappa)$ [Eq.~\eqref{eq:etastats}], so that $u_{i\to j}\simeq\mu_\eta\,\mathcal G_{i\to j}$, and at leading order the conditional occupation probability $\bar q$ reduces to $q$. A site is then occupied if $\varepsilon_i-t\kappa\le\mu_\eta E_+$, and the condition $q=1/K$ gives
\begin{eqnarray}
\label{eq:Eperc-largeK}
E_{\rm perc}(W) &\simeq&  2t\sqrt K+\frac W2
\nonumber\\
&& 
-(1-\kappa)\left(2t\sqrt K-t\kappa+\frac WK\right)\,.
\quad
\end{eqnarray}
Over the whole range shown in Fig.~\ref{fig:largeK}, this expression agrees within $1.4\,t$ with the full high-connectivity criterion $\bar q=1/K$ for $K=5$, $20$, and $100$.

\paragraph{How the two lines separate.}
Equations~\eqref{eq:Eloc-edge}, \eqref{eq:Lambda-bulk}, and \eqref{eq:Eperc-largeK} give the following picture, illustrated in Fig.~\ref{fig:largeK}.
(i) Both lines start at the same point, $(E_{\max}(W_{\min}),W_{\min})$, where $\kappa=1$.
(ii) Close to the onset, both depart from the band edge linearly in $1-\kappa$. Expanding Eq.~\eqref{eq:Lambda-edge}, one finds 
\begin{eqnarray}
&& 
E_{\max}-E_{\rm loc}\simeq
\\
&& \qquad 
(1-\kappa)[(2\sqrt K-1)(W_{\min}+2t\sqrt K-t)/K-t]  
\; , 
\nonumber
\end{eqnarray}
while Eq.~\eqref{eq:Eperc-largeK} gives 
\begin{equation}
E_{\max}-E_{\rm perc}\simeq(1-\kappa)(2t\sqrt K-t+W_{\min}/K)
\; . 
\end{equation} 
Their ratio is about $(W_{\min}+2t\sqrt K)/(Kt)\simeq\frac12\ln K$. The mobility edge therefore enters the band faster than the LLT line: just above $W_{\min}$, the LLT underestimates the extent of the localized region close to the edge.
(iii) For $W\gg W_{\min}$, $E_{\rm loc}-W/2=O(t)$, while $E_{\rm perc}-W/2\simeq2t\sqrt K\kappa-(1-\kappa)W/K$, with $\kappa\simeq(Kt/W)\ln(W/2t\sqrt K)$. The first term decreases as $1/W$ and the second grows as $W$, so the two lines cross when $E_{\rm perc}\simeq W/2+t$. With the full high-connectivity LLT line, the crossing occurs at $W\simeq11$, $125$, and $1150$ for $K=5$, $20$, and $100$. Beyond it, $E_{\rm perc}<E_{\rm loc}$: the LLT predicts a non-percolating phase at energies where the states are extended.
(iv) At strong disorder the mobility edge is re-entrant and vanishes at $W_c$, whereas $E_{\rm perc}$ keeps growing as $W(1/2-1/K)$. The two lines cross again, at $W\simeq59$, $310$, and $2000$ for $K=5$, $20$, and $100$. Beyond this second crossing, the LLT predicts extended states around the band center, where all the states are localized. At $W=W_c$, it still predicts a mobility edge at $E_{\rm perc}\simeq W_c(1/2-1/K)\simeq2Kt\ln K$.

This is the same structure as in the phase diagram for $K=2$ of Fig.~\ref{fig:phase-diagram}. There, the two lines start at the same point. The mobility edge first lies at lower energies than the LLT line, then at higher energies after the crossing at $W\simeq1.5$ (inset), and finally at lower energies again after the crossing at $W\simeq15$; the LLT line does not capture the re-entrance. The large-$K$ analysis shows that this structure is generic. It also shows that the agreement between the two lines is exact at one point only, the onset $(E_{\max}(W_{\min}),W_{\min})$. Finally, the first crossing occurs at $W_\times\simeq K^{5/4}[2\ln(W_\times/2t\sqrt K)]^{1/2}\,t$, which grows faster than $W_c\simeq4Kt\ln K$. According to Eqs.~\eqref{eq:Eloc-edge}, \eqref{eq:Lambda-bulk}, and \eqref{eq:Eperc-largeK}, the two crossings merge for $K$ of the order of $10^3$. Beyond that, the LLT line lies at larger energies than the mobility edge for all $W>W_{\min}$. Since the two lines are then only $O(t)$ apart, this estimate is only indicative.

\subsection{Distributions and extreme value statistics
}
\label{subsec:ExVal}

In this Section, we summarize the results for the distributions of the
fundamental variables that determine the critical properties of the LLT
percolation problem, and we analyze their dependence on the energy shift. We consider the general shifting convention introduced in Sec.~\ref{subsec:localization-landscape}, where $E_{\rm sh}=E_{\rm edge}-X=-2t\sqrt K -W/2-X$.

Moreover, we derive the tail behavior 
of these distributions. In Sec.~\ref{subsubsec:gtails} we focus on the cavity Green's functions' marginal and its upper-edge tail, in Sec.~\ref{subsubsec:etatails} we work with the cavity rescaled fields' marginal and derive a lower bound for its right tail, and in Sec.~\ref{subsubsec:utails}, after reviewing the features of the Localization Landscape variables' marginal, we show that its right tail is controlled by the same lower bound asymptotic behavior.  

\subsubsection{Cavity Green's functions' marginal}
\label{subsubsec:gtails}

We consider the closed recursion of Eq.~\eqref{eq:cavg}.
As the cavity Green's functions are real and positive, the denominator of Eq.~\eqref{eq:cavg} must always be positive, this means that $t^2\sum_{l \in \partial k\setminus i}\mathcal{G}_{l\to k} \le \varepsilon_i$ for any draw of $\varepsilon_k$ and $\{\mathcal{G}_{l\to k}\}_{l \in \partial k\setminus i}$ from their respective distributions. Since the support of the distribution of $\varepsilon_k$ is bounded, it follows that also $P_g$ is supported on a compact interval
\([m_g,M_g]\subset(0,+\infty)\).

The endpoints of the support are obtained by choosing, respectively, the values of the random variables that maximize or minimize the recursion for the cavity Green's functions. Thus, the upper and lower edges of the support satisfy
\begin{eqnarray}
\label{eq:Gmaxmin}
M_g
&=&
\frac{1}{\varepsilon_{\min}-Kt^2 M_g}\,,
\\
m_g
&=&
\frac{1}{\varepsilon_{\max}-Kt^2 m_g}\,,
\end{eqnarray}
where
\begin{equation}
    [\varepsilon_{\min},\varepsilon_{\max}]
    =
    [2t\sqrt K+X\,,\,2t\sqrt K+W+X]\,.
\end{equation}
Solving these equations gives
\begin{eqnarray}
\label{eq:Mg1}
M_g
&=&
\frac{1}{t\sqrt K}\,
U\left(\frac{X}{t\sqrt K}\right)\,,
\\
\label{eq:mg1}
m_g
&=&
\frac{1}{t\sqrt K}\,
U\left(\frac{W+X}{t\sqrt K}\right)\,,
\end{eqnarray}
with
\begin{equation}
    U(x)
    \equiv
    1-\frac{2}{1+\sqrt{1+\frac{4}{x}}}\,.
\end{equation}
The asymptotics of $U$ are
\begin{eqnarray}
U(x)
&=&
\begin{cases}
   1-\sqrt{x}
   +\dfrac{x}{2}
   +O\left(x^{3/2}\right),
   &
   x\to0^+\,,
   \\[2mm]
   \dfrac{1}{x}
   -\dfrac{2}{x^2}
   +O\left(\dfrac{1}{x^3}\right),
   &
   x\to+\infty\,.
\end{cases}
\end{eqnarray}
Therefore, at fixed $W$, in the limit $X\to+\infty$ one finds
\begin{eqnarray}
\mu_g\equiv \mathbb E[\mathcal G_{i\to j}]
&\sim&
M_g
\sim
m_g
\sim
\frac{1}{X}
+
O\left(\frac{1}{X^2}\right)\,,
\\
M_g-m_g
&=&
\frac{W}{X^2}
+
O\left(\frac{1}{X^3}\right)\,.
\end{eqnarray}
Thus, in this limit, the cavity Green's functions concentrate around the value $1/X$.

For $X=0$ and $W\to0^+$, the cavity Green's functions instead concentrate around
\begin{equation}
\label{eq:g0}
    g_0
    \equiv
    \frac{1}{t\sqrt K}\,.
\end{equation}
It is then convenient to rewrite Eqs.~\eqref{eq:Mg1} and~\eqref{eq:mg1} as
\begin{eqnarray}
\label{eq:Mg}
M_g
&=&
g_0\,U\left(g_0X\right)\,,
\\
\label{eq:mg}
m_g
&=&
g_0\,U\left(g_0(W+X)\right)\,.
\end{eqnarray}

The typical bulk of $P_g$ can be very well approximated for any value of $K$ by the high-connectivity marginal of Eq.\eqref{eq:Pgacc}, obtained by assuming that $\sum_{l \in \partial k\setminus i}\mathcal{G}_{l\to k}$ has typical fluctuations around its mean, i.e.
\begin{equation}
    \sum_{l \in \partial k\setminus i}\mathcal{G}_{l\to k}\sim \mathcal{N}\left(K\mu_g\,,K\sigma_g^2\right)\,,
\end{equation}
and determining self-consistently the values of $\mu_g$ and $\sigma^2_g$ from Eqs.~\eqref{eq:SCgind1} and \eqref{eq:SCgind2}.
However, this approximation is not sufficient to describe the upper-edge tail of the distribution, whose functional form is determined by the atypical realizations of the on-site disorder $\varepsilon_k$ and of the sum $\sum_{k\in\partial i\setminus j}\mathcal{G}_{k \to i}$.

In order to access the extreme values close to the upper-edge tail of the distribution, we start by defining
\begin{eqnarray}
\Delta &\equiv& M_g-\mathcal G_{k\to i}
\; ,
\label{eq:Delta}
\\
\Delta_l &\equiv& M_g-\mathcal G_{l\to k} 
\; ,
\label{eq:Delat-l}
\\
\lambda &\equiv& \varepsilon_i-\varepsilon_{\min} 
\; ,
\label{eq:lambda}
\end{eqnarray}
so that \(\lambda\in[0,W]\). Substituting these expressions 
into Eq.~\eqref{eq:cavg}, and using Eq.~\eqref{eq:Gmaxmin}, one derives
\begin{equation}
\Delta
=
M_g\left(1-\frac{1}{1+M_g\!\left(\lambda+t^2\sum_{l=1}^K \Delta_l\right)}\right) 
\; .
\label{eq:delta_recursion}
\end{equation}
We define the cumulative distribution of \(\Delta\) 
\begin{equation}
F_\Delta(\delta)\equiv\Pr\{\Delta<\delta\} 
\; .
\end{equation}
Since the function
\begin{equation}
\phi(z)= M_g\left(1-\frac{ 1}{1+M_g z}\right)
\end{equation}
is strictly increasing for \(z\ge 0\), Eq.~\eqref{eq:delta_recursion} can be inverted.
Setting
\begin{equation}
\phi^{-1}(\Delta)\equiv\frac{\Delta}{M_g(M_g-\Delta)}
\; ,
\label{eq:invphidef}
\end{equation}
one has
\begin{equation}
\Delta<\delta
\quad \Longleftrightarrow \quad
\lambda+t^2\sum_{l=1}^K \Delta_l < \phi^{-1}(\delta)
\; .
\end{equation}
Therefore,
\begin{equation}
F_\Delta(\delta)
=
\Pr\!\left\{\lambda+t^2\sum_{l=1}^K \Delta_l < \phi^{-1}(\delta)\right\}.
\end{equation}
As \(\lambda\) is uniform on \([0,W]\) and independent of the \(\Delta_l\)'s,
\begin{equation}
    F_\Delta(\delta)=\frac{1}{W}\int_0^{W} d\lambda \; 
    F_{\Sigma\Delta_l}\left(\frac{\phi^{-1}(\delta)-\lambda}{t^2}\right)
\, ,
\end{equation}
where
\begin{equation}
F_{\Sigma\Delta_l}(x)\equiv \Pr\!\left\{\sum_{l=1}^K \Delta_l < x\right\}
\; .
\end{equation}
Then, since the \(\Delta_l\)'s are strictly positive, $F_{\Sigma\Delta_l}(x)>0$ only for $x>0$, thus $\phi^{-1}(\delta)>\lambda$. Moreover, for $\delta\ll1$ we have
\begin{equation}
\phi^{-1}(\delta)
=
\frac{\delta}{M_g^2}+\frac{\delta^2}{M_g^3}+O(\delta^3)\ll 1\,,
\end{equation}
so we can safely substitute the upper bound of the integral with $\phi^{-1}(\delta)$. By changing variables we have
\begin{equation}
F_\Delta(\delta)
=
\frac{t^2}{W}\int_0^{\phi^{-1}(\delta)/t^2} dx \;   F_{\Sigma\Delta_l}(x)
\,.
\label{eq:Fdelta_exact}
\end{equation}

We assume that the probability density of \(\Delta\) has the form
\begin{equation}
P_\Delta(\delta)=L(\delta)e^{-h(\delta)},
\qquad
\delta\to0^+,
\label{eq:ansatz_Pdelta}
\end{equation}
with \(h(\delta)\to+\infty\) as \(\delta\to0^+\), and \(L(\delta)\) subexponential.
At the same exponential scale, the cumulative distribution can be written as
\begin{equation}
F_\Delta(\delta)=\widetilde L(\delta)e^{-h(\delta)},
\qquad
\delta\to 0^+
\; . 
\end{equation}
In order to evaluate \(F_{\Sigma\Delta_l}(x)\) for \(x\to0^+\), one has to estimate the probability that
the sum of \(K\) i.i.d. positive variables is very small, i.e.
\begin{eqnarray}
    F_{\Sigma\Delta_l}(x) 
    &=& 
    \int\prod_l[d\Delta_lL(\Delta_l)]e^{-\sum_lh(\Delta_l)}
    \nonumber\\
    && 
    \qquad \times \, \theta\left(x-\sum_l \Delta_l\right)
    \; . 
\end{eqnarray}
The leading exponential
contribution is given by the configuration $\{\Delta_1^*,\dots\,\Delta^*_K\}$ such that
\begin{equation}
    \sum_{l=1}^K h(\Delta_l^*) =\min\left\{\sum_{l=1}^K h(\Delta_l)\,\Bigg|\,\sum_{l=1}^K \Delta_l<x\right\}
    \; .
\end{equation}

Assuming that \(h\) is convex near \(0\), Jensen's inequality shows that the minimum
is reached at the democratic point
\begin{equation}
\Delta_1^*=\cdots=\Delta_K^*=\frac{x}{K}
\; .
\end{equation}
Therefore, at leading exponential order,
\begin{equation}
F_{\Sigma\Delta_l}(x)\sim \exp\!\left[-K\,h\!\left(\frac{x}{K}\right)\right]
\; .
\label{eq:FK_asymptotic}
\end{equation}
Substituting Eq.~\eqref{eq:FK_asymptotic} into Eq.~\eqref{eq:Fdelta_exact}, and using
again the fact that the integral is dominated by its upper endpoint, i.e. applying the Laplace (or steepest-descent) approximation, one obtains
\begin{equation}
\label{eq:Fasymp}
F_\Delta(\delta)\sim \exp\!\left[-K\,h\!\left(\frac{\phi^{-1}(\delta)}{Kt^2}\right)\right].
\end{equation}
The interpretation of the latter equation is that an extreme value of $\mathcal{G}_{k\to i}>M_g-\delta$ is going to be typically obtained from a set of neighboring cavity Green's functions $\{\mathcal{G}_{l\to k}\}_{l \in \partial k\setminus i}$ all deviating from $M_g$ at least by the same amount $\frac{\phi^{-1}(\delta)}{Kt^2}
$. 

Taking the logarithm of Eq.~\eqref{eq:Fasymp} and multiplying by $(-1)$, 
gives the asymptotic functional equation
\begin{equation}
h(\delta)\sim K\,h\!\left(\frac{\phi^{-1}(\delta)}{Kt^2}\right),
\qquad
\delta\to0^+\, .
\label{eq:h_functional}
\end{equation}

Now, depending on the value of $X$ there are two possible scenarios. 

If $X=0$, 
\begin{equation}
\frac{\phi^{-1}(\delta)}{Kt^2}
=\frac{\delta}{1-\delta/M_g}
=
\delta+\frac{\delta^2}{M_g}+O(\delta^3)\,,
\label{eq:invphi_expansion_critical}
\end{equation}
where we have used $Kt^2M_g^2=Kt^2g_0^2=1$ in Eq.~\eqref{eq:invphidef}.
Substituting Eq.~\eqref{eq:invphi_expansion_critical} into
Eq.~\eqref{eq:h_functional} and expanding the right-hand-side to first order gives
\begin{equation}
h(\delta)
\sim
K\left[h(\delta)+\frac{\delta^2}{M_g}h'(\delta)\right].
\end{equation}
Hence,
\begin{equation}
(K-1)h(\delta)\sim -\,K\,\frac{\delta^2}{M_g}h'(\delta)
\; ,
\end{equation}
that is,
\begin{equation}
\frac{h'(\delta)}{h(\delta)}
\sim
-\frac{M_g(K-1)}{K}\,\frac{1}{\delta^2}.
\end{equation}
Integrating, one finds
\begin{equation}
\ln h(\delta)\sim \frac{B}{\delta}+\mathrm{const},
\qquad
B\equiv\frac{M_g(K-1)}{K}.
\end{equation}
Therefore,
\begin{equation}
h(\delta)\sim A\,e^{B/\delta},
\label{eq:h_final}
\end{equation}
with \(A>0\) a constant. Finally, the cumulative distribution of \(\Delta\) has the
double-exponential tail
\begin{equation}
F_\Delta(\delta)\sim
\exp\!\left[-A\,e^{B/\delta}\right],
\qquad
\delta\to 0^+,
\label{eq:F_final}
\end{equation}
with
\begin{equation}
B=\frac{M_g(K-1)}{K}.
\end{equation}

This implies that the upper-edge tail of the marginal of the cavity Green's
functions is
\begin{equation}
\!\! 
P_g(g)
\sim
\frac{AB}{(M_g-g)^2}
\exp\!\left[\frac{B}{M_g-g}-A\,e^{B/(M_g-g)}\right]
\end{equation}
for $g$ close to $M_g^-$.

The case $X=0$ is marginal, in fact for any $X>0$ we have $Kt^2M_g^2<1$, therefore the equation to solve is
\begin{equation}
    h(\delta)
\sim
Kh\left(\frac{\delta}{\rho_X}\right)\,,
\end{equation}
where
\begin{equation}
    \rho_X =Kt^2M_g^2=U(g_0X)^2<1 
\end{equation}
which is solved by power law ansatz
\begin{equation}
    h(\delta)=A_X\delta^{-\gamma_X}
\end{equation}
with
\begin{equation}
    \gamma_X=-\frac{\ln K}{\ln \rho_X}=-\frac{\ln K}{2\ln U(g_0X)}>0\,.
\end{equation}
Therefore,
\begin{equation}
F_\Delta(\delta)\sim
\exp\!\left[-A_X\delta^{-\gamma_X}\right],
\qquad
\delta\to0^+,
\end{equation}
and
\begin{equation}
    P_g(g)\sim \frac{A_X\gamma_X}{(M_g-g)^{\gamma_X+1}}\exp\left[-\frac{A_X}{({M_g-g})^{\gamma_X}}\right]\,.
\end{equation}
Note, that as $X\to0^+$, $\gamma_X\to+\infty$, compatibly with the double exponential solution for $X=0$.

\subsubsection{Cavity rescaled fields' marginal}
\label{subsubsec:etatails}

The probability density function of the cavity rescaled fields can have either bounded or unbounded support, depending on the value of the additional shift $X$. Indeed, if the normal and cavity rescaled fields had finite upper bounds $\widetilde M_\eta$ and $M_\eta$, respectively, then the coupled equations
\begin{eqnarray}
M_\eta &=& 1 + t K M_g M_\eta\,,\\
\widetilde M_\eta &=& 1 + t (K+1) M_g M_\eta\,,
\end{eqnarray}
would admit finite positive solutions. This is possible only when
\begin{equation}
M_g < \frac{1}{K t}\iff X> t(\sqrt K-1)^2\equiv X_c\,,
\end{equation}
according to the solutions in Eqs.~\eqref{eq:Mg} and \eqref{eq:mg}.
Similarly, the lower bound of the cavity rescaled fields is finite only if
\begin{equation}
    m_g<\frac{1}{Kt} \iff W+X>X_c\,.
\end{equation}
Thus,
\begin{eqnarray}
\label{eq:Meta}
M_\eta &=&
\begin{cases}
+\infty\,, & \text{if } X < X_c\,,\\[0.2cm]
\displaystyle \frac{1}{1 - K t M_g}\,, & \text{if } X > X_c\,,
\end{cases}\\
\label{eq:meta}
m_\eta
&=&\begin{cases}
+\infty\,, & \text{if } \quad W+X < X_c\,,\\[0.2cm]
\displaystyle\frac{1}{1-Kt m_g}\,, & \text{if}\quad W+X > X_c\,.
\end{cases}
\end{eqnarray}
It is important to observe that for $X\le X_c$ it exists a critical value of the disorder
\begin{equation}
    W_{\rm AW}(X)\equiv X_c-X<W_{\rm min}\,,
\end{equation}
such that both $m_\eta$ and $M_\eta$ diverge.
The implication of this for the Localization Landscape is explained in Sec.~\ref{subsubsec:utails}, and the physical meaning of the critical disorder $W_{\rm AW}(X)$ is discussed in Sec.~\ref{subsec:EisoWmin}. 

The typical bulk of $P_\eta$ cannot be approximated as accurately as that of $P_g$. Indeed, within the high-connectivity Ansatz, the sum
\begin{equation}
\sum_{l \in \partial k \setminus i} \mathcal{G}_{l\to k}\eta_{l\to k}
\end{equation}
is replaced by a Gaussian random variable, leading to a self-consistent Gaussian marginal for $P_\eta$. This approximation does not reproduce well the histogram obtained through population dynamics, since it removes the asymmetry inherited from $P_g$. Nevertheless, as discussed in Sec.~\ref{subsec:hicon}, it provides an accurate estimate for the evaluation of $P_u$ through the product of $P_g$ and $P_\eta$ [see Eq.~\eqref{eq:P_u}].

To study the tails of $P_\eta$, one must distinguish between the regimes in which the support of $P_\eta$ is bounded ($X > X_c$) or unbounded ($X < X_c$).

We begin with the case $X < X_c$. To analyze the asymptotic behavior of the distribution of the rescaled fields, it is convenient to work directly with the tree representation given in Eq.~\eqref{eq:etatree}, following the same strategy as in Sec.~\ref{subsubsec:lowW}.
We introduce
\begin{equation}
\label{eq:etatree_tail}
\eta_{i\to j}
=
\sum_{r=0}^{\infty} y^{(r)}_{i\to j}\,,
\qquad
y^{(0)}_{i\to j}=1
\; .
\end{equation}
Here
\begin{equation}
\label{eq:yrdef_tail}
y^{(r)}_{i\to j}
=
t^r\sum_{\rho\in\mathcal{P}(r)}
\prod_{s=1}^{r}\mathcal{G}_{\rho_s\to\rho_{s-1}}\,,
\end{equation}
where $\mathcal{P}(r)$ denotes the set of all directed paths of length $r$ starting from $i$ and going away from the root.
Introducing the quantity
\begin{equation}
    S_r(\rho) = \sum_{s=1}^r\ln \mathcal{G}_{\rho_{s}\to\rho_{s-1}}\,,
\end{equation}
the $y$-variables can be rewritten as
\begin{equation}
   y^{(r)}_{i\to j}= t^r\sum_{\rho\in\mathcal{P}(r)}e^{S_r(\rho)}\,,
\end{equation}
or, equivalently, in terms of intermediate $y$-variables,
\begin{equation}
    y^{(r)}_{i\to j} = t^L\sum_{\rho \in \mathcal{P}(L)}e^{S_{L}(\rho)}y_{\rho_L \to \rho_{L-1}}^{(r-L)}\,, \qquad r>L\,.
\end{equation}
For a finite truncation depth $R$, one has
\begin{equation}
\label{eq:etaTtree_tail}
\eta^{(R)}_{i\to j}
\equiv
\sum_{r=0}^{R}y^{(r)}_{i\to j}\,.
\end{equation}

Now, as shown in Sec.~\ref{subsubsec:lowW}, the typical value of the exponential $e^{S_r(\rho)}$ along a generic path of length $r$ on the cavity subtree is
\begin{equation}
    e^{\mathbb E[S_r(\rho)]} = (\mu_g^{\rm typ})^r\,,
\end{equation}
where $\mu_g^{\rm typ}=e^{\mathbb E[\ln \mathcal G_{i\to j}]}$.
This suggests that an atypical realization of the infinite sequence $\{y_{i \to j}^{(r)}\}_{r\ge0}$ can be constructed by considering an atypical bubble of radius $L$ and defining
\begin{eqnarray}
    y_{i \to j}^{(r)}=\begin{cases}
       \displaystyle y_{\rm atyp}^{(r)}=t^r\sum_{\rho\in\mathcal{P}(r)}e^{S^{\rm atyp}_r(\rho)} \quad &\text{if} \quad r\le L\,, \\
    \displaystyle t^L\sum_{\rho \in \mathcal{P}(L)}e^{S_{L}^{\rm atyp}(\rho)}y_{\rho_L \to \rho_{L-1}}^{(r-L)} \quad &\text{if} \quad r> L\,.
    \end{cases}
\end{eqnarray}
The second expression, assuming that $\{y_{\rho_L \to \rho_{L-1}}^{(r-L)}\}_{r >L}$ are typical in the sense that
\begin{equation}
    y_{\rho_L \to \rho_{L-1}}^{(r-L)} \sim  (Kt)^{r-L}e^{\mathbb E [S_{r-L}(\rho)]}=(Kt\mu_g^{\rm typ})^{r-L}\,,
\end{equation}
scales as
\begin{eqnarray}
    \sum_{\rho \in \mathcal{P}(L)}e^{S_{L}^{\rm atyp}(\rho)}y_{\rho_L \to \rho_{L-1}}^{(r-L)} \sim y_{\rm atyp}^{(L)}\,(Kt\mu_g^{\rm typ})^{r-L}\,.
\end{eqnarray}
Therefore,
\begin{widetext}
\begin{equation}
    \eta_{\rm atyp}(L)=\lim_{R \to \infty} \left[ \sum_{r=0}^L y_{\rm atyp}^{(r)} +\sum_{r=L+1}^Rt^L\sum_{\rho \in \mathcal{P}(L)}e^{S_{L}^{\rm atyp}(\rho)} y_{\rho_L \to \rho_{L-1}}^{(r-L)}\right]\sim \sum_{r=0}^L y_{\rm atyp}^{(r)} + y_{\rm atyp}^{(L)} \frac{Kt\mu_g^{\rm typ}}{1-Kt\mu_g^{\rm typ}} \equiv E_L\,.
\end{equation}
\end{widetext}
The problem of determining the right tail of $P_\eta$ therefore amounts to studying the probability that $\eta_{i \to j}$ takes anomalously large values. To this end, we study the probability of obtaining an atypical realization producing the class of values $\{E_L\}_{L\ge 1}$.
In the previous subsection, the upper-edge tail of $P_g$ was obtained by
introducing the variable $\Delta$ as the distance from the upper edge $M_g$ of a cavity Green's function. The leading term in the tail was given by the event in which all branches are simultaneously close to the optimal configuration. We assume in the following that the same optimal-sharing approximation holds in the determination of the tail of the distribution of the rescaled fields, as the extreme values of the cavity rescaled fields depend directly on the extreme values of the cavity Green's functions. Taking the variable $\mathcal{G}_{i\to j}$ to be at a distance \(\Delta<\delta\) from $M_g$ generates a favorable cascade of upper-edge deficits \(\delta_r\) along the descendants. Equation~\eqref{eq:invphi_expansion_critical} gives
\begin{equation}
\delta_{r+1}
=
\frac{\delta_r}{\rho_X(1-\delta_r/M_g)}\,,
\label{eq:delta_iter_eta}
\end{equation}
where we denoted
\begin{equation}
    \rho_X= Kt^2M_g^2= U(g_0X)^2\,.
\end{equation}
Again, one has to distinguish between the case $X=0$ which is marginal, and the case $X>0$.

For $X=0$ we have $\rho_X=1$, and writing
\begin{equation}
r_{\max}(\delta) \equiv \frac{M_g}{\delta}-1,
\end{equation}
Eq.~\eqref{eq:delta_iter_eta} can be solved exactly, yielding
\begin{equation}
    \delta_r=\frac{M_g}{r_{\max}+1-r}, \qquad 0\le r\le \lfloor r_{\max}\rfloor\,.
    \label{eq:delta_r_solution}
\end{equation}
Hence, along the favorable cascade, all Green's functions at generation \(r\) satisfy the deterministic lower bound
\begin{equation}
\mathcal G_{\rho_r\to \rho_{r-1}}
\ge M_g-\delta_r
\; .
\label{eq:G_lower_along_path}
\end{equation}
Here, $r_{\max}$ should be interpreted as the maximum distance from the root for which the optimal deficit cascade is defined. However, the solution~\eqref{eq:delta_r_solution} loses self consistency before \(r\) reaches $r_{\max}$, since \(\delta_r\) becomes of order \(O(M_g)\) already for sufficiently large \(r<r_{\max}\). Indeed, configurations with a democratic distribution of deficits provide the leading contribution to the probability only asymptotically, that is, in the regime \(\delta_r \ll 1\). Nevertheless, we can use the quantity defined in the following, truncating the sequence~\eqref{eq:delta_iter_eta} at a cutoff distance $r^*$ such that
\begin{equation}
\label{eq:rstar}
    r_{\max}(\delta)-r^*(\delta) \gg 1\,,
\end{equation}
as a tool to derive a lower bound for $\bar F_\eta(x)\equiv\Pr\{\eta\ge x\}$.
In order to satisfy Eq.~\eqref{eq:rstar} we can take a general form 
\begin{equation}
    r^*(\delta)=\omega(\delta)\,r_{\max}(\delta)\,,
\end{equation}
with $\omega(\delta)<1$ and $(1-\omega(\delta))r_{\max}(\delta)\to+\infty$ for $\delta\to 0^+$. 
We start by observing that with this choice we get
\begin{equation}
    y_{\rm atyp}^{(r)}(\delta)
\equiv
(Kt)^{r}
\prod_{s=1}^{r}(M_g-\delta_s)
\; ,
\qquad
1\le r\le r^*
\; .
\end{equation}
Using Eq.~\eqref{eq:delta_r_solution},
\begin{align}
& \prod_{s=1}^{r}(M_g-\delta_s)
=
\prod_{s=1}^{r}
M_g\left(1-\frac{1}{r_{\max}+1-s}\right)
\nonumber\\
& \qquad 
=
M_g^r
\prod_{s=1}^{r}\frac{r_{\max}-s}{r_{\max}+1-s}
=
M_g^r\,\frac{r_{\max}-r}{r_{\max}}\,,
\end{align}
where in the last step we used that the product is telescopic.
Writing $r^*=\omega r_{\max}$ with $\omega<1$, and $Q= KtM_g$, one has 
\begin{equation}
 \sum_{r=1}^{r^*} Q^r\left(1-\frac{r}{r_{\max}}\right)\sim
Q^{r^*}\frac{(1-\omega)Q}{Q-1} 
\end{equation}
at the leading order
Therefore, substituting $y_{\rm atyp}^{(r)}$ into the definition of $E_L$ with $L=r^*$, and neglecting the additive constant $1$, one obtains
\begin{equation}
E_{r^*}(\delta)\sim
Q^{r^*}(1-\omega)
\left[
\frac{Q}{Q-1}
+
\frac{Kt\mu_g^{\rm typ}}{1-Kt\mu_g^{\rm typ}}
\right]\,,
\end{equation}
hence
\begin{eqnarray}
\label{eq:eta_asymptotic_R}
E_{r^*}(\delta)
&\sim&
C_\omega Q^{r^*},
\\
\label{eq:Comega}
C_\omega&\equiv&(1-\omega)\left[
\frac{Q}{Q-1}
+
\frac{Kt\mu_g^{\rm typ}}{1-Kt\mu_g^{\rm typ}}\right]\,.
\end{eqnarray}

For $X>0$ instead, $\rho_X<1$, and the solution of Eq.~\eqref{eq:delta_iter_eta} is given by
\begin{equation}
   \frac{M_g}{\delta_r}={\rho_X}^{r}\left(\frac{M_g}{\delta}+\frac{\rho_X}{1-\rho_X}\right)-\frac{\rho_X}{1-\rho_X}\,,
\end{equation}
with
\begin{equation}
    r_{\max}(\delta) \sim \frac{\ln \left(\rho_X+\right(1-\rho_X)\frac{M_g}{\delta})}{\ln(1/\rho_X)}\,,
\end{equation}
obtained by imposing $\delta_{r_{\rm max}}/M_g\sim 1$.
Following a similar derivation one finds
\begin{eqnarray}
E_{r^*}(\delta)
&\sim&
C_{X} Q_X^{r^*},\\
\label{eq:Cx}
C_{X}&\equiv&
\frac{Q_X}{Q_X-1}
+
\frac{Kt\mu_g^{\rm typ}}{1-Kt\mu_g^{\rm typ}}\,,\\
Q_X &=& KtM_g\,,
\end{eqnarray}
where $r^*$ is again $r^*(\delta)=\omega(\delta) r_{\max}(\delta)$. Therefore, the prefactor becomes independent of $\omega$, and most importantly the scaling of $r^*$ switches qualitatively from $O(1/\delta)$ to $O(\ln\left(1/\delta\right))$.

It is worth noting that the typical continuation, approximated by the term
$y_{\rm atyp}^{(L)}Kt\mu_g^{\rm typ}/(1-Kt\mu_g^{\rm typ})$, contributes only through the constant prefactor (that we denoted as $C_\omega$ or $C_X$ in Eqs.~\eqref{eq:Comega} and \eqref{eq:Cx} respectively).
Therefore, it does not modify the leading tail behavior of $P_\eta$.

Since $E_{r^*}(\delta)$ is monotonically increasing and diverges as $\delta\to0^+$, for every $x\gg1$ there exists a unique value $\delta(x)$ such that
\begin{equation}
E_{r^*}(\delta(x))=x\,,
\end{equation}
which also identifies a unique optimal cascade $\{\delta_r(x)\}_{r\le r^*}$.

Finally, since any truncated variable $\eta_{i \to j}^{(R)}$ represents a lower bound for the full rescaled field, we have
\begin{widetext}
\begin{eqnarray}
   \bar F_\eta(x)&\equiv& \Pr\{\eta\ge x\} \ge\Pr\{\eta^{(r^*)}\ge x\}=\Pr\left\{1+\sum_{r=1}^{r^*}t^{r}  \sum_{\rho \in \mathcal{P}(r)} \prod_{s=1}^r\mathcal{G}_{\rho_{s}\to\rho_{s-1}}\ge x\right\}
   \nonumber\\
   &\ge&\Pr\left\{
\mathcal G_{\rho_{r}\to \rho_{r-1}}
\ge M_g-\delta_r(x)
\;\;\text{for all paths and generations }1\le r\le r^*
\right\}\equiv \Pr\{\mathcal E_{\delta(x)}\}\,.
\label{eq:PrBoundeta}
\end{eqnarray}
\end{widetext}
The event \(\mathcal E_{\delta(x)}\) corresponds to the realization of the optimal series of deficits down to the depth $r^*(\delta(x))$.

At the same time, one may also view $\Pr\{\mathcal E_{\delta(x)}\}$ as a heuristic approximation of $\bar F_\eta(x)$, rather than only as a lower bound, provided that the typical continuation is retained through the term $Kt\mu_g^{\rm typ}/(1-Kt\mu_g^{\rm typ})$ appearing only in the constant prefactor. 
In this interpretation, one approximates the full rescaled field by completing the atypical partial sum with its typical tail. Since this contribution affects only the prefactor and not the leading exponential dependence, it is natural to expect that the event $\mathcal E_{\delta(x)}$ captures the correct asymptotic scaling of the right tail of $P_\eta$.
 
In Sec.~\ref{subsubsec:gtails} we derived that given a value of $\delta$, the scaling of $F_\Delta(\delta)=\Pr\{\Delta<\delta\}$ is given by
\begin{equation}
    F_\Delta(\delta)\sim e^{-h(\delta)}\,,
\end{equation}
where
\begin{equation}
\label{eq:hexplicit}
    h(\delta)\sim Kh\left(\frac{\delta}{\rho_X(1-\delta/M_g)}\right)\equiv Kh(\delta_1)\,,
\end{equation}
coming from the fact that also
\begin{eqnarray}
\label{eq:Ffactorization}
    F_\Delta(\delta) \sim \Pr\left\{\sum_l \Delta_l\le\delta_1\right\}
    \sim e^{-Kh\left(\delta_1\right)}\,.
\end{eqnarray}
Thus, the function $h$ is the one that maximizes the probability of obtaining the deficit $\delta$. Equation~\eqref{eq:hexplicit} is the one defining the recursion for the optimal deficits in Eq.~\eqref{eq:delta_iter_eta}.
Iterating the same procedure of Sec.~\ref{subsubsec:gtails} for an $n$-generation tree,
for $n\ge 1$, 
\begin{eqnarray}
h(\delta_r) &\sim& K\,h(\delta_{r+1})
\; ,
\\
h(\delta) &\sim&  K^n h(\delta_n)
\; . 
\label{eq:iterated_h_rec}
\end{eqnarray}

We now define the favorable cascade event \(\mathcal E_\delta^{(n)}\) up to depth \(n\) as
the event in which, at every vertex of generation \(r\le n-1\), the \(K\) descendants lie in
a small neighborhood of the democratic minimizer \(\delta_{r+1}\). Since the branches are
independent, its probability factorizes recursively. Iterating Eq.~\eqref{eq:Ffactorization} implies
\begin{equation}
-\ln \Pr\{\mathcal E_\delta^{(n)}\}
\sim
K^{n} h(\delta_n)\sim
h(\delta) \qquad \forall\,n\ge1
\; .
\end{equation}
Therefore, 
\begin{equation}
-\ln \Pr\{\mathcal E_{\delta(x)}\}\sim h(\delta(x))
\; .
\label{eq:Edelta_cost_final}
\end{equation}
Note that, if instead of studying the cavity rescaled fields we wanted to study the variables $\eta_i$, the only difference would have been the presence of a prefactor $(K+1)/K$ multiplying $h$ in Eq.~\eqref{eq:Edelta_cost_final}.

In words, the reason why \(\mathcal E_{\delta(x)}\) has the same leading exponential cost as
the one-point event \(\{\Delta<\delta(x)\}\) is that the latter is realized at leading
order, by the unique democratic minimizer of the recursive large-deviation problem.

Now, $\delta(x)$ has to be obtained enforcing $E_{r^*}(\delta(x))=x$, where $r^* = \omega(\delta)r_{\max}(\delta)$. However, in principle one has many choices for the function $\omega(\delta)$, and the slower the fraction $(1-\omega(\delta))r_{\max}$ diverges as $\delta\to 0^+$, the tighter the lower bound for the probability is going to be, as more and more terms will be included in the sum defining $E_{r^*}(\delta)$.
 A particularly simple choice is to take $\omega(\delta)$ to be a constant,
\begin{equation}
\omega(\delta)\equiv \omega\in(0,1).
\label{eq:omega_choice}
\end{equation}

Distinguishing again between the two cases, we first consider $X=0$. Using the corresponding solution for $h(\delta)$ in Eq.~\eqref{eq:h_final}, one obtains
\begin{equation}
\bar F_\eta(x)
\gtrsim
\exp\!\left[- A e^{B/\delta(x)}\right]\,.
\label{eq:eta_tail_intermediate}
\end{equation}
Then, with the choice of constant $\omega$, one has
\begin{equation}
r^*=\omega\,r_{\max}=\omega\frac{M_g}{\delta}\,,
\end{equation}
so that for all $0\le r\le r^*$,
\begin{equation}
\delta_r=\frac{M_g}{r_{\max}-r}\le \frac{M_g}{r_{\max}-r^*}
=\frac{\delta}{1-\omega}.
\end{equation}
Thus the whole truncated cascade remains in the regime $\delta_r\ll 1$ as $\delta\to 0^+$. 
Inverting $E_{r^*}(\delta(x))=x$ gives
\begin{equation}
\frac{1}{\delta(x)}
\sim
\frac{1}{\omega M_g\ln Q}
\ln\!\left(\frac{x}{C_\omega}\right),
\qquad x\to+\infty.
\label{eq:delta_inverse_eta_precise}
\end{equation}
Therefore,
\begin{eqnarray}
e^{B/\delta(x)}
&\sim&
\exp\!\left[
\frac{B}{\omega M_g\ln Q}
\ln\!\left(\frac{x}{C_\omega}\right)
\right]
\nonumber\\
&=&
\left(\frac{x}{C_\omega}\right)^{\beta_{0,\omega}}.
\label{eq:expBdelta}
\end{eqnarray}

For $X>0$, instead, the Green-function upper-edge cost is no longer marginal. In the regime $0<X\le X_c$, one has
\begin{equation}
\bar F_\eta(x)
\gtrsim
\exp\!\left[- A_X \delta(x)^{-\gamma_X}\right]\,.
\label{eq:eta_tail_intermediate_2}
\end{equation}
Taking again $\omega(\delta)\equiv\omega\in(0,1)$, one obtains
\begin{eqnarray}
r^*
&=&
\omega r_{\max}
\nonumber\\
&=&
\frac{\omega}{
\ln(1/\rho_X)
}
\ln\left[
\rho_X+(1-\rho_X)\frac{M_g}{\delta}
\right]\,.
\end{eqnarray}
This choice still keeps the truncated cascade in the small-deficit regime. Indeed, since $(1-\omega)r_{\max}\to+\infty$ for $\delta\to0^+$, all $\delta_r$'s remain small for $r<r^*$.

Enforcing $E_{r^*}(\delta(x))=x$ gives
\begin{equation}
\ln\left(\frac{x}{C_X}\right)
\sim
\frac{\omega\ln Q_X}{\ln(1/\rho_X)}
\ln\left[
\rho_X+(1-\rho_X)\frac{M_g}{\delta(x)}
\right]\,.
\end{equation}
Hence,
\begin{equation}
\rho_X+(1-\rho_X)\frac{M_g}{\delta(x)}
\sim
\left(\frac{x}{C_X}\right)^{
\frac{\ln(1/\rho_X)}{\omega\ln Q_X}
}\,,
\end{equation}
and therefore, at leading order,
\begin{equation}
\frac{1}{\delta(x)}
\sim
\frac{1}{(1-\rho_X)M_g}
\left(\frac{x}{C_X}\right)^{
\frac{\ln(1/\rho_X)}{\omega\ln Q_X}
} \; ,
\label{eq:delta_inverse_eta_X}
\end{equation}
for $x\to+\infty$.

Both cases can now be written in the common form
\begin{equation}
\bar F_\eta(x)
\gtrsim
\exp\!\left[
- A_{X,\omega}
x^{\beta_{X,\omega}}
\right],
\qquad x\to+\infty\,.
\label{eq:eta_tail_final_precise}
\end{equation}
The exponent is
\begin{equation}
\beta_{X,\omega}
=
\begin{cases}
\displaystyle
\frac{B}{\omega M_g\ln Q}
=
\frac{K-1}{\omega K\ln(KtM_g)},
\;\; & X=0\,,
\\[8pt]
\displaystyle
\frac{\gamma_X\ln(1/\rho_X)}{\omega\ln Q_X}
=
\frac{\ln K}{\omega\ln(KtM_g)},
\;\; & X>0\,,
\end{cases}
\label{eq:beta_eta_X}
\end{equation}
while the corresponding prefactor is
\begin{equation}
A_{X,\omega}
=
\begin{cases}
\displaystyle
A\,C_\omega^{-\beta_{0,\omega}},
\quad & X=0\,,
\\[8pt]
\displaystyle
A_X
\left[(1-\rho_X)M_g\right]^{-\gamma_X}
C_X^{-\beta_{X,\omega}},
\quad & X>0\,.
\end{cases}
\label{eq:A_X_omega}
\end{equation}
Thus, the tail lower bound remains a stretched exponential, 
but the exponent changes discontinuously when $X$ takes a non-vanishing value. In particular,
\begin{equation}
\frac{\beta_{0^+,\omega}}{\beta_{0,\omega}}
=
\frac{K\ln K}{K-1}\,.
\end{equation}

As already mentioned, 
for $X>X_c$, the rescaled fields are bounded, and there is no asymptotic right tail as $x\to+\infty$.

\subsubsection{Localization Landscape variables' marginal and effective potentials}
\label{subsubsec:utails}

We now study the marginal distribution of the Localization Landscape variables.
From Eq.~\eqref{eq:etau} one has
\begin{equation}
u_i=\mathcal G_{ii}\eta_i
\; .
\end{equation}
The diagonal Green's functions are positive and bounded in the interval
$[\widetilde m_g,\widetilde M_g]$, where
\begin{eqnarray}
\label{eq:tildeMg}
\widetilde M_g&=&\frac{1}{\varepsilon_{\min}-(K+1)t^2 M_g}
\; ,\\
\label{eq:tildemg}
\widetilde m_g&=&\frac{1}{\varepsilon_{\max}-(K+1)t^2 m_g}
\; . 
\end{eqnarray}
Here $M_g,\,m_g$ are the bounds of the cavity Green's functions, Eqs.~\eqref{eq:Mg},\eqref{eq:mg}.
Accordingly,
\begin{equation}
\widetilde m_g\,\eta_i\le u_i\le \widetilde M_g\,\eta_i
\; .
\label{eq:u_eta_bounds}
\end{equation}

Moreover, as explained in Sec.~\ref{subsubsec:etatails}, the support of $P_\eta$ changes at
$X_c=t(\sqrt K-1)^2$. For $X\le X_c$, the rescaled fields are unbounded, while for
$X>X_c$ they have compact support. In the latter case, the normal rescaled fields are bounded in
$[\widetilde m_\eta,\widetilde M_\eta]$, with
\begin{eqnarray}
\label{eq:tildeMetaX}
\widetilde M_\eta &=& 1+t(K+1)M_gM_\eta
\; ,\\
\label{eq:tildemetaX}
\widetilde m_\eta &=& 1+t(K+1)m_gm_\eta
\; ,
\end{eqnarray}
where $M_\eta,\,m_\eta$ are the bounds of the cavity rescaled fields, Eqs.~\eqref{eq:Meta},\eqref{eq:meta}.

It follows that the upper edge of the support of $P_u$ is
\begin{equation}
\widetilde M_u
=
\begin{cases}
+\infty 
\; ,
\quad &\text{if} \quad X\le X_c 
\; ,\\[6pt]
\displaystyle
\widetilde M_g\widetilde M_\eta
=
\frac{1}{X-X_c} 
\; ,
\quad &\text{if} \quad X>X_c
\; ,
\end{cases}
\end{equation}
where the last equality follows from Eqs.~\eqref{eq:tildeMetaX}
and \eqref{eq:tildemetaX}
after some algebra. 
Similarly,  at $W+X=X_c$, also the lower edge of the support of the cavity rescaled fields diverge. Consequently,
\begin{equation}
\widetilde m_u
= 
\begin{cases}
+\infty 
\; ,
\quad &\text{if} \quad W+X\le X_c 
\; ,\\[6pt]
\displaystyle
\widetilde m_g\widetilde m_\eta
=
\frac{1}{W+X-X_c} 
\; ,
\quad &\text{if} \quad W+X>X_c
\; .
\end{cases}
\end{equation}
In order to understand the physical meaning of these results it is convenient to translate them in terms of effective potential bounds. Defining the effective potentials $v_i=1/u_i$, one immediately sees that their distribution $P_{v}$ is supported on $[\widetilde m_v,\widetilde M_v]$, where
\begin{eqnarray}
\label{eq:tildeMv}
\widetilde M_v
&=&
\begin{cases}
0
\; ,
\quad &\text{if} \quad W+X\le X_c 
\; ,\\[6pt]
W+X-X_c
\; ,
\quad &\text{if} \quad W+X>X_c
\; .
\end{cases}\\
\label{eq:tildemv}
\widetilde m_v
&=&
\begin{cases}
0
\; ,
\quad &\text{if} \quad X\le X_c 
\; ,\\[6pt]
X-X_c
\; ,
\quad &\text{if} \quad X>X_c
\; ,
\end{cases}
\end{eqnarray}
Note that, the effective potentials concentrate on the values
\begin{eqnarray}
\label{eq:vconcentr}
    \mu_v&=&
        \begin{cases}
        0
        \; ,
        \quad &\text{if} \quad W<X_c-X\,,\, X\le X_c 
        \; ,\\[6pt]
        X-X_c
        \; ,
        \quad &\text{if} \quad W\to 0^+\,,\,\qquad X>X_c
        \; .
    \end{cases}
\end{eqnarray}
This has a remarkable consequence on the phase diagram that we explain more in detail in Sec.~\ref{subsec:dependence-shift}.

The typical bulk of the distribution of $u_i$ is well approximated by the high-connectivity
expression of Eq.~\eqref{eq:P_u}, also at low connectivity, after replacing $K$ by $K+1$ to account for the number of neighbors of a root site. However, as for $P_g$, the tails must be studied separately from the bulk, since they are controlled by rare realizations of both $\mathcal G_{ii}$ and $\eta_i$.

We now restrict to the case $X\le X_c$, where the support of $P_u$ is unbounded.
Equation~\eqref{eq:u_eta_bounds} immediately implies
\begin{equation}
\bar F_\eta\left( \frac{x}{\widetilde m_g}\right)
\le
\bar F_u(x)
\le
\bar F_\eta\left( \frac{x}{\widetilde M_g}\right),
\label{eq:Pu_from_Peta_bounds}
\end{equation}
where $\bar F_u(x)\equiv\Pr\{u\ge x\}$. Indeed, Eq.~\eqref{eq:u_eta_bounds}
gives the deterministic implications
\begin{eqnarray}
\left\{\eta_i\ge \frac{x}{\widetilde m_g}\right\}
&\Longrightarrow&
\left\{u_i\ge x\right\},\\
\left\{u_i\ge x\right\}
&\Longrightarrow&
\left\{\eta_i\ge \frac{x}{\widetilde M_g}\right\}.
\end{eqnarray}
Therefore,
\begin{eqnarray}
\Pr\left\{\eta_i\ge \frac{x}{\widetilde m_g}\right\}
&\le&
\Pr\{u_i\ge x\},\\
\Pr\{u_i\ge x\}
&\le&
\Pr\left\{\eta_i\ge \frac{x}{\widetilde M_g}\right\}.
\end{eqnarray}
Combining the two inequalities gives Eq.~\eqref{eq:Pu_from_Peta_bounds}.

Since the tail behavior of the normal rescaled fields differs from that of the cavity
rescaled fields only through the factor $(K+1)/K$ multiplying the leading cost $h(\delta)$,
we can use the estimate derived in Sec.~\ref{subsubsec:etatails}. In the whole unbounded regime
$0\le X\le X_c$, this gives
\begin{equation}
\label{eq:Pu_tail_lower}
\bar F_u(x)
\gtrsim
\exp\!\left[
-\widetilde A_{X,\omega}x^{\beta_{X,\omega}}
\right],
\qquad x\to+\infty,
\end{equation}
where $\beta_{X,\omega}$ is the exponent defined in Eq.~\eqref{eq:beta_eta_X}, and the prefactor defined in Eq.~\eqref{eq:A_X_omega} is given by 
\begin{equation}
\label{eq:tildeAomega}
\widetilde A_{X,\omega}
=
\frac{K+1}{K}
\widetilde m_g^{-\beta_{X,\omega}}
A_{X,\omega}\; . 
\end{equation}
Therefore, in the unbounded regime, the right tail of the marginal distribution of the
Localization Landscape variables is controlled by the same lower-bound asymptotic scale as the tail of the
rescaled fields. 

For $X>X_c$, instead, both $\eta_i$ and $u_i$ have compact support, and no
asymptotic right tail exists as $x\to+\infty$.

\subsubsection{Summary of the tail results}
\label{subsubsec:tails_summary}

We summarize here the main asymptotic results obtained for the right tails of the cavity Green's functions, the rescaled fields, and the Localization Landscape variables.

First, the marginal distribution of the cavity Green's functions is supported on a compact interval $[m_g,M_g]\subset(0,+\infty)$ in presence of an extra shift $X>0$. 

Defining the upper-edge deficit $\Delta=M_g-\mathcal G_{i\to j}$, for $X=0$ the cumulative distribution of $\Delta$ 
has the double-exponential behavior
\begin{equation}
F_\Delta(\delta)\sim
e^{-A\,e^{B/\delta}} 
\qquad
\delta\to0^+,
\end{equation}
with
\begin{equation}
B=\frac{M_g(K-1)}{K}\,.
\end{equation}
Equivalently, the upper-edge tail of the cavity Green's function distribution is
\begin{equation}
P_g(g)
\sim
\frac{AB}{(M_g-g)^2} \; 
e^{\frac{B}{M_g-g}-A\,e^{B/(M_g-g)}}
\end{equation}
for $g\to M_g^-$.

For any $X>0$, the tail behavior switches to stretched exponential:
\begin{equation}
F_\Delta(\delta)\sim
\exp\!\left[-A_X\delta^{-\gamma_X}\right]
\; ,
\qquad
\delta\to0^+,
\end{equation}
with
\begin{equation}
    \gamma_X=-\frac{\ln K}{2\ln U(g_0X)}>0
    \; , 
\end{equation}
and
\begin{equation}
    P_g(g)\sim \frac{A_X\gamma_X}{(M_g-g)^{\gamma_X+1}}\exp\left[-\frac{A_X}{({M_g-g})^{\gamma_X}}\right]
    \; .
\end{equation}

Second, the cavity rescaled fields are unbounded for $X\le X_c$, and their right tail can be studied through the tree decomposition
in Eq.~(\ref{eq:etatree_tail}).
Assuming the same optimal-sharing mechanism that controls the upper-edge tail of $P_g$, and truncating the favorable cascade at depth $r^*=\omega r_{\max}$ with fixed $\omega\in(0,1)$, one obtains, for $X\ge 0$, the lower-bound asymptotic
\begin{equation}
\bar F_\eta(x)\equiv \Pr\{\eta\ge x\}
\gtrsim
\exp\!\left[
- A_{X,\omega} x^{\beta_{X,\omega}}
\right] 
\end{equation}
for $x\to+\infty$,
where $\beta_{X,\omega}$ and $A_\omega>0$ are given in Eqs.~\eqref{eq:beta_eta_X} and \eqref{eq:A_X_omega}. 
Thus, within this approximation, the right tail of the cavity rescaled fields is bounded 
from below by a stretched exponential.

For $X>X_c$, the cavity rescaled fields are bounded, therefore no $x\to +\infty$ asymptotic exist.

Finally, since the Localization Landscape variables satisfy
\begin{equation}
u_i=\mathcal G_{ii}\eta_i
\end{equation}
with $\mathcal G_{ii}\in[\widetilde m_g,\widetilde M_g]$ being the upper and lower bounds for the distribution of a Green's function [Eqs.~\eqref{eq:tildeMg} and \eqref{eq:tildemg}], one has
\begin{equation}
\bar F_\eta\!\left(\frac{x}{\widetilde m_g}\right)
\le
\bar F_u(x)
\le
\bar F_\eta\!\left(\frac{x}{\widetilde M_g}\right).
\end{equation}
As a consequence, for $X\le X_c$ the right tail of the marginal distribution of the Localization Landscape variables is controlled by the same asymptotic scale as the tail of the rescaled fields. In particular, for $X=0$
\begin{equation}
\bar F_u(x)\equiv \Pr\{u\ge x\}
\gtrsim
\exp\!\left[
-\widetilde A_{X,\omega} x^{\beta_{X,\omega}}
\right]\; , 
\end{equation}
for $x\to+\infty$,
where $\beta_{X,\omega}$ and $\widetilde A_{X,\omega}$ are given in Eqs.~\eqref{eq:beta_eta_X} and \eqref{eq:tildeAomega}.
For $X>X_c$, as both cavity rescaled fields and cavity Green's functions are bounded also the Localization Landscape variables are, 
and $\bar F_u$ has  no $x\to+\infty$ tail.

In summary, the three marginal distributions exhibit two qualitatively different tail behaviors: the cavity Green's functions have a bounded support with a double-exponential ($X=0$) or stretched exponential ($X>0$) upper-edge tail. Both cavity rescaled fields 
and Localization Landscape variables switch from being unbounded for $0\le X\le X_c$, with stretched-exponential 
lower-bound asymptotics at large values, to being bounded for $X>X_c$. 

The tail behavior of the Localization Landscape distribution controls the 
behavior of the LLT estimate of the DoS close to the edge of the spectrum, which, however, 
also depends on the choice of the shift, as discussed in Sec.~\ref{subsec:density-of-states}.

\section{Conclusions}
\label{sec:Conclusions}

In this work we extended the analysis of Ref.~\cite{Tonetti2026} and provided a detailed study of the Localization Landscape Theory on the Bethe lattice. This geometry is particularly useful because both Anderson localization and the LLT percolation problem can be formulated exactly in terms of cavity equations. It therefore provides a controlled setting in which one can separate the genuinely quantum aspects of Anderson localization from the classical geometrical information encoded in the effective potential $1/u_i$.

Our first conclusion confirms and strengthens the result of Ref.~\cite{Tonetti2026}. The LLT percolation transition does not coincide with the Anderson localization transition on the Bethe lattice. Although the two critical lines are close at weak disorder, they separate at stronger
disorder and exhibit different critical properties. In particular, the inverse participation ratio of Anderson eigenstates and the inverse average cluster size of the LLT percolation problem have distinct critical behaviors. Similarly, although the localization length and the percolation correlation length both diverge with exponent $\nu=1$ on the Bethe lattice, their amplitudes differ substantially 
(and the asymptotic long-distance behavior of the corresponding correlation functions is also slightly different). This shows that the regions selected by the condition $1/u_i\leq E_+$ do not reproduce the spatial extent of the true quantum eigenstates.

This mismatch can be understood from the different nature of the two transitions. Anderson localization is driven by quantum interference, whereas the LLT transition is a classical correlated percolation problem. The correlations induced by the Localization Landscape are not sufficient to transform this percolation transition into the Anderson one. In this sense, the Bethe lattice provides an exact benchmark showing that the LLT does not constitute an exact theory of the Anderson transition in arbitrary geometries.

Concomitantly, our analysis also reveals that the LLT contains nontrivial and useful information about the onset of localization. The 
Anderson model on the Bethe lattice has an isolated eigenvalue $E_{\rm iso}(W)$ below the continuous spectrum at weak disorder. We showed that, within the LLT framework, the expected value of  the Localization Landscape variables $u_i$ diverges precisely when this isolated eigenvalue reaches the lower edge of the continuous spectrum. This gives a natural interpretation of the disorder scale $W_{\min}$ at which, according to the LLT, the mobility edge first appears. The high-connectivity solution gives an analytic prediction for this scale as well as for $E_{\rm iso}(W_{\rm min})$, in excellent agreement with the exact cavity solution.  We derived the exact bound $W_{\min}\ge2t(\sqrt K-1)^2$. For $K=2$, our numerical data for the mobility edge, which do not extend below $W=0.5$, are compatible with an onset of localization at $W_{\min}\simeq0.357$, but cannot establish it.

At large connectivity, however, the question can be settled analytically. The isolated eigenvalue and the mobility edge are fixed by the same condition, the largest eigenvalue of the Anderson stability kernel at $\beta=1/2$ being equal to one, below and inside the spectrum, respectively. Following Abou-Chacra, Anderson, and Thouless, we computed the mobility edge for $K\gg1$ close to the band edge, $E_{\rm loc}=t+(W/2)\coth(W/2Kt)$, which is the equation of the isolated eigenvalue continued above $W_{\min}$. The mobility edge therefore emerges from the edge of the spectrum exactly at $W_{\min}$, at the same point as the LLT percolation line. The comparison with the large-$K$ LLT percolation line, which we also obtained in closed form, shows that the agreement between the two lines is exact at this point only. Just above $W_{\min}$ the mobility edge enters the band faster, by a factor of about $\frac12\ln K$, so that the LLT underestimates the localized region close to the edge. At stronger disorder the two lines cross twice, as for $K=2$, and the LLT line does not capture the re-entrance of the mobility edge and the complete localization of the spectrum at $W_c\simeq4Kt\ln(W_c/2t)$.

In addition,
we showed that the analytic continuation of the LLT 
cavity equations below $W_{\min}$ (bulk-edge shift $X=0$) recovers the Aizenman--Warzel lower bound~\cite{warzel2012,aizenman2011absence} for the disorder below which 
 the states at the spectral edges are delocalized,
\begin{equation}
    W_{\rm AW}=t(\sqrt K-1)^2 .
\end{equation}
This result follows from the distributional analysis of the Localization Landscape variables and their effective potentials. Below this threshold, the effective potentials $v_i=1/u_i$ vanish identically, so the LLT percolation problem is necessarily in the percolating phase. Since in this regime the shifted Hamiltonian is not positive definite, this is a property of the analytically continued cavity equations, and in particular of the edges of the support of the distributions of the cavity variables, rather than a prediction of the standard LLT. Its origin is transparent: as in the proof of Aizenman and Warzel, the threshold is controlled by the extreme configuration in which all the on-site energies take their maximal value. The two scales $W_{\min}$ and $W_{\rm AW}$ therefore have different meanings. At $W_{\min}$, a typical Localization Landscape variable diverges, while rare finite values may still exist. At $W_{\rm AW}$, the whole support of the distribution is pushed to infinity, so no finite value of $u_i$ remains possible. This distinction clarifies the relation between the isolated eigenvalue, the LLT percolation threshold, and rigorous lower bounds for localization.

We then studied the dependence of the LLT percolation line on the spectral shift. On the Bethe lattice, the distinction between the lower edge of the continuous spectrum and the true bottom of the spectrum is essential, because the latter is controlled by the isolated eigenvalue below $W_{\min}$. We introduced a family of reference energies $E_{\rm sh}=E_{\rm edge}-X$ and compared the resulting percolation thresholds. Increasing $X$ does not improve the agreement between the LLT percolation line and the Anderson mobility edge for $W\lesssim15$. In the limit $X\to+\infty$, the cavity Green functions and the rescaled fields become deterministic, and the LLT percolation problem reduces to an effective independent-site percolation problem. For the purpose of estimating the mobility edge from the LLT percolation threshold, the bulk-edge prescription $X=0$ is therefore the natural choice, although no first-principle criterion fixes the shift. A criterion based on the LLT bound on the eigenstate amplitudes, which holds for every admissible shift, selects $X=0$ in the weak-disorder regime, where the LLT percolation line and the mobility edge are close.

Finally, we investigated the spectral predictions of the LLT through its Weyl-type approximation for the integrated Density of States~\cite{arnold2019computing,david2021landscape,Fefferman1983theUP}. This provides an additional and independent test of the theory. Below $W_{\min}$, the effective potentials vanish either typically or identically, depending on the shift regime. As a consequence, any approximation based only on $1/u_i$ has very limited sensitivity to the presence of disorder in this region. Above $W_{\min}$, the LLT prediction does not reproduce the exact Anderson Density of States. 
We derived the Weyl-type estimate from an exact decomposition of the shifted Hamiltonian into an effective kinetic operator and the effective potential $1/u_i$. For no finite value of the shift does it recover the statistical symmetry and the shape of the Anderson spectrum, even when its support coincides with the exact one ($X\ge X_c$). Increasing the shift interpolates between the LLT estimate and the naive Weyl estimate built with the bare disorder, which is recovered for $X\to\infty$; the LLT improves on the latter only at weak disorder. Since the estimate relies on two uncontrolled approximations (a slowly varying landscape and the use of the one-site distribution of $u_i$), this is a limitation of the Weyl-type construction rather than necessarily of the LLT itself. This shows that the double translation of energies, which is often used to apply LLT to non-positive Hamiltonians, does not provide a reliable way to reconstruct the spectral properties of statistically symmetric Hamiltonians.

 The tails of the LLT estimate close to the spectral boundary are controlled by rare large values of the Localization Landscape variables, whose extreme-value statistics we have analyzed. Their precise form, however, depends on the choice of the shift: for $X<X_c$ the LLT--Weyl estimate has a gap above the edge of the spectrum, while for $X\ge X_c$ we have not determined its asymptotic form. We therefore do not draw quantitative conclusions on the Lifshitz tails. Thus, while the LLT captures some qualitative information about the location of the spectral boundary and the onset of localization, it fails to reproduce the correct spectral measure and the correct critical behavior.

Overall, the Bethe lattice reveals the limitations but also the strengths of the Localization Landscape Theory. On the one hand, the LLT does not reproduce the Anderson transition as a quantum critical phenomenon, nor does its Weyl-type estimate give the correct Density of States. On the other hand, it encodes remarkable information about the isolated eigenvalue, the disorder at which localized states are expected to first appear (which, at large connectivity, we showed to be exactly the onset of localization), and the Aizenman--Warzel lower bound. 
This suggests that the effective potential $1/u_i$ correctly encodes where the spectrum can host localized states at weak disorder, but not the interference mechanism that controls the critical properties of the Anderson transition.

The framework presented here, in which both Anderson localization and the localization landscape are analytically tractable, provides a robust playground for systematically testing and refining the approximations of the LLT, potentially enhancing its predictive accuracy.
In this light, our work suggests
several directions for future research. 

One promising direction involves extending the comparison between Anderson localization and the LLT percolation transition to the loop-less Cayley tree. While this structure resembles the Bethe lattice, it possesses several fundamental differences that persist in the thermodynamic limit. Notably, the Cayley tree lacks isolated eigenvalues and is genuinely bipartite. Furthermore, the spectral properties of the metallic phase on the Cayley tree differ strikingly from those of the Bethe lattice~\cite{biroli2020anomalous,sonner2017multifractality,tikhonov2016fractality}. It would therefore be compelling to determine whether the localization landscape is similarly modified and whether it remains sensitive to the presence of boundaries. 

Another fruitful direction would be to evaluate the LLT on more general classes of hierarchical lattices where Anderson localization remains exactly solvable, yet loops and short-range disorder correlations can be naturally incorporated -- such as the graphs recently introduced in \cite{q3bz-97v6}. A distinct feature of these graphs is their non-symmetric Density of States, which could provide further insight into the limitations and strengths of the LLT framework.

It would be equally illuminating to determine which features of the Bethe lattice analysis persist in finite-dimensional systems -- such as the failure of the energy-shift prescription to restore spectral symmetry. Furthermore, alternative formulations of the LLT, specifically designed to treat non-positive-definite Hamiltonians without shifting the spectrum, should be rigorously tested within the same exactly solvable framework. 

Finally, on a more ambitious note, an exact or semi-analytical study of LLT percolation in finite dimensions ($d=2$ and $d=3$) would be invaluable. Such an investigation could clarify why the LLT provides remarkably accurate numerical estimates of the mobility edge in specific physical regimes, despite the fundamental limitations exposed by our Bethe lattice results.

\vspace{0.75cm}

{\it Acknowledgments.} This work is supported by a grant from ``Fondation CFM pour la recherche''. 
Numerical calculations were performed at the LPTMC cluster.
We acknowledge funding from the ANR research
grant ManyBodyNet ANR-24-CE30-5851. We thank M. Filoche and M. Vrech for very useful discussions, 
C. Texier for letting us know Ref.~\cite{texier2020comment} and the suggestion to study the Density of States, 
and an anonymous referee for the proposal to extend the analysis of~\cite{Tonetti2026} and publish these results as its companion paper.

\bibliography{references.bib}

\appendix
\begin{widetext}
\newpage
\section{Notation table}
\label{sec:notation-table}
\newcommand{\symbcell}[1]{\parbox[t]{0.15\textwidth}{\raggedright #1}}
\newcommand{\meancell}[1]{\parbox[t]{0.25\textwidth}{\raggedright #1}}
\newcommand{\commcell}[1]{\parbox[t]{0.50\textwidth}{\raggedright #1}}
\begin{table*}[h!]
\caption{Summary of the main notation used throughout the paper.}
\label{tab:notation1}
\footnotesize
\setlength{\tabcolsep}{4pt}
\renewcommand{\arraystretch}{1.25}
\begin{ruledtabular}
\begin{tabular}{lll}
\symbcell{\textbf{Symbol}} & \meancell{\textbf{Meaning}} & \commcell{\textbf{Convention / Comment}} \\
\hline
\symbcell{$\hat {\mathcal {H}}$} 
& \meancell{Original Anderson Hamiltonian} 
& \commcell{Defined on the Bethe lattice, or as the thermodynamic limit of a random regular graph.} \\
\symbcell{$E$} 
& \meancell{Energy in the spectrum of $\hat {\mathcal {H}}$} 
& \commcell{Physical Anderson energy. For the symmetric disorder distribution, the spectrum is statistically symmetric around $E=0$.} \\
\symbcell{$W$} 
& \meancell{Disorder strength} 
& \commcell{The on-site energies $\epsilon_i$ are drawn uniformly in $[-W/2,W/2]$.} \\
\symbcell{$t$} 
& \meancell{Hopping amplitude} 
& \commcell{Usually set to $t=1$ in the numerical results.} \\
\symbcell{$K+1$} 
& \meancell{Connectivity of the Bethe lattice} 
& \commcell{Each site has $K+1$ nearest neighbours, while cavity branches have $K$ descendants.} \\
\symbcell{$E_{\rm edge}(W)$} 
& \meancell{Lower edge of the continuous spectrum} 
& \commcell{Defined as $E_{\rm edge}(W)=-2t\sqrt{K}-\frac{W}{2}$.} \\
\symbcell{$E_{\rm iso}(W)$} 
& \meancell{Isolated eigenvalue} 
& \commcell{Isolated eigenvalue below the continuous spectrum. It exists for $W\le W_{\min}$ and reaches the bulk edge at $W=W_{\min}$.} \\
\symbcell{$W_{\min}$} 
& \meancell{Disorder where $E_{\rm iso}$ reaches the bulk edge} 
& \commcell{Defined by $E_{\rm iso}(W_{\min})=E_{\rm edge}(W_{\min})$.} \\
\symbcell{$E_{\min}(W)$} 
& \meancell{True bottom of the spectrum} 
& \commcell{Equals $E_{\rm edge}(W)$ for $W>W_{\min}$ and $E_{\rm iso}(W)$ for $W\leq W_{\min}$.} \\
\symbcell{$X$} 
& \meancell{Additional spectral shift} 
& \commcell{Used to study a family of shifted Hamiltonians.} \\
\symbcell{$E_{\rm sh}(W,X)$} 
& \meancell{Spectral reference used to define $\hat{\mathcal H}_+$} 
& \commcell{In the bulk-edge convention, $E_{\rm sh}(W,X)=E_{\rm edge}(W)-X$.} \\
\symbcell{$\hat{\mathcal H}_+(X)$} 
& \meancell{Shifted Hamiltonian} 
& \commcell{Defined by $\hat{\mathcal H}_+(X)=\hat {\mathcal {H}}-E_{\rm sh}(W,X)\hat {\mathcal {I}}$.} \\
\symbcell{$E_+$} 
& \meancell{Energy associated with $\hat{\mathcal H}_+$} 
& \commcell{Related to the original energy by $E_+=E-E_{\rm sh}(W,X)$, equivalently $E=E_{\rm sh}(W,X)+E_+$.} \\
\symbcell{$E_{\rm loc}(W)$} 
& \meancell{Anderson mobility edge} 
& \commcell{Critical energy of the original Hamiltonian $\hat {\mathcal {H}}$.} \\
\symbcell{$E_{+,\rm perc}(W,X)$} 
& \meancell{LLT percolation threshold} 
& \commcell{Critical value of $E_+$ at which $\Omega_{E_+}=\{i:\,1/u_i\leq E_+\}$ first percolates.} \\
\symbcell{$E_{\rm perc}(W,X)$} 
& \meancell{LLT threshold in the original spectrum} 
& \commcell{Defined by $E_{\rm perc}(W,X)=E_{\rm sh}(W,X)+E_{+,\rm perc}(W,X)$.} \\
\symbcell{$\epsilon_i$} 
& \meancell{Original on-site disorder} 
& \commcell{Uniformly distributed in $[-W/2,W/2]$.} \\
\symbcell{$\varepsilon_i$} 
& \meancell{Shifted on-site energy} 
& \commcell{Defined as $\varepsilon_i=\epsilon_i-E_{\rm sh}(W,X)$.} \\
\symbcell{$\hat {\mathcal {G}}(z)$} 
& \meancell{Resolvent of the original Hamiltonian} 
& \commcell{We use the convention $\hat {\mathcal {G}}(z)=(\hat {\mathcal {H}}-z\hat {\mathcal {I}})^{-1}$.} \\
\symbcell{$\mathcal G_{ii}(z)$} 
& \meancell{Diagonal (normal) Green's function} 
& \commcell{Its imaginary part determines the local Density of States.} \\
\symbcell{$\mathcal G_{k\to i}(z)$} 
& \meancell{Cavity Green's function} 
& \commcell{Green's function on the subtree rooted at $k$ when site $i$ is removed.} \\
\symbcell{$u_i$} 
& \meancell{Localization Landscape} 
& \commcell{Defined by $\hat{\mathcal H}_+\mathbf u=\mathbf{1}$. Equivalently, $\mathbf u=\hat{\mathcal H}_+^{-1}\mathbf{1}= \hat{\mathcal G}(E_{\rm sh}) \mathbf 1$.} \\
\symbcell{$1/u_i$} 
& \meancell{LLT effective potential} 
& \commcell{The classically allowed LLT region is $\Omega_{E_+}=\{i:\,1/u_i\leq E_+\}$.} \\
\symbcell{$\eta_i$} 
& \meancell{Auxiliary field used to compute $u_i$} 
& \commcell{In the cavity construction, $u_i=\mathcal G_{ii}\eta_i$.} \\
\symbcell{$\eta_{k\to i}$} 
& \meancell{Cavity auxiliary field} 
& \commcell{Defined on the subtree rooted at $k$ with site $i$ removed.} \\
\symbcell{$p_i$} 
& \meancell{Infinite-cluster probability at site $i$} 
& \commcell{Probability that site $i$ belongs to the infinite LLT cluster. It includes the occupation condition $1/u_i\leq E_+$.} \\
\symbcell{$p_{k\to i}$} 
& \meancell{Cavity percolation probability} 
& \commcell{Probability that the cavity root $k$ connects to the infinite cluster without using site $i$.} \\
\symbcell{$\bar p_{k\to i}$} 
& \meancell{Conditional cavity percolation probability} 
& \commcell{Used when separating geometric connectivity from the local occupation condition.} \\
\symbcell{$S$} 
& \meancell{Average finite cluster size} 
& \commcell{$1/S$ is the LLT analogue of the Anderson inverse participation ratio.} \\
\symbcell{$C_{\rm perc}(r)$} 
& \meancell{LLT percolation correlation function} 
& \commcell{Probability that two sites at distance $r$ belong to the same finite cluster.} \\
\symbcell{$\xi_{\rm perc}$} 
& \meancell{LLT percolation correlation length} 
& \commcell{Diverges at $E_{+,\rm perc}$.} \\
\symbcell{$\xi_{\rm loc}$} 
& \meancell{Anderson localization length} 
& \commcell{Extracted from the decay of eigenstate correlation functions.} \\
\symbcell{$\rho(E)$} 
& \meancell{Anderson Density of States} 
& \commcell{Density of eigenvalues of $\hat {\mathcal {H}}$ at energy $E$.} \\
\symbcell{$\mathcal N(E)$} 
& \meancell{Integrated Density of States} 
& \commcell{Cumulative distribution associated with $\rho(E)$.} \\
\symbcell{$\rho_{\rm LLT}(E)$} 
& \meancell{LLT prediction for the Density of States} 
& \commcell{Depends on the chosen shift convention.} \\
\symbcell{$\mathcal N_{\rm LLT}(E)$} 
& \meancell{LLT prediction for the integrated Density of States} 
& \commcell{Used to compare LLT spectral predictions with the exact Anderson result.} \\
\end{tabular}
\end{ruledtabular}
\end{table*}
\end{widetext}

\end{document}